\documentclass{hepthesis2}
\newcommand*{\theauthor}{Felix Hekhorn}
\newcommand*{\thesubject}{PhD thesis}
\newcommand*{\thetitle}{Next-to-Leading Order QCD Corrections to Heavy-Flavour Production in Neutral Current DIS}

\usepackage{xspace}
\usepackage{tikz}
\usepackage{morefloats,afterpage}
\usepackage{mathrsfs} %
\usepackage{verbatim}

\usepackage[utf8]{inputenc}
\usepackage[ngerman,english,italian,main=english]{babel}
\usepackage{amssymb}
\usepackage{dsfont}
\usepackage[backend=biber,style=numeric-comp,sorting=none,maxcitenames=4]{biblatex}
\DeclareSourcemap{
 \maps[datatype=bibtex,overwrite=true]{
  \map{
    \step[fieldsource=Collaboration, final=true]
    \step[fieldset=usera, origfieldval, final=true]
  }
 }
}

\renewbibmacro*{author}{%
  \iffieldundef{usera}{%
    \printnames{author}%
  }{%
    \textbf{\printfield{usera}}, \printnames{author}%
  }%
}%
\PassOptionsToPackage{english}{hyperref}
\usepackage{bookmark} %
\hypersetup{%
	colorlinks = true,
	bookmarksnumbered = false,
	bookmarksopen = false,
	breaklinks = false,
	citecolor = black,
	filecolor = black,
	linkcolor = black,
	urlcolor = black,
	pdfborder = {0 0 0},
	pdffitwindow = true,
	pdfnewwindow = true,
	pdfstartpage = 1,
	pdfstartview = FitH,
	setpagesize = false,
	plainpages = false,
	linktocpage = true,
	pdfauthor = {\textcopyright\ \theauthor},
	pdfsubject = \thesubject,
	pdftitle = \thetitle,
	pdfkeywords = {perturbative quantumchromodynamics, pQCD, deep inelastic scattering, DIS, heavy flavour, parton distribution function, PDF, charm, bottom},
	unicode = true,
	pdfdisplaydoctitle = true,
}

\usepackage{hepnames}
\usepackage{siunitx} %

\usepackage{slashed}

\usepackage{subcaption}
\usepackage{tabularx}
\usepackage{multirow}

\DeclareMathOperator{\Ld}{\mathcal L}
\DeclareMathOperator{\Md}{\mathcal M}

\providecommand{\Derive}[1]{\DeriveN{#1}{}}
\providecommand{\DeriveN}[2]{\DeriveNF {#1}{#2}{}}
\providecommand{\DeriveF}[2]{\DeriveNF {#1}{}{#2}}
\providecommand{\DeriveNF}[3]{\frac {d^{#2}#3} {d {#1}^{#2}}}

\DeclareMathOperator{\EqualClaim}{\stackrel{!}{=}}

\providecommand{\abs}[1]{\left|#1\right|}

\DeclareMathOperator{\MSbar}{\overline{\text{MS}}}
\DeclareMathOperator{\dPSTwo}{{d\text{PS}_2}}
\DeclareMathOperator{\dPSThree}{{d\text{PS}_3}}
\DeclareMathOperator{\dPSThreegp}{{d\text{PS}^'_{3,\Pg}}}
\DeclareMathOperator{\dPSThreeqp}{{d\text{PS}^'_{3,\Pq}}}
\DeclareMathOperator{\LQCD}{{\Lambda_{\text{QCD}}}}

\DeclareMathOperator{\DiLog}{\text{Li}_2}
\DeclareMathOperator{\acosh}{\text{arcosh}}

\DeclareMathOperator{\atanh}{\text{artanh}}
\DeclareMathOperator{\tV}{\text{V}}
\DeclareMathOperator{\tA}{\text{A}}
\DeclareMathOperator{\lVVF2}{(\tV,\tV,\hat F_2)}
\DeclareMathOperator{\lAAF2}{(\tA,\tA,\hat F_2)}
\DeclareMathOperator{\lVVFL}{(\tV,\tV,\hat F_L)}
\DeclareMathOperator{\lAAFL}{(\tA,\tA,\hat F_L)}
\DeclareMathOperator{\lVVx2g1}{(\tV,\tV,2z\hat g_1)}
\DeclareMathOperator{\lAAx2g1}{(\tA,\tA,2z\hat g_1)}
\DeclareMathOperator{\lVAxF3}{(\tV,\tA,z\hat F_3)}
\DeclareMathOperator{\lVAg4}{(\tV,\tA,\hat g_4)}
\DeclareMathOperator{\lVAgL}{(\tV,\tA,\hat g_L)}

\DeclareMathOperator{\tNC}{\text{NC}}
\DeclareMathOperator{\tOK}{\text{OK}}
\DeclareMathOperator{\tQED}{\text{QED}}
\DeclareMathOperator{\tThr}{\text{thr}}

\DeclareRobustCommand{\PQ}{\HepGenParticle{Q}{}{}\xspace} %
\DeclareRobustCommand{\PaQ}{\HepGenAntiParticle{Q}{}{}\xspace} %
\DeclareRobustCommand{\PZx}{\HepGenParticle{Z}{}{*}\xspace} %

\DeclareRobustCommand{\DSSV}{\texttt{DSSV2014}\xspace}
\DeclareRobustCommand{\MSTW}{\texttt{MSTW2008nlo90cl}\xspace}
\DeclareRobustCommand{\NNPDFpol}{\texttt{NNPDFpol11\_100}\xspace}
\DeclareRobustCommand{\NNPDF}{\texttt{NNPDF23\_nlo\_as\_0119}\xspace}
\def\MMa{{\texttt{Mathematica}}}
\def\HEPMath{\texttt{HEPMath}}
\def\FeynCalc{\texttt{FeynCalc}}
\def\LoopTools{\texttt{LoopTools}}
\def\QCDLoop{\texttt{QCDLoop}}

\begin{document}
\begin{frontmatter}
\thispagestyle{empty}
	\begin{center}

	{\usekomafont{title}\huge Next-to-Leading Order\\[.1cm] QCD Corrections\\[.1cm] to Heavy-Flavour Production\\[.5cm] in Neutral Current DIS}\\[1.5cm]
	{\usekomafont{subject}Dissertation}\\[1.5cm]
	{\usekomafont{titlehead}
		der Mathematisch-Naturwissenschaftlichen Fakultät\\
		der Eberhard Karls Universität Tübingen\\
		zur Erlangung des Grades eines\\
		Doktors der Naturwissenschaften\\
		(Dr.~rer.~nat.)\\
	}

	\vfill
	{\usekomafont{author}
		vorgelegt von\\
		\textbf{\href{mailto:felix.hekhorn@uni-tuebingen.de}{Felix Hekhorn}}\\
		aus Münsingen\\
	}

	\vspace{2cm}
	{\usekomafont{date}
		Tübingen\\[0.3em]
		2019
	}
	\end{center}
\newpage

\thispagestyle{empty}
\null
\vfill
\noindent Gedruckt mit Genehmigung der Mathematisch-Naturwissenschaftlichen Fakultät der Eberhard Karls Universität Tübingen.\\
\vspace{10pt}\\
\begin{minipage}{0.482\linewidth}
	Tag der mündlichen Qualifikation: \\
	Dekan:\\
	1. Berichterstatter:\\
	2. Berichterstatter:
\end{minipage}
\begin{minipage}{0.482\linewidth}
	25.09.2019\\
	\href{mailto:dekan@mnf.uni-tuebingen.de}{Prof.~Dr.~Wolfgang Rosenstiel}\\
	\href{mailto:werner.vogelsang@uni-tuebingen.de}{Prof.~Dr.~Werner Vogelsang}\\
	\href{mailto:thomas.gutsche@uni-tuebingen.de}{Prof.~Dr.~Thomas Gutsche}
\end{minipage}

\newpage

\chapter{Summary} \label{chap:Summary}
In the last few decades Quantum Chromo Dynamics (QCD) became a major field for high energy physics. Large experiments have been and will be built to investigate all implications that are predicted by its master formula, the Lagrangian density. The Large Hadron Collider (LHC), the \enquote{largest machine of mankind}, was built to test the predictions of the standard model (SM) to a very high accuracy and so we finally arrived in the era of high precision physics. The demand for precision requires more and more effort to compute all necessary pieces. In this PhD thesis we close one of the last missing pieces in the set of next-to-leading order (NLO) QCD calculations.

We discuss the production of a heavy quark pair in deeply inelastic scattering (DIS) and compute all parts that are required at NLO accuracy. We settle the needed frameworks and notations before computing the matrix elements first at leading order (LO) and then for all contributions at NLO. We compute two different decompositions of the required phase space to realize two different numerical codes, which in turn specialize in different experimental observables. Along the way, we give useful tips and tricks and highlight some of the mathematical challenges on the road of NLO calculations.

Finally, we discuss the possibilities to apply these calculations to the planned spin physic program at a future Electron-Ion Collider (EIC). The discussed charm quark production can improve the determination of polarized parton distribution functions (pPDFs). On the other hand, we discuss the possibilities to obtain a further improved calculation of heavy quark structure functions by including the $\PZ$-boson exchange in neutral current (NC) DIS.

\chapter{\foreignlanguage{ngerman}{Zusammenfassung}} \label{chap:Zusammenfassung}
\begin{otherlanguage}{ngerman}
In den letzten Jahrzehnten wurde die Quantum Chromo Dynamik (QCD) eines der wichtigsten Felder in der Hochenergie-Physik. Große Experimente wurden und werden gebaut um all Vorhersagen ihrer Master-Formel, der Lagrange-Dichte, zu überprüfen. Der Large Hadron Collider (LHC), die \enquote{größte Maschine der Menschheit}, wurde gebaut um die Vorhersagen des Standardmodell der Teilchenphysik mit großer Genauigkeit zu überprüfen und so wurde letztendlich das Zeitalter der Präzisions-Physik eingeläutet. Die Nachfrage nach Genauigkeit fordert immer stärkere Anstrengungen um alle nötigen Teile bereitzustellen. Diese Dissertation schließt eine der letzten Lücken in der Liste der nächst-führenden Ordnung (NLO) QCD Berechnungen.

Wir diskutieren die Erzeugung eines schweren Quarkpaars in tief-inelastischer Streuung (DIS) und berechnen alle nötigen Bestandteile in NLO Genauigkeit. Wir legen alle nötigen Grundlagen dar und fixieren die Notation, bevor wir zuerst die Matrixelemente in führender Ordnung (LO) und anschließend in NLO berechnen. Wir geben zwei verschiedene Zerlegungen des zugehörigen Phasenraums an für je zwei verschiedene numerischen Anwendungen, die jeweils auf unterschiedliche experimentelle Beobachtungen ausgelegt sind. Parallel zu den Berechnungen geben wir hilfreiche Kniffe und Tricks und zeigen einige mathematische Herausforderungen auf, die für NLO Rechnungen nötig sind.

Abschließend diskutieren wir die Möglichkeit diese Rechnung für das Spinphysik-Programm am zukünftigen Elektronen-Ionen-Beschleuniger (EIC) anzuwenden. Die diskutierte Charmquark-Produktion kann genutzt werden um eine verbesserte Bestimmung von polarisierten Partonverteilungsfunktionen (pPDFs) zu erhalten. Andererseits diskutieren wir die Möglichkeit diese Rechnung für eine verbesserte Bestimmung von massiven Beiträgen zu den Strukturfunktionen zu nutzen, indem wir den Austausch von $\PZ$-Bosonen im Fall von neutralen Strömen (NC) in DIS berücksichtigen.
\end{otherlanguage}

\chapter{List of Publications} \label{chap:Papers}
\subsubsection{Journal papers}
\begin{description}
\item[\cite{Hekhorn:2018ywm}] \fullcite{Hekhorn:2018ywm}
\item[\cite{paper2}] \fullcite{paper2}
\end{description}

\subsubsection{Conference Proceedings}
\begin{description}
\item[\cite{Hekhorn:2018aio}] \fullcite{Hekhorn:2018aio}
\item[\cite{proceedings2}] \fullcite{proceedings2}
\end{description}

\tableofcontents

\end{frontmatter}
\begin{mainmatter}
\pagenumbering{arabic}
\chapter{Introduction} \label{chap:Intro}
\chapterquote{It's got three keyboards and a hundred extra knobs, including twelve with '?' on them.}{Organ of Physics}

It is now exactly 50 years ago that Feynman\cite{Feynman:1969wa}, Bjorken, and Paschos\cite{Bjorken:1968dy,PhysRev.185.1975} set the cornerstones of Deeply Inelastic Scattering (DIS) by bringing up the idea of the parton model and \enquote{Bjorken scaling}. With the fresh results from the Stanford Linear Accelerator Center (SLAC)\cite{Bloom:1969kc} at hand, they came up with a set of ideas that turned out to be very wide looking. Since then we checked, validated and improved their predictions a million times and now consider them as an integral part in the search for the \enquote{world formula}. In the following years we established the framework of Quantum Field Theory (QFT), that joins the ideas of classical field theory, special relativity and \enquote{classical} quantum mechanics. Meanwhile, the theory of Quantum Chromo Dynamics (QCD) replaced the parton model using the strict and formal language of QFT and its use as a perturbative theory (pQCD) was demonstrated (and awarded with the Nobel Prize in 2004) by \citeauthor{Politzer:1973fx}\cite{Politzer:1973fx,Politzer:1974fr}, Gross and Wilczek\cite{Gross:1973id,Gross:1973ju} in the 1970s. An other important mathematical step was the work by \citeauthor{tHooft:1972qbu}\cite{Veltman:1968ki,tHooft:1972qbu,tHooft:1972tcz,tHooft:1971qjg} (also awarded with a Nobel Prize in 1999), who showed the mathematical consistency (i.e.\ the renormalizability) of QCD.

The framework of DIS plays a key role in studying the effects and implication of QCD. It has been and still will be a major field of research in high energy particle physics. In order to study DIS we need large collider facilities that smash two different types of particles into each other. In 2007 the HERA facility at DESY was shutdown after delivering an enormous amount of data that is still under evaluation up to this late day. Now, the scientific community awaits eagerly the construction of a new machine, called the Electron Ion Collider (EIC)\cite{Accardi:2012qut,Boer:2011fh}, that we expect to reveal even deeper levels of understanding. The political and technological development of the project has already reached quite far and the discussed proposals plan to start taking data in about ten years from now at either the Jefferson Lab (JLab) or the Brookhaven National Laboratory (BNL). A major part of this work is dedicated to the new physical possibilities that might be studied at a future EIC.

We expect the EIC not only to provide more and more accurate data to the already established observables, but also to shed new light on other interesting physical aspects, such as nuclear effects or spin physics. Especially, the EIC is the first machine that comes with a dedicated spin program in the layout, allowing scientists to reconsider recent problems such as the \enquote{proton spin puzzle}: we know that the proton is a fermion, i.e.\ its spin is $1/2$, yet, we also know that the proton is not an elementary particle, so the total spin must be accounted for by its constituents. A common choice for this decomposition (commonly referred to as Jaffe-Manohar-decomposition\cite{Jaffe:1989jz}) is
\begin{align}
\frac 1 2 &= \frac 1 2\Delta \Sigma +\Delta G + L_{\Pg} + L_{\Pq}
\end{align}
where $\Delta \Sigma$ accounts for the intrinsic quark spin, $\Delta G$ accounts for the intrinsic gluon spin, and $L_{\Pg/\Pq}$ account for the orbital momenta of the respective partons. In one part of this work we will try to reduce the large uncertainties to the gluon contributions $\Delta G$, which we can write as
\begin{align}
\Delta G &= \int\limits_0^1\! \Delta \Pg(x)\,dx
\end{align}
so we need to reduce the uncertainties of the associated polarized gluon distribution $\Delta \Pg(x)$.

At the same time, this work also provides one of the last missing pieces in the set of next-to-leading order (NLO) QCD calculations and thus can play an important role for high precision physics. High precision physics does not live just for itself, but is considered one of the major ingredients in the search of Beyond Standard Model (BSM) physics. Although the Standard Model (SM) has proven its power of prediction over several orders of magnitudes, we do know for sure that it will not have the last say, as it cannot answer correctly all question in physics. For example, the challenges in the search for dark matter and the consistent description of neutrino masses are pressing issues that might be traced back to small deviation of the SM, in a similar way the Higgs particle was discovered\cite{Aad:2012tfa,Chatrchyan:2012xdj}. Furthermore, heavy quark production is sensitive to the strange sea quark distributions at NLO which is currently under discussion\cite{Gao:2017kkx,Harland-Lang:2014zoa,Dulat:2015mca,Butterworth:2015oua,Sato:2019yez,Ball:2017nwa}.

During the calculations for this work, we were able to build extensively on previous work by others: most importantly, \citeauthor{Laenen1993162}\cite{Laenen1993162,Laenen:1992xs} computed the case of unpolarized electro-production of heavy quarks, which we extended as a first step to the polarized case in our first papers\cite{Hekhorn:2018ywm,Hekhorn:2018aio}. The paper of \citeauthor{Riedl:2012qc}\cite{Riedl:2012qc} covers the corner case of photo-production, which we used to cross check our results in a non-trivial limit. A thorough introduction into many of the technical details is provided by \citeauthor{Bojak:2000eu}\cite{Bojak:2000eu} that was very handy and instructive. The extension to the fully differential case of unpolarized electro-production was performed by \citeauthor{Harris:1995tu}\cite{Harris:1995tu,Harris:1995pr} and again we did follow closely their ideas and methods (see our papers \cite{paper2,proceedings2}). The extension to the full neutral current case is covered for the unpolarized case at leading order (LO) by \citeauthor{Leveille:1978px}\cite{Leveille:1978px}.

The rest of this work is organized as follows: in \ChapterRef{chap:QCD} we outline the basic elements of pQCD, the underlying theory, and highlight the basic principles of regularization and renormalization. In \ChapterRef{chap:DIS} we describe the fundamental setup of DIS on the various levels, fix all needed notations and set up the two fundamental use cases, charm quark production and bottom quark production. In \ChapterRef{chap:ME} we compute all needed scattering matrix elements from their respective Feynman diagrams (or matrix amplitudes). In \ChapterRef{chap:PS} we compute the needed phase space expressions using two different decompositions that provide complementary representations. In \ChapterRef{chap:Partonic} we present some of the central results of this work, the partonic coefficient functions, focusing on results that are not covered yet by our first paper \cite{Hekhorn:2018ywm}. In \ChapterRef{chap:Hadronic} we give some first phenomenological applications on a hadronic level for the introduced use cases. Finally, in \ChapterRef{chap:Conclusions} we give some concluding comments.

Furthermore, we provide a set of appendices that collect additional materials which might be useful for future reference: in \AppendixRef{sec:Appendix.V} we give some relation for the computation of virtual contributions. \AppendixaRef{sec:Appendix.s4}{sec:Appendix.s5} collect intermediate expressions for the computations of the phase space and related objects. In \AppendixRef{sec:Appendix.Partonic} we collect all analytic partonic coefficient functions, that are known by now. Finally, in \AppendixRef{sec:Appendix.Tools} we give an overview of our used programs and tools as well as a brief and interesting discussion on our program code structure.

\chapter{Perturbative Quantum Chromo Dynamics} \label{chap:QCD}
\chapterquote{And thirdly, the code is more what you'd call 'guidelines' than actual rules}{QFT seen as phenomenologist}

The SM of particle physics joins two out of the three fundamental forces: the strong nuclear force described in the framework of Quantum Chromo Dynamics (QCD) and the electro-weak (EW) force. A possible inclusion of the third force, gravity, is still subject to ongoing research and neglected for the rest of this work. The introduction of \enquote{massive particles} does affect the Lagrangian, to be defined in the following \SectionRef{sec:QCD.Lagrangian}, but will not trigger any gravitational interactions\footnote{or, at least, they can be neglected}. While we mostly deal with QCD here we will still need a small set of the electro-weak theory, as we are interested in the full neutral current (NC) interaction. The present case is too complex for an analytical solution so we rely instead on the framework of perturbative QCD (pQCD).

\section{QCD Lagrangian} \label{sec:QCD.Lagrangian}
\begin{figure}[ht]
\centering
\begin{subfigure}[t]{.5\linewidth}
	\centering
	\includegraphics[width=.4\textwidth]{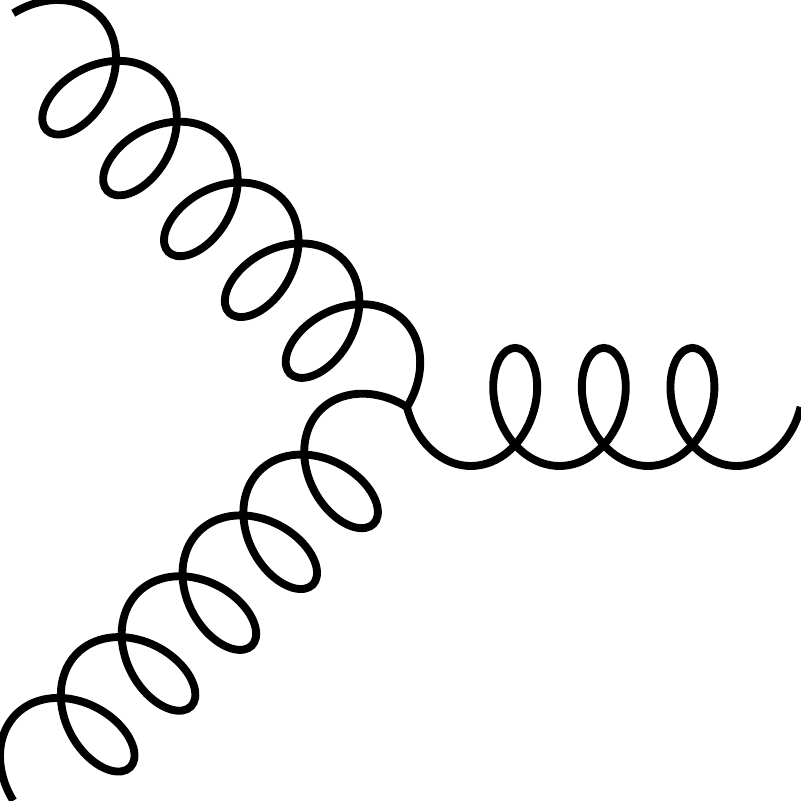}
	\caption{$3\Pg$-vertex}
	\label{fig:gSI3}
\end{subfigure}%
\begin{subfigure}[t]{.5\linewidth}
	\centering
	\includegraphics[width=.4\textwidth]{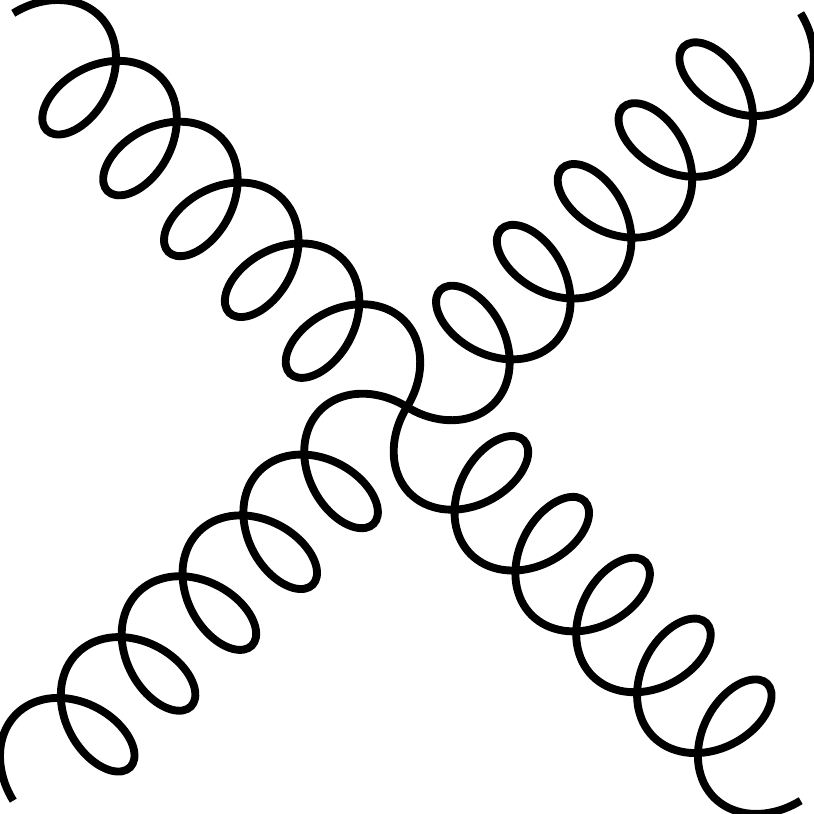}
	\caption{$4\Pg$-vertex}
	\label{fig:gSI4}
\end{subfigure}
\caption{Gluon self-interactions}\label{fig:gSI}
\end{figure}

The full derivation of the QCD Lagrangian is text book knowledge by now (see e.g.\ \cite{QFT,Leader:1996hm,Peskin:1995ev,Halzen:1984mc,Kronfeld:2010bx}), so we restrict the theoretical discussion here to some central aspects and list all necessary ingredients. The Lagrangian can be derived starting from a free, relativistic, fermionic Lagrangian
\begin{align}
\Ld_{free} &= \sum_q \bar \Psi_q(i \slashed{\partial} - m_q) \Psi_q \label{eq:Lfree}
\end{align}
where we will split the sum of quarks in the following into two parts: we will assume $n_f=n_{lf}+1$ quark flavours with $n_{lf}$ being massless ($m_q=0$) and one being massive. Up until today we know about six quark flavours: up ($\Pqu$), down ($\Pqd$), strange ($\Pqs$), charm ($\Pqc$), bottom ($\Pqb$), and top ($\Pqt$). Next, we demand invariance of \Eqref{eq:Lfree} under local transformations of $\Psi$ from the non-Abelian color gauge group $\text{SU}(N_C)$ with $N_C=3$ colors. This symmetry requires the introduction of $N_C^2-1=8$ bosonic gluon fields $A_\mu^a$ with a kinetic term of
\begin{align}
\Ld_{gluon} &= -\frac 1 4 G^a_{\mu\nu}G_a^{\mu\nu} \label{eq:Lgluon}
\end{align}
where $G_{\mu\nu}^a = \partial_\mu A_\nu^a-\partial_\nu A_\mu^a + g_s f_{abc}A_\mu^b A_\nu^c$ refers to the corresponding field tensor with $f_{abc}$ the anti-symmetric structure constants of the color space. $\Ld_{gluon}$ not only describes the propagation of the gluon field, but also contains the gluon self-interactions. QCD includes a $3\Pg$-vertex (\FigureRef{fig:gSI3}) with a coupling strength $\sim g_s$ and a $4\Pg$-vertex (\FigureRef{fig:gSI4}) with a coupling strength $\sim g_s^2$. These contributions manifest the non-Abelian nature of the gauge group $\text{SU}(3)$.

The coupling of the fermionic and the bosonic sector is mediated by
\begin{align}
\Ld_{int} &= g_s \bar\Psi\gamma^\mu T_a\Psi A_\mu^a
\end{align}
with $T_a$ the generator of the color gauge group. Following the standard prescription of QFT we now have to enhance the fields to field operators and impose the canonical quantization. For the case of QCD this is only achieved consistently in the path integral formalism. This procedure requires two important additions: first, we need to fix the gauge (we will use the Feynman gauge here), and, second, we need to add Faddeev–Popov ghost fields\cite{Faddeev:1967fc} $\eta^a$ which cancel unphysical gluon polarisations in \Eqref{eq:Lgluon}. We then get an additional term with
\begin{align}
\Ld_{gauge+ghost} &= -\frac 1 2 (\partial^\mu A_\mu^a)^2 - \bar \eta \Box \eta - g_s f_{abc} \bar c^a \partial^\mu(A_\mu^b \eta^c)\,.
\end{align}
Finally, the full QCD Lagrangian is obtained as the sum of all parts
\begin{equation}
\Ld_{\text{QCD}} = \Ld_{free} + \Ld_{gluon} + \Ld_{int} + \Ld_{gauge+ghost}
\end{equation}
from which all Feynman rules, required for in a pQCD calculation, can be obtained in a standard way. In practice, we will use the normalisation and rules from \cite{Leader:1996hm}.

The corresponding Lie algebra $\mathfrak{su}(3)$ of the gauge group $\text{SU}(3)$ has two Casimir constants: $C_F=(N_C^2-1)/(2N_C)$ in the fundamental representation and $C_A=N_C$ in the adjoint representation. The pQCD calculations generate traces of a small number of generators $T_a$ in color space which can be expressed in these group constants and, more important, can be obtained beforehand using standard methods. A more thoroughly handling of the color algebra, such as a decomposition into color multiplets\cite{Keppeler:2017kwt,Cvitanovic:2008zz}, is not necessary for the present case.

Furthermore, we need to specify the coupling of the electro-weak bosons $b\in\{\Pgg,\PZ\}$ to a fermion $f$. In order to use a common notation we denote the coupling as $-i e \Gamma^\mu_{b,f}$ with $e$ the universal electric charge unit and
\begin{align}
\Gamma^\mu_{b,f} &= g^V_{b,f} \Gamma^\mu_{V} + g^A_{b,f} \Gamma^\mu_{A} = g^V_{b,f} \gamma^\mu + g^A_{b,f} \gamma^\mu\gamma^5, \quad f\in\{\Pl,\Pq,\PQ\} \label{eq:couplingPhZ}
\end{align}
i.e.\ we split the interaction into a vector-like coupling (V), mediated by $\gamma_\mu$, and a axial-vector-like coupling (A), mediated by $\gamma_\mu\gamma_5$. The comparison to our papers \cite{Hekhorn:2018ywm,Hekhorn:2018aio,paper2}, that only considered electro-production, is obtained by observing that the photon $\Pgg$ is a purely vectorial boson ($g^A_{\Pgg,f}=0$) and its coupling is given by $g^V_{\Pgg,f}=Q_f$ with the electric charge fraction
\begin{align}
Q_f &= \left\{\begin{array}{ll}
\pm1 & f=\Plpm\\
-\frac 1 3 & f\in\{\Pqd,\Pqs,\Pqb\}\\
+\frac 2 3 & f\in\{\Pqu,\Pqc,\Pqt\}
\end{array}\right.\,.
\end{align}
For the $\PZ$-fermion couplings we find $g^V_{\PZ,f} = I_{3}^{f} - 2Q_f\sin^2(\theta_w)$ and $g^A_{\PZ,f} = I_{3}^{f}$ with the third component of the weak isospin
\begin{align}
 I_3^{f} &= \left\{\begin{array}{ll}
-\frac 1 2 & f\in\{\Plm,\Pqd,\Pqs,\Pqb\}\\
+\frac 1 2 & f\in\{\Plp,\Pqu,\Pqc,\Pqt\}
\end{array}\right.\,.
\end{align}
Furthermore, we use the PDG values\cite{Tanabashi:2018oca} as our default set of electro-weak parameters: $\sin^2(\theta_w) = 0.2315$, $M_{\PZ} = \SI{91.18}{\GeV}$ and $\alpha_{em}^{-1} = \num{137.04}$.

We define the common short-cuts $\alpha_s = g_s^2/(4\pi)$ and $\alpha_{em}=e^2/(4\pi)$ to simplify many expressions and from which our perturbation order is defined.

\section{Regularization and Renormalization} \label{sec:QCD.Renorm}
\begin{figure}[ht]
\centering
\includegraphics[width=.4\textwidth]{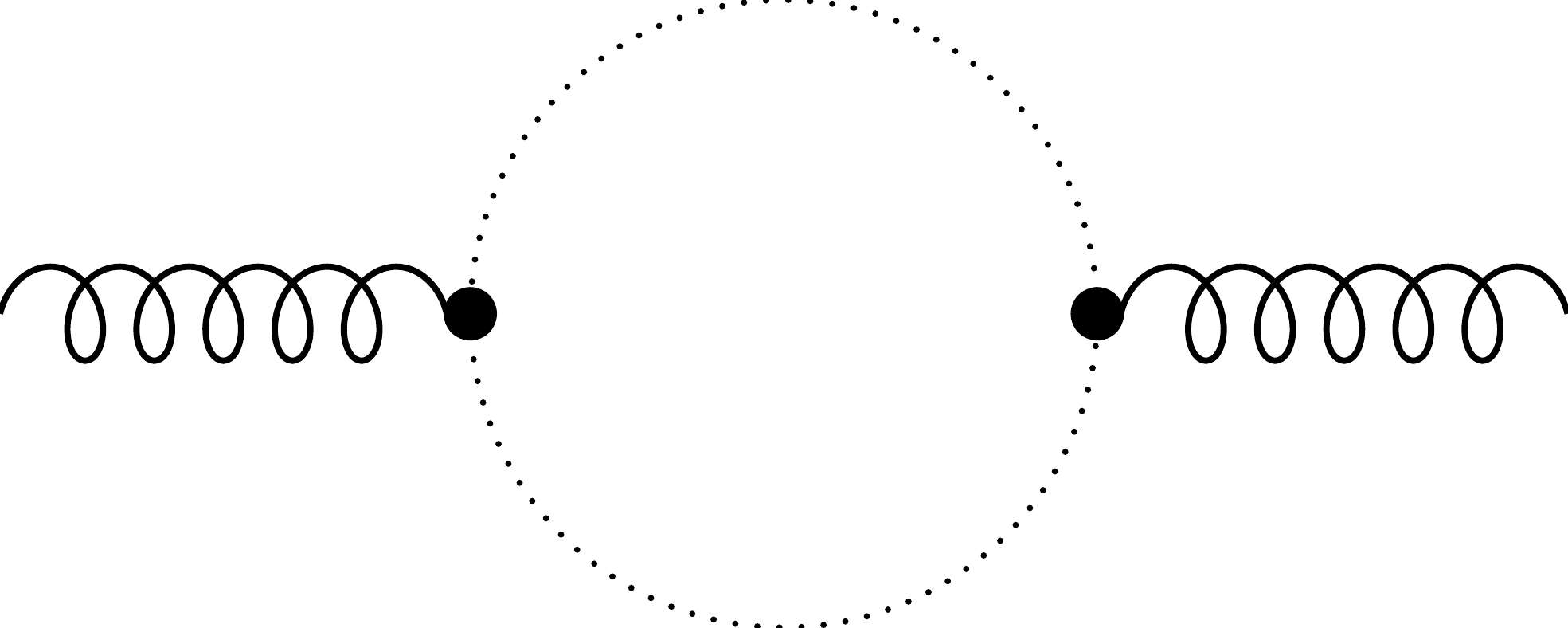}
\caption{The Feynman diagram for gluon self energy - the dotted line may refer to gluons, ghosts, light or heavy quarks}\label{fig:FeynSEg}
\end{figure}
The default QCD Lagrangian defined in a naive way includes in any higher order study infinite results (as e.g.\ the class of diagrams shown in \FigureRef{fig:FeynSEg}) that make a physical interpretation impossible. Therefore, one has to introduce a consistent \textit{regularization} that keeps intermediate results well-behaved but allows in a final step to return to a physical interpretation. While different regularization schemes\cite{QFT} exist such as the Pauli-Villars regularization or Zeta function regularization, we stick to the most common choice: dimensional regularization. It extends the space time by a small parameter $\epsilon$ and thus we work in $n=4+\epsilon$ dimensions\footnote{Note that other authors often choose $4-2\epsilon'$ dimensions}. By taking the limit $\epsilon\to 0$ in the end, i.e.\ before computing physical results, we return to the regular space time of one time dimension and three spatial dimensions. Extending the space time immediately affects the parameters in the QCD Lagrangian and thus we introduce an arbitrary mass parameter $\mu_D$ to keep the coupling $g_s$ dimensionless. Within the framework of dimensional \textit{regularization} we have to perform additionally the \textit{renormalization} of certain theory parameters.

Higher order corrections include virtual contributions where an internal loop appears. As the loop momenta is not restricted by any external boundary condition this may result potentially in a divergence. The class of diagrams shown in \FigureRef{fig:FeynSEg} can (e.g.\ for an intermediate gluon loop) be proportional to a term that behaves as
\begin{align}
\int\frac{d^n l}{l^2(k+l)^2}
\end{align}
which diverges in the ultra-violet (UV) limit, i.e.\ for large momenta, logarithmically. In dimensional regularization we perform the integral in $n$ dimensions and the divergence manifests itself as a term $1/\epsilon$. We can subtract these terms by a redefinition of the strong coupling $g_s$ or, equivalently, $\alpha_s$ and our observables will become UV safe. This \textit{renormalization} is mirrored in the Lagrangian by the inclusion of counter terms. We use the same $\MSbar_m$ scheme as defined in \cite{Bojak:2000eu,Hekhorn:2018ywm} and so the renormalization of all UV poles can be achieved by the replacement
\begin{align}
\alpha_s &\to \alpha_s(\mu_R^2)\left( 1 + 4\pi\alpha_s(\mu_R^2)C_\epsilon\left(\frac{\mu_D^2}{m^2}\right)^{-\epsilon/2} \left[\left(\frac 2 \epsilon +\ln(\mu_R^2/m^2)\right)\beta_0^f +\frac 2 3 \ln(\mu_R^2/m^2)\right] \right) \label{eq:asrep}
\end{align}
with $C_\epsilon = \exp(\epsilon/2(\gamma_E-\ln(4\pi)))/(16\pi^2)$, $\gamma_E=0.577\ldots$ the Euler-Mascheroni constant, $\beta_0^f = (11C_A- 2n_{f})/3$ the first coefficient of the beta function, and $\mu_R$ the new introduced renormalization scale. The running of the (renormalized) coupling is governed by renormalization group equation (RGE), also known as the beta function of QCD
\begin{align}
\mu\DeriveF{\mu}{\alpha_s(\mu))} &= \beta(\alpha) = -\frac{\beta_0^f}{4\pi}\alpha_s^2 + O(\alpha_s^3)
\end{align}
which is currently known up to five-loop $O(\alpha_s^6)$\cite{Herzog:2017ohr,Luthe:2016ima,Baikov:2016tgj}. At leading order (LO) this differential equation may be solved analytically, yielding
\begin{align}
\alpha(\mu_R^2) &= \frac{\alpha_s(\mu_0^2)}{1 + (\beta_0^f/4\pi) \alpha(\mu_0^2)\ln(\mu_R^2/\mu_0^2)} \label{eq:runas}
\end{align}
with $\mu_0^2$ the (arbitrary) boundary condition. The most common choice of $\mu_0^2 = m_{\PZ}^2$ leaves only its corresponding coupling strength $\alpha_s(m_{\PZ}^2)$ as a free parameter. In practice, we use the more evolved solutions given by the \texttt{LHAPDF}\cite{LHAPDF6} interface as it provides higher order prescriptions.

As can be seen from \Eqref{eq:runas}, the strong coupling has two distinct features at either limit of the choice of $\mu_R$: if the renormalization scale becomes small the value of $\alpha_s(\mu_R^2)$ increases - we refer to this as \textbf{confinement}. This explains why quarks cannot be observed as free particles, but they get bound immediately to one or several other quarks and form hadrons and baryons. Eventually, if $\mu_R$ becomes small enough, i.e.\ of the order $\LQCD=\mu_0\exp(-2\pi/(\alpha_s(\mu_0^2)\beta_0^f))$, the coupling reaches the Landau pole and becomes infinite. On the other side, if the renormalization scale becomes large the value of $\alpha_s(\mu_R^2)$ decreases - we refer to this as \textbf{asymptotic freedom}. As $\mu_R$ is to be chosen to a \enquote{typical} scale of the process we need large (i.e.\ energetic) collider experiments. So this is why heavy quarks are interesting in perturbative QCD: we define \enquote{heavy} quarks by putting the constraint that their mass is larger than this critical point, i.e.\ $m_{\PQ} > \LQCD$. So by observing heavy quark in the final state we can ensure that the measurement is a perturbative regime and thus it is safe to use pQCD. For common values of $\LQCD\sim\SI{1}{\GeV}$ we may consider the charm quark ($m_{\Pqc}\sim\SI{1.5}{\GeV}$), the bottom quark ($m_{\Pqb}\sim\SI{4.75}{\GeV}$), or the top quark ($m_{\Pqt}\sim\SI{175}{\GeV}$) as massive particles. The strong mass hierarchy ensures that all lighter quarks can be considered as massless.

Note that in \Eqref{eq:asrep} we did not treat the poles and the scaling logarithm equally to each other, but added explicitly again one of the logs. This is consistent with our choice to not renormalize the mass of the heavy quark, but instead we use the pole mass scheme\cite{Hekhorn:2018ywm,Bojak:2000eu}\footnote{the subscript in $\MSbar_m$ refers to this choice - Note that in principle the pole mass of a quark is not a physical observable and so in a strict environment a running mass is needed}, thus any mass $m$ always refers to its pole mass definition. However, there is a close relationship between the pole mass scheme and the \enquote{true} $\MSbar$ scheme\cite{Alekhin:2010sv} and we leave it to a future project to study the effect of introducing a running mass $m(\mu_m^2)$.

\chapter{Deeply Inelastic Scattering and Factorization} \label{chap:DIS}
\chapterquote{Lumos!}{The Electron}

The class of diagrams shown in \FigureRef{fig:FeynSEg} develop not only poles in the UV limit but also in the infra-red (IR) limit of the loop momenta, i.e.\ when they become small. Similar poles are created when soft, massless particles (such as gluons) are radiated. We address the issue of these IR poles by the choice to compute \textit{inclusive} observables only, i.e.\ we observe in the final state at most the produced heavy quark pair and do not observe anything else. This makes the result insensitive to any soft or collinear radiation in this unobserved part, commonly denoted as $X$, and we achieve a finite definition for our observables. In the following we either observe a specific geometric relation between the produced heavy quark pair or the momentum of the produced heavy \textit{anti}-quark $\PaQ$ or nothing at all, i.e.\ the fully inclusive case. Note that the observables we will call \enquote{fully inclusive} are \textit{not truly} fully inclusive as we still require the fact that a heavy quark (pair) is produced, we simply do not observe any properties thereof. We will denote the unobserved part in most cases as $X[\PQ]$ indicating that it has to contain at least one unobserved heavy quark $\PQ$ due to quantum number conservation.

Collider experiments can be separated into three classes: First, the scattering of two charged leptons (e.g.\ electrons $\Pem$ and positrons $\Pep$ as performed at LEP\cite{Jersak:1981sp})
\begin{equation}
\HepProcess{\Pem(l_1) + \Pep(l_2) \to \PaQ(p_2) + X[\PQ]}
\end{equation}
commonly referred to as single inclusive annihilation (SIA). Due to the confinement it is experimentally not possible to observe a final state quark. Instead one usually measures its decay product, e.g.\ a $\APDzero$-meson consisting of $(\Paqc\Pqu)$-pair for the case of $\PaQ = \Paqc$
\begin{equation}
\HepProcess{\Pem(l_1) + \Pep(l_2) \to \Paqc(p_2) + X[\Pqc]} \to \APDzero + X
\end{equation}
or, even more precisely, the decay products of the $\PD$-mesons, such as $\PK$-mesons, as $\PD$-mesons themselves are short lived. This process can be considered as the most simple case seen from a QCD perspective as there are only strong interacting particles in the final state, but not in the initial state. These experiments can be used to obtain the fragmentation functions (FF), which will be explained in more detail in the next section.

The second class of experiments is the scattering of a charged lepton $\Plpm$ off a hadron $\Ph$ (e.g.\ as performed at HERA\cite{H1:2018flt})
\begin{equation}
\HepProcess{\Plpm(l_1) + \Ph(P) \to \Plpm(l_2) + \PaQ(p_2) + X[\PQ]}
\end{equation}
commonly referred to as deeply inelastic scattering (DIS). The discussion in this work will focus on all the details of the observables that are related to DIS. This process can be considered as an intermediate step of QCD complexity as a single strong interacting particle appears in the initial state. These experiments can be used to obtain the parton distribution functions (PDF), which will be explained in the next section, as well.

The last class of experiments is the scattering of two hadrons (e.g.\ as performed at the LHC\cite{Riedl:2009ye})
\begin{equation}
\HepProcess{\Ph(P_1) + \Ph(P_2) \to \PaQ(p_2) + X[\PQ]}
\end{equation}
These processes are most challenging from a pQCD point of view as strongly interacting particles appear in the initial and in the final state.

\section{Hadronic Setup} \label{sec:DIS.Hadronic}
\begin{figure}
\includegraphics[width=.3\textwidth]{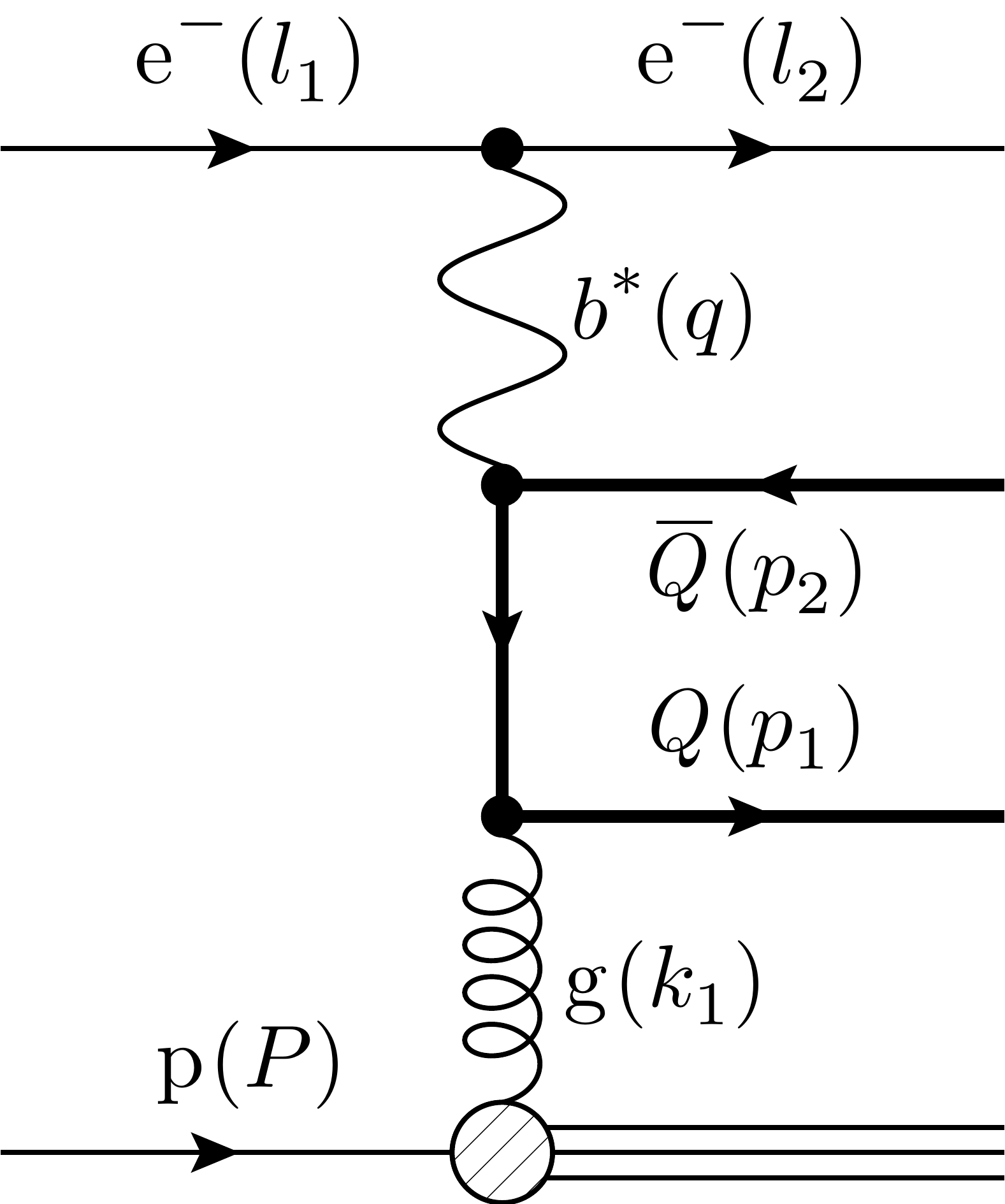}
\caption{Leading order diagram for heavy quark production in neutral-current DIS by the exchange of an electrically neutral, virtual boson $b^*=\{\Pggx,\PZx\}$. The splitting of a gluon from a proton is described by parton distributions function (PDF) in the adopted framework of collinear factorization.}
\label{fig:DIS}
\end{figure}

In the following we study the DIS reaction of a lepton scattering off a proton. As the lepton is not a strongly interacting particle it can only couple through an electro-weak exchange to the proton. The lepton will emit either a electro-magnetic photon $\Pgg$ or a weak gauge boson $\PZ,\PWpm$. The electro-magnetic charged bosons $\PWpm$ convert charged leptons into their corresponding neutrinos (e.g.\ $\HepProcess{\Pem\to\PWm+ \Pgne}$) and change the flavour of the coupling quark (e.g.\ $\HepProcess{\PWm \to \Paqc + \Pqd}$). In this work, we will neglect the case where the incoming lepton emits a charged boson, resulting in the flow of a charged current (CC) as it results in a \textit{single} heavy (anti-)quark. Instead we focus on the neutral current (NC) sector that is described by the exchange of the electrically neutral bosons $\Pgg$ and $\PZ$ (schematically depicted in \FigureRef{fig:DIS}). This way we have to study the reaction
\begin{equation}
\HepProcess{\Plpm(l_1) + \Ph(P) \to \Plpm(l_2) + \PaQ(p_2) + X[\PQ]}
\end{equation}
where $l_1(l_2), P$ and $p_2$ denote the four-momentum of the incoming (scattered) lepton, hadron and produced heavy anti-quark respectively. In the following, we will assume both, the lepton and proton to be massless $m_{\Pl}^2 = 0 = m_{\Ph}^2$ as their actual values are typically neglectable compared to other scales involved\cite{Moffat:2019qll}.

To fix the notation of the kinematical setup, we define the usual set of variables, i.e.\ the transferred momentum $q$ and the Bjorken variables $x$ and $y_{Bj}$ by
\begin{align}
q &= l_1 - l_2, &Q^2 &= -q^2, &x &= \frac{Q^2}{2 q\cdot P}, &y_{Bj} &=\frac{q\cdot P}{l_1\cdot P}\,.
\end{align}
Note that there is an additional relation $S_{\Pl} = (l_1+P)^2 = Q^2/(x \cdot y_{Bj})$ where the leptonic invariant energy $S_{\Pl}$ is typically fixed for a given experiment (e.g.\ for HERA $\sqrt{S_{\Pl}} = \SI{314}{\GeV}$). The ultimate task for theoretical physicists is then to predict the probability of the DIS process which is encoded in the leptonic cross section $\sigma_{\Pl}$.  As the coupling of the leptonic and the hadronic system is mediated by the exchange of a single, virtual vector boson, we can write the leptonic cross section $\sigma_{\Pl}$ as a contraction of a leptonic tensor $L_{\mu\mu'}$ with a hadronic tensor $W_{\mu\mu'}$
\begin{align}
\frac{d^2 \sigma_{\Pl}}{dx dy_{Bj}} &= \frac{2\pi y_{Bj} \alpha_{em}^2}{Q^4} \sum_{j} \eta_j L^{\mu\mu'}_j W_{\mu\mu'}^j\,. \label{eq:sigl}
\end{align}
Here, the sum runs over all contributing bosons $j\in\{\Pgg\Pgg,\Pgg\PZ,\PZ\PZ\}$ where we decided for the sake of unification to \textit{always} denote two bosons. The pure photon contributions ($j=\Pgg\Pgg$) are commonly referred to as electro-production and this approximation is covered in our papers \cite{Hekhorn:2018ywm,Hekhorn:2018aio,paper2}. In this work, we will not restrict the sum and thus rederive all formulae in \cite{Hekhorn:2018ywm,Hekhorn:2018aio,paper2}. The factors
\begin{align}
\eta_{\Pgg\Pgg} &= 1, &\eta_{\Pgg\PZ} &=\left(\frac{G_F M_{\PZ}^2}{2\sqrt 2 \pi \alpha_{em}}\right)\left(\frac{Q^2}{Q^2+M_{\PZ}^2}\right),  &\eta_{\PZ\PZ} =\eta_{\Pgg\PZ}^2
\end{align}
account for the difference between the $\PZ$-propagator and the photon propagator with $G_F$ Fermi's constant and we find (for the actual value of $\sin(\theta_w)$ see \SectionRef{sec:QCD.Lagrangian})
\begin{align}
\frac{G_F M_{\PZ}^2}{2\sqrt 2 \pi \alpha_{em}} &= \frac{1}{4 \sin^2(\theta_w) \cos^2(\theta_w)}\,.
\end{align}

The leptonic tensor $L_{\mu\mu'}$ can be obtained from the simple diagram of a charged lepton splitting a neutral vector boson, while averaging over the final state and keeping track of the helicity $\lambda$ of the initial lepton. We get for the photon splitting\cite{Tanabashi:2018oca}
\begin{align}
L_{\mu\mu'}^{\Pgg\Pgg} &= 2\left(k_{\mu}k_{\mu'}' + k_{\mu'}k_{\mu}' - (k\cdot k') g_{\mu\mu'} - i\lambda \epsilon_{\mu\mu'\alpha\beta}k^{\alpha}k'^{\beta}\right)\,. \label{eq:Lphph}
\end{align}
The tensor for the $\Pgg\PZ$ interference term and for the true $\PZ$ exchange ($j=\PZ\PZ$) can be obtained with the definition in \Eqref{eq:couplingPhZ} in an analog way and we get
\begin{align}
L_{\mu\mu'}^{\Pgg\PZ} &= (g^V_{\PZ,\Pl} + \lambda g^A_{\PZ,\Pl})L_{\mu\mu'}^{\Pgg\Pgg}\,, & L_{\mu\mu'}^{\PZ\PZ} &= (g^V_{\PZ,\Pl} + \lambda g^A_{\PZ,\Pl})L_{\mu\mu'}^{\Pgg\PZ}\,. \label{eq:LZZ}
\end{align}

Using Lorentz-invariance and our assumption of massless initial particles, we can decompose the hadronic tensor $W_{\mu\mu'}$ into a combination of six (hadronic) structure functions $F_1,F_2,F_3,g_1,g_4,g_5$ (using the naming conventions of \cite{Tanabashi:2018oca})
\begin{align}
W_{\mu\mu'} &= \left(-g_{\mu\mu'} + \frac{q_\mu q_{\mu'}}{q^2}\right) F_1(x,Q^2) + \frac{\hat P_\mu \hat P_{\mu'}}{P\cdot q} F_2(x,Q^2) - i\varepsilon_{\mu\mu'\alpha\beta} \frac{q^\alpha P^\beta}{2 P\cdot q} F_3(x,Q^2) \nonumber\\
 &\hspace{15pt} + i\varepsilon_{\mu\mu'\alpha\beta} \frac{q^\alpha S^\beta}{P\cdot q} g_1(x,Q^2) + \frac{S\cdot q}{P\cdot q}\left[\frac{\hat P_\mu \hat P_{\mu'}}{P\cdot q} g_4(x,Q^2) + \left(-g_{\mu\mu'} + \frac{q_\mu q_{\mu'}}{q^2}\right) g_5(x,Q^2)\right] \label{eq:HadroTen}
\end{align}
with $\hat P_\mu = P_\mu - \frac{P\cdot q}{q^2}q_\mu$ and $S$ denoting the spin vector of the scattering proton. It is convenient to introduce additionally two linear combination of structure functions
\begin{align}
F_L(x,Q^2) &= F_2(x,Q^2) - 2xF_1(x,Q^2), &g_L(x,Q^2) &= g_4(x,Q^2) - 2xg_5(x,Q^2)
\end{align}
as they facilitate many expressions. Especially, we find in the parton model\cite{PhysRev.185.1975} the Callan-Gross relation\cite{Callan:1969uq} $F_L=0$ and, its polarized counterpart, the Dicus relation\cite{Dicus:1972pq} $g_L = 0$.

We can characterize the structure functions by two properties we have to consider in parallel. First, their behaviour with respect to the spin vector of the scattering proton: the structure functions $g_1,g_4,g_5,g_L$ are proportional to Lorentz objects that are linear with respect to the spin vector $S$ and are thus not measurable when the initial spin is averaged. We refer therefore to $F_1,F_2,F_L,F_3$ as \textbf{unpolarized} structure functions and to $g_1,g_4,g_5,g_L$ as \textbf{polarized} structure functions. Polarized structure functions are most conveniently accessible when the difference of the possible initial proton spin orientations is observed. As second property of structure functions, we have to consider parity, i.e.\ their behaviour with respect to the coupling Lorentz indices. As can be seen from \Eqref{eq:Lphph}, we obtain a symmetric Lorentz tensor when averaging over the lepton helicity $\lambda\in\{-1,+1\}$ and an anti-symmetric tensor by considering the anti-symmetric combination. This symmetry has to be met by the hadronic tensor $W_{\mu\mu'}$. Therefore, we can categorize the unpolarized structure functions by this property and refer to $F_3$ as \textbf{parity violating} (PV) structure functions, as it is proportional to the only anti-symmetric object, the Levi-Civita tensor $\varepsilon_{\mu\mu'\alpha\beta}$, and to $F_1,F_2,F_L$ as \textbf{parity conserving} (PC) structure functions. For the case of polarized structure functions the situation is reversed as they already need an anti-symmetrization with respect to the proton spin, as explained above. So we refer to $g_1$ as parity conserving structure functions, proportional to $\varepsilon_{\mu\mu'\alpha\beta}$ and requiring the anti-symmetric part of $L_{\mu\mu'}$, and to $g_4,g_5,g_L$ as parity violating structure functions.
The characterizations of the structure functions is summarized in \TableRef{tab:SF}.
\begin{table}[ht]
\begin{tabular}{l|l|l}
structure functions & unpolarized  & polarized \\
\hline
parity conserving & $F_1,F_2,F_L$ & $g_1$ \\
\hline
parity violating & $F_3$ & $g_4,g_5,g_L$ \\
\end{tabular}
\caption{Characterizations of structure functions}
\label{tab:SF}
\end{table}

As we assume $m_{\Pl}^2 = 0$, the coupling of the bosons $b$ to the lepton $\Pl$ renormalizes the leptonic tensor, but leaves the Lorentz structure unchanged (see \Eqssref{eq:LZZ}). These leptonic modifications can be used to decompose the (hadronic) structure functions by the participation of the scattering bosons:
\begin{align}
H^{\tNC} &= H^{\Pgg\Pgg} - \left(g^V_{\PZ,\Pl} \pm \lambda g^A_{\PZ,\Pl}\right) \eta_{\Pgg\PZ} H^{\Pgg\PZ} + \left((g^V_{\PZ,\Pl})^2 + (g^A_{\PZ,\Pl})^2 \pm 2\lambda g^{V}_{\PZ,\Pl}g^{A}_{\PZ,\Pl}\right)\eta_{\PZ\PZ} H^{\PZ\PZ} \label{eq:HPCphZ}
\end{align}
for the parity conserving structure functions $H\in\{F_2,F_L,g_1\}$ and
\begin{align}
H^{\tNC} &= - \left(g^A_{\PZ,\Pl} \pm \lambda g^V_{\PZ,\Pl}\right) \eta_{\Pgg\PZ} H^{\Pgg\PZ} + \left(2 g^{V}_{\PZ,\Pl}g^{A}_{\PZ,\Pl} \pm \lambda\left((g^V_{\PZ,\Pl})^2 + (g^A_{\PZ,\Pl})^2 \right) \right)\eta_{\PZ\PZ} H^{\PZ\PZ}\label{eq:HPVphZ}
\end{align}
for the parity violating structure functions $H\in\{F_3,g_4,g_L\}$. As before $\lambda$ denotes the helicity of the incoming lepton and $\pm$ refers to the scattering of the initial lepton $\Plpm$.

Inserting the above definitions into \Eqref{eq:sigl}, we get for the unpolarized case
\begin{align}
\frac{d^2 \sigma^{\tNC}_{\Pl}}{dx dy_{Bj}} &= \frac{2\pi \alpha_{em}^2}{x y_{Bj} Q^2} \left(Y_+F_2^{\tNC} -y^2 F_L^{\tNC} \mp Y_- xF_3^{\tNC}\right) \label{eq:d2sigLepto}
\end{align}
and for its polarized counterpart
\begin{align}
\frac{d^2 \Delta \sigma^{\tNC}_{\Pl}}{dx dy_{Bj}} &= \frac{2\pi \alpha_{em}^2}{x y_{Bj} Q^2} \left(\mp Y_- 2xg_1^{\tNC} -Y_+g_4^{\tNC} +y^2 g_L^{\tNC} \right) \label{eq:d2sigLeptoPol}
\end{align}
with $Y_{\pm} = 1 \pm (1-y_{Bj})^2$. We are left with a determination of the structure functions as they encode all relevant physical informations. Experimental data\cite{H1:2018flt} is often provided in terms of a \textit{reduced} cross section $\sigma_{red} = F_2-y^2/Y_+ F_L$. In calculations where all initial spins are fixed rather then summed all structure functions contribute.

To compute the structure functions we introduce the framework of collinear factorization. The formal proof of the collinear factorization exists for the case of \textit{true} fully inclusive DIS only\cite{Collins:1989gx}. However, its validity is still assumed here, as we only consider a well defined subset and since calculatory justification has been experienced in other works. In the following, we highlight some of the key features and give the necessary formulae. Factorization relies on the assumption that the timescales of the various parts are hierarchically ordered, i.e.\ that we can assume the participating parton to be frozen inside its parent proton when interacting with the exchanged virtual boson. This allows us to separate the whole process into three parts: first, a non-perturbative part, we call parton distribution function (PDF), which takes care of the evolution of the parton inside the proton up to the point when it is hit by the exchanged boson. Second, the interaction of the parton with the exchanged boson itself, commonly referred to as partonic matrix element and this part is accessible via pQCD. And third, the propagation of the produced partons into measurable hadrons, commonly described as fragmentation functions (FF), which again are of non-perturbative nature. The intertwining of the various parts is mediated by the convolution of two functions
\begin{equation}
(c \otimes f)(x) = \int\limits_x^1\,\frac{dy}{y}\,c(y) f(x/y)
\end{equation}
This time separation can also be linked to a separation in energy and thus a new scale, the factorization scale $\mu_F$ is introduced and all components, the PDFs, the partonic matrix elements and the FFs become scale dependent. The FFs $D_a^h(z,\mu_F^2)$ can be interpreted as the probability density that a parton $a$ decays collinearly into a hadron $h$ carrying a momentum fraction $z$. For simplicity, we will drop the concept of the FF in the following and consider our final state only to consist of \textit{raw} heavy (anti-)quarks. We leave it to a future project to add the needed additional convolution or to experimentalists to account for the respective deconvolution\cite{H1:2018flt} as quarks themselves are not observable.

\section{Parton Distribution Functions} \label{sec:DIS.PDF}
In collinear factorization we assume that the partons move collinear to their parent hadron and we can thus interpret the PDF $f_j^h(\xi,\mu_F^2)$ as the probability density that the parton $j$ gets emitted collinear from its parent hadron\footnote{we will drop the hadron superscript to the PDFs in the following and will always assume a proton $h=\Pp$} $h$ carrying a momentum fraction $\xi$. As we are interested in both the polarized and unpolarized process we have to introduce two different sets of PDFs, $\Delta f_j(\xi,\mu_F^2)$ and $f_j(\xi,\mu_F^2)$ respectively, defined by
\begin{align}
f_j(\xi,\mu_F^2)         &= f_j^+(\xi,\mu_F^2)+f_j^-(\xi,\mu_F^2)\,,
&\Delta f_j(\xi,\mu_F^2) &= f_j^+(\xi,\mu_F^2)-f_j^-(\xi,\mu_F^2) \label{eq:DeltaFjspin}
\end{align}
where $f_j^+$ ($f_j^-$) refers to the probability of finding the parton $j$ with its spin aligned (anti-aligned) to its parent proton. Note that by defining $\Delta f_j$ as a difference it does not have any longer the \textit{strict} interpretation of a probability density and thus can be negative. Most formulas however will not depend on the polarization, so we will use the notation of $(\Delta)$ to denote both, the polarized and unpolarized case.

The evolution of the PDFs with their factorization scale $\mu_F$ is given by the Dokshitzer-Gribov-Lipatov-Altarelli-Parisi (DGLAP) equations\cite{Altarelli:1977zs,Gribov:1972ri,Dokshitzer:1977sg}
\begin{align}
\mu_F^2\Derive{\mu_F^2} (\Delta)f_j(\xi,\mu_F^2) &= \sum_k \int\limits_\xi^1 \frac {d\zeta}{\zeta} (\Delta) P_{jk}(\zeta,\alpha_s(\mu_F^2)) \cdot (\Delta)f_k(\xi/\zeta,\mu_F^2) \label{eq:DGLAP}
\end{align}
where $(\Delta) P_{jk}(\zeta,\alpha_s(\mu_F^2))$ denote the Altarelli-Parisi splitting kernels\cite{Altarelli:1977zs}. At LO these kernels can be interpreted as the probability of a parton $k$ splitting a parton $j$ with a momentum fraction $\zeta$ and they can be computed by pQCD\footnote{as all calculations in this work are performed at NLO accuracy, we will drop the order label to the kernels immediately after the definition}
\begin{align}
(\Delta) P_{jk}(\zeta,\alpha_s) &= \alpha_s\sum_{l=0} \left(\frac{\alpha_s}{4\pi}\right)^l (\Delta) P_{jk}^{(l)}(\zeta)\,.
\end{align}
The form of the splitting kernels depends on the way soft poles are renormalized and we will do this in two different ways as discussed in \ChapterRef{chap:PS}.

We can use PDFs to theoretically predict a physical observable $\sigma \sim f_j \otimes c_j$ from a pQCD calculation $c_j$. On the other hand, we can reverse the procedure and apply a fitting procedure to obtain the unperturbative PDFs by comparing different input PDFs to measured data points. By using the fact that the PDFs themselves are \textit{universal}, we are able to reuse them in other processes and evaluate the quality of the fit. Most modern research groups that provide PDFs will perform a \textit{global analysis} using many different experiments as input\footnote{it has been argued recently that one should obtain PDFs and FFs at the same time - see \cite{Sato:2019yez}}. As the fitting procedure is highly non-trivial, the groups usually provide a best-fit result together with some uncertainty estimation given as error PDF set members.

The mathematics underlying the factorization theorem usually requires a consistent truncation of the perturbation parameter $\alpha_s$. But, both polarized PDF sets (pPDFs), we are using in this work, are only available at NLO accuracy. Thus, we will relax this condition with respect to the PDFs and impose rather a strict evaluation of NLO PDFs throughout. We interface the PDFs through a wrapper that covers open-source libraries of LHAPDF\cite{LHAPDF6} and the private code for \DSSV (see also \AppendixRef{sec:Appendix.Tools}).

We will demonstrate in \ChapterRef{chap:Hadronic} that heavy quarks (HQs) can be used to constrain PDFs. To this end, we set up two different scenarios: charm quark production in \SectionRef{sec:DIS.PDF.c} and bottom quark production in \SectionRef{sec:DIS.PDF.b} that highlight two different use cases within the PDF determinations of this calculation.

\subsection{Charm Quark Production} \label{sec:DIS.PDF.c}
Here, we observe a produced anti-charm $\Paqc$ (with mass $m_{\Pqc}=\SI{1.5}{\GeV}$ and $n_{lf}=3$ light flavours ($\Pqd,\Pqu,\Pqs$)) and this was the main focus of our papers \cite{Hekhorn:2018ywm,paper2,Hekhorn:2018aio} where we studied the pPDF set \DSSV published by \citeauthor{deFlorian:2014yva}\cite{deFlorian:2008mr,deFlorian:2009vb,deFlorian:2014yva}. When obtaining pPDFs an unpolarized PDF is needed as a baseline to constrain the positivity of the single spin dependent PDF $f_j^\pm(x)$ (see \Eqref{eq:DeltaFjspin}). In the case of \DSSV the corresponding unpolarized PDF is \MSTW\cite{Martin:2009iq}. The small number of experiments with a polarized beam leaves pPDFs by far less constraint than unpolarized PDF. Moreover, the data used for \DSSV from STAR\cite{Adamczyk:2014ozi,Adamczyk:2012qj}, PHENIX\cite{Adare:2014hsq,Adare:2008aa,Adare:2008qb} and COMPASS\cite{Alekseev:2010hc,Alekseev:2010ub} cover only a small range in $x$.
\begin{figure}[ht]
\includegraphics[width=.6\textwidth]{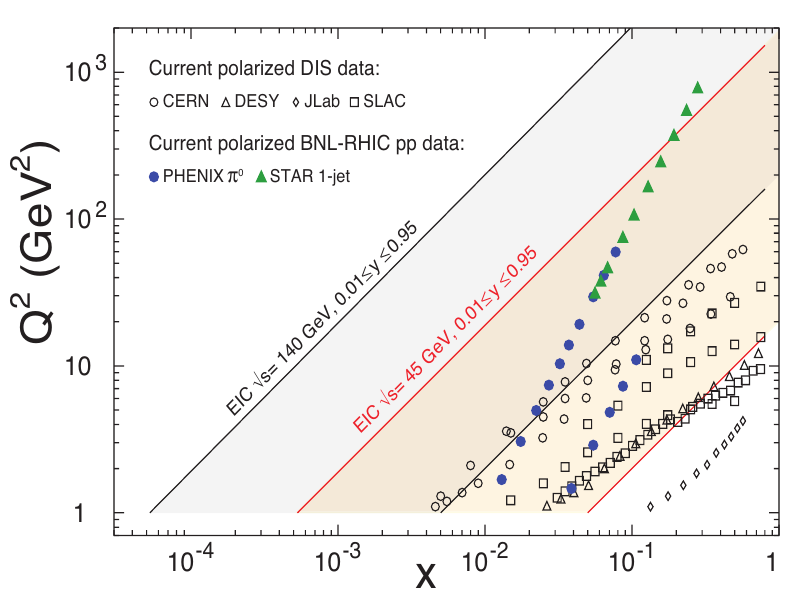}
\caption{The range in $x$ vs. $Q^2$ accessible with the EIC for different center-of-mass energies, compared to existing data - graph taken from \cite{Accardi:2012qut}}
\label{fig:EIC-xQ2}
\end{figure}
There is a dedicated project at the EIC\cite{Accardi:2012qut,Boer:2011fh} to address this issue and we show in \FigureRef{fig:EIC-xQ2} the proposed scenarios for the observable reach in $x$ and $Q^2$. The planned experiments can obviously provide an complementary measurement.
\begin{figure}[ht]
\includegraphics[width=.6\textwidth]{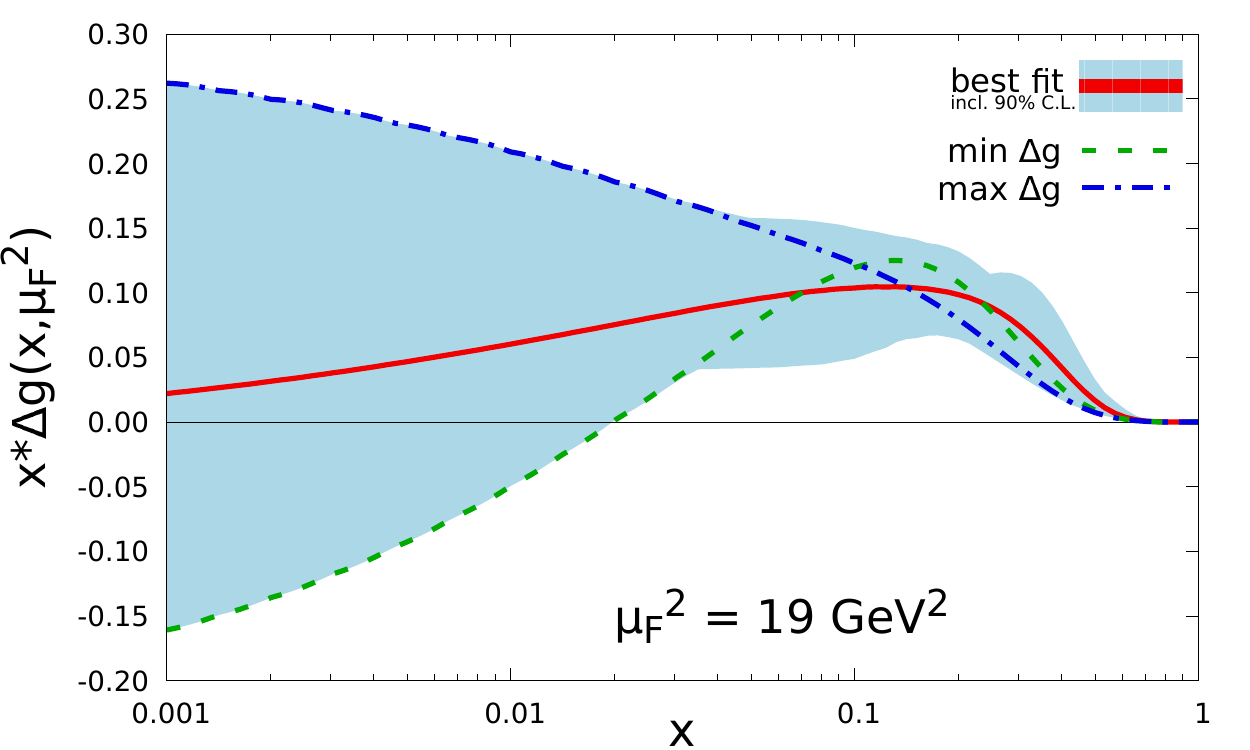}
\caption{The polarized gluon PDF $\Delta\Pg(x,\mu_F^2)$ of \DSSV\cite{deFlorian:2014yva} at $\mu_F^2 = \SI{19}{\GeV^2}$. We highlight three specific PDF set members: \enquote{best fit} representing the best fit result, \enquote{min $\Delta\Pg$} representing the variation with the smallest value for small $x$ and \enquote{max $\Delta\Pg$} representing the variation with the largest value for small $x$.}
\label{fig:DSSV-xg}
\end{figure}
The current status of the polarized gluon distribution $\Delta \Pg(x,\mu_F^2) = \Delta f_{\Pg}(x,\mu_F^2)$ is shown in \FigureRef{fig:DSSV-xg}. One can see that there are enormous uncertainties in the low $x$ region ($x \leq \num{.03}$) which is indicated by the large uncertainty bands. We will discuss in \SectionRef{sec:Hadronic.c} on how to constrain the small $x$ behaviour by a measurement of certain HQ observables. Note that we highlighted three specific variations that represent the different possible functional forms of the gluon distribution. These are the same variations as highlighted in our papers \cite{Hekhorn:2018ywm,paper2,Hekhorn:2018aio,proceedings2} and again we will keep track of them here. We will show that we can distinguish these variations for the proposed observables and thus, we can distinguish which form is favoured in a potential measurement.
\begin{figure}[ht]
\includegraphics[width=.6\textwidth]{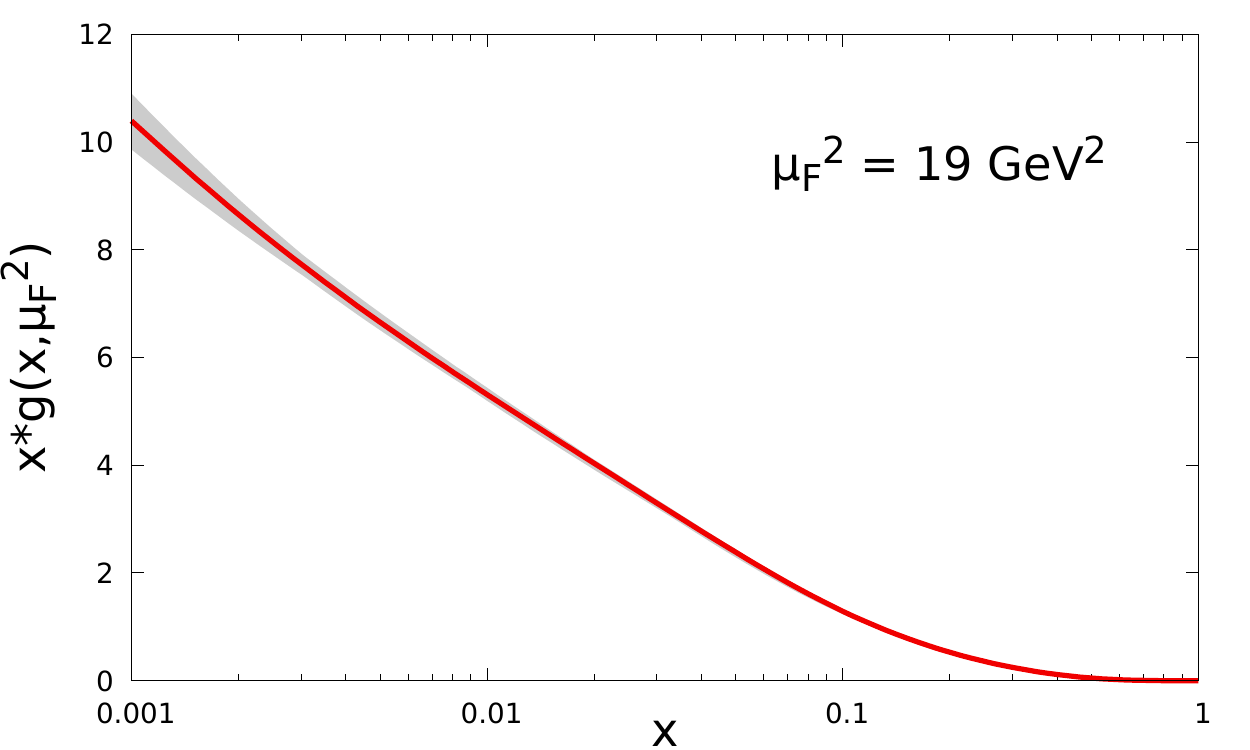}
\caption{The gluon PDF $\Pg(x,\mu_F^2)$ of \MSTW\cite{Martin:2009iq} at $\mu_F^2 = \SI{19}{\GeV^2}$. Note that for the determination of \DSSV \textit{only} the central value shown in bold, red has been used and consequently will be used in the following. The associated uncertainties shown in grey are simply shown for comparison.}
\label{fig:MSTW-xg}
\end{figure}
For comparison we also show the unpolarized gluon distribution $\Pg(x,\mu_F^2)=f_{\Pg}(x,\mu_F^2)$ of \MSTW in \FigureRef{fig:MSTW-xg} which also develops (relatively) large uncertainties in the small $x$ region, but they are several orders of magnitude smaller than their polarized counterparts in $\Delta\Pg$.

\subsection{Bottom Quark Production} \label{sec:DIS.PDF.b}
Here, we observe a produced anti-bottom $\Paqb$ ($m_{\Pqb}=\SI{4.75}{\GeV}, n_{lf}=4$) and use the PDF sets of the NNPDF collaboration, i.e.\ for the polarized case \NNPDFpol\cite{Nocera:2014gqa} and its unpolarized counterpart \NNPDF\cite{Ball:2012cx}. When focusing on bottom quark production, we need to access the polarized charm quark distribution $\Delta\Pqc(x,\mu_F^2)$ which is currently not provided by \DSSV.
\begin{figure}[ht]
\includegraphics[width=.45\textwidth]{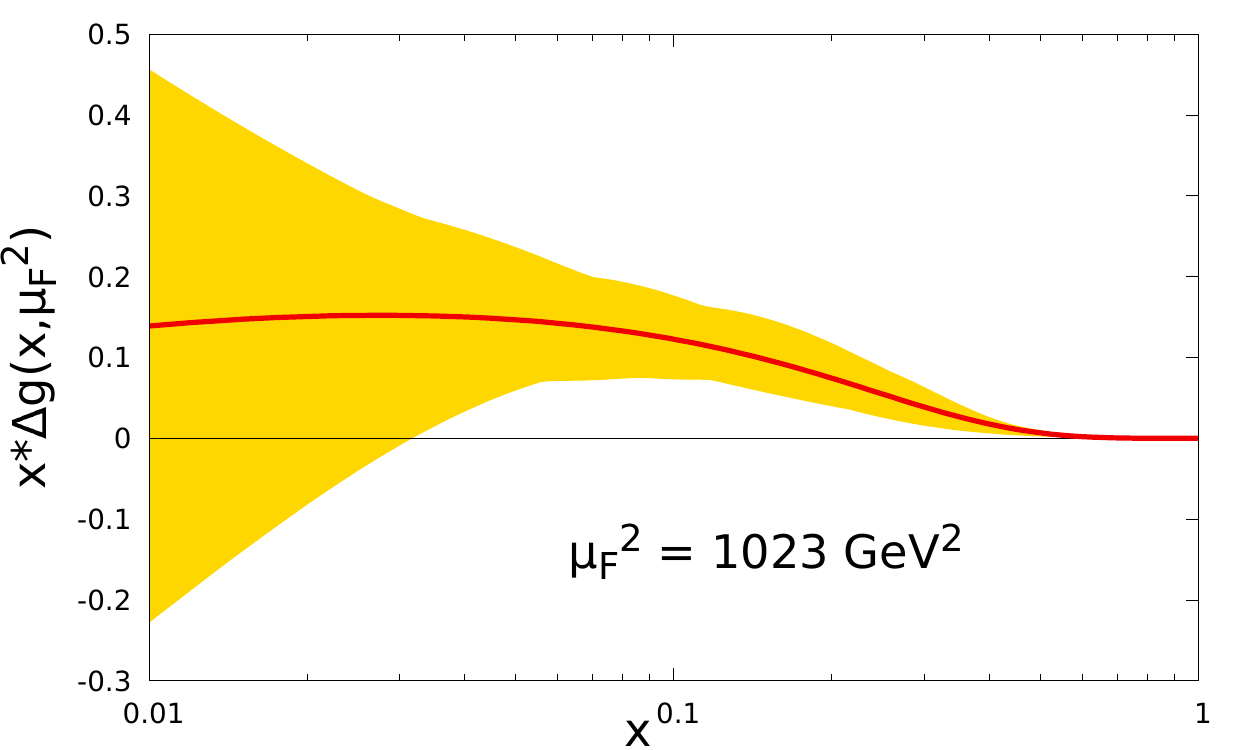}%
\includegraphics[width=.45\textwidth]{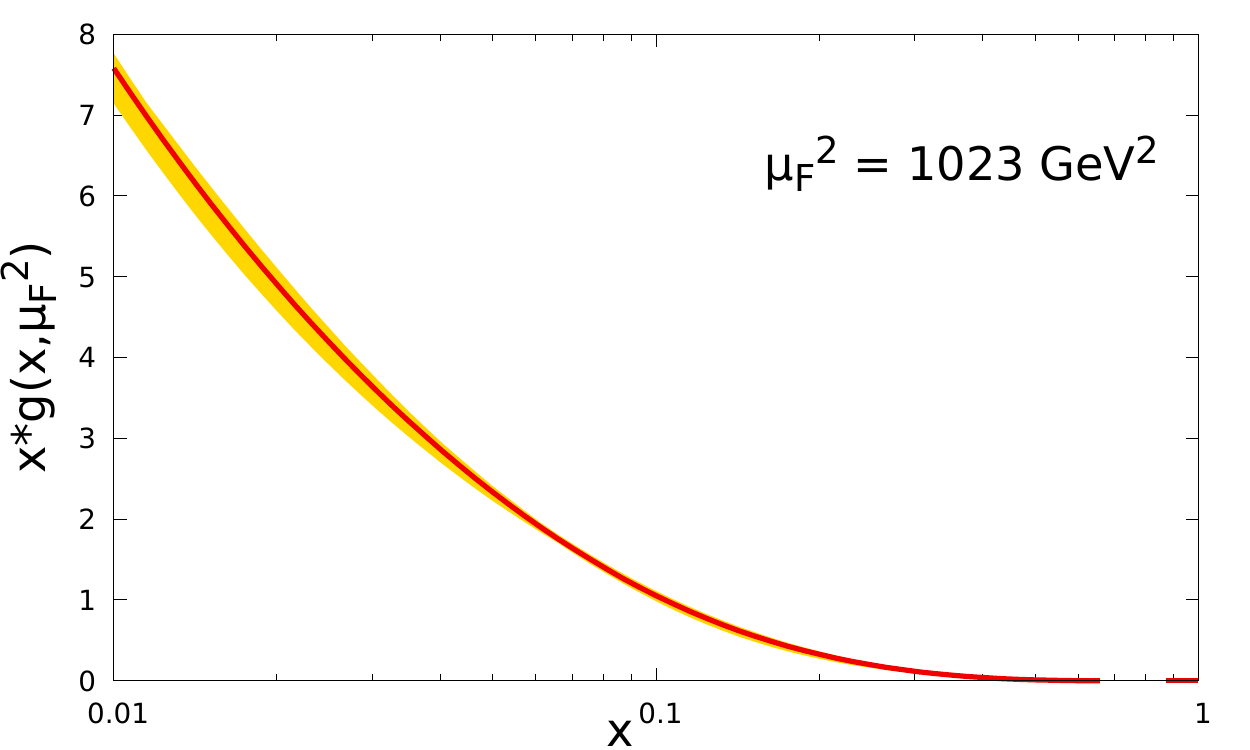}
\caption{Left: The polarized gluon PDF $\Delta\Pg(x,\mu_F^2)$ of \NNPDFpol\cite{Nocera:2014gqa} at $\mu_F^2 = 4m_{\Pqb}^2 + Q^2$ with $Q^2=\SI{1000}{\GeV^2}$. Right: The unpolarized gluon PDF $\Pg(x,\mu_F^2)$ of \NNPDF\cite{Ball:2012cx} with the very same parameters. Both show the best fit result (red solid) and the error estimate obtained by 100 Monte Carlo replicas. Note that for the determination of \NNPDFpol \textit{only} the central value has been used.}
\label{fig:NNPDF-xg}
\end{figure}
We show the polarized gluon distribution $\Delta\Pg(x,\mu_F^2)$ and its unpolarized counterpart $\Pg(x,\mu_F^2)$ in \FigureRef{fig:NNPDF-xg}. One can see mostly the same shape and error bands as in the case of charm production which confirms the universal character of PDFs. The case of bottom production is suppressed in electro-production due to the smaller electric charge fraction ($Q_{\Pqb}^2 < Q_{\Pqc}^2$) but it can be used to study electro-weak corrections (see \SectionRef{sec:Hadronic.b}). The mass ratio $m_{\Pqb}/M_{\PZ}$ towards the ultimate NC scale, the $\PZ$ mass $m_{\PZ}$, is much larger than in the case of charm production ($m_{\Pqb} > m_{\Pqc}$). The case of top quark production would join both points, but it is out of reach for now as the threshold for $m_{\Pqt}\sim\SI{175}{\GeV}$ requires much larger colliders.

\section{Partonic Setup} \label{sec:DIS.Partonic}
\subsubsection{Partonic Tensor}
As discussed in the previous section, we now assume that the proton $\Pp(P)$ emits a collinear parton $j(k_1)$ with a momentum fraction $\xi$, i.e.\ $k_1=\xi P$. So we have to consider the partonic reactions
\begin{equation}
\HepProcess{b^*(q) + j(k_1) \to \PaQ(p_2) + X[Q]},\quad b^*\in\{\Pggx,\PZx\}, j\in\{\Pg,\Pq\}
\end{equation}
where the gluonic channel $j=\Pg$ starts to contribute at $\alpha_{em}\alpha_s$ defining the leading order (LO). At next-to-leading order (NLO) $\alpha_{em}\alpha_s^2$ also initial light quarks $j=\Pq$ can contribute. The relevant partonic kinematic variables are given by
\begin{align}
z &= \frac{Q^2}{2 q\cdot k_1} = \frac {x}{\xi}, &s &= (k_1+q)^2, &s' &= s-q^2\,. \label{eq:partonicVars}
\end{align}

The partonic tensor $\hat w_{\mu\mu'}$ can be obtained from the hadronic tensor $W_{\mu\mu'}$ in \Eqref{eq:HadroTen} and is given by
\begin{align}
\frac 1 {2z} \hat w_{\mu\mu'}(S) &= \left[-g_{\mu\mu'} + \frac{q_\mu q_{\mu'}}{q^2}\right] \hat F_1(z,Q^2) + \frac{\hat k_{1,\mu} \hat k_{1,\mu'}}{k_1\cdot q} \hat F_2(z,Q^2) \nonumber\\
 &\hspace{15pt} - i\varepsilon_{\mu\mu'\alpha\beta} \frac{q^\alpha k_1^\beta}{2 k_1\cdot q} \hat F_3(z,Q^2) \nonumber\\
 &\hspace{15pt} + \frac{q_\mu q_{\mu'}}{q^2} \hat F_4(z,Q^2) + \frac{q_\mu k_{1,\mu'} + q_{\mu'}k_{1,\mu}}{2 k_1\cdot q} \hat F_5(z,Q^2)  \nonumber\\
 &\hspace{15pt} + i\varepsilon_{\mu\mu'\alpha\beta} \frac{q^\alpha S^\beta}{k_1\cdot q} \hat g_1(z,Q^2) \nonumber\\
 &\hspace{15pt} + \frac{S\cdot q}{k_1\cdot q}\left[\frac{\hat k_{1,\mu} \hat k_{1,\mu'}}{k_1\cdot q}\hat g_4(z,Q^2) + \left(-g_{\mu\mu'} + \frac{q_\mu q_{\mu'}}{q^2}\right) \hat g_5(z,Q^2)\right] \nonumber\\
 &\hspace{15pt} + \frac{S\cdot q}{k_1\cdot q}\left[\frac{q_\mu q_{\mu'}}{q^2} \hat g_6(z,Q^2) + \frac{q_\mu k_{1,\mu'} + q_{\mu'}k_{1,\mu}}{2 k_1\cdot q} \hat g_7(z,Q^2)  \right]
\label{eq:PartonTen}
\end{align}
with $\hat k_{1,\mu} = k_{1,\mu} - \frac{k_1\cdot q}{q^2}q_\mu$. Again we introduce two additional linear combinations of structure functions by
\begin{align}
\hat F_L(z,Q^2) &= \hat F_2(z,Q^2) - 2z\hat F_1(z,Q^2), &\hat g_L(z,Q^2) &= \hat g_4(z,Q^2) - 2z\hat g_5(z,Q^2)\,.
\end{align}
As can be seen from \Eqsaref{eq:d2sigLepto}{eq:d2sigLeptoPol}, it is convenient to rescale the structure functions and so, we focus in the following on the six structure functions $\hat F_2, \hat F_L, z\hat F_3, 2z\hat g_1, \hat g_4$ and $\hat g_L$ or their hadronic counterparts.

To obtain the structure functions from the partonic tensor, we define a set of projections given by
\begin{align}
\hat{\mathcal P}_{\hat F_2}^{bb',\mu\mu'} &= \frac{-g^{\mu\mu'}}{n-2} - \frac{n-1}{n-2} \cdot \frac{4 z^2 k_{1}^{\mu}k_{1}^{\mu'}}{q^2} - \frac{q^\mu q^{\mu'}}{q^2} - \frac{n-2}{n-1}\cdot \frac{2z(q^\mu k_{1}^{\mu'} + q^{\mu'}k_{1}^{\mu})}{q^2} \label{eq:PbF2} \\
\hat{\mathcal P}_{\hat F_L}^{bb',\mu\mu'} &= - \frac{4 z^2 k_{1}^{\mu}k_{1}^{\mu'}}{q^2} - \frac{q^\mu q^{\mu'}}{q^2} - \frac{2z(q^\mu k_{1}^{\mu'} + q^{\mu'}k_{1}^{\mu})}{q^2} \label{eq:PbFL} \\
\hat{\mathcal P}_{z\hat F_3}^{bb',\mu\mu'} &= -\frac{i z\varepsilon^{\mu\mu'\alpha\beta}k_{1,\alpha}q_\beta}{q^2}
\end{align}
and, due to the symmetry in the Lorentz structure of the partonic tensor, for the polarized structure functions
\begin{align}
\hat{\mathcal P}_{2z\hat g_1}^{bb',\mu\mu'} &= \hat{\mathcal P}_{z\hat F_3}^{bb',\mu\mu'}
&\hat{\mathcal P}_{\hat g_4}^{bb',\mu\mu'} &= -\hat{\mathcal P}_{\hat F_2}^{bb',\mu\mu'}
&\hat{\mathcal P}_{\hat g_L}^{bb',\mu\mu'} &= -\hat{\mathcal P}_{\hat F_L}^{bb',\mu\mu'}\,. \label{eq:LorentzSubRule}
\end{align}
For the unpolarized structure functions $\hat h\in\{\hat F_2,\hat F_L,z\hat F_3\}$ we have to average the spin of the initial parton $j$ whereas for the polarized structure functions $\hat h\in\{2z\hat g_1,\hat g_4,\hat g_L\}$ we observe the difference between the spin orientations. With this considerations at hand, we find
\begin{align}
\hat{\mathcal P}_{\hat h}^{bb',\mu\mu'}\frac 1 2\left(\hat w_{\mu\mu'}(-k_1) + \hat w_{\mu\mu'}(k_1)\right) &= \hat h &&\text{for}\,\hat h \in\{\hat F_2,\hat F_L,z\hat F_3\}\,,\\
\hat{\mathcal P}_{\hat h}^{bb',\mu\mu'}\frac 1 2\left(\hat w_{\mu\mu'}(-k_1) - \hat w_{\mu\mu'}(k_1)\right) &= \hat h &&\text{for}\,\hat h \in\{2z\hat g_1,\hat g_4,\hat g_L\}\,.
\end{align}

Note that we have to include the additional structure functions $\hat F_4,\hat F_5,\hat g_6$ and $\hat g_7$ into the partonic tensor \Eqref{eq:PartonTen}. These structure functions give rise to corrections proportional $q^\mu$ in \Eqsaref{eq:PbF2}{eq:PbFL} that are not necessary in the case of pure photon exchange (valid in the limit $Q^2\ll M_{\PZ}^2$) where Ward identities hold\cite{QFT}. However, we need these contributions for the case of the full neutral current contributions as we consider massive particles in the final state. In this case, the contraction of the axial vector current $q_\mu J^\mu_5$ does not vanish
\begin{align}
q_\mu J^{\mu}_5 &= q_\mu \bar \psi(l_1) \gamma^\mu \gamma^5\psi(l_2) = 2 m\,\bar \psi(l_1) i \gamma^5\psi(l_2)\,.
\end{align}
We assume the initial state still to be massless, thus we do not have to consider these structure functions in the hadronic tensor \Eqref{eq:HadroTen} and were able to write \Eqsaref{eq:HPCphZ}{eq:HPVphZ}.

Following the standard procedure of factorization and pQCD, we will compute the required partonic matrix elements as absolute value squared of the sum of all matrix amplitudes (see \ChapterRef{chap:ME}). Due to the Lorentz anti-symmetry of the PV structure functions $z\hat F_3,\hat g_4, \hat g_L$ in the partonic tensor $\hat w_{\mu\mu'}$, these structure functions can only receive contributions from the vectorial-axial vectorial coupling $\Gamma^\mu_V\Gamma^{\mu'}_A$, where $\mu$ refers to the Lorentz index of the boson in the matrix amplitude and the $\mu'$ to the index in the complex conjugate. Likewise, the parity conserving structure functions can only receive contributions from either the vector-vector couplings $\Gamma^\mu_V\Gamma^{\mu'}_V$ or the axial vector-axial vector couplings $\Gamma^\mu_A\Gamma^{\mu'}_A$. In the case of massless DIS, it is common to find that the vector-vector coupling gives rise to the same contributions as the axial vector-axial vector coupling\cite{Moch:2015usa}. This does no longer hold in the present case of massive contributions to the DIS structure functions.

On a partonic level the handling of the spin degree of freedom for an initial gluon $\Pg(k_1)$ is achieved by
\begin{align}
\hat{\mathcal P}_{\hat F_2}^{\Pg,\nu\nu'} = \hat{\mathcal P}_{\hat F_L}^{\Pg,\nu\nu'} = \hat{\mathcal P}_{z\hat F_3}^{\Pg,\nu\nu'} &= - g^{\nu\nu'}\\
\hat{\mathcal P}_{2z\hat g_1}^{\Pg,\nu\nu'} = \hat{\mathcal P}_{\hat g_4}^{\Pg,\nu\nu'} = \hat{\mathcal P}_{\hat g_L}^{\Pg,\nu\nu'} &= 2 i\varepsilon^{\nu\nu'\alpha\beta} \frac{k_{1,\alpha}q_\beta}{2 k_1\cdot q}
\end{align}
and by choosing just $-g^{\nu\nu'}$ for the unpolarized case, we decided to include ghosts to cancel all unphysical gluon polarizations (see \SectionRef{sec:QCD.Lagrangian}). All incoming (anti-)quarks are taken as massless partons, so the relevant projection operators onto the appropriate helicity states are given by
\begin{align}
\hat{\mathcal P}_{\hat F_2}^{\Pq,aa'} = \hat{\mathcal P}_{\hat F_L}^{\Pq,aa'} = \hat{\mathcal P}_{z\hat F_3}^{\Pq,aa'} 
&= (\slashed k_1)_{aa'}
= \hat{\mathcal P}_{\hat F_2}^{\Paq,aa'} = \hat{\mathcal P}_{\hat F_L}^{\Paq,aa'} = \hat{\mathcal P}_{z\hat F_3}^{\Paq,aa'}\\
\hat{\mathcal P}_{2z\hat g_1}^{\Pq,aa'} = \hat{\mathcal P}_{\hat g_4}^{\Pq,aa'} = \hat{\mathcal P}_{\hat g_L}^{\Pq,aa'}
&= -(\gamma_5\slashed k_1)_{aa'}
= -\hat{\mathcal P}_{2z\hat g_1}^{\Paq,aa'} = -\hat{\mathcal P}_{\hat g_4}^{\Paq,aa'} = -\hat{\mathcal P}_{\hat g_L}^{\Paq,aa'}
\end{align}
where $a$ and $a'$ refer to the Dirac-indices of the initial (anti-)quark spinors. We will add a factor of $1/2$ later on to account for the averaging of the spins.

\subsubsection{Notation}
Fully inclusive calculations will only depend on external variables, which are on a partonic level the partonic invariant energy $s$ and the momentum transfer $Q^2$ (see \Eqssref{eq:partonicVars}). We define an additional set of partonic variables that will simplify many partonic expressions
\begin{align}
0\leq&\,\rho = \frac {4m^2} s\leq 1 &0\leq&\,\beta = \sqrt{1-\rho}\leq 1 &0\leq&\,\chi = \frac{1-\beta}{1+\beta}\leq 1 \label{eq:partonicVars1}\\
0\leq&\,\rho' = \frac {4m^2} {s'}\leq 1 &0\leq&\,\beta' = \sqrt{1-\rho'}\leq 1 &0\leq&\,\chi' = \frac{1-\beta'}{1+\beta'}\leq 1 \label{eq:partonicVars2}\\
&\rho_q = \frac {4m^2} {q^2}\leq 0 &1\leq&\,\beta_q = \sqrt{1-\rho_q} &0\leq&\,\chi_q = \frac{\beta_q-1}{\beta_q+1}\leq 1 \label{eq:partonicVars3}
\end{align}
and which obey the further inequalities
\begin{align}
\rho' &< \rho, &\rho'&<\frac{\rho_q}{\rho_q-1},
&\beta &< \beta', &\beta'&<\frac 1 {\beta_q},
&\chi' &<\chi, &\chi' &<\chi_q\,.
\end{align}
As HQ production is a two-scale process, say $s/m^2$ and $Q^2/m^2$, we always can take two out of the three set of \Eqsrref{eq:partonicVars1}{eq:partonicVars3} as linearly independent.

To introduce a compact notation for the rest of this work, we write
\begin{align}
\vec \kappa =(\kappa_1,\kappa_2,\kappa_3)\quad\kappa_1\in\{\tV,\tA\},\kappa_2\in\{\tV,\tA\},\,\kappa_3\in\{\hat F_2,\hat F_L,z\hat F_3,2z\hat g_1,\hat g_4,\hat g_L\}
\end{align}
where $\kappa_1$ ($\kappa_2$) refer to the type of coupling in the (complex conjugated) matrix amplitude ($\Gamma^{\kappa_1}_\mu\Gamma^{\kappa_2}_{\mu'}$) and $\kappa_3$ to the inspected structure function\footnote{Note that we will use this notation for both, the hadronic and partonic framework}. We find that we can use just $\vec \kappa$ in all important equations following below as they will not depend explicitly on the specific projection, except for some trivial factors related to the couplings or the beam polarization. This observation actually ensures the renormalization of the considered DIS processes. We extend this notation also to the PDFs to distinguish the polarized and the unpolarized setup, i.e.\ we write ($j\in\{\Pg,\Pq\}$)
\begin{align}
f_{\kappa_3,j}(\xi,\mu_F^2) &= f_j(\xi,\mu_F^2) &\text{for}\,\kappa_3\in\{\hat F_2,\hat F_L,z\hat F_3\}\,\,\\
f_{\kappa_3,j}(\xi,\mu_F^2) &= \Delta f_j(\xi,\mu_F^2) &\text{for}\,\kappa_3\in\{2z\hat g_1,\hat g_4,\hat g_L\}\, .
\end{align}
With this notation at hand and the above considerations to the possible couplings in mind, we can decompose the hadronic structure functions further. In \Eqsaref{eq:HPCphZ}{eq:HPVphZ} we split them by the scattered bosons. Now, we decompose them again by their initial channel, i.e.\ whether a gluon $\Pg$ or a light quark $\Pq$ is hit in the proton, and by their coupling structure $\kappa_1,\kappa_2$. We write
\begin{align}
F_2^{bb'} &= H^{bb'}_{\lVVF2, \Pg} + H^{bb'}_{\lAAF2, \Pg} + \sum_{\Pq}^{n_{lf}} \left( H^{bb'}_{\lVVF2, \Pq} + H^{bb'}_{\lAAF2, \Pq} \right)
\end{align}
and similar for all other parity conserving (PC) structure functions ($F_L,2xg_1$) (see \Eqref{eq:HPCphZ}) and
\begin{align}
xF_3^{bb'} &= H^{bb'}_{\lVAxF3, \Pg} + \sum_{\Pq}^{n_{lf}} H^{bb'}_{\lVAxF3, \Pq}
\end{align}
and similar for all other parity violating (PV) structure functions ($g_4,g_L$) (see \Eqref{eq:HPVphZ}) where we symmetrize the couplings when necessary.

\subsubsection{Factorization}
To counteract the evolution of the PDFs (\Eqref{eq:DGLAP}) we have to renormalize the partonic matrix elements. This allows us to remove all collinear poles that arise in the NLO corrections with an initial state gluon $\hat h_{\vec\kappa,\Pg}^{bb',(1)}$ or an initial state light quark $\hat h_{\vec\kappa,\Pq}^{bb',(1)}$. We apply a standard mass factorization procedure
\begin{align}
d\hat h_{\vec \kappa,j}^{bb',(1),fin}(\mu_F^2)= \lim_{\epsilon\rightarrow 0}\left[d\hat h_{\vec \kappa,j}^{bb',(1)}(\mu_D^2,\epsilon) -\int\limits_0^1dx\,\Gamma_{\kappa_3,\Pg j}^{(1)}(x,\mu_F^2,\mu_D^2,\epsilon) d\hat h_{\vec \kappa,\Pg}^{bb',(0)}(x k_1,\epsilon) \right] \label{eq:PartonicFact}
\end{align}
for $j=\{\Pg,\Pq\}$ with $\Gamma_{\kappa_3,ij}^{(1)}$ the first order correction to the transition functions $\Gamma_{\kappa_3,ij}$ for \textit{incoming} particle $j$ and \textit{outgoing} particle $i$ in projection $\kappa_3$. These transition functions are to first order given by the Altarelli-Parisi splitting kernels
\begin{align}
\Gamma_{\kappa_3,ij}^{(1)}(x,\mu_F^2,\mu_D^2,\epsilon) &= \frac{\alpha_s}{2\pi}\left(P_{\kappa_3,ij}^{(0)}(x)\frac{2}{\epsilon} + f_{\kappa_3,ij}^{(1)}(x,\mu_F^2,\mu_D^2)\right)
\end{align}
and depend in addition on an arbitrary function $f_{\kappa_3,ij}^{(1)}(x,\mu_F^2,\mu_D^2)$ which defines the PDF scheme. In the adopted $\MSbar$-scheme, which is the most common, the $f_{\kappa_3,ij}^{(1)}$ take their usual form and we find
\begin{align}
\Gamma_{\kappa_3,ij}^{(1),\MSbar}(x,\mu_F^2,\mu_D^2,\epsilon) &= \frac{\alpha_s}{2\pi}P_{\kappa_3,ij}^{(0)}(x)\left(\frac{2}{\epsilon} +\gamma_E - \ln(4\pi) + \ln(\mu_F^2/m^2) - \ln(\mu_D^2/m^2)\right)\\
 &= 8\pi\alpha_s P_{\kappa_3,ij}^{(0)}(x) C_\epsilon \left(\frac{\mu_D^2}{m^2}\right)^{-\epsilon/2} \left(\frac{2}{\epsilon} + \ln(\mu_F^2/m^2)\right)
\end{align}
with $C_\epsilon = \exp(\epsilon/2(\gamma_E-\ln(4\pi)))/(16\pi^2)$, covering the $\MSbar$-scheme in a prefactor.

\chapter{Computing the Matrix Elements} \label{chap:ME}
\chapterquote{It does not do to leave a live dragon out of your calculations, if you live near him.}{$\gamma_5$}

Although regularization is, strictly speaking, only needed at next-to-leading order (discussed in \SectionRef{sec:ME.NLO}), we also apply the rules to the leading order (discussed in \SectionRef{sec:ME.LO}) as this is eventually needed later on (see discussion in \ChapterRef{chap:PS}).

Dimensional regularization affects the treatment of the Dirac matrices by their defining relation
\begin{equation}
\{\gamma_{\mu},\gamma_{\nu}\} = 2g_{\mu\nu} {\mathds{1}}_{4\times 4}
\end{equation}
as $g_{\mu\nu}$ encodes the information of the space time. Especially, the definition of $\gamma_5$ is affected as the two relations
\begin{align}
[\gamma_{\mu},\gamma_5] &= 0\,, \label{eq:G5a}\\
\gamma_5 &= \frac i {4!}\varepsilon_{\mu\nu\rho\sigma}\gamma^{\mu}\gamma^{\nu}\gamma^{\rho}\gamma^{\sigma} \label{eq:G5b}
\end{align}
cannot be true simultaneously in $n$ dimensions and so a consistent prescription has to be chosen to deal with any occurrence of $\gamma_5$ (or equivalently the anti-symmetric Levi-Civita-tensor $\varepsilon_{\mu\nu\rho\sigma}$). While a wide set of regularization schemes exist\cite{tHooft:1972tcz,breitenlohner1977,Moch:2015usa,Larin:1993tq,Gnendiger:2017pys,Kreimer:1993bh,Korner:1989is,Ferrari:1994ct,Korner:1991sx,West:1991xv,Gnendiger:2017rfh} we will only discuss briefly the algorithm of the 't-Hooft-Veltman-Breitenlohner-Maison scheme\cite{tHooft:1972tcz,breitenlohner1977} (HVBM) that was used in our papers \cite{Hekhorn:2018ywm,Hekhorn:2018aio}, but which we do not use in this work. We highlight the prescriptions of the Moch-Vermaseren-Vogt scheme\cite{Moch:2015usa} (MVV) we use instead and which is a variant of the Larin scheme\cite{Larin:1993tq}.

The basic idea of the HVBM-scheme\cite{tHooft:1972tcz,breitenlohner1977} is to split the $n$-dimensional space into a $4$-dimensional space (denoted by objects with a double hat), which behaves just as before, and the remaining $n-4$ dimensions (denoted by objects with a single hat) receive a separate definition of the Dirac matrices:
\begin{equation}
\gamma_{\mu,n\times n} = \hat{\hat{g}}_{\mu,4\times 4}\otimes \hat{\mathds{1}}_{(n-4)\times (n-4)} + \hat{\hat{\mathds{1}}}_{4\times 4}\otimes \hat{g}_{\mu,(n-4)\times (n-4)}
\end{equation}
In the $4$ dimensional space \Eqref{eq:G5a} still holds while for the counter part in hat space the commutator is replaced by an \textit{anti}-commutator. This split behaviour explicitly breaks \Eqref{eq:G5b}, but keeps the cyclicity of Dirac traces intact. We did use this scheme in our papers \cite{Hekhorn:2018ywm,Hekhorn:2018aio} but the prescription becomes impractically for the present case of full neutral current calculations as we will end up with up to four instances of $\gamma_5$ which may be distributed over two separate fermion traces.

So instead, we will employ the MVV-scheme\cite{Moch:2015usa} here which extends the Larin-scheme\cite{Larin:1993tq} that in contrast to HVBM sticks to \Eqref{eq:G5b} and defines
\begin{equation}
\gamma_{\mu}\gamma_5 = \frac i {3!} \varepsilon_{\mu\nu\rho\sigma}\gamma^{\nu}\gamma^{\rho}
\quad\text{and}\quad
\varepsilon^{\mu\nu\rho\sigma}\varepsilon_{\alpha\beta\kappa\lambda} = 
\begin{vmatrix}
\delta^\mu_\alpha & \delta^\mu_\beta & \delta^\mu_\kappa & \delta^\mu_\lambda \\ 
\delta^\nu_\alpha & \delta^\nu_\beta & \delta^\nu_\kappa & \delta^\nu_\lambda \\
\delta^\rho_\alpha & \delta^\rho_\beta & \delta^\rho_\kappa & \delta^\rho_\lambda \\
\delta^\sigma_\alpha & \delta^\sigma_\beta & \delta^\sigma_\kappa & \delta^\sigma_\lambda
\end{vmatrix}
\end{equation}
with $\delta_{\mu\nu}$ being a $n$-dimensional object. The MVV-scheme defines additionally the equivalent of a reading point, breaking the cyclicity of Dirac traces, and defines
\begin{align}
\text{Tr}\left[\gamma_{\nu_1}\gamma_{\nu_2}\cdots\gamma_{\nu_{2m-1}}\gamma_{\mu}\gamma_5\right] &= 
-4i g_{\nu_1\nu_2}\cdots g_{\nu_{2m-5}\nu_{2m-4}}\varepsilon_{\nu_{2m-3}\nu_{2m-2}\nu_{2m-1}\mu}\nonumber\\
 &\hspace{20pt}\pm \text{permutations of } \gamma_{\nu_1}\cdot\gamma_{\nu_{2m-1}}\,.
\end{align}
As the authors of \cite{Moch:2015usa} claim, this scheme provides an efficient implementation in FORM\cite{Vermaseren:2000nd} and it was the only algorithm we succeeded to get a result in less then half a day CPU time. To clarify the statement of the authors \enquote{special care is needed for more than two $\gamma_5$}, we note that in the present case it was correct to, first, mark \textit{all} occurrences of a Levi-Civita-tensor separately and, second, to rejoin them according to their fermion trace.

\section{Leading Order $O(\alpha_s\alpha_{em})$} \label{sec:ME.LO}
\begin{figure}[ht]
\centering
\begin{subfigure}[t]{.5\linewidth}
	\centering
	\includegraphics[width=.6\textwidth]{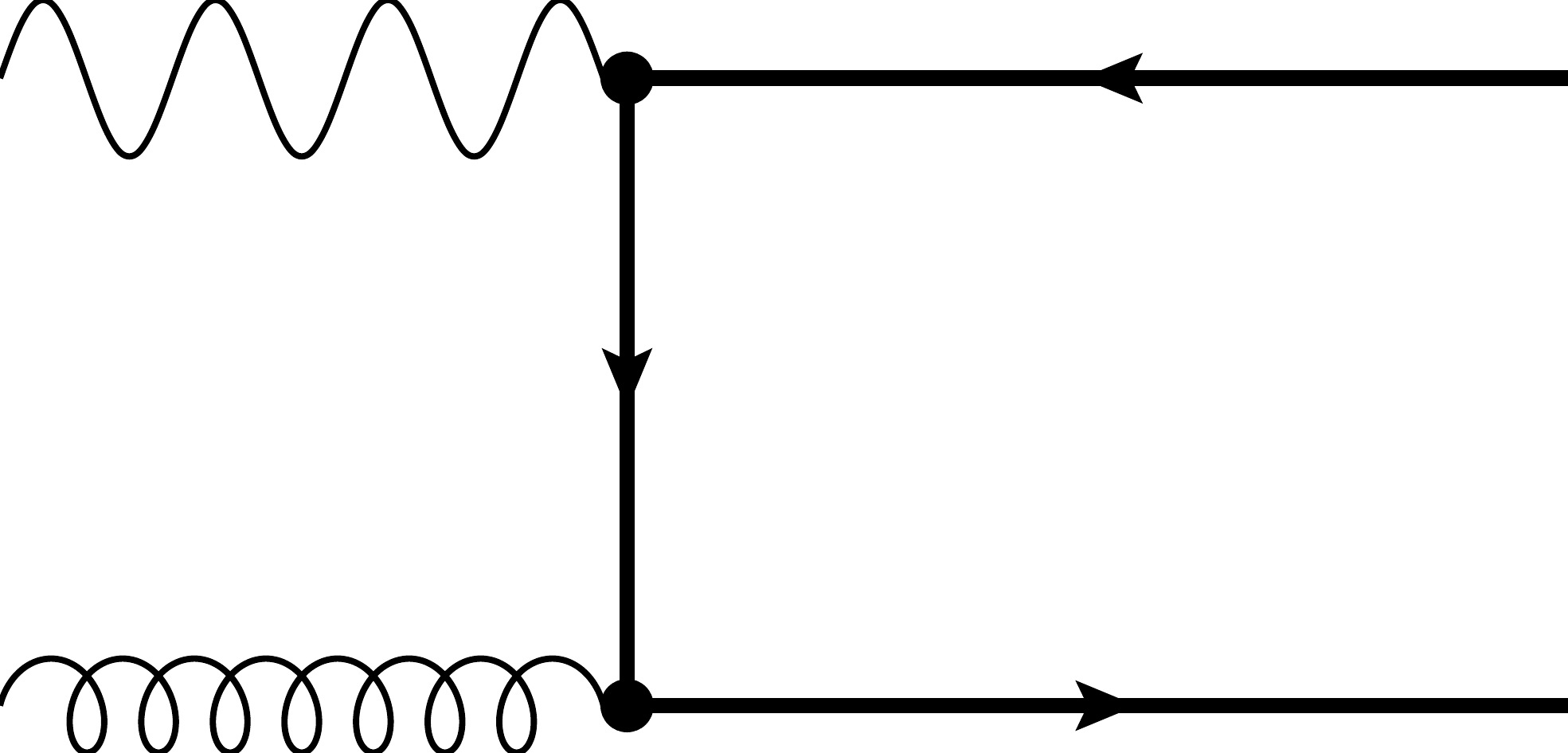}
	\caption{$i\varepsilon^{\mu}_{b}(q)\varepsilon^{\nu}_{\Pg}(k_1)\Md^{(0),1}_{b,\kappa_1,\mu\nu}$}
\end{subfigure}%
\begin{subfigure}[t]{.5\linewidth}
	\centering
	\includegraphics[width=.6\textwidth]{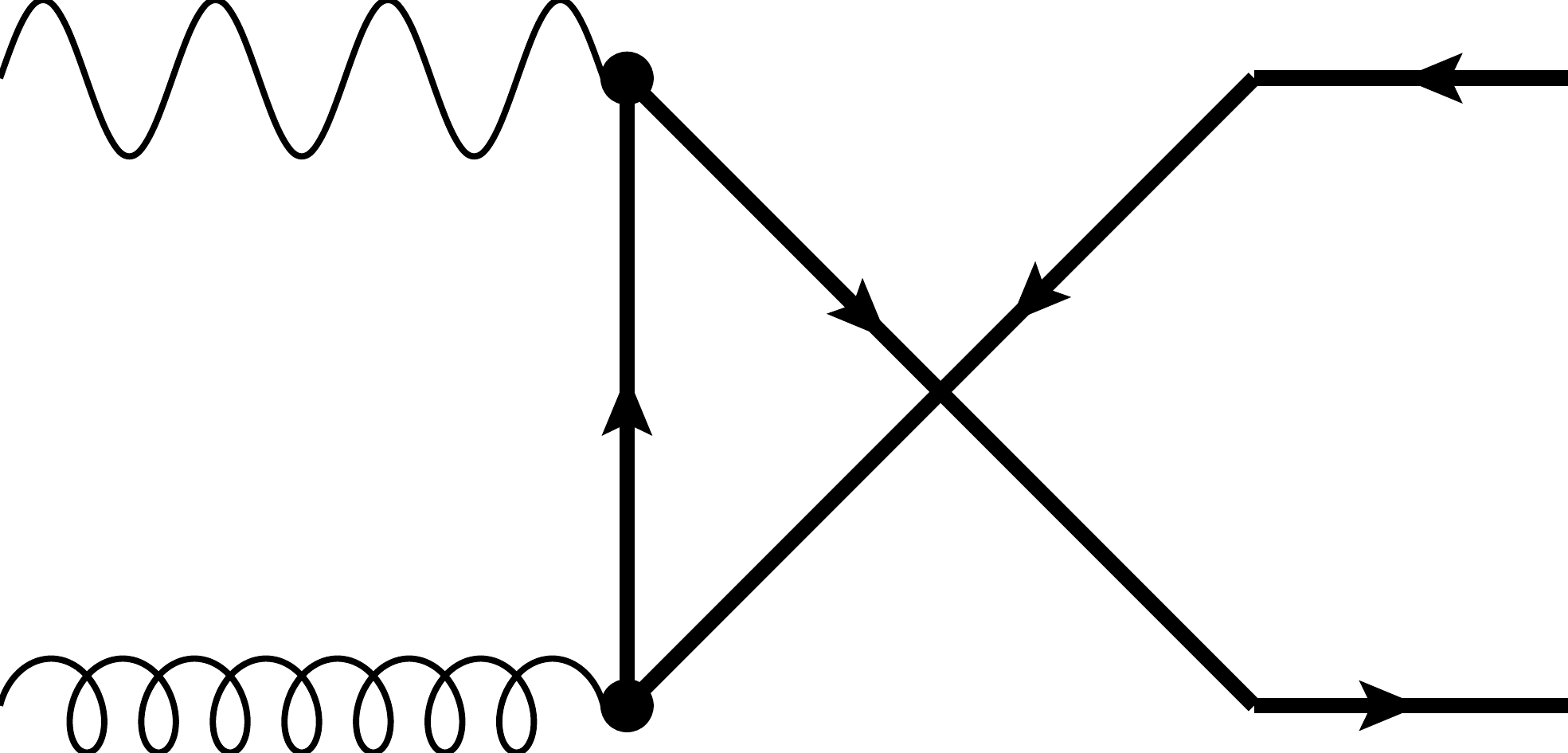}
	\caption{$i\varepsilon^{\mu}_{b}(q)\varepsilon^{\nu}_{\Pg}(k_1)\Md^{(0),2}_{b,\kappa_1,\mu\nu}$}
\end{subfigure}
\caption{The leading order Feynman diagrams}\label{fig:FeynLO}
\end{figure}

In LO we have to consider boson-gluon-fusion (BGF)
\begin{equation}
\HepProcess{b^*(q) + \Pg(k_1) \to \PQ(p_1)+\PaQ(p_2)}\quad b^*\in\{\Pggx,\PZx\}\,,
\end{equation}
with two contributing diagrams depicted in figure \ref{fig:FeynLO}.

In practice this leads already in LO to a total of $6\cdot 2^2 \cdot 2^2 = 96$ fermion traces for $6$ different projections on structure functions, $2^2=4$ combinations of vectorial and axial-vectorial couplings\footnote{Note that we may reduce this to $3=|\{\tV\tV,\tV\tA,\tA\tA\}|$, but we decided to run also $\tA\tV$ and to cross check it to $\tV\tA$ which indeed holds for all diagrams at all orders} and $2^2=4$ combinations of the $2$ LO diagrams. However, with the symmetry considerations in \SectionRef{sec:DIS.Partonic}, this reduces to only $9$ different results as given below.

The relevant kinematical invariants are given by their usual definitions
\begin{align}
s &=(k_1+q)^2, &&&s' &= s-q^2,\nonumber\\
t_1 &= (k_1-p_2) - m^2, &u_1 &= (q-p_2)^2 - m^2, &u_1' &= u_1-q^2\,.\label{eq:LOMandelstam}
\end{align}
Note that we defined $u_1$ with respect to $p_2$, thus making it independent of the initial parton momentum $k_1$. The color structure associated to the matrix elements can be solved trivially and give a single factor $N_C C_F$.

The squared matrix amplitudes can be written as
\begin{equation}
M_{\vec \kappa,bb'}^{(0)} = \hat {\mathcal P}_{\kappa_3}^{bb',\mu\mu'}\hat {\mathcal P}_{\kappa_3}^{\Pg,\nu\nu'}\sum_{j,j'=1}^2\Md^{(0),j}_{b, \kappa_1, \mu\nu}\left(\Md^{(0),j'}_{b', \kappa_2, \mu'\nu'}\right)^* = 8g_s^2\mu_D^{-\epsilon}e^2g^{\kappa_1}_{b,\PQ}g^{\kappa_2}_{b',\PQ} N_C C_F B_{\vec \kappa,\tQED}
\end{equation}
and we get for the Born cross sections
\begin{align}
B_{\lVVF2,\tQED} &= \left[-1 - \frac{6q^2}{s'} - \frac{6q^4}{{s'}^2} + \frac{q^2(6m^2+s) +2m^2 s + {s'}^2/2}{t_1u_1} - \frac{(2m^2+q^2)m^2{s'}^2}{(t_1 u_1)^2} \right] \nonumber\\
 &\hspace{20pt}+\frac{\epsilon}{2}\left[ -1 + \frac{s^2-q^2s'}{t_1u_1} - \frac{m^2q^2{s'}^2}{t_1^2u_1^2} \right] + \epsilon^2\frac{{s'}^2}{8t_1u_1}\\
B_{\lVVFL,\tQED} &= -\frac{4q^2}{s'}\left(\frac s {s'} - \frac{m^2s'}{t_1u_1}\right)\\
B_{\lVVx2g1,\tQED} &= \left\{\! 1+ \frac{2q^2}{s'} - \frac{s'(2(2m^2+q^2)+s')}{2t_1 u_1} + \frac{m^2{s'}^3}{(t_1u_1)^2}+\epsilon\left(-\frac 1 2 + \frac{{s'}^2}{4t_1u_1}\right)\!\!\right\}(1+\epsilon)
\end{align}
\begin{align}
B_{\lAAF2,\tQED} &= \frac{{m^2} {s'}^2 (1+\epsilon ) (2+\epsilon ) (12 {m^2} (-1+\epsilon )+{q^2} (-6+(-3+\epsilon ) \epsilon ))}{12 (t_1 u_1)^2}-\nonumber\\
 &\hspace{15pt} \frac{(1+\epsilon ) }{4 {q^2} {s'}^2}\left(8 {s'}^3 \epsilon +12 {q^6} (2+\epsilon )+12 {q^4} {s'} (2+\epsilon )\right.\nonumber\\
 &\hspace{40pt}\left.+{q^2} {s'}^2 (4+\epsilon  (20-(-3+\epsilon ) \epsilon ))\right)-\nonumber\\
 &\hspace{15pt}\frac{(1+\epsilon )}{48 {q^2} (t_1 u_1)}\left({q^2} (2+\epsilon ) (-6+(-3+\epsilon ) \epsilon ) \left(4 {q^4}+4 {q^2} {s'}+{s'}^2 (2+\epsilon )\right)\right.\nonumber\\
 &\hspace{40pt} \left. +48 {m^2} \left(-{s'}^2 (-2+\epsilon )+{q^4} (-4+\epsilon ) (2+\epsilon )+{q^2} {s'} \left(-2+\epsilon +\epsilon ^2\right)\right)\right)\\
B_{\lAAFL,\tQED} &= -\frac{{m^2} {s'}^2 (1+\epsilon ) (2+\epsilon ) (12 {m^2}+{q^2} \epsilon )}{6 (t_1 u_1)^2}-\nonumber\\
 &\hspace{15pt}\frac{(1+\epsilon ) \left(4 {s'}^3 \epsilon +4 {q^6} (2+\epsilon )+4 {q^4} {s'} (2+\epsilon )+{q^2} {s'}^2 \epsilon  (6+\epsilon )\right)}{2 {q^2} {s'}^2}+\nonumber\\
 &\hspace{15pt}\frac{(1+\epsilon )}{24 {q^2} (t_1 u_1)} \left(24 {m^2} \left({s'}^2 (-2+\epsilon )+4 {q^4} (2+\epsilon )+2 {q^2} {s'} (2+\epsilon )\right)+\right.\nonumber\\
 &\hspace{40pt} \left. {q^2} \epsilon  (2+\epsilon ) \left(4 {q^4}+4 {q^2} {s'}+{s'}^2 (2+\epsilon )\right)\right)\\
B_{\lAAx2g1,\tQED} &= \frac{(1+\epsilon)(2-\epsilon)}{2}B_{\lVVx2g1,\tQED}
\end{align}
\begin{align}
B_{\lVAxF3,\tQED} &= -(1+\epsilon)(2+\epsilon)(t_1^2-u_1^2)\left\{- \frac{m^2q^2}{2(t_1u_1)^2} + \frac{4q^2(q^2+s')+{s'}^2(2+\epsilon)}{8 {s'}^2 t_1 u_1}\right\} \label{eq:BQEDVAxF3}\\
B_{\lVAg4,\tQED} &= (1+\epsilon)(t_1-u_1)\left\{- \frac{m^2{s'}^2}{(t_1u_1)^2} +  \frac{4q^2 + s'(2-\epsilon)}{4 t_1 u_1} \right\} \label{eq:BQEDVAg4}\\
B_{\lVAgL,\tQED} &= 0 \label{eq:BQEDVAgL}
\end{align}
For future reference, we decompose the Born cross section further by their dependence on the dimensional parameter $\epsilon$
\begin{align}
B_{\vec \kappa,\tQED} &= B^{(0)}_{\vec \kappa,\tQED} + \epsilon B^{(1)}_{\vec \kappa,\tQED} + O(\epsilon^2)\,.
\end{align}
Note that the parity violating matrix elements $B_{(\tV,\tA,\kappa_3),\tQED}$ with $\kappa_3\in\{z\hat F_3,\hat g_4,\hat g_L\}$ (\Eqsrref{eq:BQEDVAxF3}{eq:BQEDVAgL}) are \textit{anti-symmetric} with respect to the exchange of $t_1$ and $u_1$. Also note that we find here $B_{\lVVx2g1,\tQED}^{(1)} \neq 0$ unlike in \cite{Hekhorn:2018ywm} due to the different $\gamma_5$ schemes.

\section{Next-to-Leading Order $O(\alpha_s^2\alpha_{em})$} \label{sec:ME.NLO}
We can separate the NLO contributions into three major classes: first, virtual corrections arise (discussed in \SectionRef{sec:ME.NLO.V}) which have the same kinematical signature as the LO matrix elements. Second, we can radiate a gluon to the final state (discussed in \SectionRef{sec:ME.NLO.g}) which involves $2\to 3$-kinematics. Third and last, we have to consider a light quark in the initial state (discussed in \SectionRef{sec:ME.NLO.q}) propagating to the final state and being thus a $2\to 3$-process as well.

A common element to all NLO contribution is the more complex color structure that is present in the matrix elements. Virtual diagrams and the single gluon emission generate an appearance of the $3\Pg$-vertex (see \FigureRef{fig:gSI3}) and thus introduces a non-Abelian structure (see \SectionRef{sec:QCD.Lagrangian}). It is therefore handy and common to split the matrix elements along these structures into a non-Abelian part called OK, unique to QCD and proportional to $N_C C_F C_A$, and an Abelian part called QED, where an exact analog exists in the theory of Quantum electrodynamics (QED) and that is proportional to $N_C C_F^2$.

To describe all kinematical configurations in a $2\to 3$-processes, we define ten kinematical invariants
\begin{align}
s &= (q+k_1)^2 &t_1 &=(k_1-p_2)^2-m^2 &u_1 &=(q-p_2)^2 -m^2\nonumber\\
s_3 &= (k_2+p_2)^2-m^2 &s_4 &=(k_2+p_1)^2-m^2 &s_5 &= (p_1+p_2)^2\nonumber\\
t' &= (k_1-k_2)^2\nonumber\\
u' &= (q-k_2)^2 &u_6 &=(k_1-p_1)^2-m^2 &u_7 &=(q-p_1)^2-m^2 \label{eq:NLOMandelstam}
\end{align}
where $s,t_1,u_1$ match to their leading order definitions (\Eqssref{eq:LOMandelstam}). Only five out of the ten variables are linearly independent as can be seen from momentum conservation $k_1+q=p_1+p_2+k_2$ and so five relations connect the Mandelstam variables, out of which two important ones are
\begin{align}
s_4 &= s'+t_1+u_1 &&\text{and} & s_5 &= s'+t'+u'\,.
\end{align}
Their central role will become clear in \ChapterRef{chap:PS} and, even more so, in the \AppendixaRef{sec:Appendix.s4}{sec:Appendix.s5}.

\subsection{One Loop Virtual Contributions} \label{sec:ME.NLO.V}
\begin{figure}[ht!]
\centering
\begin{subfigure}[t]{.33\linewidth}
	\centering
	\includegraphics[width=.7\textwidth]{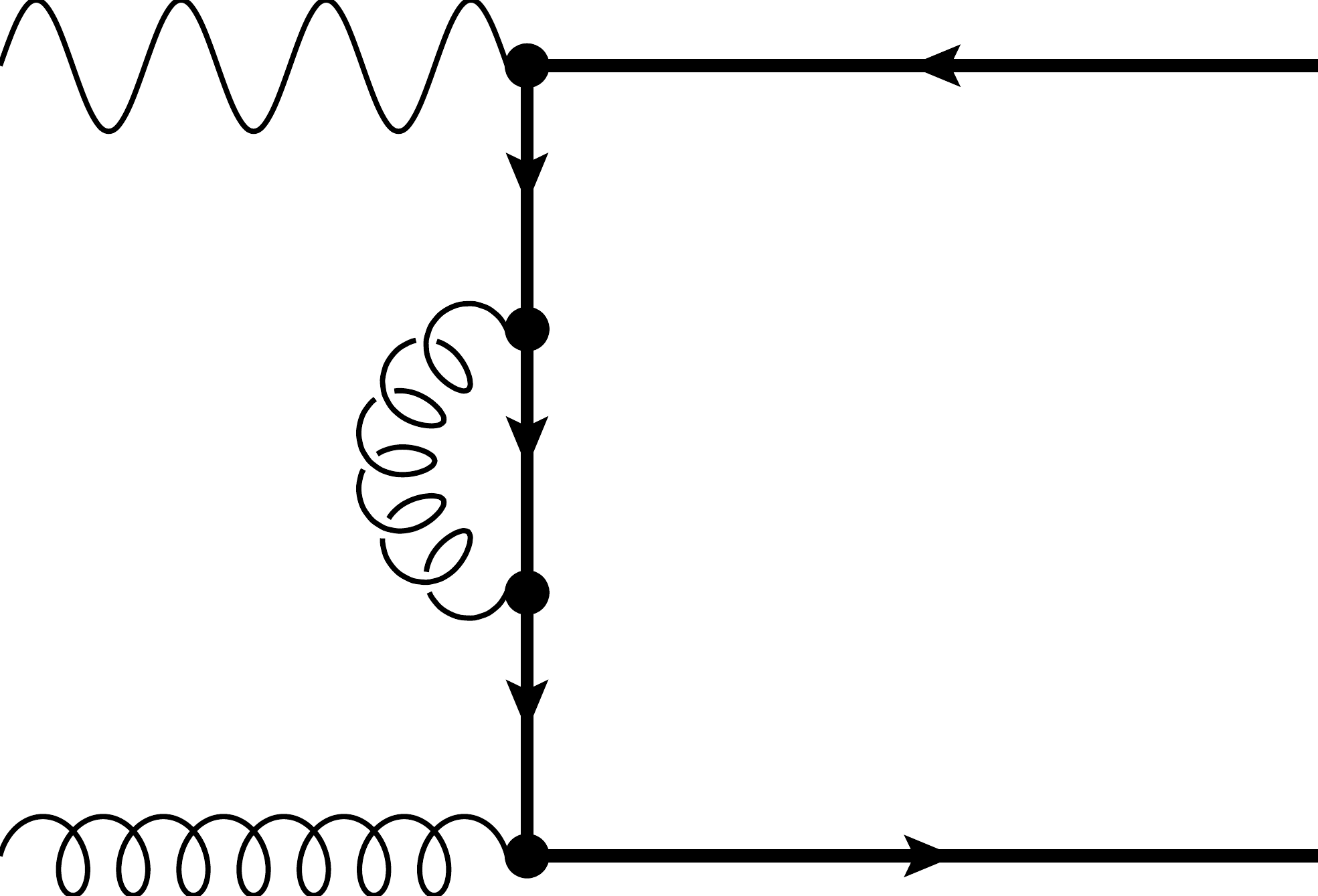}
	\caption{$i\varepsilon^{\mu}_{b}(q)\varepsilon^{\nu}_{\Pg}(k_1)\Md^{(1),\Pg,V,1}_{b,\kappa_1, \mu\nu}$}
	\label{fig:FeynNLOvm}
\end{subfigure}%
\begin{subfigure}[t]{.33\linewidth}
	\centering
	\includegraphics[width=.7\textwidth]{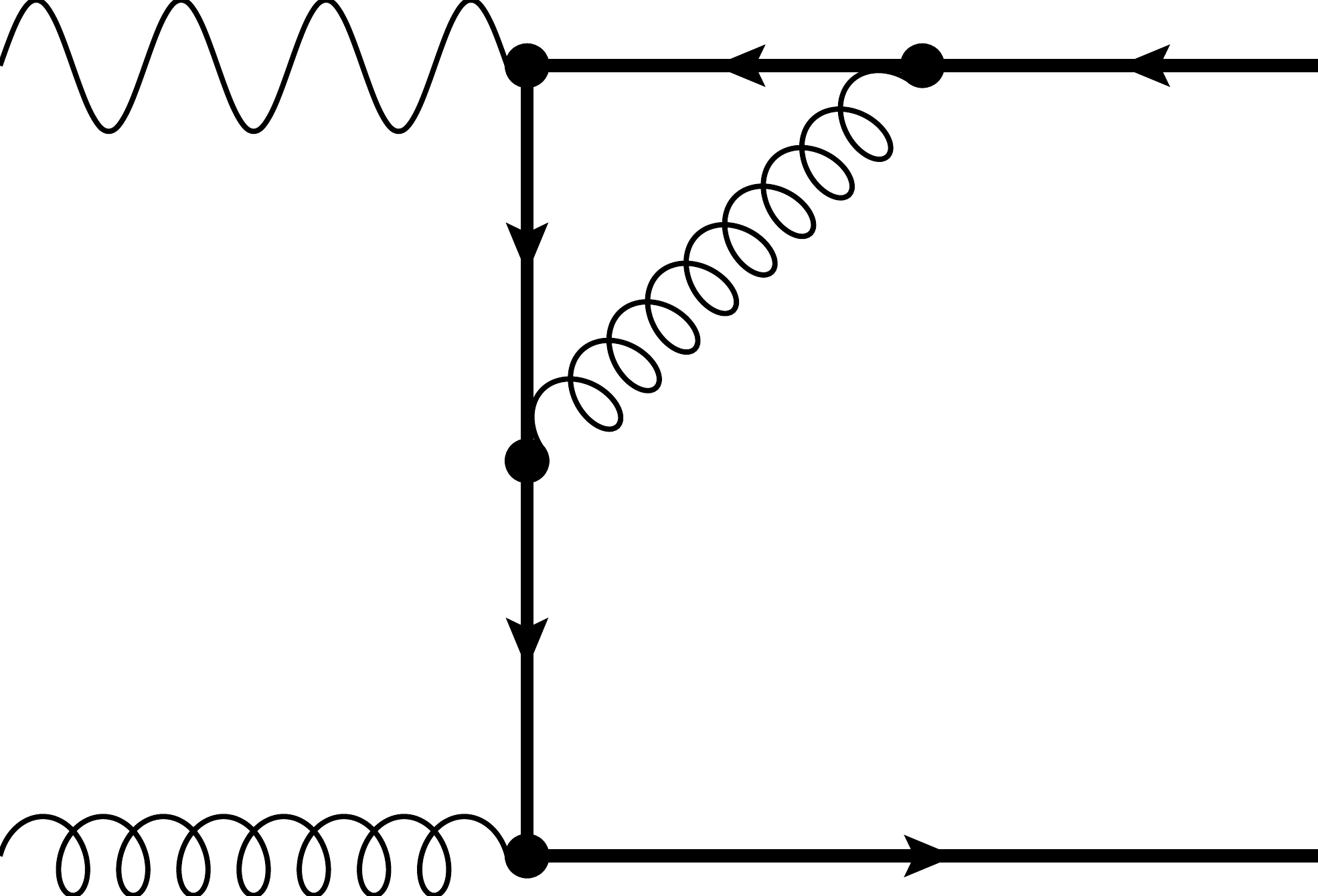}
	\caption{$i\varepsilon^{\mu}_{b}(q)\varepsilon^{\nu}_{\Pg}(k_1)\Md^{(1),\Pg,V,2}_{b,\kappa_1, \mu\nu}$}
	\label{fig:FeynNLOve}
\end{subfigure}%
\begin{subfigure}[t]{.33\linewidth}
	\centering
	\includegraphics[width=.7\textwidth]{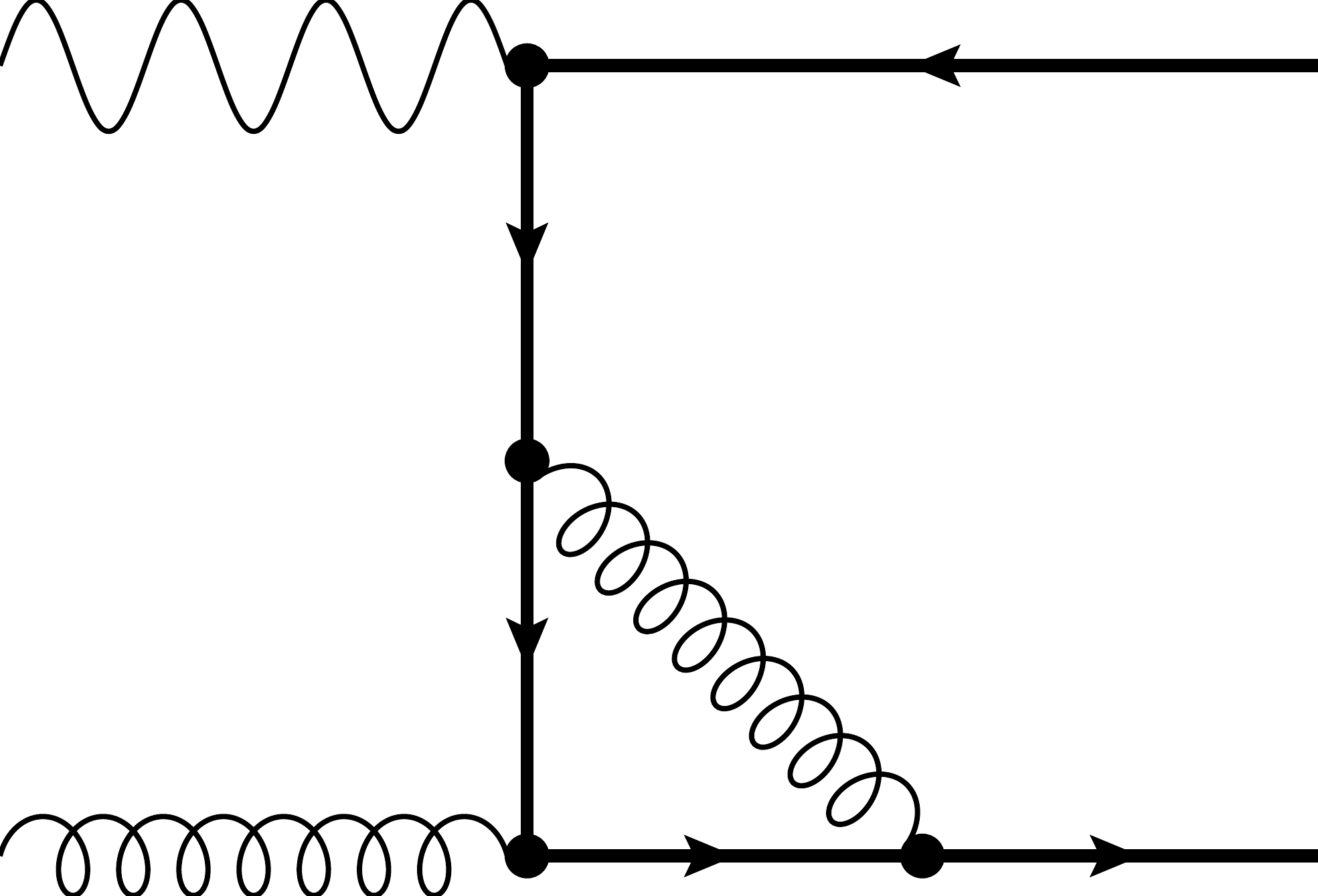}
	\caption{$i\varepsilon^{\mu}_{b}(q)\varepsilon^{\nu}_{\Pg}(k_1)\Md^{(1),\Pg,V,3}_{b,\kappa_1, \mu\nu}$}
	\label{fig:FeynNLOvga}
\end{subfigure}
\begin{subfigure}[t]{.33\linewidth}
	\centering
	\includegraphics[width=.7\textwidth]{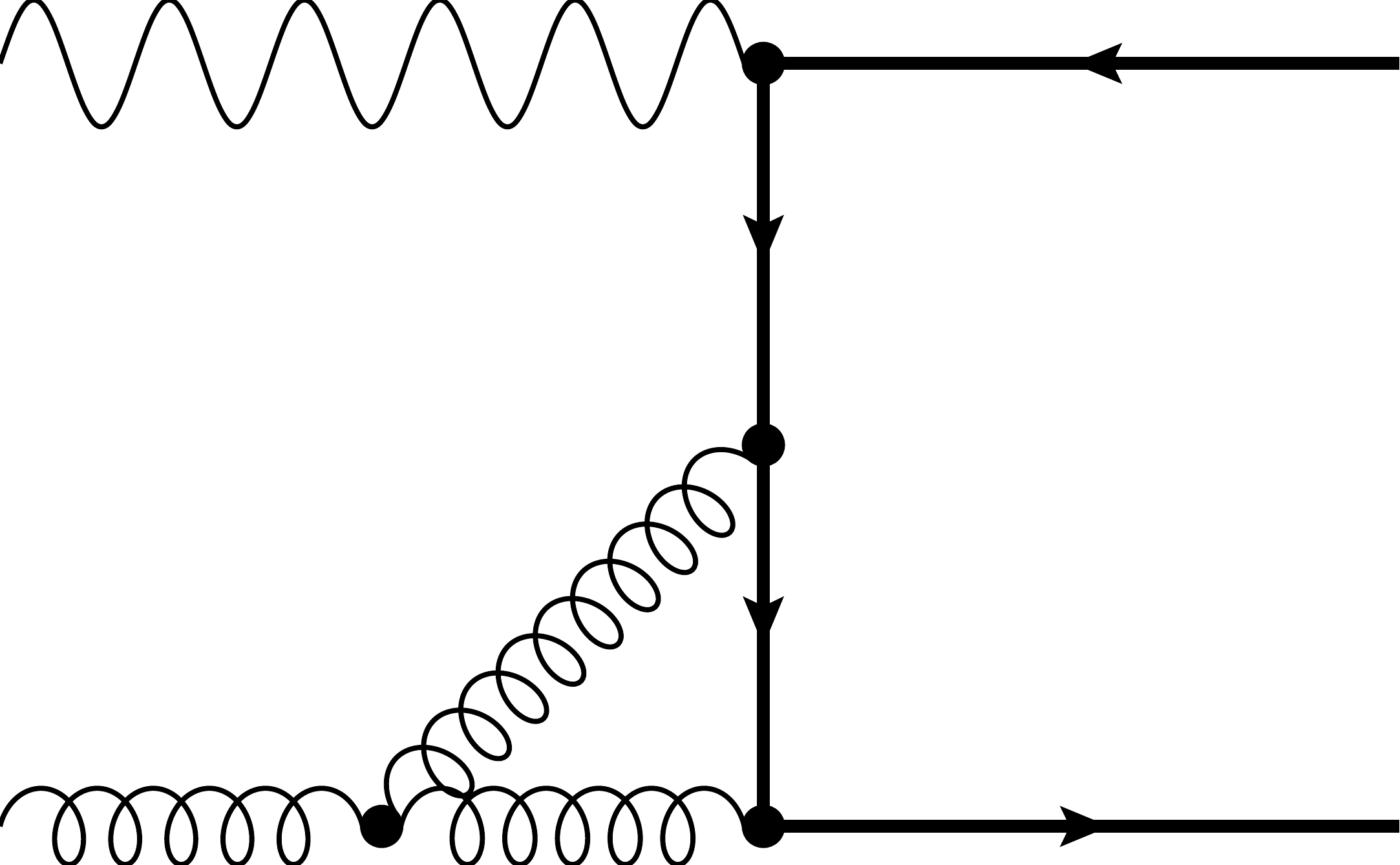}
	\caption{$i\varepsilon^{\mu}_{b}(q)\varepsilon^{\nu}_{\Pg}(k_1)\Md^{(1),\Pg,V,4}_{b,\kappa_1, \mu\nu}$}
	\label{fig:FeynNLOvgb}
\end{subfigure}%
\begin{subfigure}[t]{.33\linewidth}
	\centering
	\includegraphics[width=.7\textwidth]{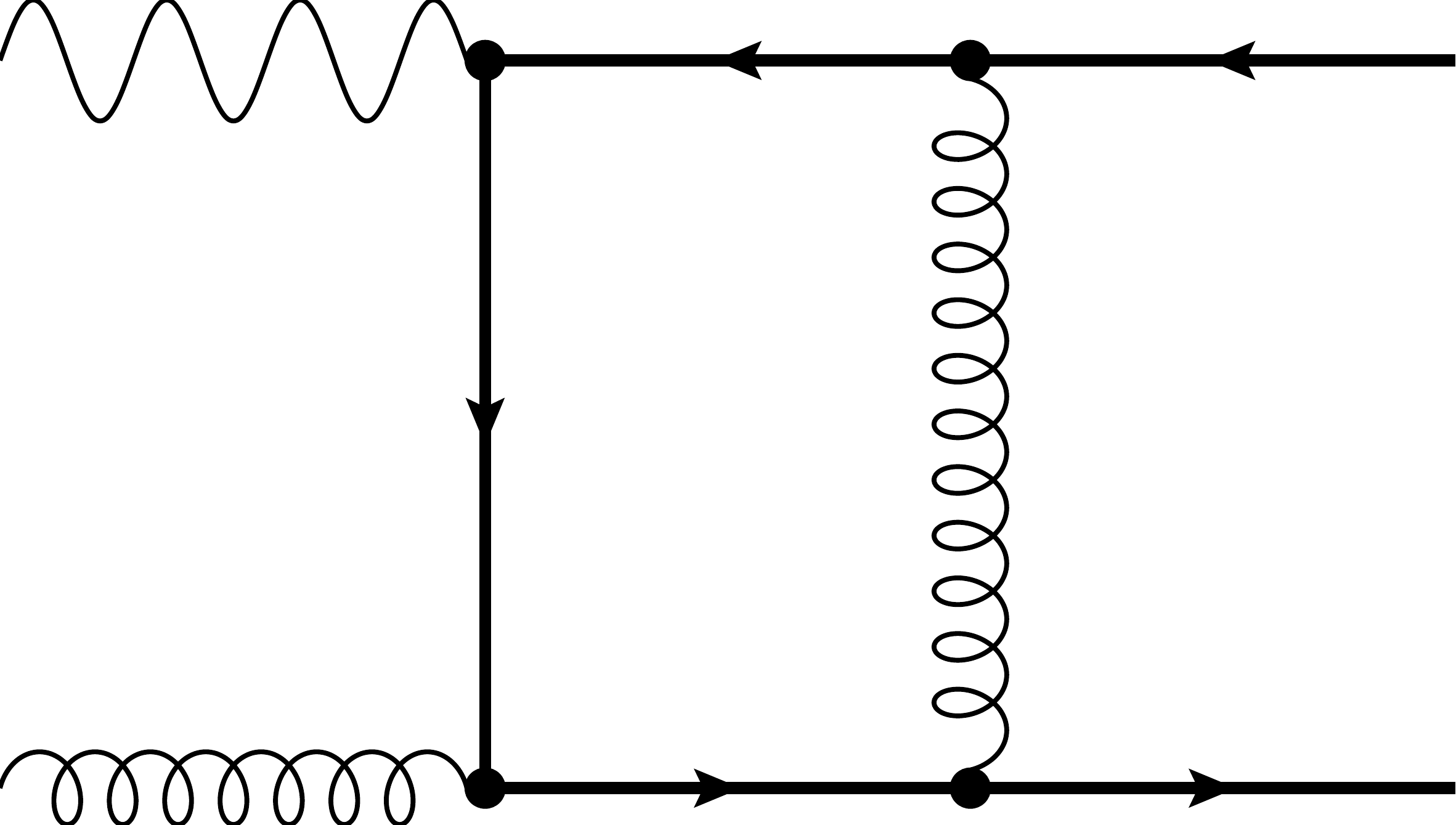}
	\caption{$i\varepsilon^{\mu}_{b}(q)\varepsilon^{\nu}_{\Pg}(k_1)\Md^{(1),\Pg,V,5}_{b,\kappa_1, \mu\nu}$}
	\label{fig:FeynNLOvboxa}
\end{subfigure}%
\begin{subfigure}[t]{.33\linewidth}
	\centering
	\includegraphics[width=.7\textwidth]{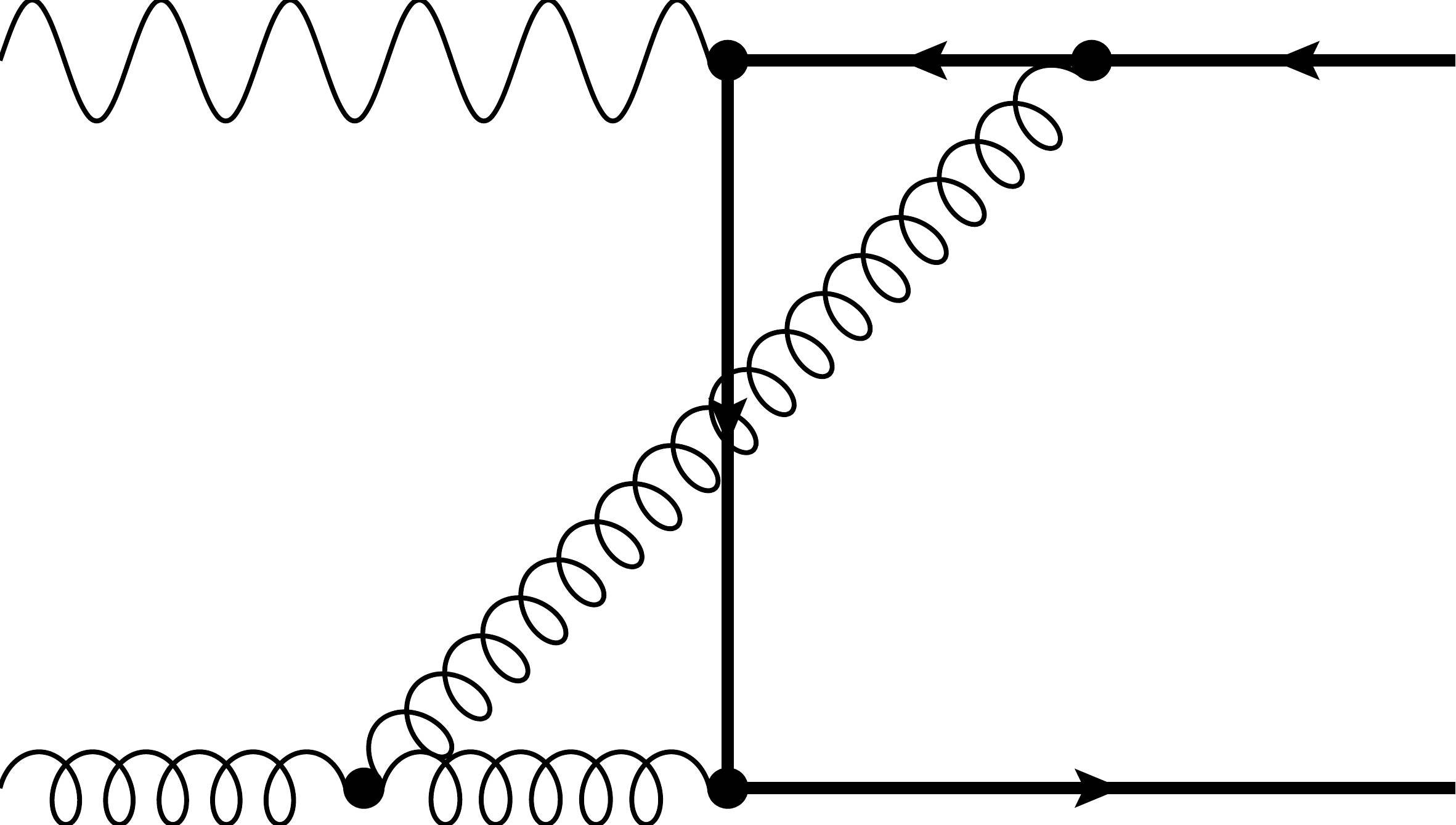}
	\caption{$i\varepsilon^{\mu}_{b}(q)\varepsilon^{\nu}_{\Pg}(k_1)\Md^{(1),\Pg,V,6}_{b,\kappa_1, \mu\nu}$}
	\label{fig:FeynNLOvboxb}
\end{subfigure}%
\caption{The next-to-leading order Feynman diagrams for the virtual corrections}\label{fig:FeynNLOv}
\end{figure}

At next-to-leading order we need to consider virtual corrections where an internal loop appears and here we find $2\cdot 5 +1 = 11$ contributing diagrams schematically depicted in \FigureRef{fig:FeynNLOv}.

The diagrams in \FigurerRef{fig:FeynNLOvm}{fig:FeynNLOvboxa} do have an additional crossed counterpart with the massive fermion current reversed ($p_1\leftrightarrow p_2$), but the diagram in \FigureRef{fig:FeynNLOvboxb} is invariant under this transformation. We can classify the diagrams by different levels of complexity as is shown in \TableRef{tab:NLOv}.
In addition to these diagrams we get self-energy diagrams on external legs (see e.g.\ \FigureRef{fig:FeynSEg}), which we treat separately. A thorough discussion of the different counter terms and their calculation may be found in \cite{Bojak:2000eu}.
\begin{table}[h]
\begin{tabular}{l|c|c|c|c|c|c}
diagram in \Figure & \ref{fig:FeynNLOvm} & \ref{fig:FeynNLOve} & \ref{fig:FeynNLOvga} & \ref{fig:FeynNLOvgb} & \ref{fig:FeynNLOvboxa} & \ref{fig:FeynNLOvboxb} \\
\hline
label & \texttt{m} & \texttt{e} & $\texttt{g1}$ & $\texttt{g2}$ & $\texttt{box1}$ & $\texttt{box2}$ \\
\hline
$n$-point function & 2 & 3 & 3 & 3 & 4 & 4 \\
\hline
color flow & x & x & $\checkmark$ & $\checkmark$ & $\checkmark$ & $\checkmark$\\
\hline
non-Abelian & x & x & x & $\checkmark$ & x & $\checkmark$ \\
\hline
non-planar & x & x & x & x & x & $\checkmark$
\end{tabular}
\caption{Classification of next-to-leading order virtual diagrams}
\label{tab:NLOv}
\end{table}

We write the matrix elements for the virtual diagrams as
\begin{align}
M_{\vec \kappa,bb'}^{(1),V}&=\hat {\mathcal P}_{\kappa_3}^{bb',\mu\mu'}\hat {\mathcal P}_{\kappa_3}^{\Pg,\nu\nu'}\sum_{j,j'}2\text{Re}\left[\Md^{(1),V,j}_{b,\kappa_1,\mu\nu}\left(\Md^{(0),j'}_{b',\kappa_2, \mu'\nu'}\right)^*\right] \nonumber\\
 &= 8g_s^4\mu_D^{-\epsilon}e^2 g^{\kappa_1}_{b,\PQ}g^{\kappa_2}_{b',\PQ} N_C C_FC_\epsilon\left( C_A V_{\vec \kappa,\tOK} + 2C_F V_{\vec \kappa,\tQED}\right)
\end{align}
where $C_\epsilon = \exp(\epsilon/2(\gamma_E-\ln(4\pi)))/(16\pi^2)$ and $\gamma_E$ is the Euler-Mascheroni constant. Note that $C_\epsilon$ effectively defines the $\MSbar$-scheme.

All virtual diagrams contain a closed, internal loop which we regularize with the adopted dimensional regularisation. The diagrams contain soft poles that arise when the momentum flow in the loop reaches the IR limit, that is when they become small, as well as UV poles, as discussed in \SectionRef{sec:QCD.Lagrangian}. The arising loop integrals can be classified by the number of internal propagators ($n$-point functions) and the arising numerator structure in the matrix elements with respect to the loop momenta. This leftover of the Dirac-algebra can be reduced by the application of the Passarino-Veltman reduction\cite{Passarino:1978jh} to a list of \textit{scalar} integrals. So, an extensive list of scalar loop integrals is needed, which can be found in \cite{Hekhorn:2018ywm,Laenen1993162,Bojak:2000eu,PhysRevD4054,Denner:1991kt,Ellis:2011cr,Ellis:2007qk}. Note that we had to correct a specific integral in \cite{Hekhorn:2018ywm} that was misprinted in \cite{Laenen1993162}. Advanced reduction methods, such as the Ossola-Papadopoulos-Pittau scheme\cite{Ossola:2006us,Ellis:2011cr} (OPP), that account for specific corner cases of the phase space, are not needed at NLO. The loop integrals involve higher order functions, such as the dilogarithm, and we use a set of relations\cite{Zagier:2007knq} to simplify these expressions. In \AppendixRef{sec:Appendix.V} we provide additional material that we found to be useful in the computation of virtual diagrams.

Although we adopt the MVV-scheme here to deal with the subtleties of $\gamma_5$ instead of the HVBM-scheme in \cite{Hekhorn:2018ywm}, the arising poles can be cast in an identical form, i.e.\
\begin{align}
V_{\vec \kappa,\tOK} &= -2B_{\vec \kappa,\tQED}\left(\frac 4 {\epsilon^2} + \frac 2 \epsilon\left[\ln(-t_1/m^2) + \ln(-u_1/m^2) +\frac{s-2m^2}{s\beta}\ln(\chi)\right] \right) + O(\epsilon^0)\,,\\
V_{\vec \kappa,\tQED} &= -2B_{\vec \kappa,\tQED}\left(1-\frac{s-2m^2}{s\beta}\ln(\chi)\right)\frac 2 \epsilon + O(\epsilon^0)\,.
\end{align}
Note that these results completely factorize and the $B_{\vec \kappa,\tQED}$ carry the only dependence on the projection $\vec\kappa$. The above results do not include self-energies on external legs. We renormalize the HQs \textit{on-shell} and $m$ thus refers to the pole mass of the HQ (see \SectionRef{sec:QCD.Lagrangian}). The double poles in $V_{\vec \kappa,\tOK}$ originate from diagrams where soft and collinear singularities can coincide. Unfortunately, the changed $\epsilon$ dependence of the Born cross section $B_{\vec \kappa,\tQED}$ as compared to \cite{Hekhorn:2018ywm}, does not allow to compare intermediate results between the two computations.

All UV poles get removed by the renormalization prescription of \Eqref{eq:runas} (introduced in \SectionRef{sec:QCD.Renorm})
\begin{align}
\frac{d^2\sigma_{\vec \kappa,bb',\Pg}^{(1),V}}{dt_1du_1} &=\left.\frac{d^2\sigma_{\vec \kappa,bb',\Pg}^{(1),V}}{dt_1du_1} \right|_{\text{bare}} \nonumber\\
 &\hspace{10pt} + 4\pi\alpha_s(\mu_R^2)C_\epsilon\left(\frac{\mu_D^2}{m^2}\right)^{-\epsilon/2} \left[\left(\frac 2 \epsilon +\ln(\mu_R^2/m^2)\right)\beta_0^f +\frac 2 3 \ln(\mu_R^2/m^2)\right]\frac{d^2\sigma_{\vec \kappa,bb',\Pg}^{(0)}}{dt_1du_1} \label{eq:d2sigmaV}
\end{align}
and in what follows, we will often drop the scale dependence of the strong coupling, i.e.\ $\alpha_s$ has to be understood as $\alpha_s(\mu_R)$. The subscript \enquote{bare} in \Eqref{eq:d2sigmaV} refers to the complete, unrenormalized virtual contributions.

\subsection{Single Gluon Radiation} \label{sec:ME.NLO.g}
\begin{figure}[ht]
\centering
\begin{subfigure}[t]{.5\linewidth}
	\centering
	\includegraphics[width=.6\textwidth]{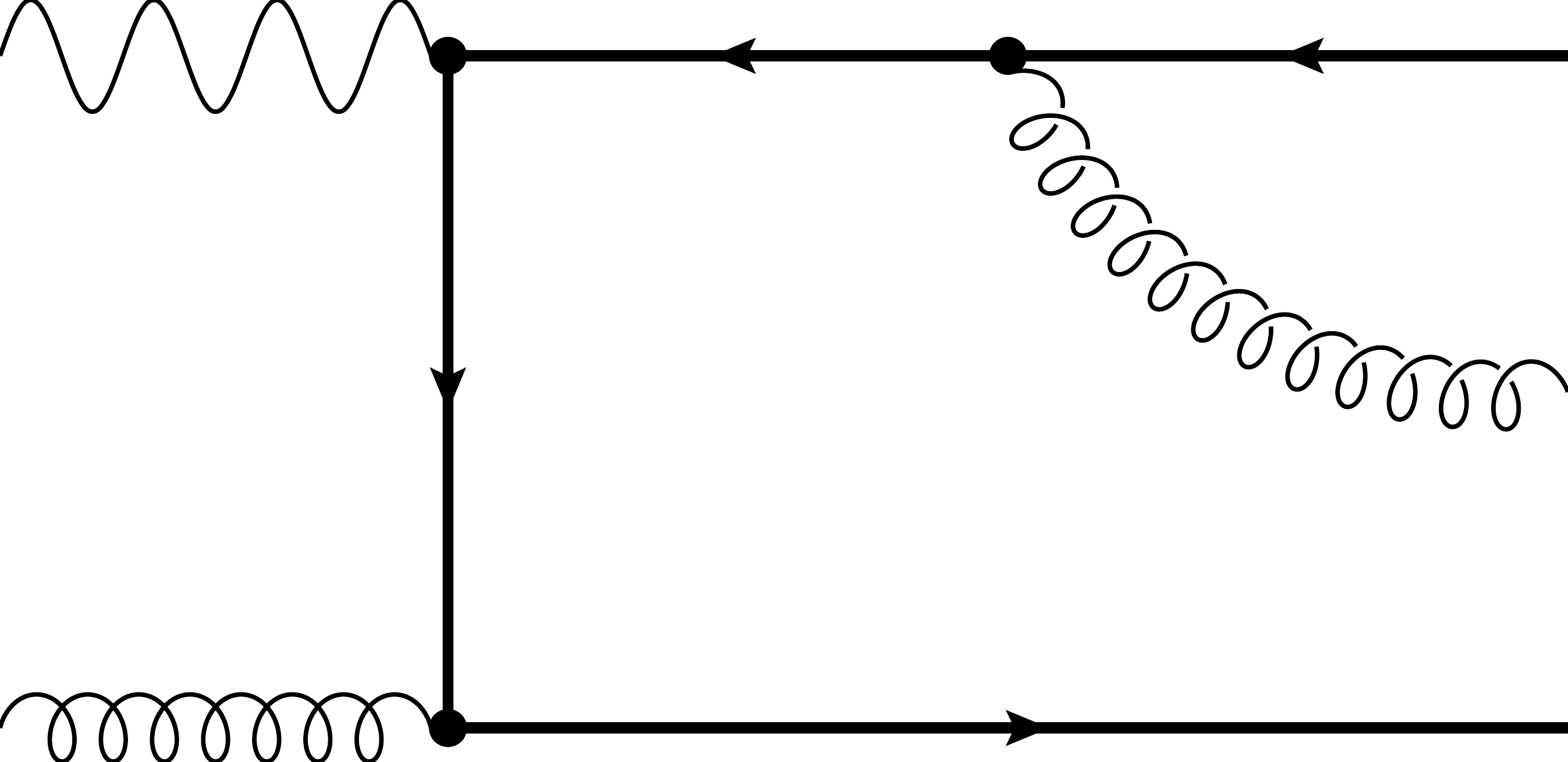}
	\caption{$i\varepsilon^{\mu}_{b}(q)\varepsilon^{\nu}_{\Pg}(k_1)\Md^{(1),\Pg,1}_{b,\kappa_1,\mu\nu}$}
\end{subfigure}%
\begin{subfigure}[t]{.5\linewidth}
	\centering
	\includegraphics[width=.6\textwidth]{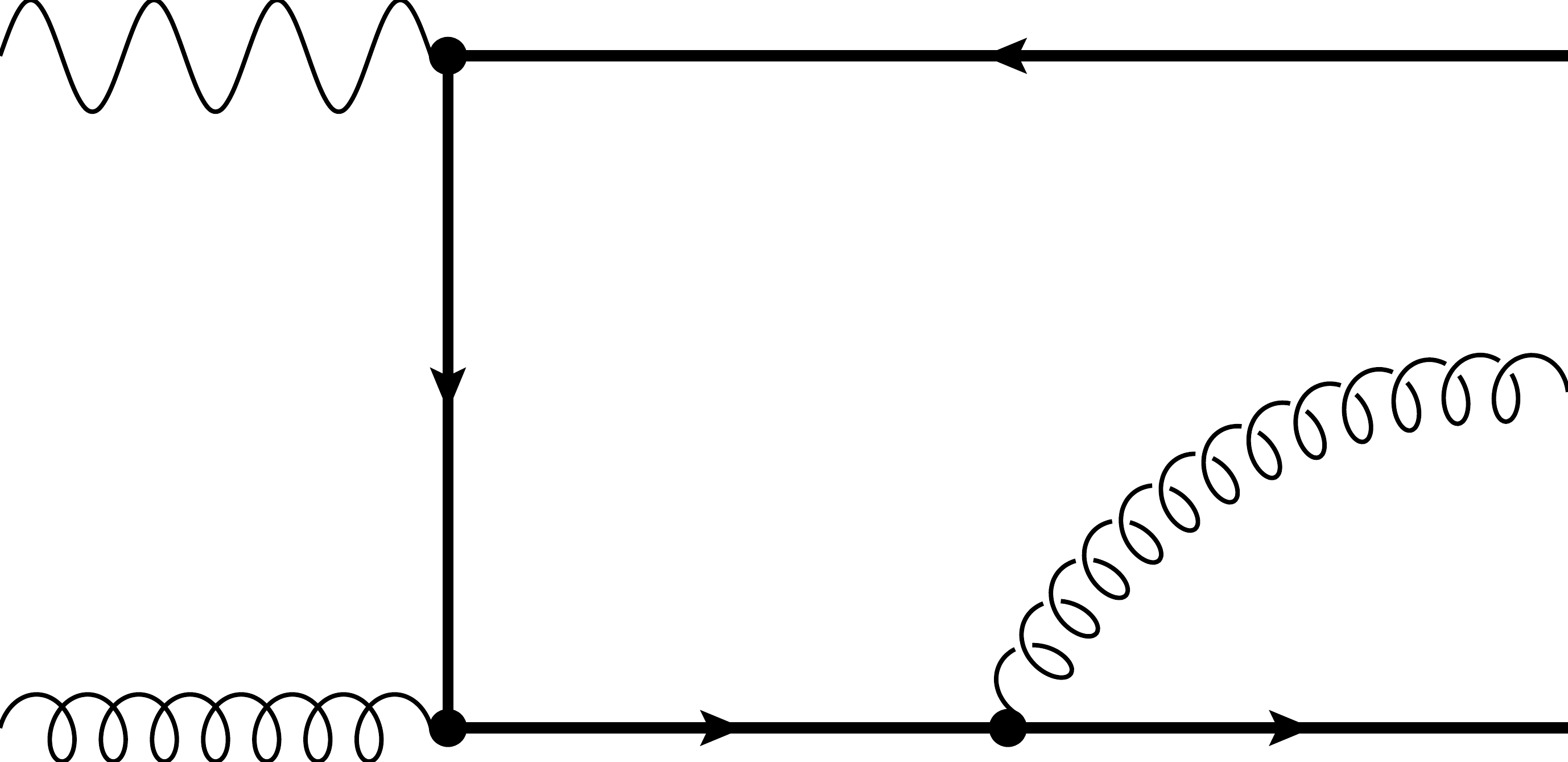}
	\caption{$i\varepsilon^{\mu}_{b}(q)\varepsilon^{\nu}_{\Pg}(k_1)\Md^{(1),\Pg,2}_{b,\kappa_1,\mu\nu}$}
\end{subfigure}
\begin{subfigure}[t]{.5\linewidth}
	\centering
	\includegraphics[width=.4\textwidth]{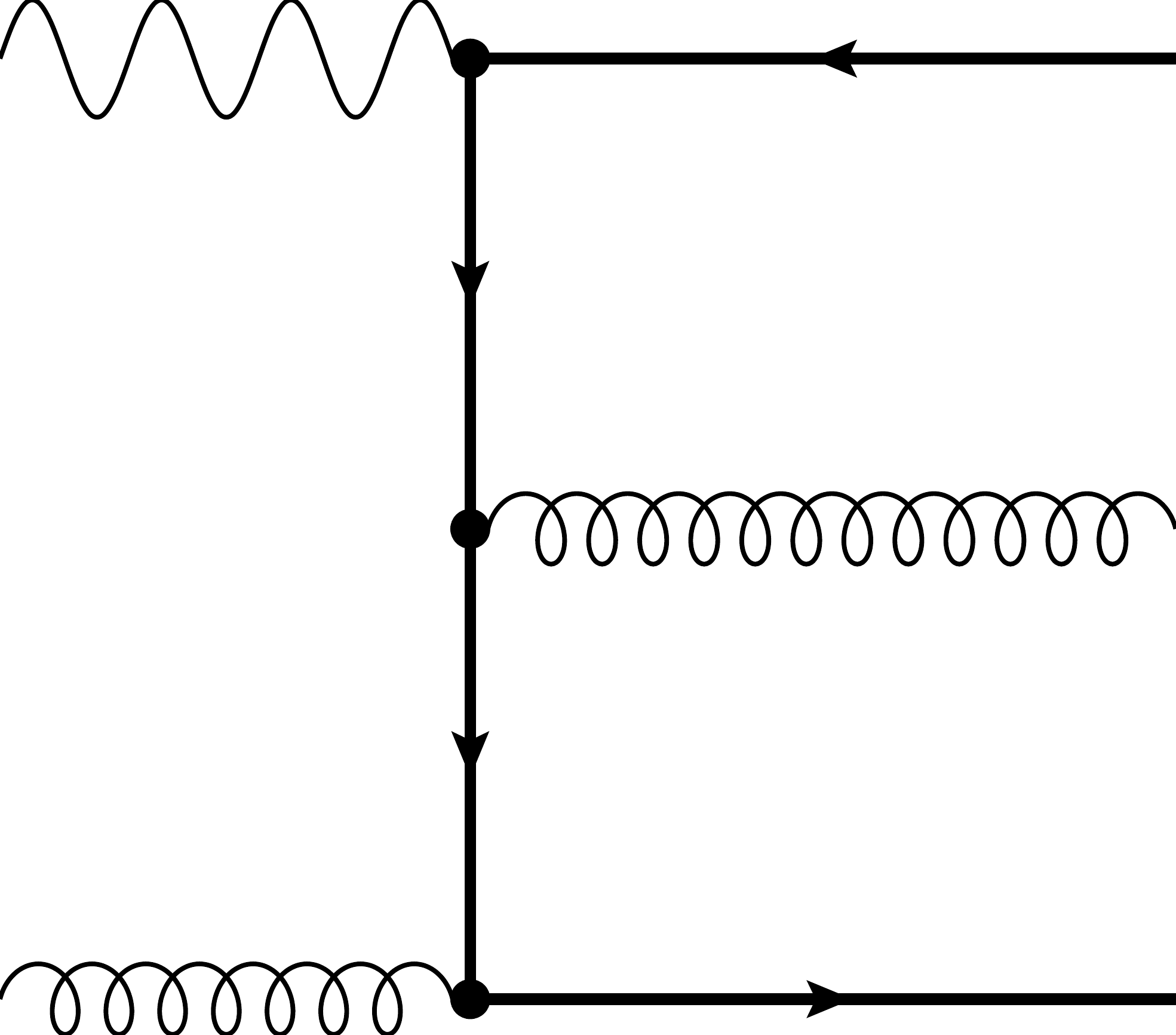}
	\caption{$i\varepsilon^{\mu}_{b}(q)\varepsilon^{\nu}_{\Pg}(k_1)\Md^{(1),\Pg,3}_{b,\kappa_1,\mu\nu}$}
\end{subfigure}%
 \begin{subfigure}[t]{.5\linewidth}
  \centering
  \includegraphics[width=.4\textwidth]{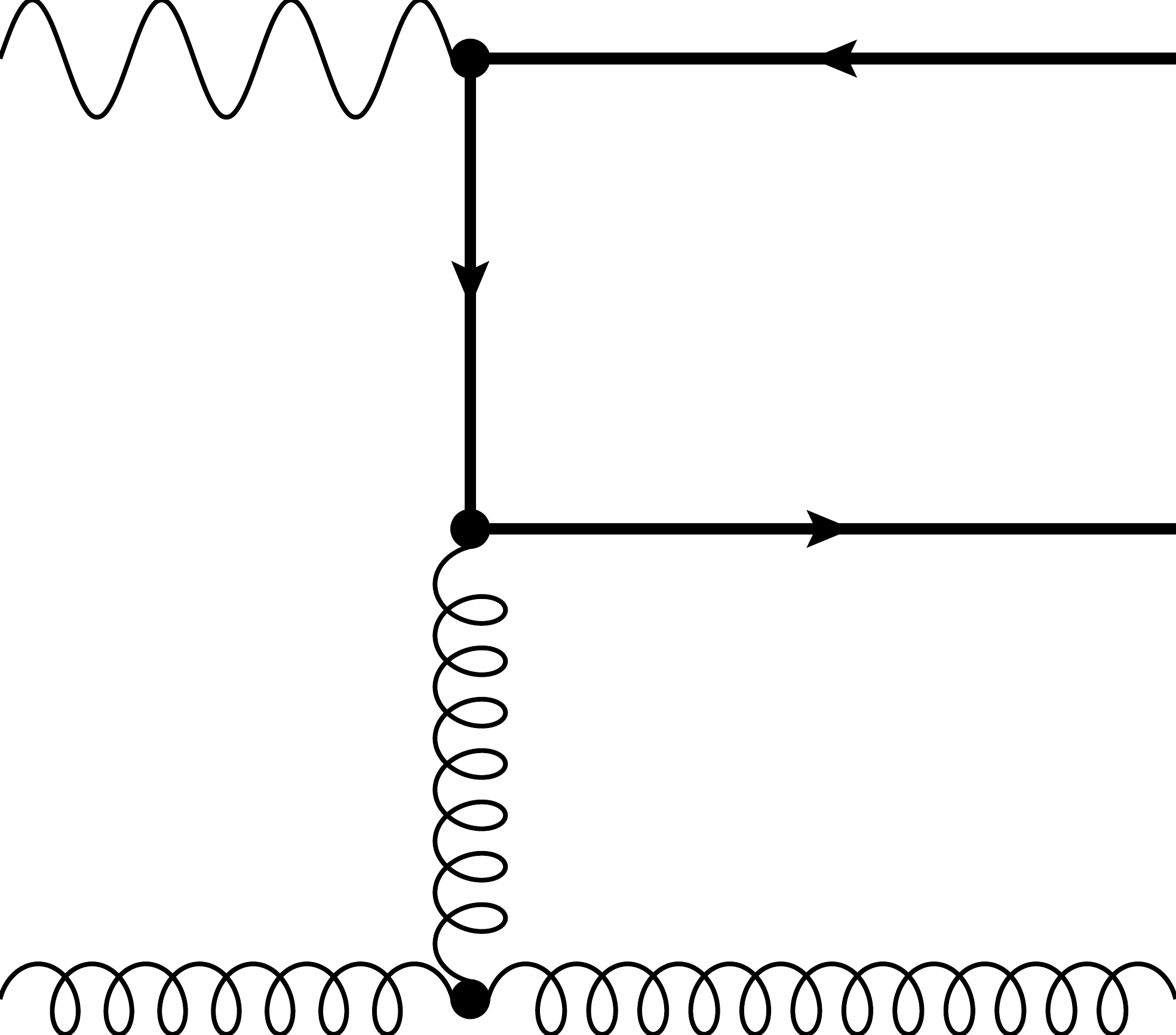}
  \caption{$i\varepsilon^{\mu}_{b}(q)\varepsilon^{\nu}_{\Pg}(k_1)\Md^{(1),\Pg,4}_{b,\kappa_1,\mu\nu}$}
  \label{fig:FeynNLOg4}
 \end{subfigure}
\caption{The next-to-leading order Feynman diagrams for an initial gluon.}\label{fig:FeynNLOg}
\end{figure}

In addition to the virtual corrections we have to consider the radiation of an additional gluon at next-to-leading order, i.e.\ the $2\to 3$ process
\begin{equation}
\HepProcess{b^*(q) + \Pg(k_1) \to \PQ(p_1)+\PaQ(p_2) + \Pg(k_2)},\quad b^*\in\{\Pggx,\PZx\}
\end{equation}
from which we find $2\cdot 4 = 8$ contributing diagrams schematically depicted in \FigureRef{fig:FeynNLOg}.

All diagrams have an additional crossed counterpart with the massive fermion current reversed ($p_1\leftrightarrow p_2$). The diagram in \FigureRef{fig:FeynNLOg4} has to be treated with special care: first, we need an additional diagram (not shown) with an initial ghost to cancel the unphysical polarisations of the gluon as we decided to work in Feynman gauge. Second, as the final state gluon in the diagram of \FigureRef{fig:FeynNLOg4} is massless, it can generate soft and/or collinear poles. Third, this diagram introduces the non-Abelian OK part as it does contain the $3\Pg$-vertex (see \FigureRef{fig:gSI3}). Therefore, it is convenient to keep track of this diagram and it interferences throughout the calculations.

We write the matrix elements for the single gluon radiation as
\begin{align}
M_{\vec \kappa,bb'}^{(1),\Pg} &= \hat {\mathcal P}_{\kappa_3}^{b,\mu\mu'}\hat {\mathcal P}_{\kappa_3}^{\Pg,\nu\nu'}\sum_{j,j'=1}^4\Md^{(1),\Pg,j}_{b,\kappa_1,\mu\nu}\left(\Md^{(1),\Pg,j'}_{b',\kappa_2,\mu'\nu'}\right)^*\\
 &= 8g_s^4\mu_D^{-2\epsilon}e^2 g^{\kappa_1}_{b,\PQ}g^{\kappa_2}_{b',\PQ} N_C C_F\left( C_A R_{\vec \kappa,\tOK} + 2C_F R_{\vec \kappa,\tQED}\right)\,. \label{eq:MER}
\end{align}

In the following we will need the explicit limit of the matrix elements when the radiated gluon becomes soft ($k_2\to 0$) and as can be seen from \Eqssref{eq:NLOMandelstam} $s_3,s_4$ and $t'$ will vanish in this limit (keep in mind that $u'$ does not vanish as we only consider true DIS $Q^2>0$). We obtain
\begin{align}
\lim_{k_2\rightarrow 0}\left(C_A R_{\vec \kappa,\tOK} + 2C_F R_{\vec \kappa,\tQED}\right) &= \left(C_A S_{\vec \kappa,\tOK} + 2C_F S_{\vec \kappa,\tQED}\right) + O(1/s_4,1/s_3,1/t')
\end{align}
with $S_{\vec \kappa,\tOK} = E_{\tOK}B_{\vec \kappa,\tQED}, S_{\vec \kappa,\tQED} = E_{\tQED}B_{\vec \kappa,\tQED}$ and
\begin{align}
E_{\tOK} &= \frac{(k_1\cdot p_2)_{k_2=0}}{2(k_2\cdot k_1)(k_2\cdot p_2)} + \frac{(k_1\cdot p_1)_{k_2=0}}{2(k_2\cdot k_1)(k_2\cdot p_1)} - \frac{(p_1\cdot p_2)_{k_2=0}}{2(k_2\cdot p_2)(k_2\cdot p_1)} \nonumber\\
 &=2\left(\frac{t_1}{t's_3} + \frac{u_1}{t's_4}-\frac{s-2m^2}{s_3s_4}\right), \label{eq:EOK}\\
E_{\tQED} &= \frac{(p_1\cdot p_2)_{k_2=0}}{2(k_2\cdot p_2)(k_2\cdot p_1)} - \frac{(p_1\cdot p_1)}{(k_2\cdot p_1)^2} - \frac{(p_2\cdot p_2)}{2(k_2\cdot p_2)^2}\nonumber\\
 &= 2\left(\frac{s-2m^2}{s_3s_4} - \frac{m^2}{s_3^2} - \frac{m^2}{s_4^2}\right). \label{eq:EQED}
\end{align}
Note that the eikonal factors $E_{\tOK}$ and $E_{\tQED}$ neither depend on the photon's virtuality $q^2$ nor on the projection $\vec \kappa$, ensuring the factorization of the process.

\subsection{Initial Light Quark Contribution} \label{sec:ME.NLO.q}
\begin{figure}[ht]
\centering
\begin{subfigure}[t]{.5\linewidth}
	\centering
	\includegraphics[width=.4\textwidth]{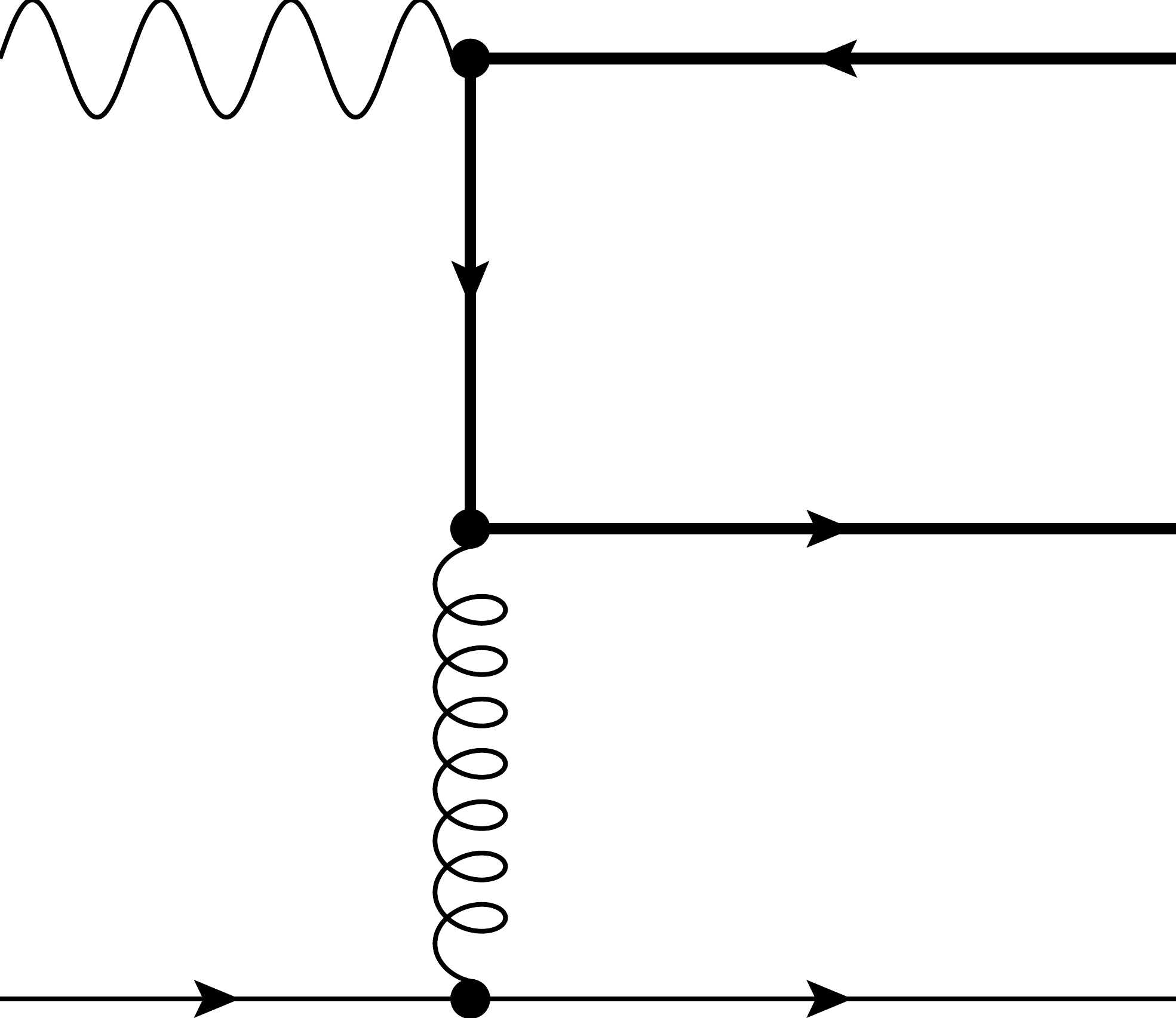}
	\caption{$i\varepsilon^{\mu}_{b}(q)\bar u_a(k_1)\Md^{(1),\Pq,1}_{b,\kappa_1,\mu a}$}
	\label{fig:FeynNLOq1}
\end{subfigure}%
\begin{subfigure}[t]{.5\linewidth}
	\centering
	\includegraphics[width=.4\textwidth]{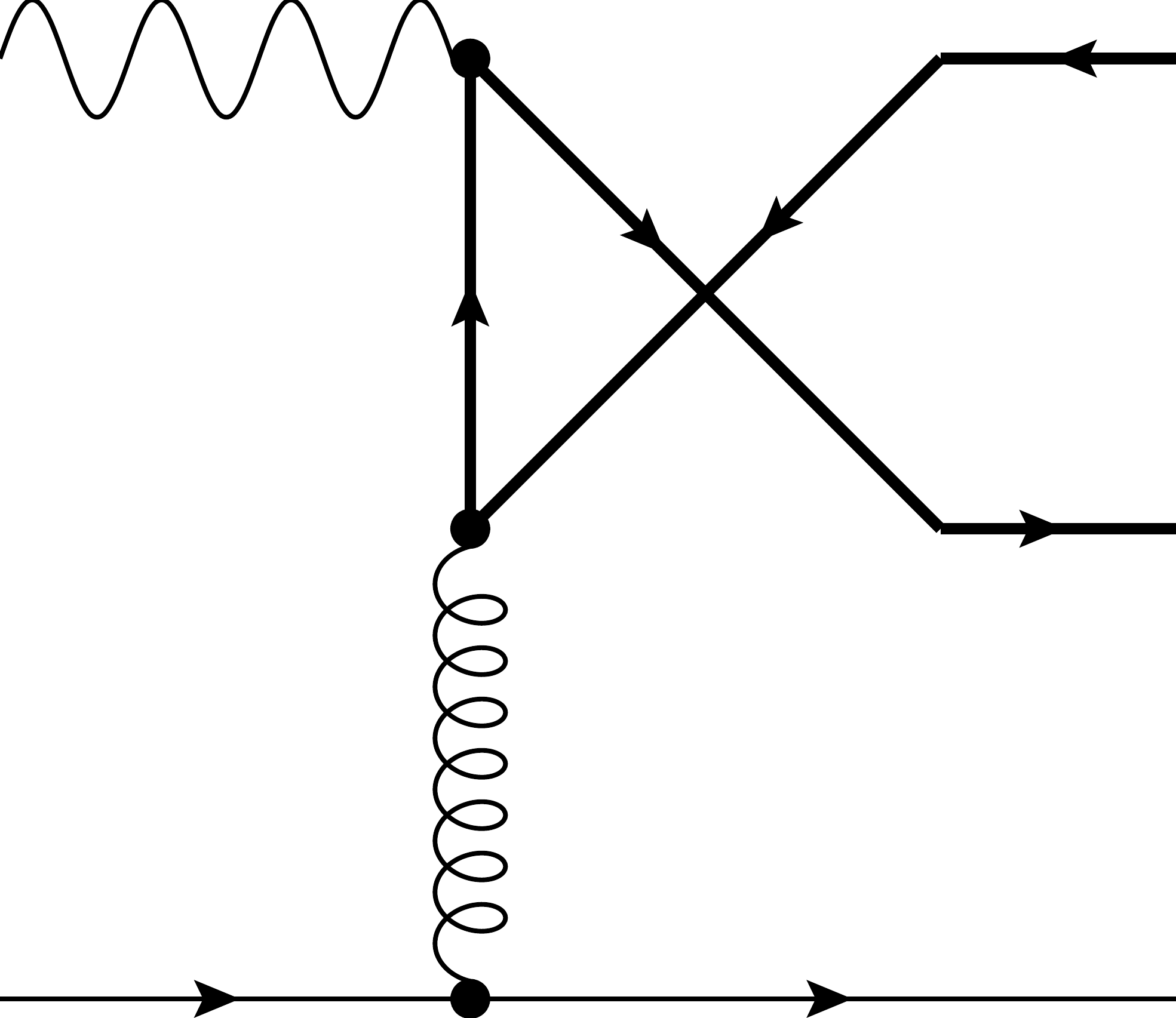}
	\caption{$i\varepsilon^{\mu}_{b}(q)\bar u_a(k_1)\Md^{(1),\Pq,2}_{b,\kappa_1,\mu a}$}
	\label{fig:FeynNLOq2}
\end{subfigure}
\begin{subfigure}[t]{.5\linewidth}
	\centering
	\includegraphics[width=.6\textwidth]{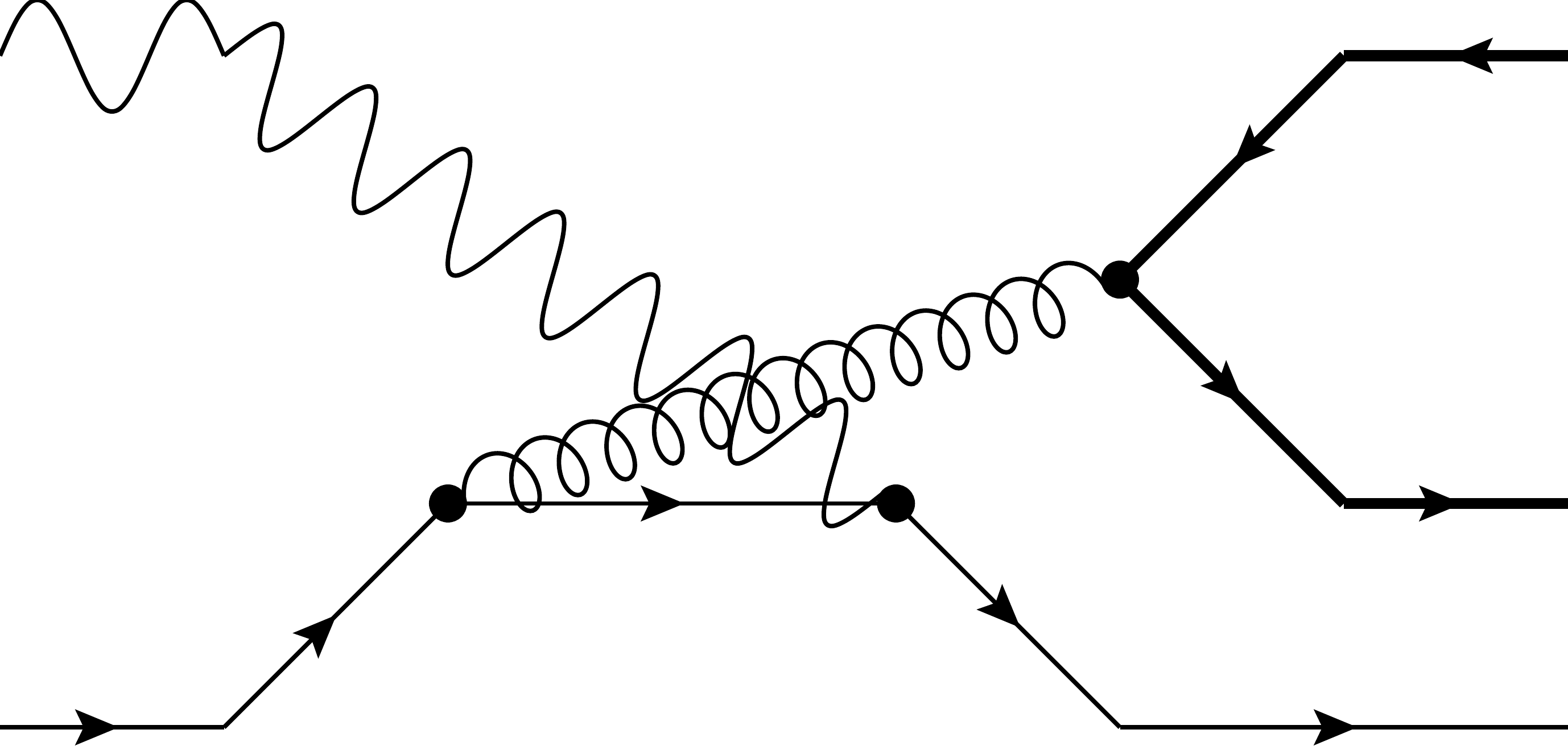}
	\caption{$i\varepsilon^{\mu}_{b}(q)\bar u_a(k_1)\Md^{(1),\Pq,3}_{b,\kappa_1,\mu a}$}
	\label{fig:FeynNLOq3}
\end{subfigure}%
\begin{subfigure}[t]{.5\linewidth}
	\centering
	\includegraphics[width=.6\textwidth]{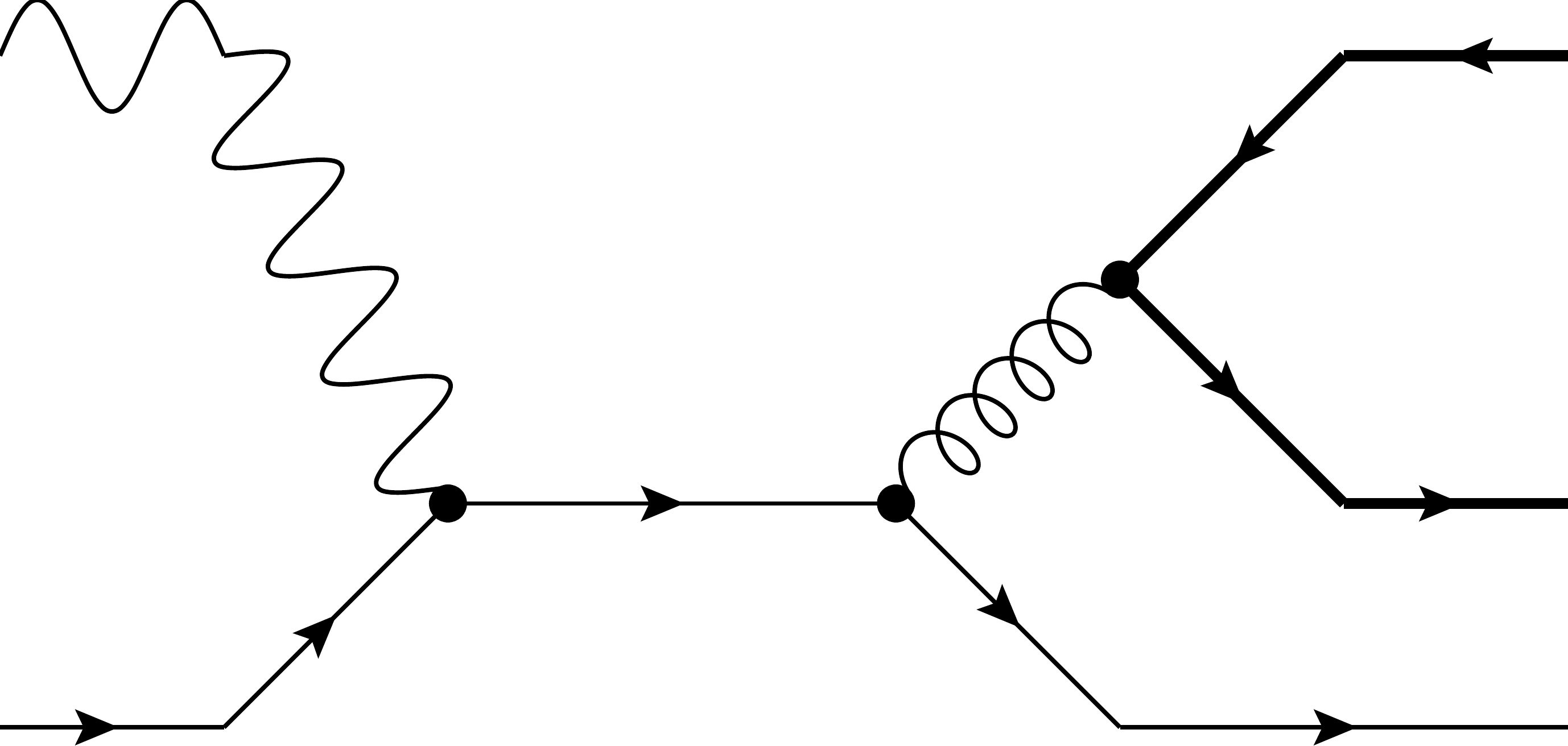}
	\caption{$i\varepsilon^{\mu}_{b}(q)\bar u_a(k_1)\Md^{(1),\Pq,4}_{b,\kappa_1,\mu a}$}
	\label{fig:FeynNLOq4}
\end{subfigure}
\caption{The next-to-leading order Feynman diagrams for an initial light quark}\label{fig:FeynNLOq}
\end{figure}

At NLO we also have to consider a light quark as initial parton, so we have to consider the process
\begin{equation}
\HepProcess{b^*(q) + \Pq(k_1) \to \PQ(p_1)+\PaQ(p_2) + \Pq(k_2)},\quad b^*\in\{\Pggx,\PZx\}
\end{equation}
from which we find four contributing Feynman diagrams depicted in \FigureRef{fig:FeynNLOq}.

Note that the diagram in \FigureRef{fig:FeynNLOq2} is the crossed version of the diagram in \FigureRef{fig:FeynNLOq1} and that the diagrams in \FigureaRef{fig:FeynNLOq3}{fig:FeynNLOq4} are invariant under the crossing transformation ($p_1\leftrightarrow p_2$). Note that in the diagrams in \FigureaRef{fig:FeynNLOq3}{fig:FeynNLOq4} the initial bosons and the final heavy quarks are not coupled directly and thus employ a simpler analytic structure.

We write the matrix elements for the light quark contributions as
\begin{align}
M_{\vec \kappa,bb'}^{(1),\Pq} &= \hat {\mathcal P}_{\kappa_3}^{bb',\mu\mu'}\hat {\mathcal P}_{\kappa_3}^{\Pq,aa'}\sum_{j,j'}\Md^{(1),\Pq,j}_{b,\kappa_1,\mu a}\left(\Md^{(1),\Pq,j'}_{b',\kappa_2,\mu' a'}\right)^*\\
 &= 8g_s^4\mu_D^{-2\epsilon}e^2 N_C C_F\left( g^{\kappa_1}_{b,\PQ}g^{\kappa_2}_{b',\PQ} A_{\vec \kappa,1} +  g^{\kappa_1}_{b,\Pq}g^{\kappa_2}_{b',\Pq} A_{\vec \kappa,2} +  g^{\kappa_1}_{b,\PQ}g^{\kappa_2}_{b',\Pq} A_{\vec \kappa,3} \right) \label{eq:MEA}
\end{align}
where we adopt the common decomposition by the coupling of the bosons. We also adopt the usual language and refer to $A_{\vec\kappa,1}$ as Bethe-Heitler subprocess and to $A_{\vec\kappa,2}$ as Compton subprocess.

\chapter{Computing the Phase Space} \label{chap:PS}
\chapterquote{You Shall Not Pass!}{The Phase Space Boundary Condition}

As we decided to apply dimensional regularization, we need to be careful \textit{if} we average over initial state bosons and thus define
\begin{align}
E_{\hat F_2}(\epsilon) = E_{\hat F_L}(\epsilon) = E_{z\hat F_3}(\epsilon) &= \frac{1}{1+\epsilon/2}\,,\\
E_{2z\hat g_1}(\epsilon) = E_{\hat g_4}(\epsilon) = E_{\hat g_L}(\epsilon) &= 1
\end{align}
to account for additional degrees of freedom in $n=4+\epsilon$ dimensions.

At leading order (LO) we require the $2\to 2$ $n$-dimensional phase space given by
\begin{align}
\dPSTwo &= \!\int\!\!\frac{d^{n}p_1}{(2\pi)^{n-1}}\frac{d^{n}p_2}{(2\pi)^{n-1}}(2\pi)^n\delta^{(n)}(k_1+q-p_1-p_2) \nonumber\\
 &\hspace{50pt}\Theta(p_{1,0})\delta(p_1^2-m^2)\Theta(p_{2,0})\delta(p_2^2-m^2) \label{eq:PS2}
\end{align}
involving the three Mandelstam variables
\begin{align}
s &= (q+k_1)^2 &t_1 &=(k_1-p_2)^2-m^2 &u_1 &=(q-p_2)^2 -m^2
\end{align}
out of which only two are linearly independent (Note, that we defined additionally $s' =s-q^2$ and $u_1' = u_1-q^2$).

At next-to-leading order (NLO) we require in addition the $2\to 3$ $n$-dimensional phase space given by
\begin{align}
\dPSThree &= \!\int\!\!\frac{d^{n}p_1}{(2\pi)^{n-1}}\frac{d^{n}p_2}{(2\pi)^{n-1}}\frac{d^{n}k_2}{(2\pi)^{n-1}}(2\pi)^n\delta^{(n)}(k_1+q-p_1-p_2-k_2) \nonumber\\
 &\hspace{50pt}\Theta(p_{1,0})\delta(p_1^2-m^2)\Theta(p_{2,0})\delta(p_2^2-m^2)\Theta(k_{2,0})\delta(k_2^2) \label{eq:PS3}
\end{align}
involving the ten Mandelstam variables of the \Eqssref{eq:NLOMandelstam}. The expression can be reduced by writing it as product of a $2\to 2$ decay and a subsequent $1\to 2$ decay\cite{PhysRevD4054}.

The processing of the general phase space expressions (\Eqsaref{eq:PS2}{eq:PS3}) to a minimal set of integrations is not a unique procedure. Furthermore, it is often advantageous to apply different decomposition in different regions of phase space or to compute specific observables. We will apply two different phase space decompositions here: the $s_4$-centered decomposition (discussed at LO in \SectionRef{sec:PS.LO} and at NLO in \SectionRef{sec:PS.s4}), which will turn $s_4$ into a good variable\footnote{the term \enquote{good variable} will become obvious in the following \SectionrRef{sec:PS.LO}{sec:PS.s5} and by the arguments in \AppendixRef{sec:Appendix.s4}}, and the $s_5$-centered decomposition (discussed at LO in \SectionRef{sec:PS.LO} and at NLO in \SectionRef{sec:PS.s5}), which will turn $s_5$ into a good variable.

We will use the $s_4$-centered decomposition for our analytic approach, which aims at doing as many integrals as possible in an analytic way. The factorization actually holds on a differential level
\begin{align}
\frac{dH_{\vec \kappa,j}^{bb'}}{dT_1 dU_1} &= \frac{Q^2}{4\pi^2\alpha}\int\limits_{\xi_{min}}^1\!d\xi\,\xi f_{\kappa_3,j}(\xi,\mu_F^2) \frac{d\sigma_{\vec\kappa,j}^{bb'}}{dt_1 du_1}\quad j\in\{\Pg,\Pq\} \label{eq:DiffFact}
\end{align}
with $T_1=(p-p_2)^2 - m^2$ and $U_1 = (q-p_2)^2-m^2$ referring to the \textit{hadronic} Mandelstam variables, $\xi_{min} = x(4m^2+Q^2)/Q^2$ the minimum momentum fraction to justify the production threshold and $\sigma_{\vec\kappa,j}^{bb'}$ the partonic cross sections. In this approach we will split the five\footnote{to be exact: $2n-3=5+2\epsilon$ integrations - the angular parts will cover $2+2\epsilon$ integrations} integrals at NLO resulting from \Eqref{eq:PS3} into two angular integrals that will be performed analytically and into three integrals related to the momentum fraction $\xi$, the angle $t_1$ and $s_4(=s'+t_1+u_1)$ which effectively measures the softness of the radiation. The further regularization procedure requires us to split $s_4$-integral into a hard ($s_4>\Delta$) and a soft ($s_4<\Delta$) part - a method that is commonly known as phase space slicing\cite{Hekhorn:2018ywm,Laenen1993162,epub29831,Bojak:2000eu}. In this approach we gain access to the inclusive distributions of the structure functions, i.e.\ the distributions towards the transverse momentum of the heavy anti-quark $p_T$ and/or its rapidity $y$ as we can transform the differentials in \Eqref{eq:DiffFact} into these variables (see \AppendixRef{sec:Appendix.s4.Hadronic}).

We will use the $s_5$-centered decomposition for our Monte-Carlo approach, which aims at getting reliable access to \textit{any} differential of the structure functions. In this approach, we will always compute the full structure functions
\begin{align}
H^{bb'}_{\vec \kappa, j}(x,Q^2,m^2) &= \frac{Q^2}{4\pi^2\alpha} \int\limits_x^{z_{max}}\frac {dz}{z} f_{\kappa_3,j}\left(\xi=\frac x z,\mu_F^2\right) \cdot \sigma^{bb'}_{\vec \kappa,j}(s'=Q^2/z),\quad j\in\{\Pg,\Pq\}
\end{align}
while collecting all distributions \textit{on-the-fly} in histograms. Here, we will choose a specific phase space parametrisation of the final state momenta turning all soft and collinear poles explicit. This prescription allows to subtract the poles in a meaningful way from other regions of the phase space - a method that is commonly known as subtraction method\cite{paper2,Harris:1995tu,epub29831,Mangano:1991jk,Frixione:1993dg}. We will solve all occurring five integrals by sampling the integration kernel at a given number of random points (Monte Carlo integration) and thus, we get access to the four-momenta of all particles. This informations allow to observe any distribution and in especially, correlated distributions that are not accessible in the inclusive approach.

As was mentioned in \ChapterRef{chap:ME}, we will not repeat the calculation of hat-momenta here, but we compare how this additional level of complexity is dealt with in the different approaches:
\begin{description}
\item [hat momenta] they generate additional terms $O({\hat k_1}^2)$ in the NLO matrix elements $R_{\vec\kappa,\tOK}$ (\Eqref{eq:MER}) and $A_{\vec\kappa,1}$ (\Eqref{eq:MEA}) and a correction to the $2\to 3$ phase space $\dPSThree$ (for details see \cite{Hekhorn:2018ywm,Bojak:2000eu})
\item [$s_4$-centered] matrix elements are affected at all orders: additional terms $O(\epsilon)$ are generated in the $B_{\vec\kappa,\tQED}$ entering at NLO through factorization and additional terms $O(\epsilon)$ are generated in the NLO matrix elements $R_{\vec\kappa,\tOK}$ and $A_{\vec\kappa,1}$ resulting in finite contributions when joined by collinear poles (see \SectionRef{sec:PS.s4})
\item [$s_5$-centered] the $O(\epsilon)$ contributions to the splitting kernels $P_{\kappa_3,jk}$ become important in NLO when performing the phase space subtractions, but, instead, all matrix elements can be performed in just $n=4$ dimensions (see \SectionRef{sec:PS.s5})
\end{description}
We verified that up to the considered accuracy (NLO) all three approaches give the same results and thus are compatible. In fact, we did use this to carefully cross check our results.

\section{Leading Order Phase Space} \label{sec:PS.LO}
\subsubsection{$s_4$-centered}
At leading order we may still define $s_4=s'+t_1+u_1$ but due to momentum conservation it is manifest $s_4=0$. Nevertheless, we stick to the idea of setting up a $t_1$ integration and find\cite{Hekhorn:2018ywm,Laenen1993162}
\begin{align}
\dPSTwo &= \frac {2\pi S_\epsilon}{s'\Gamma(1+\epsilon/2)}\left(\frac{(t_1u_1'-s'm^2)s' - q^2t_1^2}{s'^2}\right)^{\epsilon/2}\delta(s'+t_1+u_1)\,dt_1du_1 \label{eq:dps2}
\end{align}
with $S_\epsilon = (4\pi)^{-2+\epsilon/2}$ and $\Gamma$ representing Euler's Gamma function. One typically solves the $u_1$ integration trivially by using the $\delta$-distribution, leaving only a single integration in $t_1$ with\footnote{recall the definition of the partonic variables in \Eqsrref{eq:partonicVars1}{eq:partonicVars3}}
\begin{align}
-\frac{s'}{2}(1+\beta)<t_1<-\frac{s'}{2}(1-\beta)\,.
\end{align}
The three fundamental integrals for the representation in \SectionRef{sec:ME.LO} are
\begin{align}
\int \dPSTwo &= \frac{\beta}{8\pi}\\
\int\frac{s'^2}{t_1 u_1} \,\dPSTwo &= -\frac{\ln\chi}{4\pi}\\
\int\left(\frac{s'^2}{t_1 u_1}\right)^2 \,\dPSTwo &= \frac{1-\chi^2-2\chi\ln\chi}{4\pi\chi} \label{eq:PS2s4}
\end{align}
We can additionally reduce the occurring phase space integrals by using partial fractioning and by applying Laporta's algorithm\cite{Laporta:2001dd}. We then only need the following master integral
\begin{align}
\int \left(\frac{s'}{t_1}\right)^x \dPSTwo &= \frac{1}{2^{4-x}\pi(1-x)}\left((1+\beta)\left(-\frac{1-\beta}{\rho}\right)^x-(1-\beta)\left(-\frac{1+\beta}{\rho}\right)^x\right)
\end{align}
from which the correct result for $x=1$ can be obtained by taking the appropriate limit.

\subsubsection{$s_5$-centered}
At leading order we may still define $s_5=(p_1+p_2)^2$ but due to momentum conservation it is manifest $s_5=s$. Nevertheless, we stick to the idea of picking the four-momenta of the final state particles in their center-of-mass system ($\vec p_1 +\vec p_2 = \vec 0$). Thus, we define an angle $\theta_1$ by\cite{paper2,Harris:1995tu}
\begin{align}
k_1 &= \frac{s'}{2\sqrt s}(1,0,0,1)\,, &p_2 &= \frac{\sqrt s}{2}(1,0,\beta \sin\theta_1,\beta\cos\theta_1)\,.
\end{align}
and write the $2\to 2$ phase space as
\begin{align}
\dPSTwo &= \frac{\beta\sin\theta_1}{16\pi\Gamma(1+\epsilon/2)} \left(\frac{s\beta^2\sin^2(\theta_1)}{16\pi}\right)^{\epsilon/2} d\theta_1
\end{align}

We can relate the two decomposition by observing
\begin{align}
t_1 &= -2k_1\cdot p_2 = -\frac{s'}{2}(1-\beta\cos\theta_1) \,.
\end{align}

\subsubsection{Partonic Cross Sections}
The actual physics of course is unaltered from the representation of phase space and we find for the spin and color averaged LO partonic cross section
\begin{align}
d\sigma_{\vec \kappa,bb',\Pg}^{(0)} &= \frac 1 {2s'}\frac {E_{\kappa_3}(\epsilon)K_{\Pg\Pgg}} 2 M_{\vec \kappa,bb'}^{(0)} \dPSTwo \label{eq:LOxs}
\end{align}
At NLO the virtual corrections will share the same kinematics as the LO process and so we can give the bare partonic one-loop virtual BGF cross section by
\begin{align}
\left.d\sigma_{\vec \kappa,bb',\Pg}^{(1),V}\right|_{\text{bare}} &= \frac 1 {2s'}\frac {E_{\kappa_3}(\epsilon) K_{\Pg\Pgg}} 2 M_{\vec \kappa,bb'}^{(1),V} \dPSTwo
\end{align}
We have defined
\begin{align}
K_{\Pg\Pgg} &= \frac 1 {N_C^2 -1} &K_{\Pq\Pgg} = \frac{1}{N_C}
\end{align}
to account for the color averaging in the gluonic or light quark initial state respectively (Note that $K_{\Pq\Pgg}/K_{\Pg\Pgg} = 2C_A$).

\section{Next-to-Leading Order $s_4$-centered Phase Space} \label{sec:PS.s4}
We move to the center-of-mass system of the two unobserved particles with momenta $k_2$ and $p_1$ and write\cite{Laenen1993162}
\begin{align}
k_2 &= (\omega_2,\ldots,+k_{2,x},+\omega_2\sin\theta_1\cos\theta_2,+\omega_2\cos\theta_1)\\
p_1 &= (E_1,\ldots,-k_{2,x},-\omega_2\sin\theta_1\cos\theta_2,-\omega_2\cos\theta_1)
\end{align}
where $k_{2,x}=k_{2,x}(\omega_2,\theta_1,\theta_2,\hat k_2)$ is such, that $k_2^2=0$. We have then the additional freedom on where to align the $z$-axis and we can associate this with different sets. In \textit{Set I} we align $k_1$ to the $z$-axis and this is our default choice. In \textit{Set II} we align $q$ to the $z$-axis and a potential Set III, with $p_2$ aligned, is actually not needed for the calculation as will be explained in the \AppendixRef{sec:Appendix.s4.Decomposition}, where we also explicitly give the decomposition of the vectors.

The required $n$-dimensional phase space $\dPSThree$ is, of course, independent of the definition of the sets and we find\cite{Hekhorn:2018ywm,Laenen1993162}
\begin{align}
\dPSThree &= \frac{S_\epsilon^2}{\Gamma(1+\epsilon)s'} \frac{s_4^{1+\epsilon}}{(s_4+m^2)^{1+\epsilon/2}}\left(\frac{(t_1u_1'-s'm^2)s' - q^2t_1^2}{s'^2}\right)^{\epsilon/2}\! dt_1 du_1 d\Omega_n \label{eq:PS3s4}
\end{align}
with $d\Omega_n = \sin^{n-3}(\theta_1)d\theta_1\sin^{n-4}(\theta_2)d\theta_2$ carrying all angular dependence. In practice, we write $du_1 = ds_4$ making the label \enquote{$s_4$-centered} even more obvious, but the representation given here allows an easy comparison to the LO case in \Eqref{eq:PS2s4}. Note that we do not consider hat momenta\cite{Hekhorn:2018ywm} here, as we use the MVV-scheme in this work.

For solving the angular integrals $d\Omega_n$, we insert the definitions of the different sets and classify the angular-dependent variables $s_3,s_5,t',u',u_6,u_7$ into two categories: $(ab)$-variables, which have the general form $a+b\cos\theta_1$, and $[ABC]$-variables, which have the general form $A+B\cos\theta_1+C\sin\theta_1\cos\theta_2$. They may contain collinear poles if either $a=-b$ or $A^2 = B^2+C^2$. The parameters $a,b,A,B,C$ are given by a combination of the angular-independent variables $s',t_1,u_1,s_4$. Soft poles are encoded by terms proportional $1/s_4$ implicitly and only appear explicitly after a potential integration in the small $s_4$ region. In practice, all integration have to be reduced to a product of two angular-dependent variables, one being an $(ab)$-variable and one being an $[ABC]$-variable. This is why we need two different sets of phase space decomposition as they interchange the role of the variable type. We are left with the solution of the generic integrals
\begin{align}
I^{(k,l)}_n &=\int\!d\Omega_n\,(a+b\cos(\theta_1))^{-k}(A+B\cos\theta_1+C\sin\theta_1\cos\theta_2)^{-l} \label{eq:Ikln}
\end{align}
We split them by their collinear structure:
\begin{description}
\item[$I_{0,n}^{(k,l)}$] with no collinear poles and at NLO accuracy we need them only at $O(\epsilon^0)$. Most of the higher order integrals can be obtained from the master integral $I_{0,n}^{(1,1)}$ by differential equations.
\item[$I_{a,n}^{(k,l)}$] with the $(ab)$-variable being collinear ($a=-b$) and at NLO accuracy we need them up to $O(\epsilon)$ (which is potentially three terms as they may start at $1/\epsilon$). We give a new general solution to a certain range in parameter space in \AppendixRef{sec:Appendix.s4.PSInt}.
\item[$I_{A,n}^{(k,l)}$] with the $[ABC]$-variable being collinear ($A^2=B^2+C^2$). These integrals can be related to the case of $I_{a,n}^{(k,l)}$ by a rotation which leaves the integral measure unchanged, furthermore, we do not need these integrals here.
\item[$I_{aA,n}^{(k,l)}$] with both the $(ab)$- and $[ABC]$-variable being collinear. These integrals are not needed here as only $t'$ can become collinear.
\end{description}
An extensive list for many cases of all types of integrals can be found in \cite{PhysRevD4054} and \cite{Bojak:2000eu} where also several technical details are covered.

The matrix elements for the Bethe-Heitler subprocess $A_{\vec\kappa,1}$ and the Compton subprocess $A_{\vec\kappa,2}$ involve each only two angular-dependent variables: $t',u_7$ and $s_5,u'$ respectively. So, we achieve the decomposition of the matrix elements into pairs of variables by substituting all other Mandelstam variables by expressions that contain the two desired variables along with some of the angular-independent variables $s',t_1,u_1,s_4$, depending on the chosen phase space decomposition set. For all other matrix elements, i.e.\ $R_{\vec\kappa,\tOK}$, $R_{\vec\kappa,\tQED}$, and, $A_{\vec\kappa,3}$, we have to perform the decomposition more carefully: on some individual terms we may apply the procedure described above, but on most terms we have to apply partial fractioning leading to more terms. Partial fractioning relies only on the relations connecting the Mandelstam variables and not on the chosen phase space decomposition set. The product of $s_3$ and $u_7$ cannot be simplified by partial fractioning, but instead we have to solve these terms in our \textit{Set II} (see \AppendixRef{sec:Appendix.s4.Decomposition}).

As said in the introduction to this chapter, we will split the $s_4$ integral into two pieces: a soft part $s_4<\Delta$ which will be treated separately and a hard part $s_4>\Delta$ which will enter the numerical codes. We can link this separation to the momentum fraction $\xi$ by observing
\begin{align}
s_4 &= s' + t_1 + u_1 = \xi S_h' + \xi T_1 + U_1 \sim \Delta\,.
\end{align}
So, the hard region is mapped by $1-\xi>\delta$ with $\delta=\Delta/(S_h'+T_1)$ and the soft limit refers to $\xi\to 1$. With this definition at hand we can now give the gluon-to-gluon Altarelli-Parisi splitting kernels with
\begin{align}
P^{\delta,H}_{\hat F_2,\Pg\Pg}(x) &= P^{\delta,H}_{\hat F_L,\Pg\Pg}(x) = P^{\delta,H}_{z\hat F_3,\Pg\Pg}(x)= P^{\delta,H}_{\Pg\Pg}(x) \nonumber\\
 &= 2C_A\left(\frac x {1-x} + \frac {1-x} x + x(1-x)\right) \label{eq:PFgg}\\
P^{\delta,H}_{2z\hat g_1,\Pg\Pg}(x) &= P^{\delta,H}_{\hat g_4,\Pg\Pg}(x) = P^{\delta,H}_{\hat g_L,\Pg\Pg}(x) = \Delta P^{\delta,H}_{\Pg\Pg}(x) \nonumber\\
 &= 2C_A\left(\frac 1 {1-x} - 2x + 1\right) \label{eq:Pggg}
\end{align}
for the hard part and the full kernel is then given by
\begin{align}
P_{\kappa_3,\Pg\Pg}^{\delta}(x) &= \Theta(1-x-\delta)P_{\kappa_3,\Pg\Pg}^{\delta,H}(x)+\delta(1-x)\left(2C_A\ln(\delta)+\frac{\beta_0^{lf}}{2}\right)
\end{align}
with $\beta_0^{lf} = (11C_A - 2n_{lf})/3$. Note that the hard part of the polarized and unpolarized splitting kernels do have the \textit{same} soft limit
\begin{align}
\lim\limits_{x\to 1}(1-x)P_{\Pg\Pg}^{\delta,H}(x) &= 2C_A = \lim\limits_{x\to 1}(1-x)\Delta P_{\Pg\Pg}^{\delta,H}(x)\,.
\end{align}
For the quark-to-gluon kernels we have
\begin{align}
P^{\delta}_{\hat F_2,\Pg\Pq}(x) = P^{\delta}_{\hat F_L,\Pg\Pq}(x) = P^{\delta}_{z\hat F_3,\Pg\Pq}(x) = P^{\delta}_{\Pg\Pq}(x) &= C_F\left(\frac {1+(1-x)^2} x\right) \label{eq:PFgq}\,, \\
P^{\delta}_{2z\hat g_1,\Pg\Pq}(x) = P^{\delta}_{\hat g_4,\Pg\Pq}(x) = P^{\delta}_{\hat g_L,\Pg\Pq}(x)  = \Delta P^{\delta}_{\Pg\Pq}(x) &= C_F\left(2-x\right)\,. \label{eq:Pggq}
\end{align}

\subsection{Single Gluon Radiation} \label{sec:PS.s4.g}
The diagram shown in \FigureRef{fig:FeynNLOg4} introduces collinear poles as it depends on the propagator $t'$. The poles can be cast in a compact form by
\begin{align}
\frac{s_4}{2\pi(s_4+m^2)}\int\!d\Omega_n \,C_A R_{\vec \kappa,\tOK} &=-\frac 2 {u_1}B_{\vec \kappa,\tQED}\left(x_1k_1\right) P^{\delta,H}_{\kappa_3,\Pg\Pg}(x_1)\frac 2 \epsilon + O(\epsilon^0) \label{eq:ROKPoles}
\end{align}
with $x_1 = -u_1/(s'+t_1)$. By denoting $B_{\vec \kappa,\tQED}\left(x_1k_1\right)$ we mean the shift of momentum $k_1\to x_1k_1$ in the LO matrix element which effectively shifts $s'\to x_1s'$ and $t_1\to x_1 t_1$ (Note that we have to replace every $u_1$ \textit{inside} $B_{\vec\kappa,\tQED}$ with its definition in terms of $s',t_1$, i.e.\ $u_1=-s'-t_1$). The hard Abelian QED part $R_{\vec \kappa,\tQED}$ does not contain any collinear poles.

To proceed with the soft contributions ($s_4<\Delta$), we need to treat the soft limit of the matrix elements (\Eqsaref{eq:EOK}{eq:EQED}) with the soft limit of $\dPSThree$ (see \cite{Hekhorn:2018ywm}). After the angular integrations the soft variables $s_3$ and $t'$ reveal explicitly their soft behaviour, i.e.\ they become proportional to $1/s_4$. Actually, all soft terms behave in general as $1/s_4^2$ and together with the phase space expression \Eqref{eq:PS3s4} we get a dependence $s_4^{-1+\epsilon}$. Next, we can replace this dependence with a standard regulator\cite{Hekhorn:2018ywm,Bojak:2000eu}
\begin{align}
s_4^{-1+\epsilon}= \frac{\Delta^\epsilon}{\epsilon} \delta(s_4) + \left[s_4^{-1+\epsilon}\right]_\Delta\,.
\end{align}
We find for the poles\cite{Hekhorn:2018ywm,Bojak:2000eu}
\begin{align}
\nonumber
\lim_{s_4\to 0}\,&\frac{s_4}{2\pi(s_4+m^2)} \left[1-\frac 3 8 \zeta(2)\epsilon^2\right] \frac{s_4}{\epsilon}
\int\!d\Omega_n \,S_{\vec \kappa,\tOK} \\
\nonumber
&=2B_{\vec \kappa,\tQED}\left\{ \frac 4 {\epsilon^2} + \frac 2 \epsilon \left[\ln\left(\frac{t_1}{u_1}\right)
+\frac{s-2m^2}{s\beta} \ln(\chi)\right]
-\ln^2(\chi) - \frac 3 2 \zeta(2)+\frac 1 2\ln^2\left(\frac{u_1\chi}{t_1}\right)\right. \\
&\left. \hspace{20pt} +\DiLog\left(1- \frac{t_1}{u_1\chi}\right) 
- \DiLog\left(1-\frac{u_1}{t_1\chi}\right) + \frac{s-2m^2}{s\beta}\left[
\DiLog(1-\chi^2)+\ln^2(\chi)
\right] \right\} + \mathcal O(\epsilon)
\end{align}
and
\begin{align}
\lim_{s_4\to 0}\,&\frac{s_4}{2\pi(s_4+m^2)} \left[1-\frac 3 8 \zeta(2)\epsilon^2\right] \frac{s_4}{\epsilon}
\int\!d\Omega_n\,S_{\vec \kappa,\tQED}\nonumber\\
&= 2B_{\vec \kappa,\tQED}  \left\{-\frac 2 \epsilon  \left(1+\frac{s-2m^2}{s\beta}\ln(\chi)\right) +1 \right.\nonumber\\
 & \hspace{75pt} \left. - \frac{s-2m^2}{s\beta} \left[ \ln(\chi)\left[1+\ln(\chi)\right] + \DiLog(1-\chi^2) \right] \right\} + \mathcal O(\epsilon)\,.
\label{eq:soft-qed}
\end{align}
Note that we get double poles $1/\epsilon^2$ in the OK part (when soft and collinear poles coincide) that multiply the full $n$-dimensional Born amplitude $B_{\vec\kappa,\tQED}$. This leads to slightly different expressions for the \textit{single} poles compared to \cite{Hekhorn:2018ywm} as our $B_{\vec\kappa,\tQED}^{(1)}$ here differ from the expressions there.

\subsection{Initial Light Quark Contribution} \label{sec:PS.s4.q}
The diagrams shown in \FigureaRef{fig:FeynNLOq1}{fig:FeynNLOq2} introduce collinear poles as they depend on the propagator $t'$. The poles can be cast in a compact form by
\begin{align}
\frac{s_4}{4\pi(s_4+m^2)}\int\!d\Omega_n \,C_F A_{\vec \kappa,1} &=-\frac 1 {u_1}B_{\vec \kappa,\tQED}\left(x_1 k_1\right) P^{\delta}_{\kappa_3,\Pg\Pq}(x_1)\frac 2 \epsilon + O(\epsilon^0)
\end{align}
using again $x_1 = -u_1/(s'+t_1)$. For the current case of DIS ($Q^2 > 0$) the Coulomb subprocess $A_{\vec \kappa,2}$ does not contain collinear poles. Due to Furry's theorem we find $\int\!\dPSThree \,A_{\vec \kappa,3}=0$ (of course $A_{\vec \kappa,3}$ may still contribute to any more differential observable).

\subsection{Finite results} \label{sec:PS.s4.fin}
As shown above, the non-Abelian part of the gluon channel $R_{\vec\kappa,\tOK}$ and the Bethe-Heitler subprocess of the light quark channel $A_{\vec\kappa,1}$ contain collinear poles (see \FigureRef{fig:FeynNLOg4} and \FigureaRef{fig:FeynNLOq1}{fig:FeynNLOq2} respectively). We subtract these by applying the factorization formula \Eqref{eq:PartonicFact} and get finite results. 

\subsubsection{Initial State Gluon}
The final, finite partonic cross sections for the NLO gluon channel is split, as discussed, into two parts: first, the hard contributions given by
\begin{align}
{s'}^2\frac{d^2\sigma_{\vec \kappa,bb',\Pg}^{(1),H,fin}}{dt_1du_1} &=\alpha\alpha_s g^{\kappa_1}_{b,\PQ}g^{\kappa_2}_{b',\PQ} K_{\Pg\Pgg}N_CC_F\left[-\frac 2 {u_1}P^H_{\kappa_3,\Pg\Pg}(x_1)\right. \nonumber\\
 &\hspace{15pt}\left\{ B^{(0)}_{\vec \kappa,\tQED}\left(x_1 k_1\right) \left(\ln\left(\frac{s_4^2}{m^2(s_4+m^2)}\right)-\ln(\mu_F^2/m^2)\right) -2B^{(1)}_{\vec \kappa,\tQED}\left(x_1 k_1\right)\right\} \nonumber\\
 &\hspace{10pt}+C_A\frac{s_4}{2\pi(s_4+m^2)}\left(\int\!d\Omega_n\,R_{\vec \kappa,\tOK}\right)^{finite} \nonumber\\
 &\hspace{10pt}\left.+2C_F\frac{s_4}{2\pi(s_4+m^2)}\int\!d\Omega_4\,R_{\vec \kappa,\tQED}\right] \,. \label{eq:Hfinal}
\end{align}
The term proportional $B_{\vec\kappa,\tQED}^{(1)}$ together with terms in the finite part of the phase space integration that arise when a collinear pole ($1/\epsilon$) is multiplied by a $O(\epsilon)$ term in $R_{\vec\kappa,\tOK}$ contribute the additional information that can also be attributed to hat momenta.

Second, the soft plus virtual contributions given by
\begin{align}
{s'}^2\frac{d^2\sigma_{\vec \kappa,bb',\Pg}^{(1),S+V,fin}}{dt_1du_1} &=4\alpha\alpha_s g^{\kappa_1}_{b,\PQ}g^{\kappa_2}_{b',\PQ} K_{\Pg\Pgg}N_CC_FB^{(0)}_{\vec \kappa,\tQED}\delta(s'+t_1+u_1)\Bigg[C_A\ln^2(\Delta/m^2) \Bigg.\nonumber\\
 &\hspace{20pt} + \ln(\Delta/m^2)\Big\{\left(\ln(-t_1/m^2)-\ln(-u_1/m^2)-\ln(\mu_F^2/m^2)\right)C_A \Big. \nonumber\\
 &\hspace{80pt} \Big. - \frac{2m^2-s}{s\beta}\ln(\chi)(C_A-2C_F) - 2C_F\Big\} \nonumber\\
 &\hspace{20pt}\left. + \frac {\beta_0^{lf}}{4}\left(\ln(\mu_R^2/m^2)- \ln(\mu_F^2/m^2)\right) +  f^4_{\vec \kappa}(s',t_1)\right] \label{eq:SVfinal}
\end{align}
with functions $f^4_{\vec \kappa}$ that contain logarithms and dilogarithms with different, complicated arguments, but they do not depend on $\Delta,\mu_F^2,\mu_R^2$ nor $n_f$ and $\beta_0^{lf} = (11C_A - 2n_{lf})/3$.

\subsubsection{Initial State Light Quark}
The final, finite partonic cross section for the light quark initiated process is given by
\begin{align}
{s'}^2\frac{d^2\sigma_{\vec \kappa,bb',\Pq}^{(1),fin}}{dt_1du_1} &=\alpha\alpha_s K_{\Pq\Pgg}N_C\left[-\frac 1 {u_1} g^{\kappa_1}_{b,\PQ}g^{\kappa_2}_{b',\PQ} P^{\delta}_{\kappa_3,\Pg\Pq}(x_1)\right. \nonumber\\
 &\hspace{30pt}\left\{B^{(0)}_{\vec \kappa,\tQED}\left(x_1 k_1\right) \left(\ln\left(\frac{s_4^2}{m^2(s_4+m^2)}\right)-\ln(\mu_F^2/m^2)-2\partial_\epsilon E_{\kappa_3}(\epsilon=0)\right)\right.\nonumber\\
 &\hspace{100pt}\Big. -2 B^{(1)}_{\vec \kappa,\tQED}\left(x_1 k_1\right)\Big\} \nonumber\\
 &\hspace{20pt}+C_F\frac{s_4}{4\pi(s_4+m^2)}\left(\int\!d\Omega_n\, g^{\kappa_1}_{b,\PQ}g^{\kappa_2}_{b',\PQ} A_{\vec \kappa,1}\right)^{finite} \nonumber\\
 &\hspace{20pt}+C_F\frac{s_4}{4\pi(s_4+m^2)}\int\!d\Omega_4\, g^{\kappa_1}_{b,\Pq}g^{\kappa_2}_{b',\Pq} A_{\vec \kappa,2} \nonumber\\
 &\hspace{20pt}\left.+C_F\frac{s_4}{4\pi(s_4+m^2)}\int\!d\Omega_4\, g^{\kappa_1}_{b,\PQ}g^{\kappa_2}_{b',\Pq} A_{\vec \kappa,3} \right] \label{eq:qfinal}
\end{align}
where the $\partial_\epsilon E_{\kappa_3}$ part originates from the subtraction in \Eqref{eq:PartonicFact}. Again the term proportional $B_{\vec\kappa,\tQED}^{(1)}$ together with  terms $(1/\epsilon)\cdot\epsilon$ contribute the additional information that can also be attributed to hat momenta.

\section{Next-to-Leading Order $s_5$-centered Phase Space} \label{sec:PS.s5}
We move to the center-of-mass system of the produced heavy quark pair and choose the decomposition\cite{Harris:1995tu}
\begin{align}
q &= (q^0,0,0,\abs{\vec q})\\
k_1 &= k_1^0(1,0,\sin\psi,\cos\psi)\\
p_1 &= \frac {\sqrt{s_5}}{2}(1,\beta_5\sin\theta_2\sin\theta_1,\beta_5\sin\theta_2\cos\theta_1,\beta_5\cos\theta_1)\\
p_2 &= \frac {\sqrt{s_5}}{2}(1,-\beta_5\sin\theta_2\sin\theta_1,-\beta_5\sin\theta_2\cos\theta_1,-\beta_5\cos\theta_1)\\
k_2 &= (k_2^0,0,k_1\sin\psi,\abs{\vec q}+k_1^0\cos\psi)
\end{align}
with
\begin{align}
q_0 &= \frac {s+u'}{2\sqrt{s_5}},
&\abs{\vec q} &= \frac {1}{2\sqrt{s_5}}\sqrt{(s+u')^2-4s_5q^2},
&s_5' &= s_5-q^2\\
k_1^0 &= \frac{s_5-u'}{2\sqrt{s_5}},
&\cos\psi &= \frac{2k_1^0q^0-s'}{2k_1^0\abs{\vec q}},
&\beta_5 &= \sqrt{1-4m^2/s_5},\\
k_2^0 &= \frac{s-s_5}{2\sqrt{s_5}}
\end{align}
from which the name \enquote{$s_5$-centered} becomes obvious. Furthermore, we introduce\footnote{Following the notation of \cite{Harris:1995tu}, we define here a symbol $x$ which marks soft poles and is not to be confused with the Bjorken variable $x$. We also define here a symbol $y$ which marks collinear poles and is not to be confused with the Bjorken variable $y_{Bj}$ nor the rapidity of the heavy quark to be defined later on.}
\begin{align}
&\rho^* = \frac{4m^2-q^2}{s'} \leq x = \frac{s_5'}{s'} \leq 1 &&\text{and} &&-1\leq y\leq 1
\end{align}
where $y$ is the cosine of the angle between $\vec q$ and $\vec k_2$ in the system with $\vec q+\vec k_1 = \vec 0$. We find for the phase space\cite{Harris:1995tu}
\begin{align}
\dPSThree &= \frac {T_\epsilon}{2\pi} \left(\frac{{s'}^2}{s}\right)^{1+\epsilon/2}(1-x)^{1+\epsilon}(1-y^2)^{\epsilon/2}\dPSTwo^{(5)} dy \sin^{\epsilon}(\theta_2)d\theta_1d\theta_2
\end{align}
with $0\leq\theta_1\leq\pi, 0\leq \theta_2\leq \pi$ and
\begin{align}
S_\epsilon &= (4\pi)^{-2-\epsilon/2}\,,\\
T_\epsilon &= \frac{\Gamma(1+\epsilon/2)}{\Gamma(1+\epsilon)}S_\epsilon = \frac 1 {16\pi^2}\left(1 + \frac {\epsilon} 2(\gamma_E - \ln(4\pi)) + O(\epsilon^2)\right)\,,\\
\dPSTwo^{(5)} &= \frac{\beta_5 \sin(\theta_1)}{16\pi\Gamma(1+\epsilon/2)}\left(\frac{s_5\beta_5^2\sin^2(\theta_1)}{16\pi}\right)^{\epsilon/2}d\theta_1dx = \dPSTwo(s\rightarrow s_5) dx\,.
\end{align}
With the above decomposition, we find
\begin{align}
t' &= -2k_1\cdot k_2 = -\frac{s'}{2} \left(\frac{s'}{s}\right) (1-x)(1+y)\,, \\
u' &= q^2 - 2q\cdot k_2=q^2\left(\frac{s_5}{s}\right) - \frac{s'}{2}\left(\frac{s'}{s}\right) (1-x)(1-y)
\end{align}
which allow to observe and the soft ($x\to 1$) and collinear $(y\to -1)$ limit.

In the following, both, soft $(1-x)^{-1+\epsilon}$ and collinear factors $(1+y)^{-1+\epsilon}$ will appear and they can be replaced by generalized plus distributions\cite{Harris:1995tu,Mangano:1991jk,Frixione:1993dg}
\begin{align}
(1-x)^{-1+\epsilon} &\sim \left(\frac 1 {1-x}\right)_{\tilde\rho} + \epsilon \left(\frac{\ln(1-x)}{1-x}\right)_{\tilde \rho} + \delta(1-x)\left(\frac 1 \epsilon + 2\ln\tilde\beta + 2\epsilon\ln^2(\tilde\beta)\right) + O(\epsilon^2)\\
(1+y)^{-1+\epsilon} &\sim \left(\frac 1 {1+y}\right)_\omega + \delta(1+y)\left(\frac 1 \epsilon + \ln\omega\right) + O(\epsilon) \label{eq:collDist}
\end{align}
inside integration over smooth functions with $\tilde \beta = \sqrt{1-\tilde\rho}$. The distributions are defined by
\begin{align}
\int\limits_{\tilde\rho}^1\!\!dx\,f(x)\left(\frac 1 {1-x}\right)_{\tilde\rho} &= \int\limits_{\tilde\rho}^1\!\!dx\,\frac {f(x) - f(1)} {1-x}\\
\int\limits_{\tilde\rho}^1\!\!dx\,f(x)\left(\frac {\ln(1-x)} {1-x}\right)_{\tilde\rho} &= \int\limits_{\tilde\rho}^1\!\!dx\,\frac {f(x) - f(1)} {1-x}\ln(1-x)\\
\int\limits_{-1}^{-1+\omega}\!\!\!dy\,f(y)\left(\frac 1 {1+y}\right)_{\omega} &= \int\limits_{-1}^{-1+\omega}\!\!\!dy\,\frac {f(y)-f(-1)} {1+y}
\end{align}
with $\rho^*\leq\tilde\rho < 1$ and $0<\omega\leq 2$. If the integration does not include a singularity, the distribution sign ($\tilde\rho,\omega$) can be dropped and the terms become regular again. From an analytical point of view the results may not depend on the specific choice of the regularisation parameters $\tilde\rho$ and $\omega$ but for any numerical purpose they may influence the rate of convergence or stability. For numerical computations we also have to cut the poles out of the integrations
\begin{align}
&\int\limits_{\rho^*}^1\!dx \rightarrow \int\limits_{\rho^*}^{1-\tilde\delta_x}\!\!\!dx &&\text{and} &&\int\limits_{-1}^1\!dy \rightarrow \int\limits_{-1+\delta_y}^{1}\!\!dy\,.
\end{align}
If not stated otherwise we use as our default setup
\begin{align}
\tilde\rho &= \rho^* + \tilde x(1-\rho^*)\,\text{with}\, \tilde x=0.8\,, &\omega &= 1.0\,,\\
\tilde\delta_x &= \delta_x\cdot\left(\frac{s-4m^2}{s}\right)^{2/3} \,\text{with}\,\delta_x=\num{1e-6}\,, &\delta_y &=\num{7e-6}\,.
\end{align}
For the regularisation of the soft poles ($x\to 1$) we use the same dependence on external variables as \texttt{HVQDIS}\cite{Harris:1995tu,Harris:1995pr,Harris:1997zq} from which we also inherited our default choice of parameters.

With the $\tilde\rho$ prescription at hand, we can give the needed Altarelli-Parisi splitting kernels $(\Delta)P_{\Pg\Pg}^{\tilde\rho}(x)$ and $(\Delta)P_{\Pq\Pg}^{\tilde\rho}(x)$\cite{Harris:1995tu,Vogelsang:1995vh,Altarelli:1977zs}
\begin{align}
P_{  \hat F_2,\Pg\Pg}^{\tilde\rho}(x) &= P_{\hat F_L,\Pg\Pg}^{\tilde\rho}(x) = P_{z\hat F_3,\Pg\Pg}^{\tilde\rho}(x) = P_{\Pg\Pg}^{\tilde\rho}(x)\nonumber\\
 &= 2C_A\left[\left(\frac{x}{1-x}\right)_{\tilde\rho} + \frac{1-x}{x} + x(1-x) \right] + \delta(1-x)\left(\frac{b_0^{lf}}{2} + 4C_A\ln(\tilde\beta)\right),\\
P_{2z\hat g_1,\Pg\Pg}^{\tilde\rho}(x) &= P_{\hat g_4,\Pg\Pg}^{\tilde\rho}(x) = P_{ \hat g_L,\Pg\Pg}^{\tilde\rho}(x) = \Delta P_{\Pg\Pg}^{\tilde\rho}(x)\nonumber\\
 &= 2C_A\left[\left(\frac{1}{1-x}\right)_{\tilde\rho} - 2x + 1 + \epsilon(x-1) \right]\nonumber\\
 &\hspace{10pt} + \delta(1-x)\left(\frac{b_0^{lf}}{2} + 4C_A\ln(\tilde\beta) - \epsilon\frac{N_C}{6}\right),\\
P_{  \hat F_2,\Pq\Pg}^{\tilde\rho}(x) &= P_{\hat F_L,\Pq\Pg}^{\tilde\rho}(x) = P_{z\hat F_3,\Pq\Pg}^{\tilde\rho}(x) = P_{\Pq\Pg}^{\tilde\rho}(x) = C_F\left[\frac{1+(1-x)^2}{x} + \frac x 2 \epsilon\right],\\
P_{2z\hat g_1,\Pq\Pg}^{\tilde\rho}(x) &= P_{\hat g_4,\Pq\Pg}^{\tilde\rho}(x) = P_{ \hat g_L,\Pq\Pg}^{\tilde\rho}(x) = \Delta P_{\Pq\Pg}^{\tilde\rho}(x)= C_F\left[2-x + (x-1) \epsilon \right]\,.
\end{align}
For future reference we decompose the kernels into a soft and a hard part, and, at the same time, by their $\epsilon$ dependence 
\begin{align}
(\Delta)P_{jk}^{\tilde\rho}(x) &= \left((\Delta)P_{jk,H}^{\tilde\rho,(0)}(x) + \epsilon \cdot (\Delta)P_{jk,H}^{\tilde\rho,(1)}(x)\right) \nonumber\\
 &\hspace{15pt} + \delta(1-x)\left((\Delta)P_{jk,S}^{\tilde\rho,(0)} + \epsilon \cdot (\Delta)P_{jk,S}^{\tilde\rho,(1)}\right) + O(\epsilon^2)\,.
\end{align}
Note that these definitions here are closely related to their counter parts in the $s_4$-centered phase space (\Eqsrref{eq:PFgg}{eq:Pggq}) and furthermore, the quark-to-gluon splittings match exactly as they have a smooth soft limit. Nevertheless, we need them here at $O(\epsilon)$ as this information will become relevant, as discussed below.

\subsection{Single Gluon Radiation} \label{sec:PS.s5.g}
We shift the occurring soft ($x\rightarrow 1$) and collinear ($y\rightarrow -1$) poles from the gluonic matrix elements to the phase space by dividing by $t'\propto(1+y)(1-x)$ and $u'-q^2s_5/s\propto(1-x)$
\begin{align}
\dPSThreegp &= \frac{\dPSThree}{t'(u'-q^2s_5/s)} = \dPSThree \cdot \left(\frac {2s}{{s'}^2}\right)^2\frac 1 {(1-x)^2(1-y)(1+y)}\\
 &= \frac {2T_\epsilon}{\pi} \left(\frac {{s'}^2} s\right)^{-1+\epsilon/2} (1-x)^{-1+\epsilon}(1-y^2)^{-1+\epsilon/2}\dPSTwo^{(5)}dy \sin^\epsilon(\theta_2)d\theta_2\,,\\
{M_{\vec\kappa,bb'}^{(1),\Pg}}' &= t'(u'-q^2s_5/s)M_{\vec\kappa,bb'}^{(1),\Pg}\,.
\end{align}

We get for the partonic cross sections
\begin{align}
d\sigma_{\vec\kappa,bb'}^{(1),\Pg} &= \frac{1}{2s'}\frac {K_{\Pg\Pgg}E_k(\epsilon)} 2 {M_{\vec\kappa,bb'}^{(1),\Pg}}' \dPSThreegp \,.
\end{align}
which we can split with the given distributions into three pieces\cite{Harris:1995tu}
\begin{align}
d\sigma_{\vec \kappa,bb'}^{(1),\Pg} &= d\sigma_{\vec \kappa,bb'}^{(1),\Pg,S} + d\sigma_{\vec \kappa,bb'}^{(1),\Pg,\hat c-} + d\sigma_{\vec \kappa,bb'}^{(1),\Pg,f} \label{eq:PS.s5.sigg}
\end{align}
corresponding to the soft ($d\sigma_{\vec \kappa,bb'}^{(1),\Pg,S} \sim \delta(1-x)$), the collinear ($d\sigma_{\vec \kappa,bb'}^{(1),\Pg,\hat c-} \sim \delta(1+y)$) and the finite parts ($d\sigma_{\vec \kappa,bb'}^{(1),\Pg,f} \sim \left(\frac 1 {1-x}\right)_{\tilde \rho}\left(\frac 1 {1+y}\right)_{\omega}$). Note, there is only a single collinear contribution ($y\rightarrow -1$) for $q^2 < 0$.

We find for the integrated soft part
\begin{align}
d\sigma_{\vec \kappa,bb'}^{(1),\Pg,S} &= \frac 1 {2s'}\frac {K_{\Pg\Pgg}E_k(\epsilon)} 2  M_{\vec \kappa,bb'}^{(1),\Pg,S} \dPSTwo \,,\\
M_{\vec \kappa,bb'}^{(1),\Pg,S} &= 8g^4\mu_D^{-\epsilon}e^2 g^{\kappa_1}_{b,\PQ}g^{\kappa_2}_{b',\PQ} N_C C_F C_\epsilon\left( C_A \tilde S^5_{\tOK} + 2C_F \tilde S^5_{\tQED}\right) B_{\vec \kappa,\tQED}
\end{align}
where the full expressions for $\tilde S^5_{\tOK/\tQED}$ can be found in \cite{Harris:1995tu} and their poles are given by
\begin{align}
\tilde S^5_{\tOK} &= 2\left(\frac 4 {\epsilon^2} + \left(\ln(-t_1/m^2) + \ln(-u_1/m^2) - \frac{2m^2-s}{s\beta}\ln(\chi) + 4\ln(\tilde\beta)\right)\frac 2 {\epsilon}\right) + O(\epsilon^0)\,,\\
\tilde S^5_{\tQED} &=-2\cdot \left(1 - \frac{2m^2-s}{s\beta}\ln(\chi)\right)\frac 2 \epsilon + O(\epsilon^0)\,.
\end{align}

\subsection{Initial Light Quark Contribution} \label{sec:PS.s5.q}
We shift the occurring collinear ($y\rightarrow -1$) poles from the light quark matrix elements to the phase space by dividing by $t'\propto(1+y)(1-x)$
\begin{align}
\dPSThreeqp &= \frac{dPS_3}{t'} = \dPSThree \cdot \left(\frac {2s}{{s'}^2}\right)\frac 1 {(1-x)(1+y)}\\
 &= \frac {T_\epsilon}{\pi} \left(\frac {{s'}^2} s\right)^{\epsilon/2} (1-x)^{\epsilon}(1-y)^{\epsilon/2}(1+y)^{-1+\epsilon/2}\dPSTwo^{(5)}dy \sin^\epsilon(\theta_2)d\theta_2\,,\\
{M_{\vec \kappa,bb'}^{(1),\Pq}}' &= t'M_{\vec\kappa,bb'}^{(1),\Pq}\,.
\end{align}

We get for the partonic cross section
\begin{align}
d\sigma_{\vec\kappa,bb'}^{(1),\Pq} &= \frac{1}{2s'}\frac {K_{\Pq\Pgg}} 2 {M_{\vec\kappa,bb'}^{(1),\Pq}}' \dPSThreeqp
\end{align}
which we split into two pieces\cite{Harris:1995tu}
\begin{align}
d\sigma_{\vec \kappa,bb'}^{(1),\Pq} &= d\sigma_{\vec \kappa,bb'}^{(1),\Pq,\hat c-} + d\sigma_{\vec \kappa,bb'}^{(1),\Pq,f} \label{eq:PS.s5.sigq}
\end{align}
corresponding, as before, to the collinear part ($d\sigma_{\vec \kappa,bb'}^{(1),\Pq,\hat c-} \sim \delta(1+y)$) and the finite part ($d\sigma_{\vec \kappa,bb'}^{(1),\Pq,f}\sim\left(\frac 1 {1+y}\right)_\omega$). Note, there is only a single collinear contribution ($y\rightarrow -1$) in $A_{\vec \kappa,1}$ for $q^2 < 0$, i.e.\ neither $A_{\vec\kappa,2}$ nor $A_{\vec\kappa,3}$ do contain any poles.

\subsection{Finite results} \label{sec:PS.s5.fin}
Via \Eqref{eq:collDist} collinear poles enter into $\dPSThreegp$ and $\dPSThreeqp$ (see \FigureRef{fig:FeynNLOg4} and \FigureaRef{fig:FeynNLOq1}{fig:FeynNLOq2} respectively). We remove them by applying the factorization formula \Eqref{eq:PartonicFact} and we then get finite results.

\subsubsection{Initial State Gluon}
The final, finite cross sections for the gluonic part can be split into three pieces
\begin{align}
d\sigma_{\vec\kappa,bb'}^{\Pg,(1),fin} &= d\sigma_{\vec \kappa,bb'}^{\Pg,(1),s+v} + d\sigma_{\vec \kappa,bb'}^{\Pg,(1),c-} + d\sigma_{\vec \kappa,bb'}^{\Pg,(1),f}
\end{align}
analog to \Eqref{eq:PS.s5.sigg}.

The first piece
\begin{align}
d\sigma_{\vec \kappa,bb'}^{\Pg,(1),s+v} &= \alpha_s^2\alpha_{em} g_{b,\PQ}^{\kappa_1}g_{b',\PQ}^{\kappa_2}\cdot \frac 1 {2s'} \frac {K_{\Pg\Pgg}E_{\kappa_3}(0)}{2} N_C C_F B_{\vec \kappa,\tQED} \cdot 2^7 \left[ 4C_A\ln^2(\tilde\beta) \right.\nonumber\\
 &\hspace{40pt} +\ln(\tilde\beta)\Big( 	2\left(\ln(-t_1/m^2)+\ln(-u_1/m^2)-\ln(\mu_F^2/m^2)\right)C_A \Big. \nonumber\\
 &\hspace{80pt}\Big. - 2C_F + \frac{s-2m^2}{s\beta}\ln(\chi)(C_A-2C_F) \Big) \nonumber\\
 &\hspace{40pt}\left. + \frac {\beta_0^{lf}}{4}\left(\ln(\mu_R^2/m^2)- \ln(\mu_F^2/m^2)\right) + f_{\vec \kappa}^5(s,\theta_1) \right] \dPSTwo
\end{align}
collects all $2\to 2$-phase space contributions, as there is no dependence neither on $x,y$ nor $\theta_2$. These are soft contributions $d\sigma_{\vec \kappa,bb'}^{\Pg,(1),S}$, virtual contributions $d\sigma_{\vec \kappa,bb'}^{\Pg,(1),V}$, self-energies on external legs and factorization contributions. The functions $f_{\vec \kappa}^5$ contain all remaining finite contributions, mainly expressed a lot of logarithms and dilogarithms, but they do not depend on $\tilde\rho,\mu_F^2,\mu_R^2$ nor $n_f$ nor $\beta_0^{lf} = (11C_A - 2n_{lf})/3$.

The second piece
\begin{align}
d\sigma_{\vec \kappa,bb'}^{\Pg,(1),c-} &= \alpha_s^2\alpha_{em} g_{b,\PQ}^{\kappa_1}g_{b',\PQ}^{\kappa_2}\cdot \frac 1 {2xs'} \frac {K_{\Pg\Pgg}E_{\kappa_3}(0)}{2} \cdot N_C C_F B_{\vec \kappa,\tQED}(xk_1) \cdot 2^6\pi \nonumber\\
 &\hspace{20pt}\cdot \left[ (1-x)P_{\kappa_3,\Pg\Pg,H}^{\tilde\rho,(0)}(x)\left(\frac{1}{(1-x)_{\tilde\rho}}\left(\ln(s'/\mu_F^2) + \ln(s'/s) + \ln(\omega/2)\right) \right.\right.\nonumber\\
 &\hspace{120pt} \left.\left. + 2\left(\frac{\ln(1-x)}{1-x}\right)_{\tilde\rho} \right) + 2P_{\kappa_3,\Pg\Pg,H}^{\tilde\rho,(1)}(x)\right] \dPSTwo^{(5)}
\end{align}
collects all collinear contributions (note the absence of $y$) encoded by the Altarelli-Parisi splitting kernels. The $P_{\kappa_3,\Pg\Pg,H}^{\tilde\rho,(1)}(x)$ contribute the additional information that also could be attributed to hat momenta.

And the third and last piece
\begin{align}
d\sigma_{\vec \kappa,bb'}^{\Pg,(1),f} &= 2 \left(\frac 1 {4\pi}\right)^4 \frac 1 {2s'} \frac {K_{\Pg\Pgg}E_{\kappa_3}(0)}{2} \frac{s\beta_5}{(s')^2} \left(\frac 1 {1-x}\right)_{\tilde\rho} \left(\frac 1 {1+y}\right)_\omega \frac 1 {1-y} \nonumber\\
 &\hspace{40pt} \cdot {M_{\vec \kappa,bb'}^{(1),\Pg}}' dx\,dy\sin(\theta_1)d\theta_1\,d\theta_2\\
 &= \alpha_S^2\alpha_{em} g_{b,\PQ}^{\kappa_1}g_{b',\PQ}^{\kappa_2} \cdot K_{\Pg\Pgg} N_C C_F \frac{s\sin(\theta_1)}{\pi(s')^3} \frac{\beta_5}{1-y} \left(\frac 1 {1-x}\right)_{\tilde\rho} \left(\frac 1 {1+y}\right)_\omega \nonumber\\
 &\hspace{40pt} \cdot t'(u'-q^2s_5/s) \left(C_A R_{\vec\kappa,\tOK} + 2C_F R_{\vec \kappa,\tQED}\right) dx\,dy\,d\theta_1\,d\theta_2
\end{align}
collects all finite contributions from the matrix elements regularized by the simultaneous appearance of the $\tilde\rho$ and $\omega$ prescriptions. These prescriptions require soft and/or collinear counter terms which can be computed beforehand. The soft counter terms ($x\to 1$) can be obtained from \Eqsaref{eq:EOK}{eq:EQED} and we find
\begin{align}
&\lim\limits_{x\to 1} t'(u'-q^2s_5/s) E_{\tOK}\nonumber\\
&= \frac{2 s'^2 (y-1)}{s^2 \left(1-y \beta  \cos(\theta_1)+\sqrt{1-y^2} \beta  \cos(\theta_2) \sin(\theta_1)\right) }\nonumber\\
&\hspace{10pt}\cdot\frac{s y - 2m^2 (1+y) - s y \beta^2\cos^2(\theta_1) +s \beta^2 \sqrt{1-y^2} \cos(\theta_2)\sin(\theta_1)\cos(\theta_1)}{1+y \beta\cos(\theta_1)-\sqrt{1-y^2} \beta  \cos(\theta_2)\sin(\theta_1)},\\
&\lim\limits_{x\to 1} t'(u'-q^2s_5/s) E_{\tQED}\nonumber\\
&= -\frac{2(4m^2-s) s'^2 (y^2-1)}{s^2 \left(1-y \beta  \cos(\theta_1)+\sqrt{1-y^2} \beta  \cos(\theta_2) \sin(\theta_1)\right)^2 }\nonumber\\
&\hspace{5pt}\cdot\frac{-1 + y^2 \cos^2(\theta_1) -2y \sqrt{1-y^2} \cos(\theta_2)\sin(\theta_1)\cos(\theta_1) + (1-y^2)\cos^2(\theta_2)\sin^2(\theta_1)}{\left(1+y \beta\cos(\theta_1)-\sqrt{1-y^2} \beta  \cos(\theta_2)\sin(\theta_1)\right)^2}\,.
\end{align}
The collinear counter terms ($y\to -1$) related to $R_{\vec \kappa,\tQED}$ vanish (as expected) and the expressions for $R_{\vec\kappa,\tOK}$ are shown in \AppendixRef{sec:Appendix.s5}. The soft-collinear counter terms can be obtained from the expressions above and we find
\begin{align}
\lim\limits_{x\to 1}\lim\limits_{y\to -1} t'(u'-q^2s_5/s) \left(C_A R_{\vec\kappa,\tOK} + 2C_F R_{\vec \kappa,\tQED}\right) &= C_A \frac{4 s'^2}{s} B_{\vec \kappa,\tQED}\,. \label{eq:scR}
\end{align}

\subsubsection{Initial State Light Quark}
The final, finite light quark process can be split into two pieces
\begin{align}
d\sigma_{\vec\kappa,bb'}^{\Pq,(1),fin} &= d\sigma_{\vec \kappa,bb'}^{\Pq,(1),c-} + d\sigma_{\vec \kappa,bb'}^{\Pq,(1),f}
\end{align}
analog to \Eqref{eq:PS.s5.sigq}.

The first piece
\begin{align}
d\sigma_{\vec\kappa,bb',\Pq}^{(1),c-} &= \alpha_s^2\alpha_{em}g_{b,\PQ}^{\kappa_1}g_{b',\PQ}^{\kappa_2} \cdot  \frac 1 {2xs'} \frac {K_{\Pq\Pgg}} 2 \cdot N_C B_{\vec\kappa,\tQED}(xk_1) \cdot 2^5\pi \dPSTwo^{(5)} \nonumber\\
 &\hspace{20pt}\cdot \left(P_{\kappa_3,\Pg\Pq}^{(0)}(x)\left(\ln(s'/\mu_F^2) + \ln(s'/s) + \ln(\omega/2) + 2\ln(1-x)\right) + 2P_{\kappa_3,\Pg\Pq}^{(1)}(x)\right)
\end{align}
collects all collinear contributions. Again, the $P_{\kappa_3,\Pq\Pg}^{\tilde\rho,(1)}(x)$ contribute the additional information that also could be attributed to hat momenta.

The second piece
\begin{align}
d\sigma_{\vec\kappa,bb',\Pq}^{(1),f} &= - \left(\frac 1 {4\pi}\right)^4 \frac 1 {2s'} \frac {K_{\Pq\Pgg}} 2  \beta_5 \left(\frac 1 {1+y}\right)_\omega {M_{\vec\kappa,bb'}^{(1),\Pq}}' dxdy\sin(\theta_1)d\theta_1d\theta_2\\
 &= \alpha_s^2\alpha_{em} \cdot K_{\Pq\Pgg} N_C C_F \left(-\frac{1}{4\pi}\right) \frac{\beta_5\sin(\theta_1)}{s'}  \left(\frac 1 {1+y}\right)_\omega \nonumber\\
 &\hspace{40pt} \cdot t'\left(g_{b,\PQ}^{\kappa_1}g_{b',\PQ}^{\kappa_2}A_{\vec\kappa,1} + g_{b,\Pq}^{\kappa_1}g_{b',\Pq}^{\kappa_2} A_{\vec\kappa,2} + g_{b,\Pq}^{\kappa_1}g_{b',\PQ}^{\kappa_2} A_{\vec\kappa,3}\right) dxdyd\theta_1d\theta_2
\end{align}
collects all finite contributions from the matrix elements regularized by the $\omega$ prescription. The collinear counter terms ($y\to -1$) of $A_{\vec\kappa,1}$ is given in \AppendixRef{sec:Appendix.s5}.

\chapter{Partonic Results} \label{chap:Partonic}
\chapterquote{And I'm not insane, my mother had me tested.}{The Parton}

Using the partonic matrix elements of \ChapterRef{chap:ME} and the expressions for the full inclusive phase space of \ChapterRef{chap:PS}, we investigate the structure on a partonic level, i.e.\ without any convolutions of PDFs. We decompose the contributions to the total partonic cross sections by their respective order in QCD pertubation theory, their scale dependence and their electro-weak charge structure and write
\begin{align}
\sigma^{bb'}_{\vec \kappa,\Pg} &= \frac{\alpha_{em}\alpha_s}{m^2} \cdot g^{\kappa_1}_{b,\PQ}g^{\kappa_2}_{b',\PQ} \left\{ c_{\vec\kappa,\Pg}^{(0)} + 4\pi\alpha_s\left[c_{\vec\kappa,\Pg}^{(1)} + \bar c_{\vec\kappa,\Pg}^{(1),F}\ln(\mu_F^2/m^2)  + \bar c_{\vec\kappa,\Pg}^{(1),R}\ln(\mu_R^2/m^2) \right]\right\}\,,\label{eq:sigPg}\\
\sigma^{bb'}_{\vec \kappa,\Pq} &= \frac{\alpha_{em}\alpha_s}{m^2} \cdot 4\pi\alpha_s\left[g^{\kappa_1}_{b,\PQ}g^{\kappa_2}_{b',\PQ} \left(c_{\vec \kappa,\Pq}^{(1)} + \bar c_{\vec\kappa,\Pq}^{(1),F}\ln(\mu_F^2/m^2) \right) + g^{\kappa_1}_{b,\Pq}g^{\kappa_2}_{b',\Pq} d_{\vec \kappa,\Pq}^{(1)}\right]\,. \label{eq:sigPq}
\end{align}
All partonic coefficient functions $c/d_{\vec\kappa,j}$ depend on two scales, e.g.\ $c_{\vec\kappa,\Pg}^{(0)}(s/m^2,Q^2/m^2)$, for which we defined a set of transformations (see \Eqsrref{eq:partonicVars1}{eq:partonicVars3}). For the sake of readability we already have suppressed these dependencies in \Eqsaref{eq:sigPg}{eq:sigPq} and we will continue to do so. For the diagrams we define additionally the distance from threshold
\begin{align}
\eta = \frac{s - 4m^2}{4m^2} = 1/\rho -1 \,.
\end{align}
The partonic coefficient functions are a central result of this work, as they are required for the most important case of the fully inclusive structure functions where no property of the heavy quark pair is observed. However, they are not sufficient for distributions as they have all informations integrated out.

\begin{figure}[ht]
\begin{center}
\includegraphics[width=0.45\textwidth]{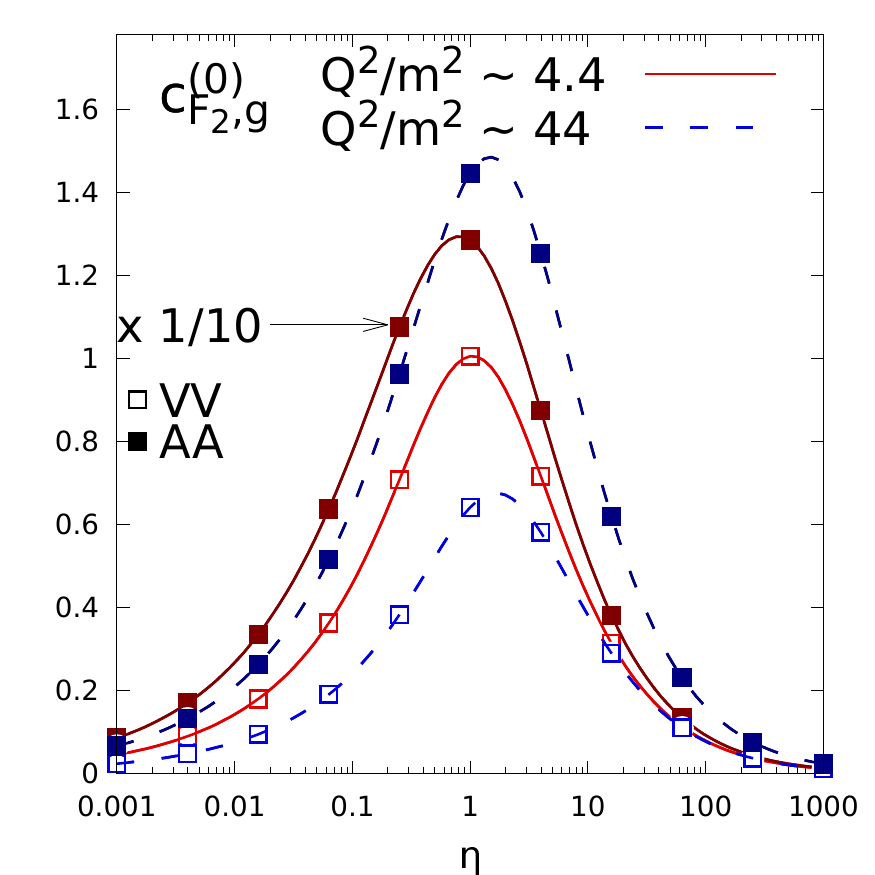}
\includegraphics[width=0.45\textwidth]{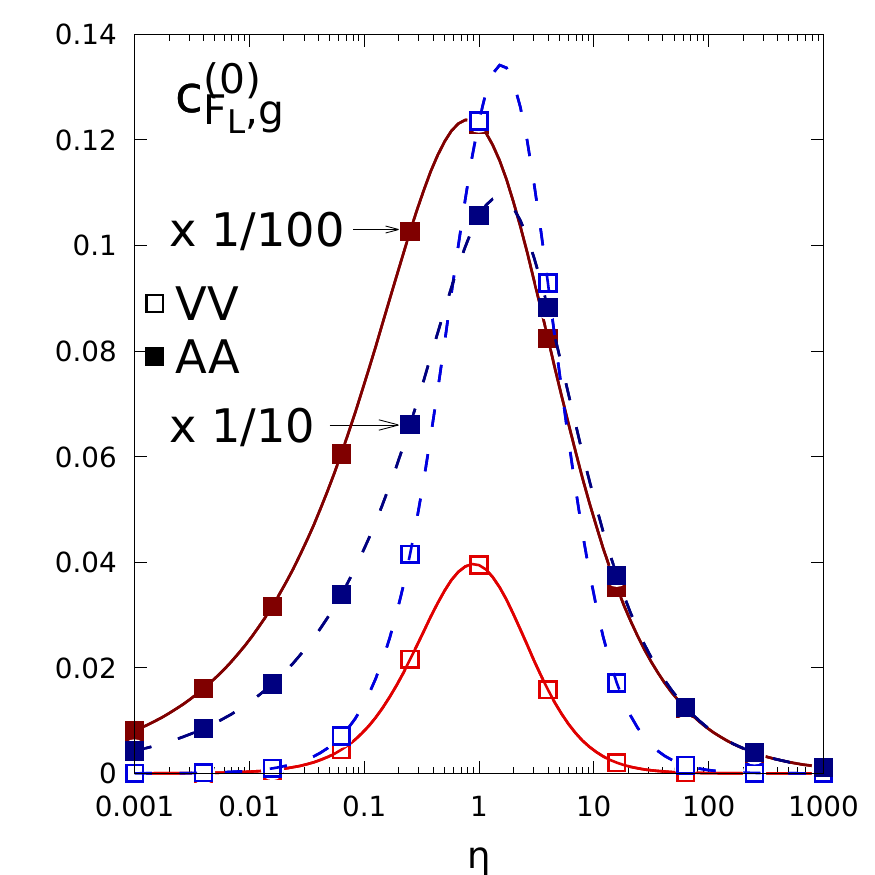}
\end{center}
\caption{LO coefficient functions $c_{\vec\kappa,\Pg}^{(0)}$ for the unpolarized structure function $\kappa_3 = \{F_2,F_L\}$. Note that $c_{\lVVx2g1,\Pg}^{(0)} = c_{\lAAx2g1,\Pg}^{(0)}$ (for a plot see \cite{Hekhorn:2018ywm}) and $c_{(\tV,\tA,\kappa_3),\Pg}^{(0)} = 0$ for $\kappa_3\in\{z\hat F_3,\hat g_4,\hat g_L\}$. Encoded by color and line type we show two different values of the ratio $Q^2/m^2$ and by open (closed) symbols the contributions from the vector-vector (axial vector-axial vector) coupling. The plots for the vector-vector parts can also be found in \cite{Laenen1993162}. Note that some of the axial vector-axial vector coefficient functions had to be rescaled to fit into a single diagram.} \label{fig:cg0-F}
\end{figure}
In \FigureRef{fig:cg0-F} we show the LO coefficient functions $c_{\vec\kappa,\Pg}^{(0)}$. They can be computed from \Eqref{eq:LOxs} with analytical tools and the corresponding expressions are given in \AppendixRef{sec:Appendix:Partonic:cg0}. From these analytical results we obtain a list of interesting properties: the parity violating $B_{(\tV,\tA,\kappa_3),\tQED}$ (\Eqsrref{eq:BQEDVAxF3}{eq:BQEDVAgL}) are anti-symmetric with respect to the symmetric $t_1$-integration of LO phase space (\Eqref{eq:PS2s4}) so they do not contribute to the full inclusive cross sections and we find
\begin{equation}
c_{(\tV,\tA,\kappa_3),\Pg}^{(0)} = 0 \quad \kappa_3\in\{z\hat F_3,\hat g_4,\hat g_L\}\,. \label{eq:cg0PV}
\end{equation}
This is in accordance with the results available in literature for both, the unpolarized\cite{Leveille:1978px} and the polarized\cite{Vogelsang:1990ka} case. We find a simple V-A-structure for the polarized coefficient function only, i.e.\ $c_{\lVVx2g1,\Pg}^{(0)} = c_{\lAAx2g1,\Pg}^{(0)}$, but the axial vector-axial vector counterpart for the unpolarized coefficient functions behave differently. Most prominently they diverge for small virtualities $c_{\lAAF2,\Pg}^{(0)},c_{\lAAFL,\Pg}^{(0)} \sim m^2/Q^2$ for $Q^2\to 0$ which is mirrored in \FigureRef{fig:cg0-F} by a rescaling of these functions. Also for the longitudinal projection $\hat F_L$ the vector-vector part and the axial vector-axial vector part do not share the same behaviour near threshold ($s\to 4m^2 \Leftrightarrow \eta\to 0$), i.e.\ $c_{\lVVFL,\Pg}^{(0),\tThr} \sim \beta^3$ but $c_{\lAAFL,\Pg}^{(0),\tThr}\sim \beta$.

Next, we turn to the NLO scaling functions $\bar c_{\vec\kappa,\Pg}^{(1),F}$, $\bar c_{\vec\kappa,\Pg}^{(1),R}$ and $\bar c_{\vec\kappa,\Pq}^{(1),F}$ given by the \Eqsrref{eq:Hfinal}{eq:qfinal}. Their simplified structure allows to give analytical results for these coefficient functions as well, see \AppendixRef{sec:Appendix:Partonic:cgBar1} and \AppendixRef{sec:Appendix:Partonic:cqBarF1} respectively. The arguments concerning the symmetry of $B_{(\tV,\tA,\kappa_3),\tQED}$, as discussed above, can also be applied to the PV NLO scaling functions and we find
\begin{align}
\bar c_{(\tV,\tA,\kappa_3),\Pg}^{(1),F} = \bar c_{(\tV,\tA,\kappa_3),\Pg}^{(1),R} = \bar c_{(\tV,\tA,\kappa_3),\Pq}^{(1),F} = 0 \quad \kappa_3\in\{z\hat F_3,\hat g_4,\hat g_L\}\,.
\end{align}
For convenience we define the common combination $\bar c_{\vec\kappa,\Pg}^{(1)} = \bar c_{\vec\kappa,\Pg}^{(1),F} + \bar c_{\vec\kappa,\Pg}^{(1),R}$, which accounts the common choice of a unified scale $\mu_F^2=\mu_R^2$.

\begin{figure}[ht]
\begin{center}
\includegraphics[width=0.33\textwidth]{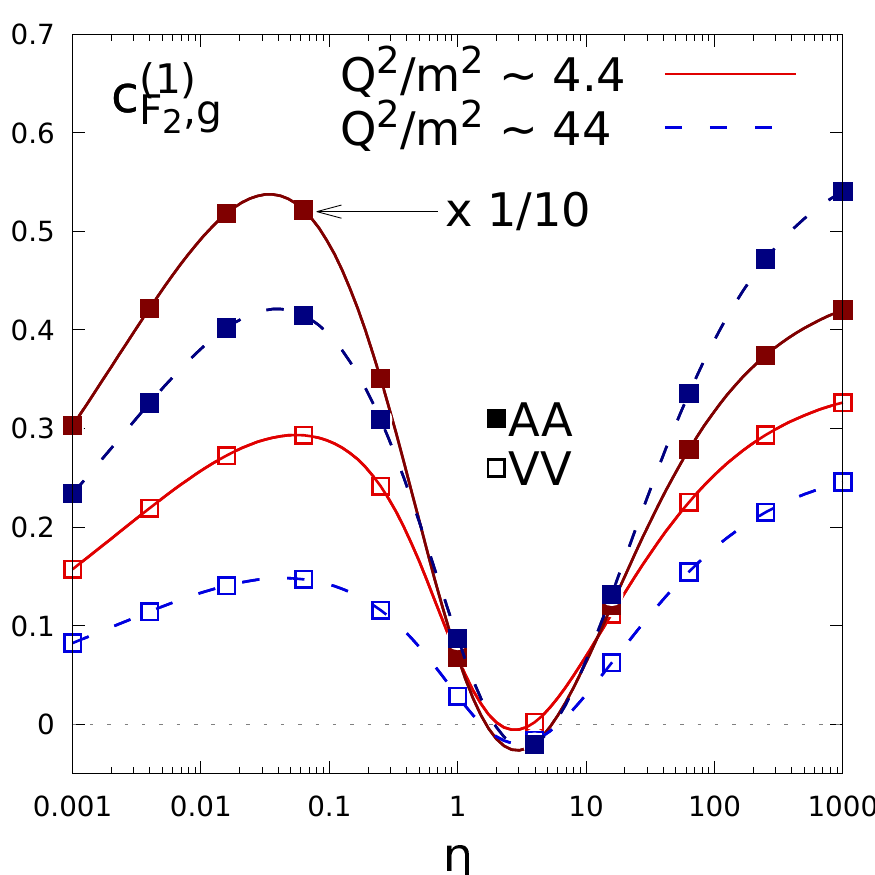}%
\includegraphics[width=0.33\textwidth]{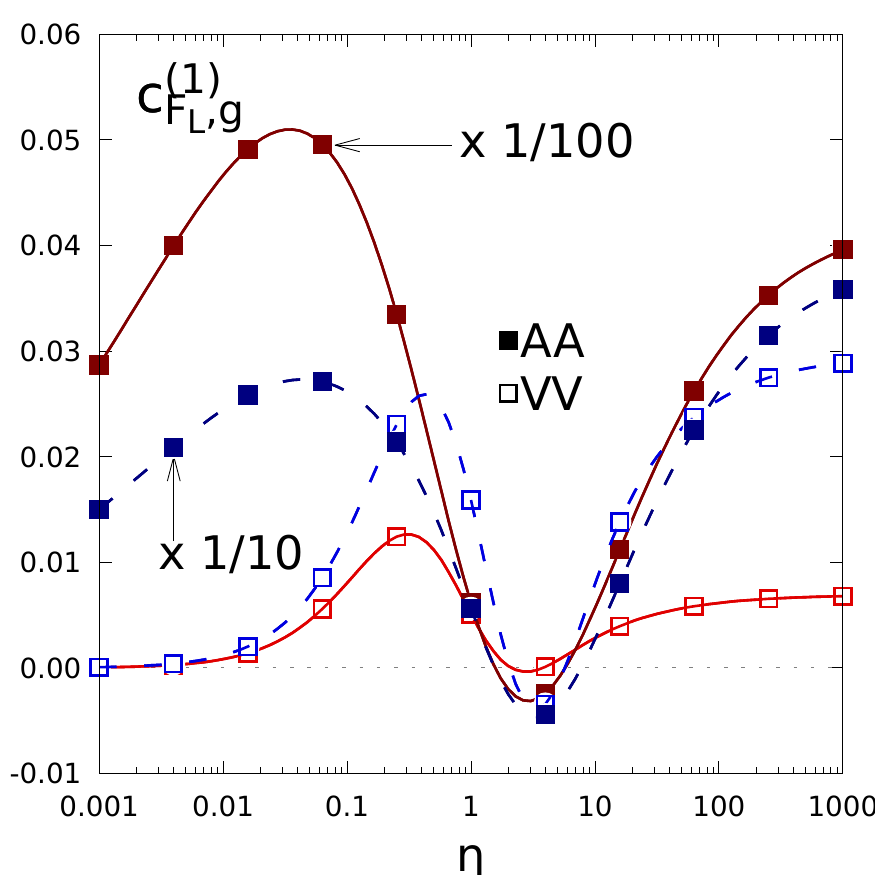}%
\includegraphics[width=0.33\textwidth]{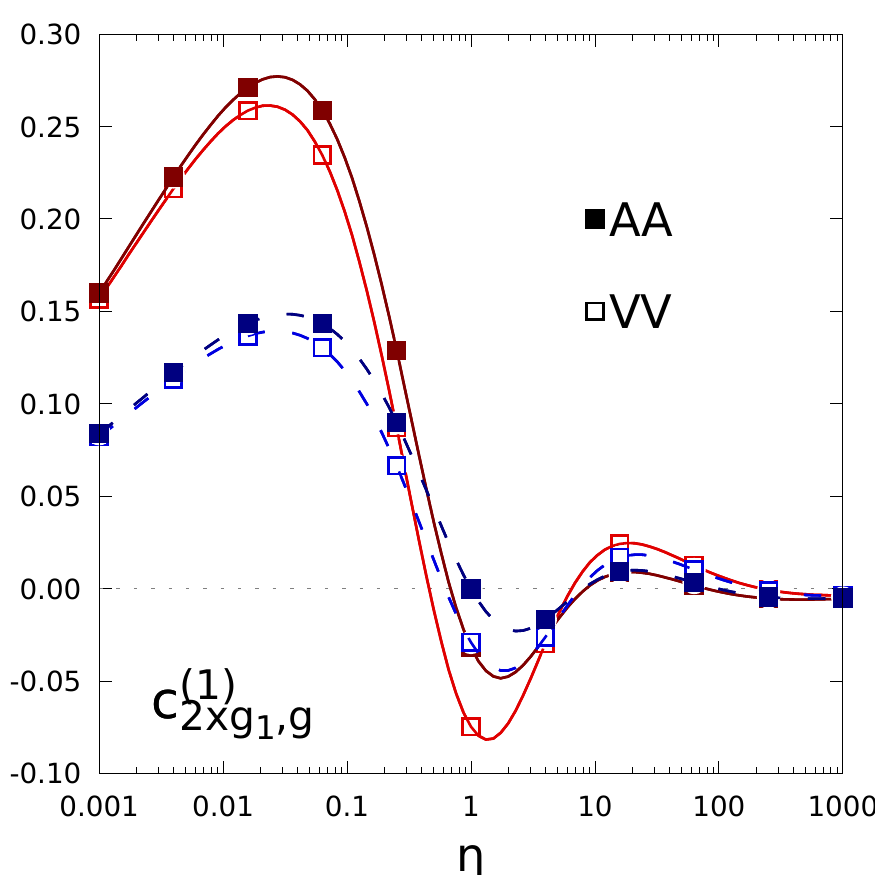}%
\end{center}
\caption{NLO gluon coefficient functions $c_{\vec\kappa,\Pg}^{(1)}$ for the parity conserving structure function $\kappa_3 = \{F_2,F_L,2xg_1\}$. Encoded by color and line type we show two different values of the ratio $Q^2/m^2$ and by open (closed) symbols the contributions from the vector-vector (axial vector-axial vector) coupling. The plots for the vector-vector parts can also be found in \cite{Hekhorn:2018ywm,Laenen1993162}. Note that some of the axial vector-axial vector coefficient functions had to be rescaled to fit into a single diagram.} \label{fig:cg1-PC}
\end{figure}
In \FigureRef{fig:cg1-PC} we show the NLO gluon coefficient functions $c_{\vec\kappa,\Pg}^{(1)}$. We can see that the described behaviour for $Q^2\to 0$ is inherited to the NLO case of the unpolarized, parity conserving structure functions $F_2$ and $F_L$. For the polarized case of $g_1$, now, the axial vector-axial vector coupling is no longer identical to the vector-vector coupling and near threshold ($s\to 4m^2$) we can quantify this breaking analytically, see \AppendixRef{sec:Appendix:Partonic:cg1}. As was pointed out in \cite{Hekhorn:2018ywm}, we found an error in the threshold approximation of $F_L$ in \cite{Laenen1993162} demonstrating once more the need for independent cross checks to improve and fix calculations\footnote{more over, this error was also inherited into \cite{Riemersma:1994hv}, from there into \texttt{OPENQCDRAD}\cite{OPENQCDRAD} and from there into \texttt{xFitter}\cite{Bertone:2017tig}}.

\begin{figure}[ht]
\begin{center}
\includegraphics[width=0.33\textwidth]{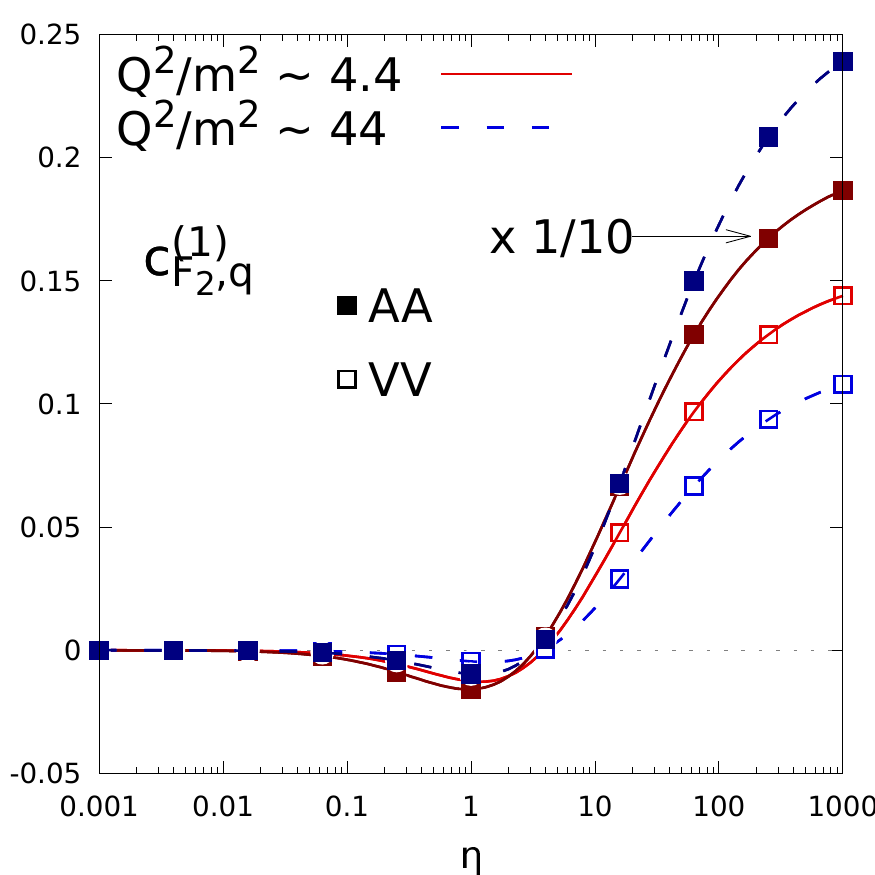}%
\includegraphics[width=0.33\textwidth]{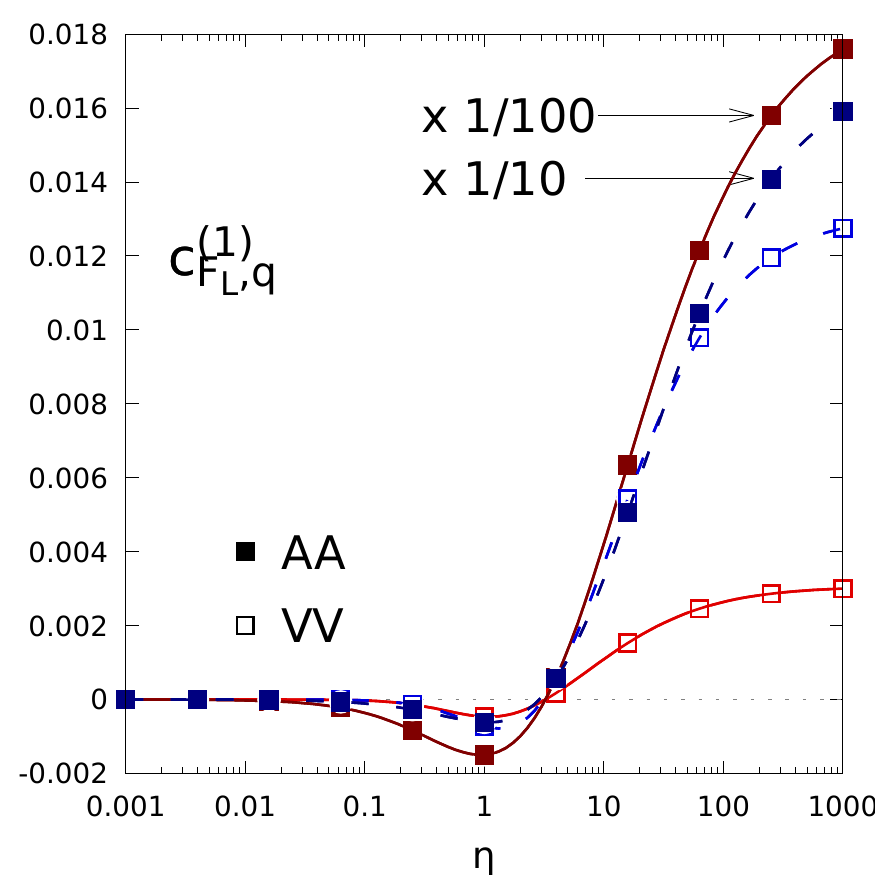}%
\includegraphics[width=0.33\textwidth]{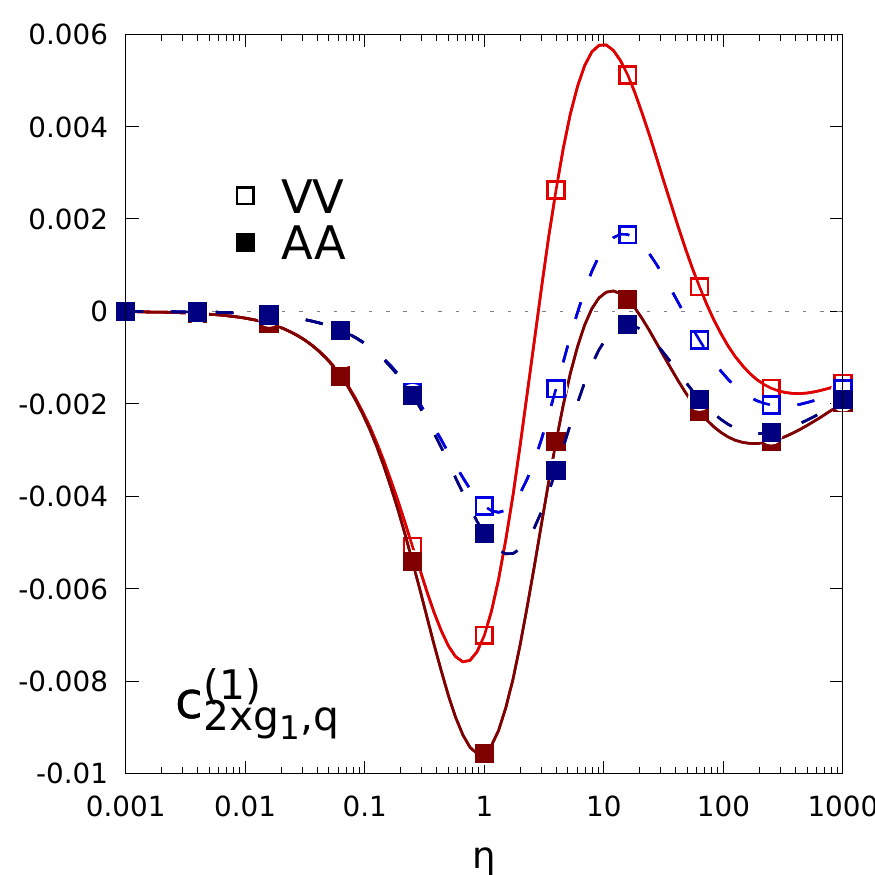}%
\end{center}
\caption{NLO Bethe-Heitler coefficient functions $c_{\vec\kappa,\Pq}^{(1)}$ for the parity conserving structure function $\kappa_3 = \{F_2,F_L,2xg_1\}$. Encoded by color and line type we show two different values of the ratio $Q^2/m^2$ and by open (closed) symbols the contributions from the vector-vector (axial vector-axial vector) coupling. The plots for the vector-vector parts can also be found in \cite{Hekhorn:2018ywm,Laenen1993162,Harris:1995tu}. Note that some of the axial vector-axial vector coefficient functions had to be rescaled to fit into a single diagram.} \label{fig:cq1-PC}
\end{figure}
In \FigureRef{fig:cq1-PC} we show the NLO Bethe-Heitler coefficient functions $c_{\vec\kappa,\Pq}^{(1)}$. All arguments stated in the discussing of \FigureRef{fig:cg1-PC} can also be applied to these coefficient functions, i.e.\ the prominent low $Q^2$ behaviour, the V-A breaking for the polarized coefficient function related to $g_1$ and the given threshold approximation in \AppendixRef{sec:Appendix:Partonic:cq1}.

The NLO Coulomb coefficient functions $d_{\vec\kappa,\Pq}^{(1)}$ have a simplified analytic structure, so we can give analytic results for them in the \AppendixRef{sec:Appendix:Partonic:dq1}. We find
\begin{align}
d_{\lVVF2,\Pq}^{(1)} &= d_{\lAAF2,\Pq}^{(1)} = -d_{\lVAg4,\Pq}^{(1)}\,,\\
d_{\lVVFL,\Pq}^{(1)} &= d_{\lAAFL,\Pq}^{(1)} = -d_{\lVAgL,\Pq}^{(1)}\,,\\
d_{\lVVx2g1,\Pq}^{(1)} &= d_{\lAAx2g1,\Pq}^{(1)} = d_{\lVAxF3,\Pq}^{(1)}\,. 
\end{align}
So, here, the coefficient functions for the parity conserving $F_2,F_L$ and $2xg_1$ do show the V-A structure of the massless case, i.e., e.g.\ $d_{\lVVF2,\Pq}^{(1)} = d_{\lAAF2,\Pq}^{(1)}$, as the exchanged boson couples to the massless quark $\Pq$. They also follow the simple substitution rule of \Eqref{eq:LorentzSubRule} given by Lorentz symmetry. For a graphical visualisation of these functions we refer the reader to \cite{Hekhorn:2018ywm,Laenen1993162,Harris:1995tu}, as we do not generate new contributions here.

\chapter{Hadronic Results} \label{chap:Hadronic}
\chapterquote{And then, of course, I've got this terrible pain in all the diodes down my left side.}{Higgs01}

We now return to the hadronic setup, i.e.\ the scattering of the bosons with the proton
\begin{equation}
\HepProcess{b(q) + \Pp(P) \to \PaQ(p_2) + X[\PQ]} \quad b\in\{\Pgg,\PZ\}
\end{equation}
that is used in a measurable observables. As outlined in \ChapterRef{chap:DIS} the factorization theorem provides the necessary link between the physical results and the pQCD calculations of \ChapterRef{chap:Partonic} by the non-perturbative PDFs. The factorization formula is given by
\begin{align}
H^{bb'}_{\vec \kappa, j}(x,Q^2,m^2) &= \frac{Q^2}{4\pi^2\alpha_{em}} \int\limits_x^{z_{max}}\frac {dz}{z} f_{\kappa_3,j}\left(\xi=\frac x z,\mu_F^2\right) \cdot \sigma^{bb'}_{\vec \kappa,j}(s'=Q^2/z),\quad j\in\{\Pg,\Pq\}\label{eq:HtoPartonic}
\end{align}
with the kinematic limit $z_{max} = Q^2/(4m^2+Q^2)$ that is approached near threshold ($z\to z_{max}\Leftrightarrow s\to 4m^2$). The recombination of the contributions from the different bosons $H^{bb'}_{\vec \kappa, j}$ to the full neutral current (NC) structure functions $H^{\tNC}$ has been given in \Eqsaref{eq:HPCphZ}{eq:HPVphZ}. We are then left with the task to obtain a detailed description of the structure functions as they encode all physical information of the scattering. Therefore, we will provide a first study of all structure functions and provide some differentials thereof.

Actually, the factorization holds on a more differential level (see \Eqref{eq:DiffFact}) and we get access in the $s_4$-centered approach to the inclusive distributions. These distributions may be differential in the transverse momentum of the heavy anti-quark $p_T$ and/or its rapidity $y$. The corresponding identities are provided in \AppendixRef{sec:Appendix.s4.Hadronic}. We can use these inclusive distributions to verify the implementation of our Monte-Carlo approach and to compare (e.g.)\ the accuracy of the rapidity distributions. In the following, we will discuss some results obtained for the two different scenarios we have set up in \SectionRef{sec:DIS.PDF}: charm quark production from \SectionRef{sec:DIS.PDF.c} is discussed in \SectionRef{sec:Hadronic.c} and bottom quark production from \SectionRef{sec:DIS.PDF.b} is discussed in \SectionRef{sec:Hadronic.b}.

To investigate the PV structure functions $xF_3$,$g_4$ and $g_L$ we have to refer to their rapidity distributions as neither, the full inclusive structure functions not their distributions towards the transverse momentum $p_T$ contribute at LO. The former can be seen directly from the vanishing of the relevant partonic coefficient functions (see \Eqssref{eq:cg0PV}) and the latter is proven in the \AppendixRef{sec:Appendix.s4.TransverseMomentum}. Recall that in the case of the PV structure functions, only the VA-coupling contributes and thus we can account for the inclusion of the $\PZ$-boson exchange by applying the simple factors of \Eqref{eq:HPVphZ}. The kinematical limit of the rapidity distributions is given by\cite{Laenen:1992xs}
\begin{align}
S_h &= (q+P)^2\,, &\beta_h &= \sqrt{1-\frac {4m^2}{S_h}}\,, &|y| &\leq \atanh(\beta_h)\,.
\end{align}

All distributions will be defined in the center-of-mass system of the exchanged boson $b(q)$ and the proton $\Pp(P)$, i.e.\ $\vec q+ \vec P =\vec 0$, and with the proton aligned along the positive $z$-axis, $\vec P \sim \hat e_z$. To fix the weak contribution we further assume an electron in the initial state ($\Pl = \Pem$) with positive helicity ($\lambda=1$). The used (p)PDF sets have been introduced in \SectionRef{sec:DIS.PDF} and we will always use the running coupling $\alpha_s(\mu_R^2)$ of the respective unpolarized PDF set as provided by LHAPDF\cite{LHAPDF6}. Our default choice, if not indicated specifically, for the renormalization and factorization scale is $\mu_F^2 = \mu_R^2 = 4m^2+Q^2$.

In \TableRef{tab:numeric} we give our default numerical setup and list the used routines and their most important input parameters. Note that for the MC approach we need an integration routine that passes \textit{explicitly} the associated MC weight as this is required to fill the produced histograms\footnote{for which we use $\texttt{gsl\_histogram}$\cite{GSL}} correctly. Whenever possible we reduce the number of integrated dimensions in the inclusive case to improve the accuracy. Furthermore, we apply to all VEGAS integrations an additional adaptation to ensure a convergence, i.e.\ $|\chi^2_{MC}-1|\leq \num{0.5}$.
\begin{table}[ht]
\begin{tabularx}{\textwidth}{|l|l|X|}
\hline
 \multirow{2}{*}{inclusive} & $H^{inc,LO}$ & doubly-adaptive general-purpose quadrature $\texttt{gsl\_integration\_cquad}$\cite{GSL} in 1 dimension\\ \cline{2-3}
 & $H^{inc,NLO}$ & adaptive VEGAS integration $\texttt{gsl\_monte\_vegas\_integrate}$\cite{GSL} in 4 dimensions with $\num{70}\si{\kilo}$ points and 5 iterations and in 3 dimensions with $\num{50}\si{\kilo}$ points\\ \hline
 \multirow{2}{*}{MC} & $H^{MC,LO}$ & adaptive VEGAS integration $\texttt{Dvegas}$\cite{Kauer:2001sp,Kauer:2002sn} in 2 dimensions with $\num{500}\si{\kilo}$ points and 20 iterations\\ \cline{2-3}
 & $H^{MC,NLO}$ & adaptive VEGAS integration $\texttt{Dvegas}$ in 5 dimensions with $\num{5}\si{\mega}$ points and 20 iterations\\ \hline
\end{tabularx}
\caption{Default numerical routines and parameters}
\label{tab:numeric}
\end{table}

\section{Charm Quark Production} \label{sec:Hadronic.c}
The most important inclusive and fully differential formulae and graphs for polarized charm quark production have already been shown and discussed in \cite{Hekhorn:2018ywm,Hekhorn:2018aio} and \cite{paper2} respectively. Therefore, the discussion here is restricted to some additional properties.

\begin{figure}[ht]
\begin{center}
\includegraphics[width=0.7\textwidth]{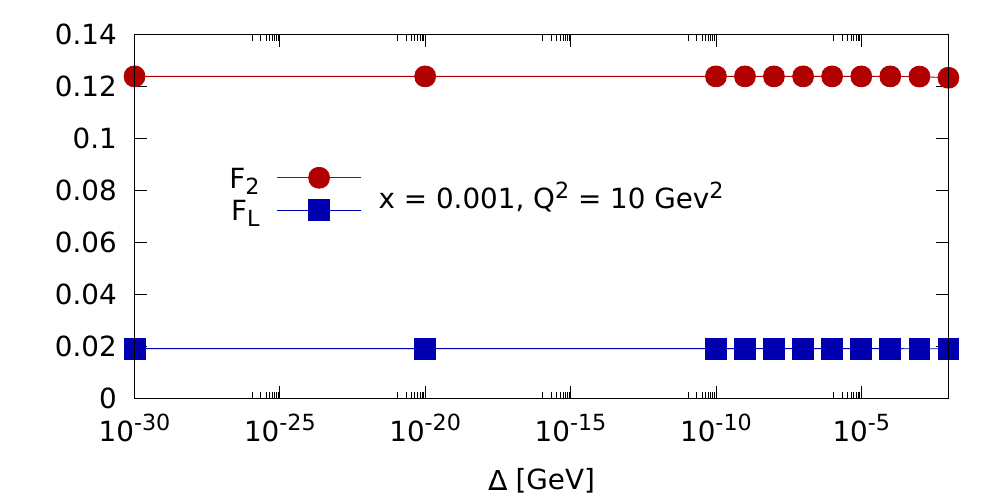}
\end{center}
\caption{Dependence of the unpolarized structure functions $H^{\Pqc}$ for $H=F_2$ (red circles) and $H=F_L$ (blue squares) on the (unphysical) phase space slicing parameter $\Delta$ at $x=0.001$ and $Q^2 = \SI{10}{\GeV^2}$} \label{fig:Fc-Delta}
\end{figure}
As a first step, we investigate the accuracy of our two phase space decompositions which are implemented in two separate numerical codes. In \FigureRef{fig:Fc-Delta} we show the dependence of the unpolarized structure functions $F_2^{\Pqc}$ and $F_L^{\Pqc}$ on the phase space slicing parameter $\Delta$ for the inclusive code. As outlined in \cite{Hekhorn:2018ywm}, we use the identity of \cite{Bojak:2000eu} to rewrite the terms of $\delta(s_4)\ln^k(\Delta/m^2)$ in \Eqref{eq:SVfinal} into terms $\Theta(s_4-\Delta) f(s_4)$ that can be merged with the integration for the hard parts. This procedure eliminates the unphysical $\Delta$-dependence in a large parameter region and thus it allows to obtain a stable result. When implementing the prescription, one has to consider the entire factorization formula \Eqref{eq:HtoPartonic} (or its differential equivalent \Eqref{eq:DiffFact}) as the parton fraction $\xi$ depends on $s_4$ as well. Our default value of $\Delta=\SI[retain-unity-mantissa = false]{1e-6}{\GeV}$ is taken over from \cite{Bojak:2000eu}.

\begin{figure}[ht]
\begin{center}
\includegraphics[width=0.99\textwidth]{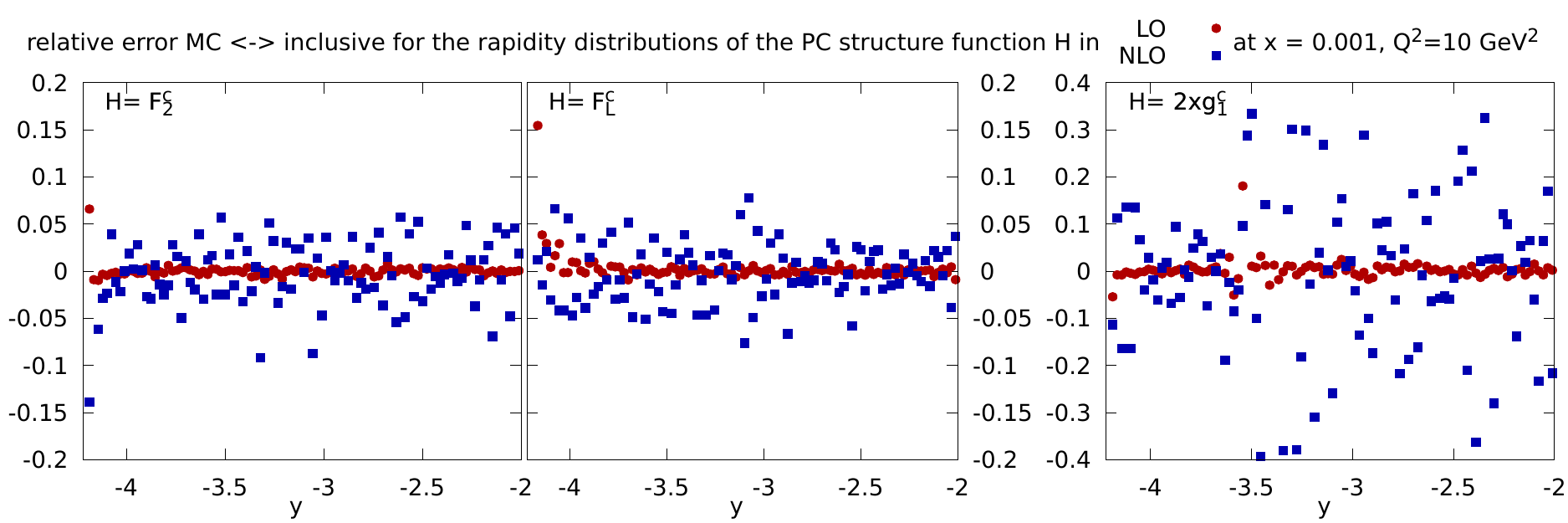}
\end{center}
\caption{Relative error for the rapidity distributions of the PC structure functions $\frac{dH^{MC}/dy - dH^{inc}/dy}{dH^{inc}/dy}$ at LO (red circles) and NLO (blue squares) accuracy at $x=\num{0.001}$ and $ Q^2=\SI{10}{\GeV^2}$ for $H=F_2^{\Pqc}$ (left), $H=F_L^{\Pqc}$ (middle), and, $H=2xg_1^{\Pqc}$(right). } \label{fig:Fc-y-PC-relE}
\end{figure}
Next, we give the relative error of the rapidity distributions obtained by our Monte-Carlo (MC) code towards the inclusive calculations in \FigureRef{fig:Fc-y-PC-relE}. For the unpolarized structure functions $dF_2^{\Pqc}/dy$ and $dF_L^{\Pqc}/dy$ we achieve mostly a $\sim 1\%$ accuracy at LO and $\sim 5\%$ at NLO. For the polarized case of $2x\cdot dg_1^{\Pqc}/dy$ the situation is more difficult as it involves a more complex integration, i.e.\ there are subtle cancellations in different phase space regions, as both, the pPDF and the coefficient functions are not strictly positive functions. Overall, we find a sufficient agreement between our routines. Comparisons like this may serve as a first error analysis for the MC code, as there is no straightforward error obtained for the correlated distributions.

\begin{figure}[ht]
\begin{center}
\includegraphics[width=0.99\textwidth]{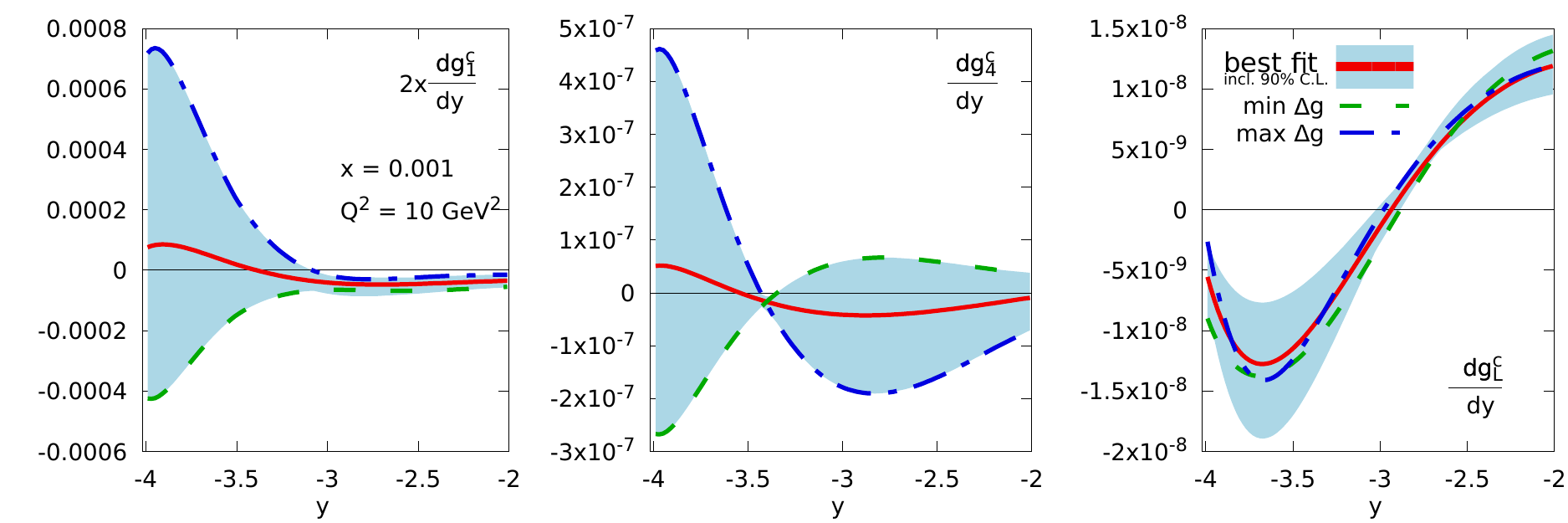}
\end{center}
\caption{Rapidity distributions of the polarized structure functions $dH/dy$ at $x=\num{0.001}$ and $Q^2=\SI{10}{\GeV^2}$ for $H=2xg_1^{\Pqc}$ (left), $H=g_4^{\Pqc}$ (middle), and, $H=g_L^{\Pqc}$(right) for the \DSSV pPDF set (see \FigureRef{fig:DSSV-xg}). } \label{fig:Fc-y-DSSV}
\end{figure}
In \FigureRef{fig:Fc-y-DSSV} we demonstrate the dependence of the rapidity distributions of the polarized structure functions on the pPDFs. We mainly find the same conclusions as in our paper \cite{paper2}: the marked variations of min/max $\Delta\Pg$ build mostly the envelope to the distributions and can be separated clearly. The contributions to $dg_L^{\Pqc}/dy$ are mostly driven by the non-singlet contributions linked to $d_{\lVAgL,\Pq}^{(1)}$ due to the vanishing of $B_{\lVAgL,\tQED}$ and thus depend stronger on the polarized quark distributions rather than the polarized gluon distribution. Note, that the absolute values of the different structure functions have a strong hierarchical ordering.

In experiments it is often convenient to observe a spin asymmetry instead of the structure functions themselves. We define the spin asymmetry $A_1^{\Pqc}$ in the fully inclusive case by
\begin{align}
A_1^{\Pqc}(x,Q^2) = \frac{g_1^{\Pqc}(x,Q^2)}{F_1^{\Pqc}(x,Q^2)} = \frac{2xg_1^{\Pqc}(x,Q^2)}{F_2^{\Pqc}(x,Q^2)-F_L^{\Pqc}(x,Q^2)} \label{eq:A1inc}
\end{align}
and it was studied extensively in our first papers \cite{Hekhorn:2018ywm,paper2,Hekhorn:2018aio}. The idea of defining such a ratio relies on the expectation, that some uncertainties are common to both terms, numerator and denominator, and cancel in the ratio. This expectation is not always fulfilled, as we did show in \cite{Hekhorn:2018ywm}. The representation in \Eqref{eq:A1inc} confirms again the usefulness of our chosen normalizations of the structure functions, i.e.\ e.g.\ to consider $2xg_1$ as fundamental structure function instead of just $g_1$. We extend the definition of the spin asymmetry to a differential case to study distributions and define
\begin{align}
A_{1,x_T}^{\Pqc}(x,Q^2,x_T) = \frac{dg_1^{\Pqc}/dx_T(x,Q^2,x_T)}{dF_1^{\Pqc}/dx_T(x,Q^2,x_T)}
\end{align}
with the transverse momentum fraction $x_T = p_T/p_T^{max}$. The kinematical limit of the transverse momentum of the heavy anti-quark $p_T^{max}$ is given by\cite{Laenen:1992xs}
\begin{align}
p_T^{max} &= \frac{\sqrt{S_h}}{2}\beta_h\,.
\end{align}
Due to the finite mass $m$, we also obtain a finite limit for the distributions in the limit $p_T\to 0$ as the relevant transverse mass $m_T$ is given by $m_T^2 = m^2+p_T^2$.

\begin{figure}[ht]
\begin{center}
\includegraphics[width=0.8\textwidth]{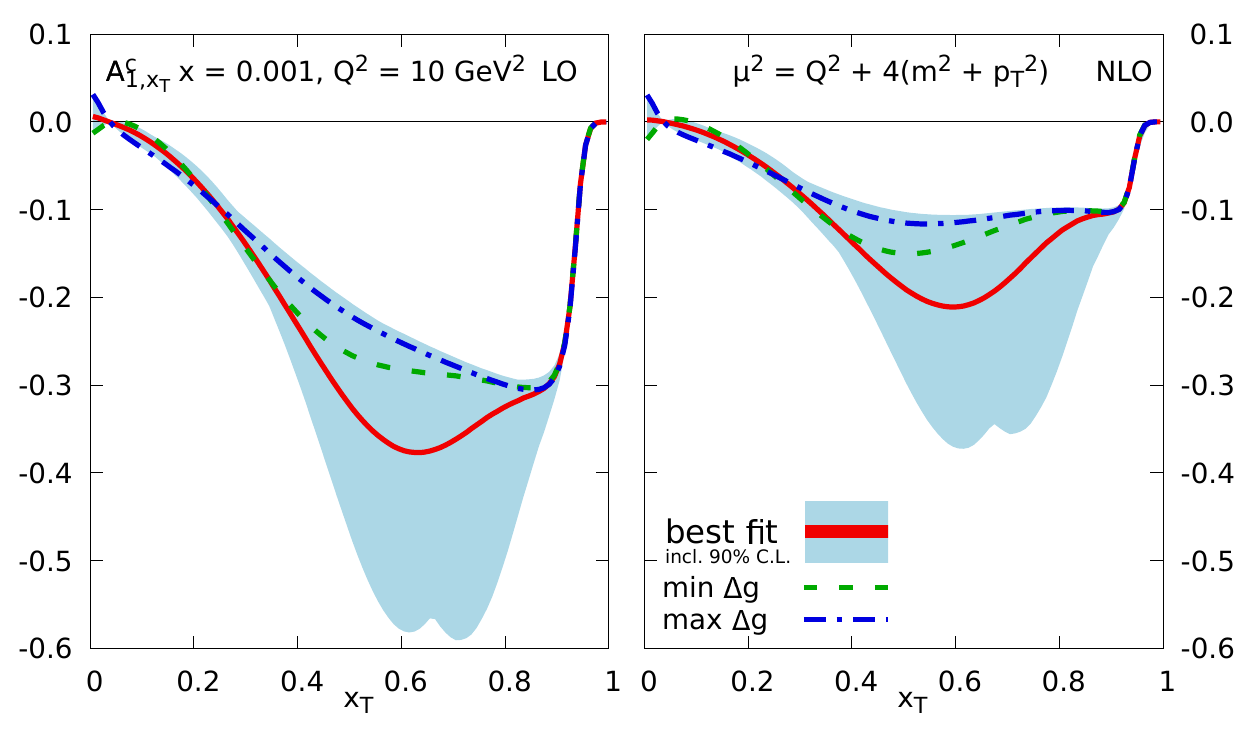}
\end{center}
\caption{Spin asymmetry $A_{1,x_T}^{\Pqc}$ dependent on the transverse momentum fraction $x_T$ at $x=\num{0.001}$ and $Q^2=\SI{10}{\GeV^2}$ at LO (left) and NLO (right) for the \DSSV pPDF set (see \FigureRef{fig:DSSV-xg}). } \label{fig:A1-xt-DSSV}
\end{figure}
In \FigureRef{fig:A1-xt-DSSV} we investigate the dependence of the transverse momentum fraction dependent spin asymmetry $A_{1,x_T}^{\Pqc}$ on the pPDFs. The case for small $p_T$ (or equivalently small $x_T$) is shown in \cite{Hekhorn:2018ywm}, where we observed the usual behaviour of the two marked variations building the envelope to the distributions. This behaviour is not visible for larger $x_T$, where light quark induced contributions become important. Note that the spin asymmetry, as shown in \FigureRef{fig:A1-xt-DSSV}, is largely enhanced in the medium $x_T$ region, but an actual measurement in this region is at the moment out of reach as both structure functions, $g_1^{\Pqc}$ and $F_1^{\Pqc}$, drop rapidly for increasing $x_T$. The unified behavior of the distributions in the large $x_T$ region is linked to the closing of the available phase space. The overall size of the uncertainties gets reduced when increasing the perturbation order as expected. We have set the renormalization and factorization scale to $\mu^2 = Q^2 + 4(m^2 +p_T^2)$ where a $p_T$ dependence is strictly required in order to get physical results in the high $x_T$ region. This behaviour hints to potentially dangerous logarithms of the type $\ln(p_T^2/m^2)$ that spoil the convergence of the perturbation theory. These contributions are typically counteracted by fragmentation functions, which we have dropped here, as they can resum such terms. By introducing a $p_T$ dependence on the renormalization scale we can mimic to some extend this resummation as the running coupling itself does resum certain logarithms\footnote{as can be seen from an expansion of \Eqref{eq:runas}}.

\section{Bottom Quark Production} \label{sec:Hadronic.b}
Using the decomposition of \Eqref{eq:HPCphZ} we define the relative contributions of the bosons to the full NC PC structure functions by
\begin{align}
R^{\Pgg\PZ}_{H}(x,Q^2) &= \frac{(g_{\PZ,\Pe}^V-g_{\PZ,\Pe}^A)   \eta_{\Pgg\PZ}   H^{\Pgg\PZ}(x,Q^2)}{H^{NC}(x,Q^2)}\,, \nonumber\\
R^{\PZ\PZ}_{H}(x,Q^2) &= \frac{(g_{\PZ,\Pe}^V-g_{\PZ,\Pe}^A)^2 \eta_{\Pgg\PZ}^2 H^{\PZ\PZ}(x,Q^2)}{H^{NC}(x,Q^2)}
&\text{with}\,H\in\{F_2,F_L,2xg_1\} \label{eq:Rbb}
\end{align}
to investigate their influence to the full NC corrections. The respective contributions are almost independent of the chosen Bjorken-$x$ if we are in safe distance to the threshold region, i.e.\ in the hadronic case $x<z_{max}$. Near threshold the different behaviour of the partonic coefficient functions, as outlined in \ChapterRef{chap:Partonic} and \AppendixRef{sec:Appendix.Partonic}, come into effect. Note, that the described small $Q^2$ behaviour of the partonic coefficient functions is cured in the hadronic setup by the factor $\eta_{\Pgg\PZ}^2$ ensuring a physical, regular $Q^2$ dependence.

\begin{figure}[ht]
\begin{center}
\includegraphics[width=0.7\textwidth]{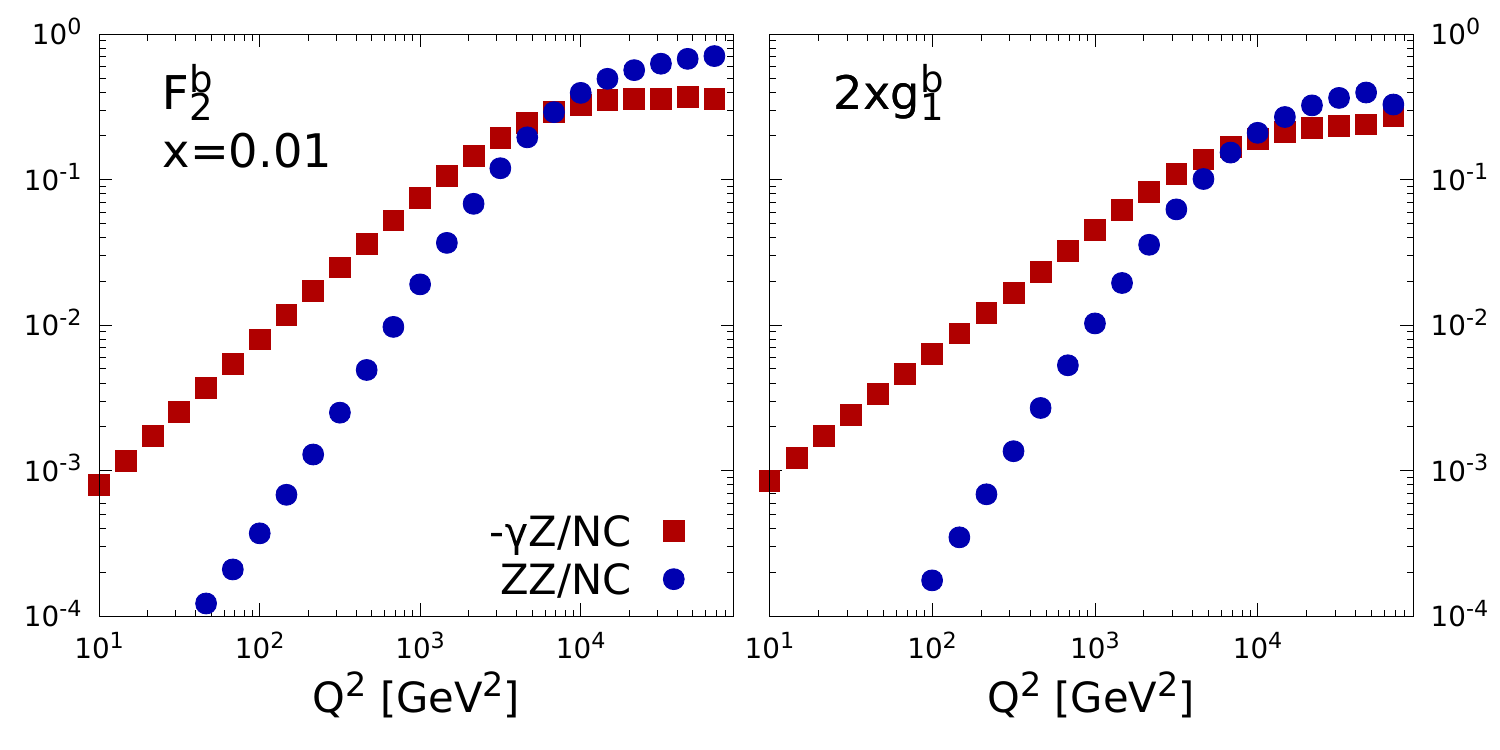}
\end{center}
\caption{Relative contributions of the bosons to the full NC structure functions (see \Eqssref{eq:Rbb}): $R^{\Pgg\PZ}_H(x=\num{0.01},Q^2)$ (red squares) and $R^{\PZ\PZ}_H(x=\num{0.01},Q^2)$ (blue circles) for $F_2^{\Pqb}$ (left) and $2xg_1^{\Pqb}$(right). Note the minus sign for the $\Pgg\PZ$ contributions.} \label{fig:Fb-PC-R}
\end{figure}
In \FigureRef{fig:Fb-PC-R} we plot the relative contribution as function of $Q^2$. Note, that we have to show the negative ratio of $R^{\Pgg\PZ}$ as the $\Pgg\PZ$-interference terms do contribute with a negative sign. Due to different $Q^2$ behaviour of the axial vector-axial vector coupling the two parts do not contribute equally and this leads to the effect that in the very high $Q^2$ region the $\PZ$-boson contributions compensate and even dominate the $\Pgg\PZ$-interference terms. Over all, we see that NC contributions cannot be neglected if $Q^2 > \SI[retain-unity-mantissa = false]{1e3}{\GeV^2}$, both, in the polarized and the unpolarized setup.

\begin{figure}[ht]
\begin{center}
\includegraphics[width=0.99\textwidth]{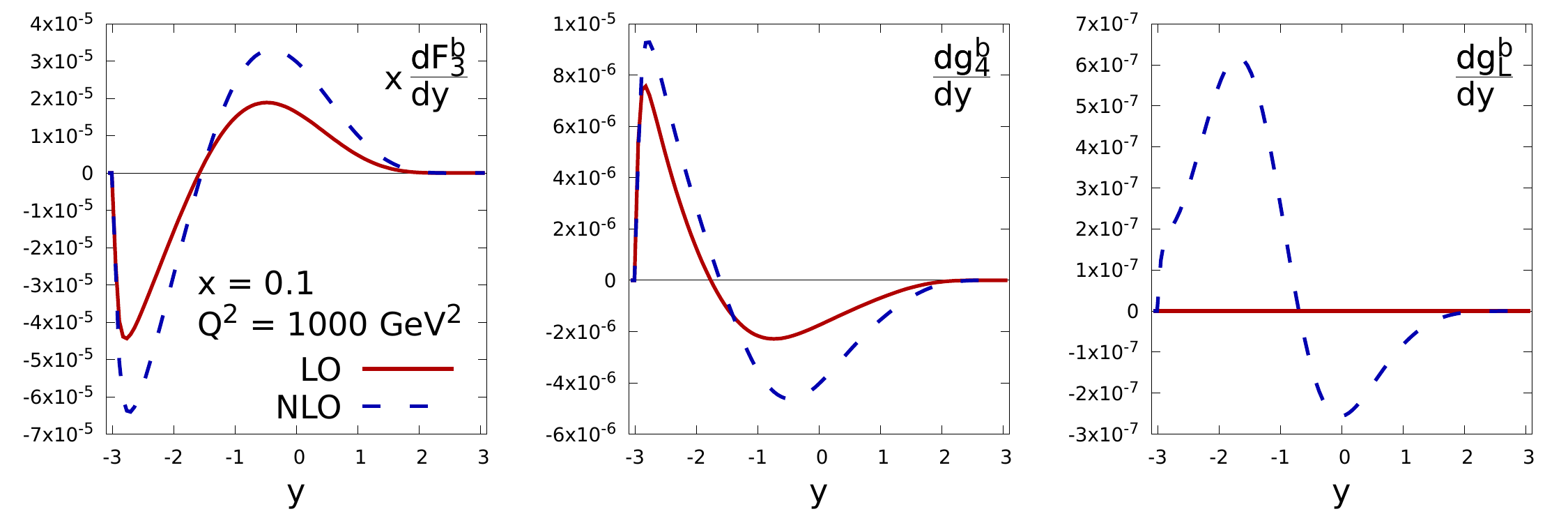}
\end{center}
\caption{Rapidity distributions of the PV structure functions: $x\DeriveF{y}{F^{\Pqb}_3}$ (left), $\DeriveF{y}{g^{\Pqb}_4}$ (middle), and, $\DeriveF{y}{g^{\Pqb}_L}$ (right) at LO (solid red) and NLO (dashed blue) accuracy at $x=\num{0.1}$ and $Q^2=\SI[retain-unity-mantissa = false]{1e3}{\GeV^2}$. } \label{fig:Fb-y-PV}
\end{figure}
With the inclusion of the $\PZ$ boson we get access to the PV structure functions $xF_3^{\Pqb}, g_4^{\Pqb}, g_L^{\Pqb}$ and we can present them here for the first time, to the best of our knowledge, at NLO accuracy. As explained in the beginning of this chapter, we have to refer to the rapidity distributions which we present in \FigureRef{fig:Fb-y-PV}. As the integral over all these functions at LO is exactly 0, all distributions have to include a node. For the case of the unpolarized $x\cdot dF^{\Pqb}_3/dy$ we can see an enhancement when increasing the perturbation order, leaving the overall shape mostly intact. As $F_3^{\Pqb}$ refers to a helicity difference it is not a strictly positive number unlike its PC counter-parts $F_2^{\Pqb}$ and $F_L^{\Pqb}$. In the case of the polarized $dg^{\Pqb}_4/dy$ we can see additionally a small shift of the node of the distribution. Due to the vanishing of the LO matrix element $B_{\lVAgL,\tQED}$ the structure function $dg^{\Pqb}_L/dy$ starts contributing only at NLO. Furthermore, this results to only a mild breaking of the Dicus relation\cite{Dicus:1972pq} which is in contrast to the unpolarized counter-part, the Callan-Gross relation\cite{Callan:1969uq}.

\begin{figure}[ht]
\begin{center}
\includegraphics[width=0.99\textwidth]{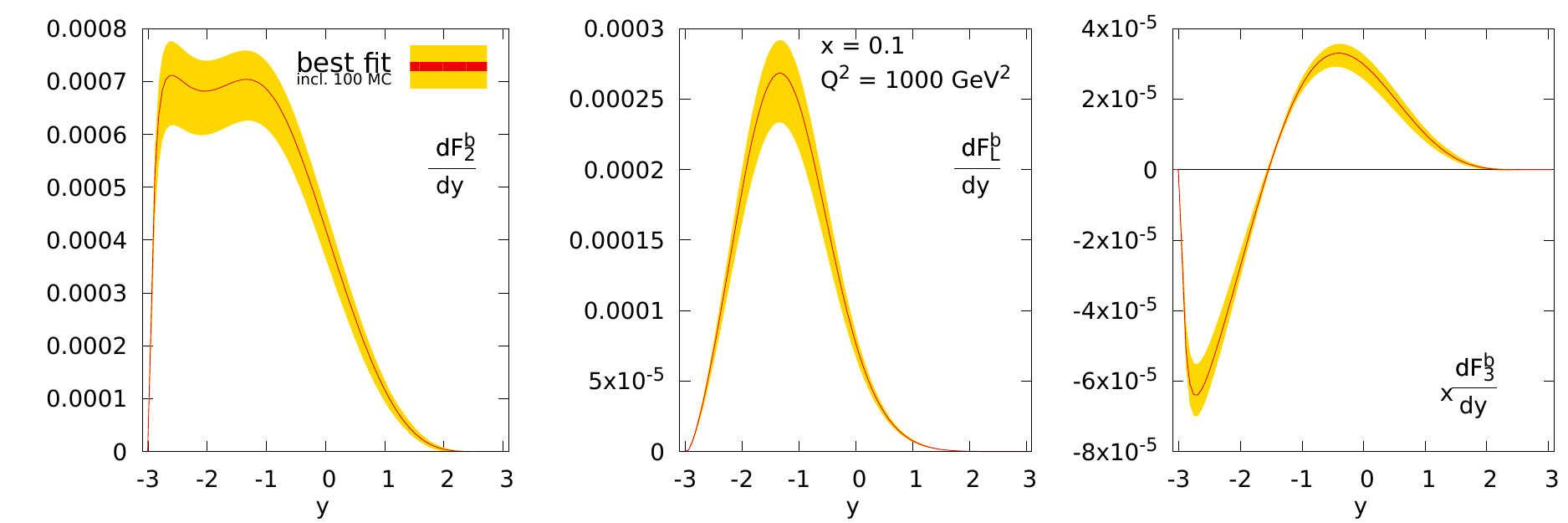}
\end{center}
\caption{Rapidity distributions of the unpolarized structure functions $dH/dy$ at $x=\num{0.1}, Q^2=\SI[retain-unity-mantissa = false]{1e3}{\GeV^2}$ for $H=F_2^{\Pqb}$ (left), $H=F_L^{\Pqb}$ (middle), and, $H=xF_3^{\Pqb}$(right) for the \NNPDF PDF set (see \FigureRef{fig:NNPDF-xg}). } \label{fig:Fb-y-NNPDF}
\end{figure}
In \FigureRef{fig:Fb-y-NNPDF} we plot the dependence of the rapidity distributions of the unpolarized structure functions on the PDFs. The related PDF uncertainties are much smaller then for the polarized counterpart in \FigureRef{fig:Fc-y-DSSV}. Note again, that the absolute value of the structure functions are hierarchically ordered which confirms a common approximation to neglect $F_L$ and $F_3$. For the PC structure functions $dF_2^{\Pqb}/dy$ and $dF_L^{\Pqb}/dy$ we find small negative, unphysical contributions in the very small and very large rapidity limit. This unphysical behavior can be probably traced back to threshold effects, that are not properly resummed over and we leave it to a future project to study resummation effects.

\begin{figure}[ht]
\begin{center}
\includegraphics[width=0.7\textwidth]{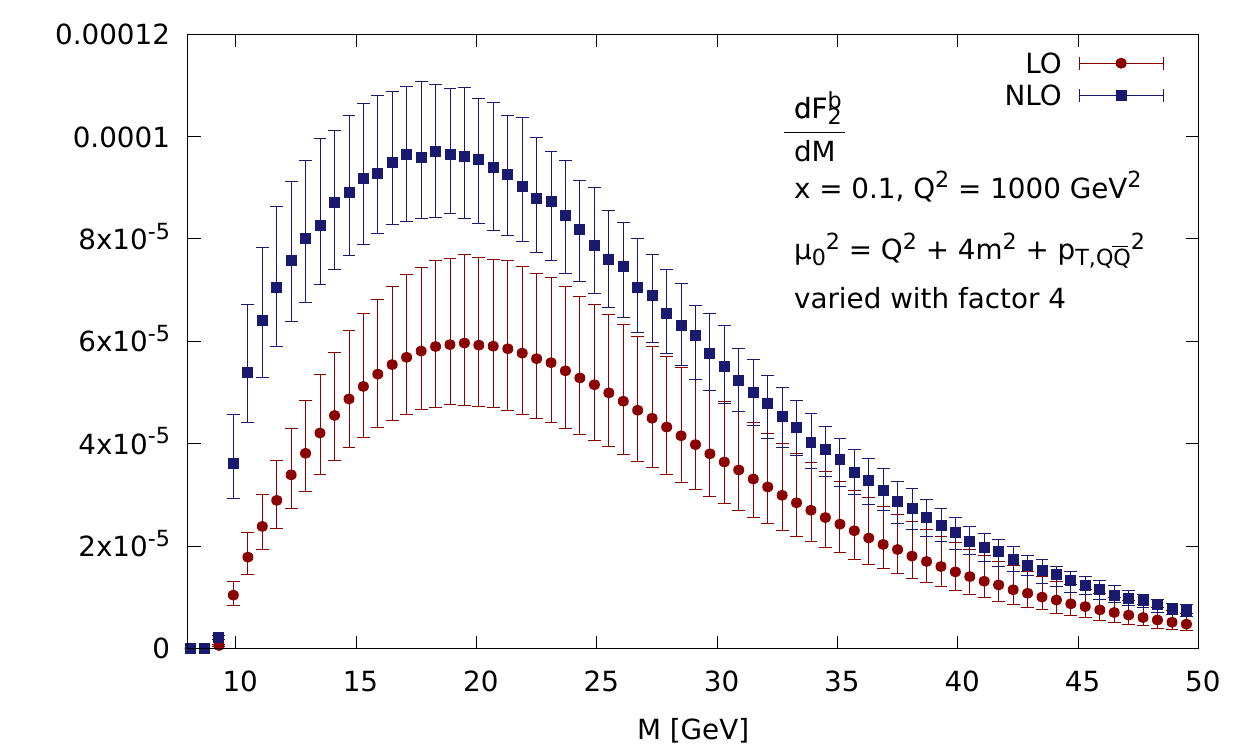}
\end{center}
\caption{Scale dependence of the pair mass distribution of the unpolarized structure functions $F_2^{\Pqb}$, i.e.\ $dF_2^{\Pqb}/dM(\mu_R^2,\mu_F^2)$ at $x=\num{0.1}$ and $Q^2=\SI[retain-unity-mantissa = false]{1e3}{\GeV^2}$ varying the renormalization scale $\mu_R^2$ and the factorization scale $\mu_F^2$ independently around the default scale $\mu_0^2 = Q^2+4m^2 + p_{T,\PQ\PaQ}^2$ with a factor 4 up and down at LO (red circles) and NLO (blue squares)} \label{fig:Fb-F2-M}
\end{figure}
As an example of our Monte-Carlo code, we now turn to the pair mass $M=\sqrt{(p_1+p_2)^2}$ that is not accessible in an inclusive computation. In \FigureRef{fig:Fb-F2-M} we investigate the scale dependence of the pair mass distribution of the unpolarized structure functions $F_2^{\Pqb}$. We can conclude two important points from the graph: first, the scale uncertainties are reduced when the perturbation order is increased, as expected. Second, the NLO curve is not covered by the scale uncertainty band at LO. This hints either to a slowly converging perturbation series or to the fact, that a variation of a factor 4 is too small. Interestingly, we see only very small difference between an 3-point, 7-point and a 9-point error estimate (shown here). Our choice for the default scale $\mu_0^2 = Q^2+4m^2 + p_{T,\PQ\PaQ}^2$ is inspired by similar plots in \cite{Harris:1995tu}, where $p_{T,\PQ\PaQ}=(p_1+p_2)_T$ labels the transverse momentum of the heavy quark pair. The complete framework of estimating higher order terms by scale variations as been reviewed recently in \cite{AbdulKhalek:2019ihb}.

As a final example and to exploit the full capabilities of our Monte-Carlo code, we investigate two geometrical correlation. First, we turn to the azimuthal angle correlation $\Delta\phi$ between the heavy quark pair. At LO this is a trivial function as, due to momentum conservation, the pair is always produced back-to-back ($\Delta\phi = \pi$) and we may thus write the distributions as
\begin{align}
\DeriveF{\Delta\phi}{H^{LO}}(x,Q^2,\Delta\phi) &= H^{LO}(x,Q^2)\cdot\delta(\Delta\phi-\pi)
\end{align}
At NLO we get the possibility to radiate an additional (potentially) hard parton and thus a smooth function can be generated. The singular point at $\Delta\phi=\pi$ will receive contributions from a potential soft gluon resummation as it corresponds to the soft (LO) configuration. Second, we discuss the heavy quark cone size\cite{Harris:1995tu} $\Delta R$, which is related to $\Delta\phi$ by its definition
\begin{align}
(\Delta R)^2 &= (\Delta \phi)^2 + (\Delta\eta)^2
\end{align}
with $\Delta\eta$ the difference in pseudo-rapidity\footnote{Note that for massive particles pseudo-rapidity and rapidity do not coincide}. With the above considerations at hand we can again give a simplified expression at LO
\begin{align}
\DeriveF{\Delta R}{H^{LO}}(x,Q^2,\Delta R) &= \DeriveF{\Delta R}{\tilde H^{LO}}(x,Q^2,\Delta R)\cdot \Theta(\Delta R-\pi)
\end{align}

\begin{figure}[ht]
\begin{center}
\includegraphics[width=0.8\textwidth]{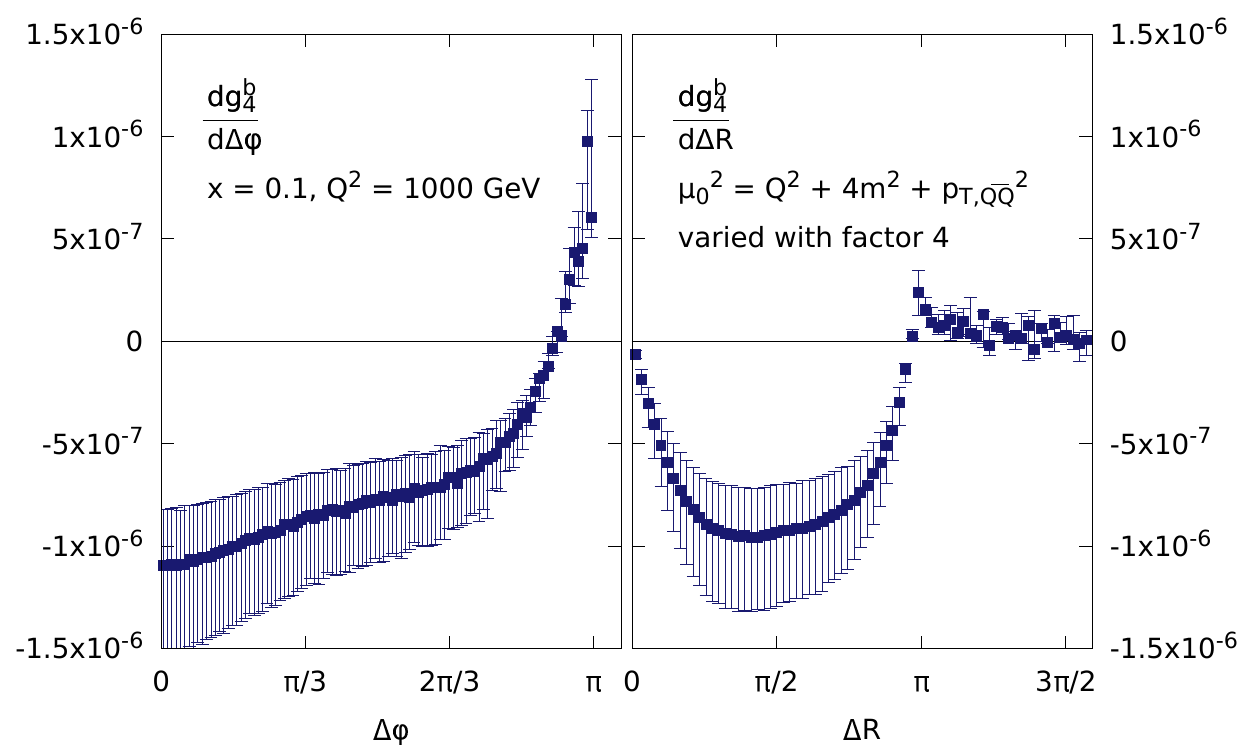}
\end{center}
\caption{Scale dependence of the $\Delta\phi$ (left) and $\Delta R$ (right) distributions of the polarized structure function $g_4^{\Pqb}$ at $x=\num{0.1}, Q^2=\SI[retain-unity-mantissa = false]{1e3}{\GeV^2}$ by varying the scales analog to \FigureRef{fig:Fb-F2-M} at NLO} \label{fig:Fb-g4-dR-dphi}
\end{figure}
In \FigureRef{fig:Fb-g4-dR-dphi} we plot the scale dependence of the $\Delta\phi$ and $\Delta R$ distribution of the parity violating polarized structure function $g_4^{\Pqb}$. The analogous plots for $2xg_1^{\Pqc}$ are shown in \cite{paper2}. The scale setup is the very same as for \FigureRef{fig:Fb-F2-M}. Note, that although the pair can be produced on the same side of the collider ($\Delta\phi=0$) it cannot be produced at the same position ($\Delta R=0$).

\chapter{Conclusions} \label{chap:Conclusions}
\chapterquote{The world is not enough}{Theorists}

The framework of deeply inelastic scattering (DIS) is one of the corner stones of Quantum Chromodynamics (QCD) and thus the Standard Model (SM) of particle physics. This work provides one of the last missing pieces in the set of next-to-leading order (NLO) perturbative QCD (pQCD) computations within in the SM. Heavy quark (HQ) production is standardized enough to be tackled with publicly available tools and yet complicated enough to require the full attention in any step of the calculation. It requires all typical tools that are needed in any higher order study and we have demonstrated their use and some of the advanced settings here. Our calculations provide an important cross check to existing codes (e.g.\ \texttt{HVQDIS}\cite{Harris:1997zq}) and give two important extensions, that were not available before: first, a polarized setup is considered, and, second, the full neutral current (NC) case is covered, that might become important for high precision measurements.

In \ChapterRef{chap:PS} we provided two different phase space decompositions by using two different mathematical prescriptions. Thus, we can provide two independent implementations for inclusive distributions that allow to compare e.g.\ accuracy or speed of the routines. However, the two numerical methods can also be seen in a complementary way: the phase space slicing in the $s_4$-centered phase space decomposition (see \SectionRef{sec:PS.s4}) requires more analytical tools but provides more accurate results. On the other side, the Monte-Carlo (MC) approach in the $s_5$-centered phase space decomposition (see \SectionRef{sec:PS.s5}) requires more numerical tools but provides more complex results. Overall, we demonstrated the consistency of the used prescription and the independence of physics from mathematical regularizations. We implemented the two numerical routines into a single program code, demonstrating the usefulness of program patterns in scientific computations.

In \ChapterRef{chap:Partonic} we provided the fundamental partonic coefficient functions (see also \AppendixRef{sec:Appendix.Partonic}) which are the main ingredients to the hadronic structure functions. Their determination can be regarded as one of the central results of this thesis as they encode all physical informations. We investigated their parameter dependence and highlighted some of their analytic features such as their $Q^2$ dependence or their threshold approximations. We successfully verified our own results against available results in literature, finding in turn one small error\footnote{in the threshold approximation of partonic structure function $\hat F_L$ - see \cite{Hekhorn:2018ywm}} in a commonly used code.

In \ChapterRef{chap:Hadronic} we have set up two different use cases for our calculations: first, we studied charm quark production (see \SectionRef{sec:Hadronic.c}), demonstrating its need and usefulness in the determination of polarized parton distribution functions (PDF). Polarized PDFs are largely unconstrained up to this late day but will offer new insights into spin physics. The polarized gluon PDF $\Delta\Pg$ is currently least constraint and HQ production offers a unique opportunity to reduce its uncertainties as it is a leading order (LO) ingredient. We demonstrated that different variations for the gluon PDF will result in different, distinguishable results and that thus a measurement of the proposed observables will help in constraining the pPDFs. This holds for the most part for both, parity conserving and parity violating structure functions, as well as their distributions. We have also shown the limit of this expectation and found it to be only a minor issue. We provided here and in our publications \cite{Hekhorn:2018ywm,paper2} a first theoretical estimate of the needed precision in order to use such a measurement at a possible future Electron-Ion Collider (EIC).

Second, we studied bottom quark production (see \SectionRef{sec:Hadronic.b}) to investigate the weak corrections to the full NC contributions. We calculated these contributions here for the first time and have shown, that they cannot be simulated by applying a $K$-factor prescription, but rather they carry their own parameter dependence. We have shown, that these contributions cannot be neglected in the high momentum transfer region ($Q^2 > \SI[retain-unity-mantissa = false]{1e3}{\GeV^2}$) where they will even dominate the full NC cross sections. We provided a first NLO computation of the parity violating structure functions $F_3,g_4,g_L$ that can be used for a high-precision determination of (un-)polarized PDFs. We studied their dependence on the rapidity of the produced heavy anti-quark and, for some of them, their dependence on some correlated variables, such as the heavy quark pair mass.

For the future, there is still some work left, on both, the theoretical and experimental aspects of phenomenology. On the experimental side, one might consider the inclusion of fragmentation functions to obtain actual measurable result. We might also include additional constraints to the phase space, motivated by experimental setups, to improve the match to the actual measurement and to reduce the need for extrapolation from unmeasured (or unmeasurable) regions. Important theoretical extensions would be the inclusion of a running mass to study the sensitivity of the DIS process to scale variations or the inclusion of resummation effects, such as soft gluon resummation. Finally, this calculations may become useful for PDF groups to obtain better fits in both, the polarized and the unpolarized framework. With the recent progress in the detector constructions, theoretical uncertainties, i.e.\ PDF uncertainties, become a limiting factor of predictions. To facilitate this use case, an exact analytical solution to \textit{all} main ingredients, the partonic coefficient functions, would be very beneficial - meanwhile the results presented here will provide a substitute.

\end{mainmatter}

\begin{appendices}
 \chapter{Details of Virtual Calculations}\label{sec:Appendix.V}
\chapterquote[ngerman]{Ich rede nicht geschwollen, ich bin geschwollen!}{Diese Arbeit}

\section{Ambiguous Definitions}
One has to be very careful when comparing results from various sources as there is quite a range of definitions for loop integrals. In the original paper of \citeauthor{Passarino:1978jh}\cite{Passarino:1978jh} they set
\begin{align}
A(m) &= \frac 1 {i\pi^2}\int d^nq\frac 1 {q^2+m^2}\\
B_0(p,m_1,m_2) &= \frac 1 {i\pi^2}\int d^nq\frac 1 {(q^2+m_1^2)((q+p)^2+m_2^2)}
\end{align}
and apart from their pole term (called $\Delta$ - see \cite[eq. D.1]{Passarino:1978jh}), they keep $n=4$.

In \cite{PhysRevD4054,Bojak:2000eu} they define:
\begin{align}
A(m) &= \mu^{-\epsilon}\int\frac{d^nq}{(2\pi)^n} \frac 1 {q^2-m^2}\\
B(q_1,m_1,m_2) &= \mu^{-\epsilon}\int\frac{d^nq}{(2\pi)^n} \frac 1 {(q^2-m_1^2)((q+q_1)^2-m_2^2)}
\end{align}
and $n=4+\epsilon$ (\cite{PhysRevD4054} writes \textquote{The notations for the one-, two-, three-, and four-point functions have been taken over from Ref. \cite{Passarino:1978jh}.} - obviously they do not).

\HEPMath\cite{Wiebusch:2014qba}, \FeynCalc\cite{Mertig:1990an,Shtabovenko:2016sxi}, \LoopTools\cite{Hahn:1998yk,LoopTools212Guide}, \QCDLoop\cite{Ellis:2007qk} and \cite{Ellis:2011cr} use the definition
\begin{align}
T_{\mu_1\ldots\mu_P}^N &=
\frac{\mu^{4 - D}}{i\pi^{D/2}\,r_\Gamma}
\int d^Dq\,
\frac{q_{\mu_1}\cdots q_{\mu_P}}
  {\bigl[q^2 - m_1^2\bigr]\,
   \bigl[(q + k_1)^2 - m_2^2\bigr] \cdots
   \bigl[(q + k_{N - 1})^2 - m_N^2\bigr]} \\[1ex]
\notag
r_\Gamma &= \frac{\Gamma^2(1 - \varepsilon)\Gamma(1+\varepsilon)}
  {\Gamma(1 - 2\varepsilon)}\,,
\quad D = 4 - 2\varepsilon\,
\end{align}

Finally, \cite{Denner:1991kt} does use the definition
\begin{align}
T_{\mu_1\ldots\mu_P}^N(p_1,\ldots,p_{N-1},m_0,\ldots,m_{N-1}) &= \frac{(2\pi\mu)^{4-D}}{i\pi^2}\int d^Dq \frac{q_{\mu_1}\cdots q_{\mu_P}}{L_0 L_1 \cdots L_{N-1}}\\
L_0 &= q^2-m_0^2 +i\varepsilon\\
L_i &= (q+p_i)^2-m_i^2+i\varepsilon \, i=1,\ldots,N-1
\end{align}

We will use the loop integral definition of \cite{Bojak:2000eu} and list the arguments (if possible) in the non-vector signature of \QCDLoop\cite{Ellis:2007qk}
\begin{align}
A(m^2), B_0(p^2,m_1^2,m_2^2), C_0(p_1^2,p_2^2,(p_1+p_2)^2,m_1^2,m_2^2,m_3^2)\nonumber\\
D_0(p_1^2,p_2^2,p_3^2,p_4^2,(p_1+p_2)^2,(p_2+p_3)^2,m_1^2,m_2^2,m_3^2,m_4^2)
\end{align}

The transformation of the \textit{analytic} results from the notation in \cite{Ellis:2007qk} is given by
\begin{equation}
\text{\cite{Ellis:2007qk}:}\,\frac{\mu^{4-n}}{i\pi^{n/2}r_\Gamma} \mathcal{I} \leftrightarrow \text{\cite{Bojak:2000eu}:}\,\frac{\mu^{4-n}}{(2\pi)^n}\mathcal{I}
\end{equation}
with $\mathcal{I}$ denoting the \textit{raw} integral. We then need to solve (B=Bojak\cite{Bojak:2000eu},E=Ellis\cite{Ellis:2007qk}):
\begin{align}
\Rightarrow\, &iC_\epsilon\left(\frac{a_{2}^B}{(n-4)^2}+\frac{a_{1}^B}{n-4}+a_0^B + O(n-4)\right)\nonumber\\
 &\EqualClaim \frac{\mu^{4-n}}{(2\pi)^n} \frac{i\pi^{n/2}r_\Gamma}{\mu^{4-n}}\left(\frac{a_{2}^E}{(n-4)^2}+\frac{a_{1}^E}{n-4}+a_0^E + O(n-4)\right)
\end{align}
We find
\begin{align}
\Rightarrow a_{2}^B &=a_{2}^E\\
a_{1}^B &= a_{1}^E-\frac 1 2 a_{2}^E\ln(m^2/\mu^2)\\
a_0^B &= a_0^E-\frac{a_{2}^E}{8}\zeta(2)+\frac{a_{2}^E}8\ln^2(m^2/\mu^2)-\frac{a_{1}^E} 2\ln(m^2/\mu^2)
\end{align}

To compare \textit{numeric} results form \LoopTools{} or \QCDLoop{} one needs to solve
\begin{align}
\Rightarrow\, &\left(\frac{b_2^B}{(n-4)^2}+\frac{b_1^B}{n-4}+b_0^B + O(n-4)\right)\nonumber\\
 &\EqualClaim \frac{\mu^{4-n}}{(2\pi)^n} \frac{i\pi^{n/2}r_\Gamma}{\mu^{4-n}}\left(\frac{b_2^E}{(n-4)^2}+\frac{b_1^E}{n-4}+b_0^E + O(n-4)\right)
\end{align}
We find
\begin{align}
\Rightarrow b_2^B &=\frac{i}{16\pi^2}b_2^E\\
b_1^B &= \frac{i}{16\pi^2}\left(b_1^E+\frac{b_2^E}2(\gamma_E-\ln(4\pi))\right)\\
b_0^B &= \frac{i}{16\pi^2}\left(b_0^E+\frac{b_1^E}2(\gamma_E-\ln(4\pi))+\frac{b_2^E}{8}\left((\gamma_E-\ln(4\pi))^2-\zeta(2)\right)\right)\\
\end{align}

\section{Momenta Decomposition}
The tensor coefficients have to be expanded by momenta and there are also different conventions: \HEPMath{} and \LoopTools{}(LT) use the \textit{internal} momenta $k_i$ and \cite{Passarino:1978jh,Bojak:2000eu,Ellis:2011cr}(E) use the \textit{external} momenta $p_i$. They are related by
\begin{equation}
k_1 = p_1, \quad k_2 = k_1+p_2=p_1+p_2, \quad k_i = k_{i-1}+p_i, \forall i>1
\end{equation}
So the decomposition is different, e.g.\ $C_{\mu}:$
\begin{align}
C_{\mu}^{LT} &= k_{1\mu}C_{1}^{LT}+k_{2\nu}C_{2}^{LT}\\
 &=p_{1\mu}(C_1^{LT}-C_2^{LT}) + p_{2\mu}C_2^{LT}\\
C_{\mu}^{E} &= p_{1\mu}C_{1}^{E}+p_{2\mu}C_{2}^{E}
\end{align}
Keep in mind, that also the labeling is not unique in literature, i.e.\ \cite{Passarino:1978jh,Bojak:2000eu} do use other naming schemes, but we will use the scheme of \cite{Ellis:2011cr} here, as it seems more natural and extensible.

Scalar Passarino-Veltman-Coefficients do not need any transformation, because they do not depend on any momenta. B-Coefficients also do not need any transformation, because $k_1=p_1$. The transformation for all other needed coefficients is given by:
\begin{equation}
\begin{pmatrix}C_1\\C_2\end{pmatrix}^{E} = \begin{pmatrix}1&-1\\0&1\end{pmatrix}\begin{pmatrix}C_1\\C_2\end{pmatrix}^{LT}
\qquad
\begin{pmatrix}D_1\\D_2\\D_3\end{pmatrix}^{E} = \begin{pmatrix}1&-1&0\\0&1&-1\\0&0&1\end{pmatrix}\begin{pmatrix}D_1\\D_2\\D_3\end{pmatrix}^{LT}
\end{equation}
\begin{equation}
\begin{pmatrix}C_{00}\\C_{11}\\C_{12}\\C_{22}\end{pmatrix}^{E} =
\begin{pmatrix}
 1 & 0 & 0 & 0 \\
 0 & 1 & -2 & 1 \\
 0 & 0 & 1 & -1 \\
 0 & 0 & 0 & 1 \\
\end{pmatrix}
\begin{pmatrix}C_{00}\\C_{11}\\C_{12}\\C_{22}\end{pmatrix}^{LT}
\end{equation}
\begin{equation}
\begin{pmatrix}D_{00}\\D_{11}\\D_{12}\\D_{13}\\D_{22}\\D_{23}\\D_{33}\end{pmatrix}^{E} =
\begin{pmatrix}
 1 & 0 & 0 & 0 & 0 & 0 & 0 \\
 0 & 1 & -2 & 0 & 1 & 0 & 0 \\
 0 & 0 & 1 & -1 & -1 & 1 & 0 \\
 0 & 0 & 0 & 1 & 0 & -1 & 0 \\
 0 & 0 & 0 & 0 & 1 & -2 & 1 \\
 0 & 0 & 0 & 0 & 0 & 1 & -1 \\
 0 & 0 & 0 & 0 & 0 & 0 & 1 \\
\end{pmatrix}
\begin{pmatrix}D_{00}\\D_{11}\\D_{12}\\D_{13}\\D_{22}\\D_{23}\\D_{33}\end{pmatrix}^{LT}
\end{equation}
\begin{equation}
\begin{pmatrix}D_{001}\\D_{002}\\D_{003}\\D_{111}\\D_{112}\\D_{113}\\D_{122}\\D_{123}\\D_{133}\\D_{222}\\D_{223}\\D_{333}\end{pmatrix}^{E} =
\left(\!\!
\begin{array}{ccccccccccccc}
 1 & -1 & 0 & 0 & 0 & 0 & 0 & 0 & 0 & 0 & 0 & 0 & 0 \\
 0 & 1 & -1 & 0 & 0 & 0 & 0 & 0 & 0 & 0 & 0 & 0 & 0 \\
 0 & 0 & 1 & 0 & 0 & 0 & 0 & 0 & 0 & 0 & 0 & 0 & 0 \\
 0 & 0 & 0 & 1 & -3 & 0 & 3 & 0 & 0 & -1 & 0 & 0 & 0 \\
 0 & 0 & 0 & 0 & 1 & -1 & -2 & 2 & 0 & 1 & -1 & 0 & 0 \\
 0 & 0 & 0 & 0 & 0 & 1 & 0 & -2 & 0 & 0 & 1 & 0 & 0 \\
 0 & 0 & 0 & 0 & 0 & 0 & 1 & -2 & 1 & -1 & 2 & -1 & 0 \\
 0 & 0 & 0 & 0 & 0 & 0 & 0 & 1 & -1 & 0 & -1 & 1 & 0 \\
 0 & 0 & 0 & 0 & 0 & 0 & 0 & 0 & 1 & 0 & 0 & -1 & 0 \\
 0 & 0 & 0 & 0 & 0 & 0 & 0 & 0 & 0 & 1 & -3 & 3 & -1 \\
 0 & 0 & 0 & 0 & 0 & 0 & 0 & 0 & 0 & 0 & 1 & -2 & 1 \\
 0 & 0 & 0 & 0 & 0 & 0 & 0 & 0 & 0 & 0 & 0 & 1 & -1 \\
 0 & 0 & 0 & 0 & 0 & 0 & 0 & 0 & 0 & 0 & 0 & 0 & 1 \\
\end{array}
\!\!\right)
\begin{pmatrix}D_{001}\\D_{002}\\D_{003}\\D_{111}\\D_{112}\\D_{113}\\D_{122}\\D_{123}\\D_{133}\\D_{222}\\D_{223}\\D_{333}\end{pmatrix}^{LT}
\end{equation}

\chapter{Details of $s_4$-centered Phase Space}\label{sec:Appendix.s4}
\chapterquote[ngerman]{Bücher? Bücher müsst ich tragen, dann könnt ich nicht mehr so schnell laufen, dann könnt ich nicht mehr durch die Luft fliegen \ldots und überhaupt!}{Richard P. Feynman}

\section{Phase Space Decomposition}\label{sec:Appendix.s4.Decomposition}
The general case for the $2\to 3$ scattering is given by
\begin{equation}
\HepProcess{b^*(q) + j(k_1) \to \PQ(p_1)+\PaQ(p_2) + j(k_2)}, \quad b^*\in\{\Pggx,\PZx\}, j\in\{\Pg,\Pq\}
\end{equation}
We use the center-of-mass system of the unobserved particles, i.e.\ the recoiling heavy quark $Q(p_1)$ and the massless particle $j(k_2)$ and write
\begin{align}
k_2 &= (\omega_2,\ldots,+k_{2,x},+\omega_2\sin\theta_1\cos\theta_2,+\omega_2\cos\theta_1)\\
p_1 &= (E_1,\ldots,-k_{2,x},-\omega_2\sin\theta_1\cos\theta_2,-\omega_2\cos\theta_1)
\end{align}
where $k_{2,x}=k_{2,x}(\omega_2,\theta_1,\theta_2,\hat k_2)$ is such that $k_2^2=0$. We are then left with the freedom on where to align the $z$-axis, which is reflected by our \textbf{Set I} (see \AppendixRef{sec:Appendix.s4.Decomposition.SetI}) and \textbf{Set II} (see \AppendixRef{sec:Appendix.s4.Decomposition.SetII}). Recall that $s_4=(k_2+p_1)^2-m^2$ which obviously is angular independent in this decomposition and justifies thus the name $s_4$-centered decomposition.

\subsection{Set I}\label{sec:Appendix.s4.Decomposition.SetI}
Align $k_1$ to $z$-axis
\begin{align}
k_1 &= (\omega_1,\ldots,0,0,\omega_1)\\
q &= (q_0,\ldots,0,\abs{\vec p_2}\sin\psi,\abs{\vec p_2}\cos\psi-\omega_1)\\
p_2 &= (E_2,\ldots,0,\abs{\vec p_2}\sin\psi,\abs{\vec p_2}\cos\psi)
\end{align}
with
\begin{align}
E_2 &= -\frac{s-s_4-2m^2}{2\sqrt{s_4+m^2}} & \omega_2 &= \frac{s_4}{2\sqrt{s_4+m^2}} \\
E_1 &= \frac{s_4+2m^2}{2\sqrt{s_4+m^2}} & q_0 &= \frac{s+u_1}{2\sqrt{s_4+m^2}} \\
\omega_1 &= \frac{s'+t_1}{2\sqrt{s_4+m^2}} = \frac{s_4-u_1}{2\sqrt{s_4+m^2}} & \abs{\vec p_2} &= \frac {\sqrt{(s-s_4)^2-4sm^2}} {2\sqrt{s_4+m^2}}
\end{align}
\begin{equation}
\cos\psi &= \frac {(u_1+m^2)(t_1-s')-(m^2-q^2-t_1)(s'+t_1)} {(s'+t_1)\sqrt{(s-s_4)^2-4sm^2}}
\end{equation}

From the above definitions we can classify the variables: $t'$ is collinear as an $(ab)$-variable, $u_6$ is a regular $(ab)$-variable and all other angular dependent variables $s_3,s_5,u_7,u'$ are regular $[ABC]$-variables. The angle $\theta_1$ defined here measures exactly the collinear pole associated to $t'$.

\subsection{Set II}\label{sec:Appendix.s4.Decomposition.SetII}
Align $q$ to $z$-axis
\begin{align}
q &= (q_0,\ldots,0,0,q_z)\\
k_1 &= (\omega_1,\ldots,0,\abs{\vec p_2}\sin\psi,\abs{\vec p_2}\cos\psi-q_z)\\
p_2 &= (E_2,\ldots,0,\abs{\vec p_2}\sin\psi,\abs{\vec p_2}\cos\psi)
\end{align}
with the same definitions as for \textbf{Set I} but
\begin{align}
q_z &= \frac{\sqrt{(s'+u_1')^2-4q^2t}}{2\sqrt{s_4+m^2}}\\
\cos\psi &= \frac{s(q^2-t-m^2)- 2q^2(s_4+m^2)+s_4 u_1}{\sqrt{((s-s_4)^2-4m^2s)((s'+u_1')^2-4q^2 t)}}
\end{align}

From the above definitions we can classify the variables: $t'$ is still the only collinear variable, but now as an $[ABC]$-variable, $u'$ and $u_7$ are regular $(ab)$-variables and all remaining angular dependent variables $s_3,s_5,u_6$ are regular $[ABC]$-variables.

In a potential Set III $s_3$ and $s_5$ become $(ab)$-variables, but our diagrams to this order are such that we never need this choice and can instead always use one of the other two sets. In fact the only product we truly need Set II for is just the combination of $s_3$ and $u_7$, because now, in Set II, we can treat $s_3$ as $(ab)$-variable and $u_7$ as $[ABC]$-variable.

\section{Collinear Phase Space Integrals}\label{sec:Appendix.s4.PSInt}
As explained in \SectionRef{sec:PS.s4} all phase space integrals has to be reduced to a simple combination of an (ab)- and an [ABC]-variable, so we then need to solve the master formula for $I_n^{(k,l)}$ in the different type of collinearity (see \Eqref{eq:Ikln}). Recall that
\begin{align}
d\Omega_n = \sin^{n-3}(\theta_1)d\theta_1\cdot\sin^{n-4}(\theta_2)d\theta_2
\end{align}
In the following we provide a new general solutions which can be applied to simple integrals with any collinearity and to semi-simple integrals with a single collinearity.

\subsubsection{Helper Integral}
We define the helper integral
\begin{equation}
\hat I^{(q)}(\nu) := \int\limits_0^\pi\!dt\,\sin^{\nu-3}(t)\cos^{q}(t)
\end{equation}
and use \cite[Eq. 5.12.6]{NIST:DLMF}
\begin{equation}
\int_{0}^{\pi}(\mathop{\sin\/}\nolimits t)^{\alpha-1}e^{i\beta t}dt=\frac{\pi}{2^{\alpha-1}} \frac{e^{i\pi \beta/2}}{\alpha\mathop{B}\left((\alpha+\beta+1)/2,(\alpha-\beta+1)/2\right)} \qquad\text{if}\,\Re(\alpha) > 0
\end{equation}
If q is odd, we find $\hat I^{(q)}=0$, due to the symmetry of the kernel. If q is even, we set $q=2p$ with $p\in\mathds N_0$ and obtain
\begin{align}
\hat I^{(2p)}(\nu) &= \frac{\sqrt{\pi}(2p)!}{2^{2p}p!} \frac {\Gamma((\nu-2)/2)} {\Gamma(\frac{\nu-1}2+p)}
\end{align}

\subsubsection{Any Collinearity and $-k,-l\in\mathds N_0$}
If $-k,-l\in\mathds N_0$ $I_{n}^{(k,l)}$ can always be reduced in a straight forward manner to combinations of $\hat I^{(q)}(n)$ and this way one finds\cite[Ch. 5]{Bojak:2000eu}\cite[App. C]{PhysRevD4054}:
\begin{align}
I^{(0,0)}_{n} &= \hat I^{(0)}(n-1) \cdot \hat I^{(0)}(n) = \frac{2\pi}{n-3} \mathrel{\mathop{\longrightarrow}_{n=4}} 2\pi\\
I^{(-1,0)}_{n} &= \hat I^{(0)}(n-1) \cdot (a\hat I^{(0)}(n)+b\hat I^{(1)}(n)) = \frac{2\pi a}{n-3}  \mathrel{\mathop{\longrightarrow}_{n=4}} 2\pi a \\
I^{(0,-1)}_{n} &= \hat I^{(0)}(n-1) \cdot (A\hat I^{(0)}(n) + B\hat I^{(1)}(n)) + C\hat I^{(1)}(n-1)\hat I^{(0)}(n)\\
 &= \frac{2\pi A}{n-3}  \mathrel{\mathop{\longrightarrow}_{n=4}} 2\pi A\\
I^{(-2,0)}_{n} &= \hat I^{(0)}(n-1) \cdot (a^2\hat I^{(0)}(n)+2ab\hat I^{(1)}(n) + b^2 \hat I^{(2)}(n))\\
 &= 2\pi\left(\frac{a^2(n-1)+b^2}{(n-1)(n-3)}\right)  \mathrel{\mathop{\longrightarrow}_{n=4}} 2\pi(a^2 + b^2/3) \\
I^{(0,-2)}_{n} &= \hat I^{(0)}(n-1) \cdot (A^2\hat I^{(0)}(n) + B^2\hat I^{(2)}(n)) + C^2\hat I^{(2)}(n-1)\hat I^{(0)}(n+2) \\
 &= 2\pi\left(\frac{A^2(n-1)+B^2+C^2}{(n-1)(n-3)}\right) \mathrel{\mathop{\longrightarrow}_{n=4}} 2\pi(A^2+(B^2+C^2)/3) \\
I^{(-1,-1)}_{n} &= \hat I^{(0)}(n-1) \cdot (a A\hat I^{(0)}(n) + b B \hat I^{(2)}(n)) = 2\pi\left(\frac{aA(n-1)+bB}{(n-1)(n-3)}\right) \\
 & \mathrel{\mathop{\longrightarrow}_{n=4}}  2\pi(aA + bB/3)
\end{align}

\subsubsection{Single Collinear $a$ and $-l\in\mathds N_0$}
We find:
\begin{align}
\hat I_{a}^{(k,q)}(\nu) &= \int\limits_0^\pi\!\frac{\sin^{\nu-3}(t)}{(1-\cos(t))^k}\cos^q(t)\, dt \\
 &= \int\limits_0^\pi\!\frac{\sin^{\nu-3}(t)}{(1-\cos^2(t))^k}\cos^q(t)(1+\cos(t))^k\, dt\\
 &=\int\limits_0^\pi\!\sin^{\nu-3-2k}(t)\cos^q(t)(1+\cos(t))^k\, dt\\
 &= \sum\limits_{l=0}^k\binom{k}{l}\hat I^{(q+l)}(\nu-2k)
\end{align}
and with such\cite[Ch. 5]{Bojak:2000eu}\cite[App. C]{PhysRevD4054}
\begin{align}
I^{(1,0)}_{a,n} &= \frac 1 a\hat I^{(0)}(n-1) \cdot \hat I^{(0)}(n-2)\\
 &= \frac {2\pi}{a(n-4)} = \frac {2\pi}{a\epsilon}\\
I^{(1,-1)}_{a,n} &= \frac 1 a\hat I^{(0)}(n-1)\cdot \left(A\hat I^{(0)}(n-2)+B\hat I^{(2)}(n-2)\right)\\
 &= \frac {2\pi}{a}\left(\frac{A}{n-4}+\frac{B}{(n-4)(n-3)}\right) = \frac {2\pi}{a}\left(\frac{A+B}\epsilon - 2B + O(\epsilon)\right)\\
I^{(1,-2)}_{a,n} &= \frac 1 a\left(\hat I^{(0)}(n-1)\cdot \left(A^2\hat I^{(0)}(n-2)+(2AB+B^2)\hat I^{(2)}(n-2)\right) \right.\nonumber\\
 &\hspace{30pt} \left. + C^2\hat I^{(2)}(n-1)\hat I^{(0)}(n)\right)\\
 &= \frac {2\pi}{a}\left(\frac {A^2}{n-4} + \frac {2AB + B^2}{(n-4)(n-3)} + \frac {C^2}{(n-3)(n-2)}\right)\\
 &= \frac {2\pi}{a}\left(\frac{(A+B)^2}{\epsilon}+\frac{C^2}{2}-2AB-B^2+O(\epsilon)\right)\\
I^{(1,-3)}_{a,n} &= \frac {2\pi} a \left(\frac{A^3}{n-4}+\frac{3A^2B+3AB^2}{(n-4)(n-3)}+\frac{3B^3}{(n-4)(n-3)(n-1)}\right.\nonumber\\
 &\hspace{40pt} \left.+\frac{3AC^2}{(n-3)(n-2)}+\frac{3BC^2}{(n-3)(n-2)(n-1)}\right)\\
 &= \frac {2\pi}{a}\left(\frac{(A+B)^3}{\epsilon} + \frac 1 6 \left( 3(3A+B)C^2 - 2B(9A^2+9AB+4B^2) \right) + O(\epsilon)\right)
\end{align}
and
\begin{align}
I^{(2,0)}_{a,n} &= \frac 1 {a^2}\hat I^{(0)}(n-1) \cdot \left(\hat I^{(0)}(n-4) + \hat I^{(2)}(n-4)\right)\\
 &= \frac {2\pi}{a^2(n-6)}
  = -\frac {\pi}{a^2} + O(\epsilon)\\
I^{(2,-1)}_{a,n} &= \frac 1 {a^2}\hat I^{(0)}(n-1) \cdot \left(A\left(\hat I^{(0)}(n-4) + \hat I^{(2)}(n-4)\right) + 2B\hat I^{(2)}(n-4)\right)\\
 &= \frac {2\pi}{a^2}\left(\frac{A}{n-6}+\frac{2B}{(n-6)(n-4)}\right)
  = -\frac{2\pi}{a^2}\left(\frac{B}{\epsilon} + \frac{A+B} 2\right) + O(\epsilon)\\
I^{(2,-2)}_{a,n} &= \frac 1 {a^2}\left(\hat I^{(0)}(n-1)\cdot \left(A^2(\hat I^{(0)}(n-4)+\hat I^{(2)}(n-4))+4AB\hat I^{(2)}(n-4) \right.\right.\nonumber\\
 &\hspace{20pt}\left.\left. + B^2(\hat I^{(2)}(n-4)+\hat I^{(4)}(n-4))\right) + C^2\hat I^{(2)}(n-1)(\hat I^{(0)}(n-2) + \hat I^{(2)}(n-2))\right)\\
 &= \frac {2\pi}{a^2}\left(\frac {A^2}{n-6}+\frac{4AB}{(n-6)(n-4)} + \frac{B^2 n}{(n-6)(n-4)(n-3)} + \frac{C^2}{(n-4)(n-3)} \right) \\
 &= \frac {2\pi}{a^2}\left(\frac{-2AB-2B^2+C^2}{\epsilon}+\frac{B^2-A^2}{2}-AB-C^2+O(\epsilon)\right)\\
\end{align}

\section{Logarithm and Friends in Phase Space Integrals}\label{sec:Appendix.s4.PSLog}
All phase space integrals involve at a given order a finite set of logarithms and eventually higher order transcendental functions. In the present case, we get logarithms and dilogarithms for the phase space which are still differential in $s_4,t_1$ and also only logarithms and dilogarithms for the full inclusive integrals of $\bar c_{\vec\kappa,\Pg},\bar c_{\vec\kappa,\Pq},d_{\vec\kappa,\Pq}$ (see \AppendixRef{sec:Appendix.Partonic}). For the full inclusive integrals to $c_{\vec\kappa,\Pg}$ and $c_{\vec\kappa,\Pq}$, which are yet to be determined, we expect higher order functions, such as trilogarithms and potentially elliptic functions\cite{Blumlein:2018cms,Ablinger:2013jta,Czakon:2008ii}.

For practical issues it is a good idea to keep track of all special functions in the phase space integrals with a separate symbol. The actual value of the symbol will depend on the pair of the $(ab)$-variable with the $[ABC]$-variable, which in turn depends on the chosen set $S$ of phase space decomposition. The list of logarithms needed in the present case is given by
\begin{align}
\texttt{psLog1}[S,(ab)]  &= \ln\left(\frac{a+b}{a-b}\right),\\
\texttt{psLog2}[S,[ABC]] &= \ln\left(\frac{(A+B)^2}{A^2-B^2-C^2}\right),\\
\texttt{psLog3}[S,[ABC]]  &= \ln\left(\frac{A+\sqrt{B^2+C^2}}{A-\sqrt{B^2+C^2}}\right),\\
\texttt{psLog4}[S,(ab),[ABC]]  &= \ln\left(\frac{Aa - Bb + X}{Aa - Bb - X}\right),\\
\texttt{psLog5}[S,[ABC]]  &= \ln\left(\frac{A-\sqrt{B^2+C^2}}{A+B}\right)
\end{align}
where $X[(ab),[ABC]] = \sqrt{(Aa-bB)^2-(A^2-B^2-C^2)(a^2-b^2)}$. The list of dilogarithms needed in the present case is given by
\begin{align}
\texttt{psDiLog1}[S,[ABC]]  &= \DiLog\left(\frac{2\sqrt{B^2+C^2}}{A+\sqrt{B^2+C^2}}\right),\\
\texttt{psDiLog2}[S,[ABC]] &= \DiLog\left(-\frac{B+\sqrt{B^2+C^2}}{A-\sqrt{B^2+C^2}}\right),\\
\texttt{psDiLog3}[S,[ABC]]  &= \DiLog\left(\frac{B-\sqrt{B^2+C^2}}{A+B}\right)
\end{align}
Note that the respective terms can only appear if they are well-defined, i.e., e.g., $\texttt{psLog1}[S,(ab)]$ will never appear in an $I_{a,n}^{(k,l)}$ expression and quite in general they will only appear if $k>0\lor l>0$. Note also that they mostly depend only on \textit{one} element of the pair and can thus be combined later on with terms from other phase space integrals. The most evolved term $\texttt{psLog4}[S,(ab),[ABC]]$ is introduced by the most complicated case of $I_{0,n}^{(1,1)}$. Casting the special functions into symbols also improves the numerical speed of the calculations, as the call to a special function (such as $\ln(x),\DiLog(x)$) are much more time consuming that just raw additions and multiplication. It also helps by comparing to other codes as, as usual, the representation of logarithms and dilogarithms is \textit{not} unique\cite{Zagier:2007knq}.

\section{Inclusive Hadronic Phase Space Integrals}\label{sec:Appendix.s4.Hadronic}
Here, we investigate the hadronic phase space integrals that are left open in the inclusive calculations and which we can rewrite in several ways to highlight various aspects. The required hadronic variables are given by $S_h = (q + P)^2$ and $S_h' =S_h-Q^2 = Q^2 / x$.

The integrals are given by\cite{Laenen:1992xs}
\begin{align}
\hat I &:= \int\!dT_1\int\!dU_1\int\!d\xi
\end{align}
with
\begin{align}
\beta_h &= \sqrt{1-4m^2/{S_h}}\\
-\frac{S_h'} 2 (1+\beta_h) &\leq T_1 \leq -\frac{S_h'} 2 (1-\beta_h)\\
-S_h' - T_1 &\leq U_1 \leq \frac{S_h'}{T_1}\left(m^2 - \frac{T_1^2 Q^2}{{S_h'}^2} - \frac{Q^2 T_1}{S_h'}\right)\\
\frac{-U_1}{S_h' + T_1} &\leq \xi \leq 1
\end{align}

We can rewrite this to\cite{Laenen:1992xs}
\begin{align}
\hat I &= S_h' \int\!dm_T^2\int\!dy\int\!d\xi \label{eq:PSmTyxi}
\end{align}
with
\begin{align}
m^2 &\leq m_T^2 \leq {S_h}/4\\
y_T &= \acosh\left(\frac {\sqrt {S_h}}{2m_T}\right)\\
-y_T &\leq y \leq y_T\\
T_1 &= - S_h' \frac{\exp(-y)}{2\cosh(y_T)}\\
U_1 &= -Q^2 - \frac 1 {2\cosh(y_T)}\left({S_h}\exp(y) - Q^2\exp(-y)\right)\\
\frac{-U_1}{S_h' + T_1} &\leq \xi \leq 1
\end{align}

We can rewrite this to\cite{Laenen:1992xs}
\begin{align}
\hat I &= S_h' \int\! dy \int\! dm_{T}^2 \int\!d\xi \label{eq:PSymTxi}
\end{align}
with 
\begin{align}
-\atanh(\beta_h) &\leq y \leq \atanh(\beta_h)\\
m^2 &\leq m_T^2 \leq \frac {S_h}{4\cosh^2(y)}\\
T_1 &= - S_h' \sqrt{\frac{m_T^2}{S_h}} \exp(-y)\\
U_1 &= -Q^2 - \sqrt{\frac{m_T^2}{S_h}}\left({S_h}\exp(y) - Q^2\exp(-y)\right)\\
\frac{-U_1}{S_h' + T_1} &\leq \xi \leq 1
\end{align}

We use either \Eqref{eq:PSmTyxi} or \Eqref{eq:PSymTxi} in our numerical code to compute the rapidity or transverse momentum distributions. Recall that $m_T^2 = m^2 + p_T^2$.

We can rewrite the operator to
\begin{align}
\hat I &= S_h' \int\! dy  \int\!d\xi \int\! d\hat m_{T}^2
\end{align}
with 
\begin{align}
-\atanh(\beta_h) &\leq y \leq \atanh(\beta_h)\\
x + (1-x)\frac{\sqrt{m^2/S_h}\exp(y)}{1-\sqrt{m^2/S_h}\exp(-y)} &\leq \xi \leq 1\\
m^2 &\leq \hat m_T^2 \leq S_h \left(\frac{\xi-x}{(1-x)\exp(y) + (\xi-x)\exp(-y)}\right)^2\\
T_1 &= - {S_h}' \sqrt{\frac{\hat m_T^2}{S_h}} \exp(-y)\\
U_1 &= -Q^2 - \sqrt{\frac{\hat m_T^2}{S_h}}\left({S_h}\exp(y) - Q^2\exp(-y)\right)
\end{align}

We can rewrite this to
\begin{align}
\hat I &= 2 S_h S_h' \int\! dy  \int\!d\xi \int\! dw\,w
\end{align}
with 
\begin{align}
-\atanh(\beta_h) &\leq y \leq \atanh(\beta_h)\\
x + (1-x)\frac{\sqrt{m^2/S_h}\exp(y)}{1-\sqrt{m^2/S_h}\exp(-y)} &\leq \xi \leq 1\\
\sqrt{m^2/S_h} &\leq w \leq \frac{\xi-x}{(1-x)\exp(y) + (\xi-x)\exp(-y)}\\
T_1 &= - S_h' w \exp(-y)\\
U_1 &= -Q^2 - w\left({S_h}\exp(y) - Q^2\exp(-y)\right)
\end{align}

We can rewrite this to
\begin{align}
\hat I &= 2 S_h  \int\! dy  \int\!d\xi \int\! ds_4\, r^2\left(\xi-x-\frac{s_4}{S_h'}\right)
\end{align}
with 
\begin{align}
-\atanh(\beta_h) &\leq y \leq \atanh(\beta_h)\\
x + (1-x)\frac{\sqrt{m^2/S_h}\exp(y)}{1-\sqrt{m^2/S_h}\exp(-y)} &\leq \xi \leq 1\\
r &= \frac 1 {(1-x)\exp(y) + (\xi-x)\exp(-y)}\\
0 &\leq s_4 \leq S_h'\left(\xi - x - \sqrt{\frac{m^2}{S_h}}\frac 1 r\right)\\
T_1 &= - S_h' r \left(\xi - x - \frac{s_4}{S_h'}\right) \exp(-y)\\
U_1 &= s_4-\xi\left(S_h' + T_1\right)
\end{align}

\section{Transverse Momentum Distributions}\label{sec:Appendix.s4.TransverseMomentum}
Recall that we can write the PV LO matrix elements $B_{(\tV,\tA,\kappa_3),\tQED}$ as a product of two functions
\begin{align}
B_{(\tV,\tA,\kappa_3),\tQED} &=\frac{t_1-u_1}{s'}\cdot g_{\kappa_3}\left(\frac{t_1 u_1}{{s'}^2}\right)
\end{align}
and that we can rewrite the phase-space delta function as 
\begin{align}
\delta(s_4) &= \delta(\xi {S_h}' + \xi T_1 + U_1) = \frac 1 {{S_h}' + T_1} \delta(\xi)\,.
\end{align}
We can insert this into \Eqref{eq:PSmTyxi} and write the full expression for the transverse momentum dependent distribution as
\begin{align}
\frac{dH_{(\tV,\tA,\kappa_3),\Pg}^{bb',(0)}}{dm_T^2} &\sim \int\limits_{-y_T}^{y_T}\!\!dy\, \frac{{S_h}'}{{S_h}'+T_1} \cdot \frac{t_1-u_1}{s'} \cdot g_{\kappa_3}\left(\frac{t_1 u_1}{{s'}^2}\right) \cdot f_{\kappa_3,\Pg}(\xi(y,m_T),\mu_F^2)\\
 &=\int\limits_{v_{min}}^{1-v_{min}}\!\!\!dv\, \frac{(1-v) - v}{v(1-v)}\cdot g_{\kappa_3}\left(\frac{t_1 u_1}{{s'}^2}=v(1-v)\right) f_{\kappa_3,\Pg}(\xi(v),\mu_F^2)\
\end{align}
with
\begin{align}
v &= \frac{-T_1}{{S_h}'} = \frac{\exp(-y)}{2\cosh(y_T)}\\
v_{min} &= v(y_T)= \frac{\exp(-y_T)}{2\cosh(y_T)}\\
\xi(v) &= x + \frac{1-x}{4\cosh^2(y_T)}\frac 1 {v(1-v)} = \xi(1-v)\,.
\end{align}
Now, we can see easily that the integral vanishes as the first term in the integral kernel is \textit{anti}-symmetric under the transformation $v\rightarrow 1-v$, but both, the truncated matrix element $g_{\kappa_3}$ and (p)PDF $f_{\kappa_3,\Pg}$, as well as the integral itself, are invariant under this transformation.

\chapter{Details of $s_5$-centered Phase Space}\label{sec:Appendix.s5}
\chapterquote[ngerman]{ne Palme! Mensch, mitten in der Ostsee ne Palme!}{die Seequarks}
\section{Collinear Counter Terms for $R_{\vec\kappa,\tOK}$ and $A_{\vec\kappa,1}$}
We define two sets of helper functions
\begin{align}
\tilde g_{\lVVF2}^{\Pg}(x) &= \tilde g_{\lAAF2}^{\Pg}(x) = \tilde g_{\lVVFL}^{\Pg}(x) = \tilde g_{\lAAFL}^{\Pg}(x) = \tilde g_{\lVAxF3}^{\Pg}(x) = 2-2x+x^2\\
\tilde g_{\lVVx2g1}^{\Pg}(x) &= \tilde g_{\lAAx2g1}^{\Pg}(x) = \tilde g_{\lVAg4}^{\Pg}(x) = \tilde g_{\lVAgL}^{\Pg}(x) = \frac{2-3x+2x^2}{x}
\end{align}
\begin{align}
\tilde g_{\lVVF2}^{\Pq}(x) &= \tilde g_{\lAAF2}^{\Pq}(x) = \tilde g_{\lVVFL}^{\Pq}(x) = \tilde g_{\lAAFL}^{\Pq}(x) = \tilde g_{\lVAxF3}^{\Pq}(x) = -2\\
\tilde g_{\lVVx2g1}^{\Pq}(x) &= \tilde g_{\lAAx2g1}^{\Pq}(x) = \tilde g_{\lVAg4}^{\Pq}(x) = \tilde g_{\lVAgL}^{\Pq}(x) = \frac{2(x-2)}{x}
\end{align}
and
\begin{align}
\tilde f_{\lVVF2} &= 2\frac{1-x}{t_1 u_1 x^2}\left(-(s')^2 + 2t_1 u_1 +s_5\beta_5^2 \cos^2(\theta_2)\sin^2(\theta_1)\left(6q^2 - \frac{(2m^2+q^2)(s')^2}{t_1 u_1} \right) \right)\\
\tilde f_{\lAAF2} &= \tilde f_{\lVVF2}+2\frac{1-x}{ x^2} \cdot \frac{4m^2(s')^2}{t_1 u_1 q^2}\\
\tilde f_{\lVVFL} &= 8\frac{(1-x)q^2s_5\beta_5^2 \cos^2(\theta_2)\sin^2(\theta_1)}{t_1 u_1 x^2}\\
\tilde f_{\lAAFL} &= 8\frac{1-x}{t_1 u_1 x^2}\left(\frac{m^2(s')^2}{q^2} +s_5\beta_5^2 \cos^2(\theta_2)\sin^2(\theta_1)\left(q^2 - \frac{m^2(s')^2}{t_1 u_1} \right) \right)\\
\tilde f_{\lVAxF3} &= -2\frac{(1-x)s'(t_1-u_1)}{t_1 u_1 x^2}\left(1 +s_5\beta_5^2 \cos^2(\theta_2)\sin^2(\theta_1)\frac{q^2}{t_1 u_1} \right)\\
\end{align}
\begin{align}
\tilde f_{\lVVx2g1} = \tilde f_{\lAAx2g1} = \tilde f_{\lVAg4} = \tilde f_{\lVAgL} = 0
\end{align}
where we used LO kinematics, i.e.\ $u_1=-s'-t_1$.

We can then give the collinear counter terms in a compact form
\begin{align}
\frac{s}{4(s')^2}\lim\limits_{y\to -1} t' (u' - q^2 s_5/s) R_{\vec\kappa,\tOK} &= \left[ \tilde g^{\Pg}_{\vec\kappa}(x) B_{\vec\kappa,\tQED} - \frac{1-x}{4} \tilde f_{\vec\kappa} \right]_{s'\to x s',t_1\to xt_{1,c}}\\
\lim\limits_{y\to -1} t' A_{\vec\kappa,1} &= \left[ \tilde g^{\Pq}_{\vec\kappa}(x) B_{\vec\kappa,\tQED} + \tilde f_{\vec\kappa} \right]_{s'\to x s',t_1\to xt_{1,c}}
\end{align}
where the right-hand-side of the Equations has to be evaluated in the collinear limit, i.e.\ with the collinear limit of $t_1$ and $u_1$ which are given by\cite{Harris:1995tu}
\begin{align}
t_{1,c} &= -\frac {s'} 2 (1-\beta_5 \cos(\theta_1)) &u_{1,c} &= -\frac {s_5'} 2 (1 + \beta_5 \cos(\theta_1))
\end{align}
In the collinear limit we find $x s' + xt_{1,c} + u_{1,c} = 0$ as $x=s_5'/s'$.

From these expressions we obtain the very same soft-collinear counter terms as in \Eqref{eq:scR}, as we have
\begin{align}
\tilde g^{\Pg}_{\vec\kappa}(1) &= 1 &\left.\tilde f_{\vec\kappa}\right|_{x=1} = 0
\end{align}

\chapter{Partonic Coefficient Functions}\label{sec:Appendix.Partonic}
\chapterquote[ngerman]{Forscht, wo ihr was zum Forschen findet! Das Unerforschbare lasst unergründet!}{Fake News}

In this Appendix we collect all analytic expressions for the partonic coefficient functions. In the Appendix of \cite{Hekhorn:2018ywm} we have already computed the expressions for $\lVVF2,\lVVFL$ and $\lVVx2g1$, but we will repeat them here for the sake of completeness. Recall that we used $F_G = F_2 - \frac 3 2 F_L$ in \cite{Hekhorn:2018ywm}. In the following we will make extensive use of the \Eqsrref{eq:partonicVars1}{eq:partonicVars3}.

\section{Leading Order Gluon Coefficient Function $c_{\vec\kappa,\Pg}^{(0)}$} \label{sec:Appendix:Partonic:cg0}
In leading order we find
\begin{align}
c^{(0)}_{\lVVF2,\Pg} &= -\frac{\pi {\rho'}^3 }{4 \rho ^2 {\rho_q}^2}\Big[2\beta\left(\rho ^2+{\rho_q}^2+\rho  {\rho_q} (6+{\rho_q})\right)\Big.\nonumber\\
 &\hspace{20pt} \Big.+\left(2 {\rho_q}^2+2 \rho  {\rho_q}^2+\rho ^2 (2-(-4+{\rho_q}) {\rho_q})\right) \ln(\chi)\Big]\\
c^{(0)}_{\lVVFL,\Pg} &= -\frac{\pi{\rho'}^3}{\rho\rho_q}\left[2\beta + \rho\ln(\chi)\right]\\
c^{(0)}_{\lVVx2g1,\Pg} &= \frac{\pi{\rho'}^2}{2\rho\rho_q}\left[\beta(\rho+3\rho_q) + (\rho+\rho_q)\ln(\chi)\right]
\end{align}
\begin{align}
c^{(0)}_{\lAAF2,\Pg} &= \frac{\pi {\rho'}^3}{4\rho^2{\rho_q}^2}\Big[ 2\beta\left(\rho ^2+{\rho_q}^2+\rho  {\rho_q} (6+{\rho_q})\right) \Big. \nonumber\\
 &\hspace{20pt} \Big. - \left(-6 \rho  {\rho_q}^2+2 (-1+{\rho_q}) {\rho_q}^2+\rho ^2 (-2+(-2+{\rho_q}) {\rho_q})\right) \ln(\chi) \Big] \\
c^{(0)}_{\lAAFL,\Pg} &=-\frac{\pi{\rho'}^3}{2\rho^2\rho_q}\left[2\beta\rho(2+\rho_q) - \left(\rho ^2 (-1+{\rho_q})-4 \rho  {\rho_q}+{\rho_q}^2\right) \ln(\chi)\right]\\
c^{(0)}_{\lAAx2g1,\Pg} &= c^{(0)}_{\lVVx2g1,\Pg}
\end{align}
and due to symmetry
\begin{align}
c^{(0)}_{\lVAxF3,\Pg} &= c^{(0)}_{\lVAg4,\Pg} = c^{(0)}_{\lVAgL,\Pg} = 0
\end{align}

Near threshold ($s\to 4m^2 \Leftrightarrow \rho\to 1 \Leftrightarrow \beta\to 0 \Leftrightarrow \chi\to 1$) we find
\begin{align}
c^{(0),\tThr}_{\lVVF2,\Pg} &= c^{(0),\tThr}_{\lVVx2g1,\Pg} = c^{(0),\tThr}_{\lAAx2g1,\Pg} = \frac{\pi\beta\rho_q}{2(\rho_q-1)}\\
c^{(0),\tThr}_{\lVVFL,\Pg} &= \frac{4\pi\beta^3\rho_q^2}{3(1-\rho_q)^3}\\
c^{(0),\tThr}_{\lAAF2,\Pg} &= \frac{\pi\beta\rho_q^2}{1-\rho_q}\\
c^{(0),\tThr}_{\lAAFL,\Pg} &= \frac{\pi\beta(1-2\rho_q)\rho_q}{2(\rho_q-1)}
\end{align}
Note that $c^{(0),\tThr}_{\lVVFL,\Pg} \sim \beta^3$, but $c^{(0),\tThr}_{\lAAFL,\Pg}\sim \beta$.

\section{Next-to-leading Order Coefficient Function $c_{\vec\kappa,\Pg}^{(1)}$} \label{sec:Appendix:Partonic:cg1}
In \cite{Hekhorn:2018ywm} we described the methods to obtain an approximation of the coefficient functions near threshold, but we use here a slightly modified definition of the helper functions $g_{1,2}(\chi_q)$ and drop $g_3(\chi_q)$. We define
\begin{align}
g_1(\chi_q) &= \frac 1 8\left[\DiLog\left(\frac{\rho_q}{2-\rho_q}\right) + \frac 1 2\ln^2\left(\frac{-\rho_q}{2-\rho_q}\right) - \frac {\pi^2} 2 - \frac 3 2 \ln^2(\chi_q) -2 \beta_q\ln(\chi_q)\right]\\
g_2(\chi_q) &= \frac 1 4\left[\ln^2\left(\frac{-\rho_q}{2(1-\rho_q)}\right) - \ln^2(\chi_q) + 9\ln^2(2)+15\ln(2)-\frac{25}{2} \right]
\end{align}
and find
\begin{align}
c_{\vec \kappa,\Pg}^{(1),\tThr} &= c_{\vec \kappa,\Pg}^{(0),\text{thr}} \frac{1}{\pi^2}\left[
     C_A\left(a_{\vec \kappa,\Pg}^{(1,2)}\ln^2(\beta) + a_{\vec \kappa,\Pg}^{(1,1)}\ln(\beta) - \frac{\pi^2}{16\beta} + a_{\vec \kappa,\Pg,\tOK}^{(1,0)}\right) \right.\nonumber\\
 &\hspace{60pt} \left. + 2C_F\left(\frac{\pi^2}{16\beta} + a_{\vec \kappa,\Pg,\tQED}^{(1,0)}\right)\right]
\end{align}
with
\begin{align}
a^{(1,2)}_{\vec \kappa,\Pg} &= 1
\end{align}
and
\begin{align}
a^{(1,1)}_{\lVVF2,\Pg} &= a^{(1,1)}_{\lVVx2g1,\Pg} = a^{(1,1)}_{\lAAF2,\Pg} = a^{(1,1)}_{\lAAFL,\Pg} = a^{(1,1)}_{\lAAx2g1,\Pg} = -\frac 5 2 + 3\ln(2)\\
a^{(1,1)}_{\lVVFL,\Pg} &= a^{(1,1)}_{\lVVF2,\Pg} - \frac 2 3
\end{align}
and for the \enquote{finite} OK parts
\begin{align}
a^{(1,0)}_{\lVVF2,\Pg,\tOK} &= g_1(\chi_q)+g_2(\chi_q)-\frac{\pi ^2 {\rho_q}}{32 (-1+{\rho_q})}-\frac{(-3+{\rho_q}) \ln \left[\frac{{\rho_q}}{2 (-1+{\rho_q})}\right]}{4 (-2+{\rho_q})}\nonumber\\
 &\hspace{20pt}+\frac{(-5+(7-2 {\rho_q}) {\rho_q}) \ln [{\chi_q}]}{8 {\beta_q} (-1+{\rho_q})}-\frac{{\rho_q} \ln [{\chi_q}]^2}{32 (-1+{\rho_q})}
\end{align}
\begin{align}
a^{(1,0)}_{\lVVFL,\Pg,\tOK} &= \frac{49}{72}+g_2(\chi_q)-\frac{\pi ^2 (-4+{\rho_q}) {\rho_q}}{24 (-1+{\rho_q})^2}+\frac{g_1(\chi_q) (1+2 {\rho_q})}{(-1+{\rho_q})^2}-\ln [2]\nonumber\\
 &\hspace{20pt}+\frac{(-1+{\rho_q} (-3-2 (-3+{\rho_q}) {\rho_q})) \ln \left[\frac{{\rho_q}}{2 (-1+{\rho_q})}\right]}{4 (-2+{\rho_q}) (-1+{\rho_q})^2}-\frac{(-4+{\rho_q}) {\rho_q} \ln [{\chi_q}]^2}{8 (-1+{\rho_q})^2}
\end{align}
\begin{align}
a^{(1,0)}_{\lVVx2g1,\Pg,\tOK} &= a^{(1,0)}_{\lVVF2,\Pg,\tOK}
\end{align}
\begin{align}
a^{(1,0)}_{\lAAF2,\Pg,\tOK} &= g_2(\chi_q)-\frac{\pi ^2 {\rho_q} (3+{\rho_q} (-25+18 {\rho_q}))}{96 (-1+{\rho_q})^2 (-1+2 {\rho_q})}+\frac{g_1(\chi_q) \left(-1-{\rho_q} \left(-4+{\rho_q}+{\rho_q}^2\right)\right)}{(-1+{\rho_q})^2 (-1+2 {\rho_q})}\nonumber\\
 &\hspace{20pt}-\frac{\left(-1+(-1+{\rho_q}) {\rho_q}^2\right) \ln \left[\frac{{\rho_q}}{2 (-1+{\rho_q})}\right]}{4 (-2+{\rho_q}) (-1+{\rho_q}) (-1+2 {\rho_q})}+\frac{(-3+2 {\rho_q} (2+{\rho_q})) \ln [{\chi_q}]}{8 {\beta_q} (-1+2 {\rho_q})}\nonumber\\
 &\hspace{20pt}-\frac{{\rho_q} (1+{\rho_q} (-19+14 {\rho_q})) \ln [{\chi_q}]^2}{32 (-1+{\rho_q})^2 (-1+2 {\rho_q})}
\end{align}
\begin{align}
a^{(1,0)}_{\lAAFL,\Pg,\tOK} &= g_2(\chi_q)+\frac{\pi ^2 (4-9 (-1+{\rho_q}) {\rho_q})}{96 (-1+{\rho_q})^2}+\frac{g_1(\chi_q) (4-{\rho_q} (1+{\rho_q}))}{2 (-1+{\rho_q})^2}\nonumber\\
 &\hspace{20pt}+\frac{\left(-4+5 {\rho_q}-{\rho_q}^3\right) \ln \left[\frac{{\rho_q}}{2 (-1+{\rho_q})}\right]}{8 (-2+{\rho_q}) (-1+{\rho_q})^2}+\frac{\left(2-3 {\rho_q}+{\rho_q}^3\right) \ln [{\chi_q}]}{8 {\beta_q} (-1+{\rho_q})^2}\nonumber\\
 &\hspace{20pt}+\frac{(4-7 (-1+{\rho_q}) {\rho_q}) \ln [{\chi_q}]^2}{32 (-1+{\rho_q})^2}
\end{align}
\begin{align}
a^{(1,0)}_{\lAAx2g1,\Pg,\tOK} &= g_2(\chi_q)+\frac{g_1(\chi_q)}{(-1+{\rho_q})^2}+\frac{\pi ^2 (11-7 {\rho_q}) {\rho_q}}{96 (-1+{\rho_q})^2}\nonumber\\
 &\hspace{20pt}+\frac{(1+{\rho_q} (-5-2 (-3+{\rho_q}) {\rho_q})) \ln \left[\frac{{\rho_q}}{2 (-1+{\rho_q})}\right]}{4 (-2+{\rho_q}) (-1+{\rho_q})^2}\nonumber\\
 &\hspace{20pt}+\frac{3 \ln [{\chi_q}]}{8 {\beta_q}}+\frac{(9-5 {\rho_q}) {\rho_q} \ln [{\chi_q}]^2}{32 (-1+{\rho_q})^2}
\end{align}
and the QED parts
\begin{align}
a^{(1,0)}_{\lVVF2,\Pg,\tQED} &= -\frac{g_1(\chi_q)}{-1+{\rho_q}}+\frac{\pi ^2 (-4+3 {\rho_q})}{96 (-1+{\rho_q})}-\frac{-9+5 {\rho_q}}{8 (-2+{\rho_q})}\nonumber\\
 &\hspace{20pt}+\frac{(-3+(5-2 {\rho_q}) {\rho_q}) \ln \left[\frac{{\rho_q}}{2 (-1+{\rho_q})}\right]}{4 (-2+{\rho_q})^2 (-1+{\rho_q})}-\frac{\ln [{\chi_q}]}{8 {\beta_q}}+\frac{(-4+{\rho_q}) \ln [{\chi_q}]^2}{32 (-1+{\rho_q})}
\end{align}
\begin{align}
a^{(1,0)}_{\lVVFL,\Pg,\tQED} &= \frac{3-2 {\rho_q}}{8 (-2+{\rho_q})}-\frac{g_1(\chi_q) (-1+6 {\rho_q})}{(-1+{\rho_q})^2}-\frac{\pi ^2 (-1+6 {\rho_q})}{24 (-1+{\rho_q})^2}\nonumber\\
 &\hspace{20pt}+\frac{(3+2 {\rho_q} (5+(-5+{\rho_q}) {\rho_q})) \ln \left[\frac{{\rho_q}}{2 (-1+{\rho_q})}\right]}{4 (-2+{\rho_q})^2 (-1+{\rho_q})}\nonumber\\
 &\hspace{20pt}+\frac{\left(-6+{\rho_q}+{\rho_q}^2\right) \ln [{\chi_q}]}{8 {\beta_q} (-2+{\rho_q})}-\frac{(-1+6 {\rho_q}) \ln [{\chi_q}]^2}{8 (-1+{\rho_q})^2}
\end{align}
\begin{align}
a^{(1,0)}_{\lVVx2g1,\Pg,\tQED} &= a^{(1,0)}_{\lVVF2,\Pg,\tQED}
\end{align}
\begin{align}
a^{(1,0)}_{\lAAF2,\Pg,\tQED} &= \frac{-5-4 (-3+{\rho_q}) {\rho_q}}{8 (-2+{\rho_q}) (-1+2 {\rho_q})}+\frac{g_1(\chi_q) (-1+(-2+{\rho_q}) {\rho_q} (-2+3 {\rho_q}))}{(-1+{\rho_q})^2 (-1+2 {\rho_q})}\nonumber\\
 &\hspace{20pt}+\frac{\pi ^2 (-4+{\rho_q} (19+{\rho_q} (-41+18 {\rho_q})))}{96 (-1+{\rho_q})^2 (-1+2 {\rho_q})}\nonumber\\
 &\hspace{20pt}+\frac{(1+{\rho_q}) (-1+{\rho_q} (7+{\rho_q} (-7+2 {\rho_q}))) \ln \left[\frac{{\rho_q}}{2 (-1+{\rho_q})}\right]}{4 (-2+{\rho_q})^2 (-1+{\rho_q}) (-1+2 {\rho_q})}\nonumber\\
 &\hspace{20pt}-\frac{(1+6 (-1+{\rho_q}) {\rho_q}) \ln [{\chi_q}]}{8 {\beta_q} (-1+2 {\rho_q})}+\frac{(-4+{\rho_q} (17+7 {\rho_q} (-5+2 {\rho_q}))) \ln [{\chi_q}]^2}{32 (-1+{\rho_q})^2 (-1+2 {\rho_q})}
\end{align}
\begin{align}
a^{(1,0)}_{\lAAFL,\Pg,\tQED} &= -\frac{-5+2 {\rho_q}}{8 (-2+{\rho_q})}+\frac{g_1(\chi_q) {\rho_q} (-7+3 {\rho_q})}{2 (-1+{\rho_q})^2}+\frac{\pi ^2 {\rho_q} (-17+9 {\rho_q})}{96 (-1+{\rho_q})^2}\nonumber\\
 &\hspace{20pt}+\frac{(12+{\rho_q} (-7+{\rho_q} (-3+2 {\rho_q}))) \ln \left[\frac{{\rho_q}}{2 (-1+{\rho_q})}\right]}{8 (-2+{\rho_q})^2 (-1+{\rho_q})}\nonumber\\
 &\hspace{20pt}+\frac{\left(-\frac{1}{2}+{\rho_q}-\frac{3 {\rho_q}^2}{8}\right) \ln [{\chi_q}]}{{\beta_q} (-2+{\rho_q})}+\frac{{\rho_q} (-15+7 {\rho_q}) \ln [{\chi_q}]^2}{32 (-1+{\rho_q})^2}
\end{align}
\begin{align}
a^{(1,0)}_{\lAAx2g1,\Pg,\tQED} &= \frac{5-2 {\rho_q}}{8 (-2+{\rho_q})}+\frac{g_1(\chi_q) (1+(-4+{\rho_q}) {\rho_q})}{(-1+{\rho_q})^2}+\frac{\pi ^2 (4+{\rho_q} (-19+7 {\rho_q}))}{96 (-1+{\rho_q})^2}\nonumber\\
 &\hspace{20pt}+\frac{(1+(-2+{\rho_q}) {\rho_q} (-3+2 {\rho_q})) \ln \left[\frac{{\rho_q}}{2 (-1+{\rho_q})}\right]}{4 (-2+{\rho_q})^2 (-1+{\rho_q})}\nonumber\\
 &\hspace{20pt}+\frac{(-2+(5-2 {\rho_q}) {\rho_q}) \ln [{\chi_q}]}{8 {\beta_q} (-2+{\rho_q})}\nonumber\\
 &\hspace{20pt}+\frac{(4+{\rho_q} (-17+5 {\rho_q})) \ln [{\chi_q}]^2}{32 (-1+{\rho_q})^2}
\end{align}
Note that we can see manifest here, that the polarized coefficient function do not have a V-A structure, i.e.\ $a^{(1,0)}_{\lVVx2g1,\Pg,\tOK/\tQED}\neq a^{(1,0)}_{\lAAx2g1,\Pg,\tOK/\tQED}$.

\section{Next-to-leading Order Gluon Scaling Coefficient Function $\bar c_{\vec\kappa,\Pg}^{(1)}$} \label{sec:Appendix:Partonic:cgBar1}
We define the very same helper functions as in \cite{Hekhorn:2018ywm} by
\begin{align}
h_1(\chi,\chi_q) &= \DiLog\!\left(\frac{1-\chi_q}{1+\chi}\right) +\DiLog\!\left(-\frac{1-\chi_q}{(1+\chi )\chi_q}\right) -\DiLog\!\left(\frac{\chi (1-\chi_q)}{1+\chi}\right) -\DiLog\!\left(-\frac{\chi (1-\chi_q)}{(1+\chi)\chi_q}\right) \nonumber \\
&\hspace{20pt}+\frac 1 2 \ln^2 (\chi ) -\zeta(2)-2\DiLog(-\chi )+\ln (\chi )\ln\left[\frac{\chi_q}{(\chi +\chi_q)(1+\chi  \chi_q)} \right]\\
h_2(\chi,\chi_q) &= - \zeta(2)+2\DiLog(\chi)+2\DiLog(-\chi)-\frac 1 2 \ln(\chi) -\ln(\chi)\ln\left[\frac{\chi_q}{(\chi+\chi_q)(1+\chi\chi_q)}\right]\\
h_3(\chi,\chi_q) &= \ln(\beta)-\ln(\rho)+\ln(2)+\frac 1 2 \ln(\rho')
\end{align}
and find
\begin{align}
\bar c_{\lVVF2,\Pg}^{(1)} &= \frac{\rho_q}{128\pi(1-\rho_q)(\rho-\rho_q)^3}\left\{ 96 h_3(\chi,\chi_q) \beta  \rho  (-1+{\rho_q}) \left(\rho ^2+{\rho_q}^2+\rho  {\rho_q} (6+{\rho_q})\right)\right.\nonumber\\
 &\hspace{0pt}+48 h_1(\chi,\chi_q) \rho  (\rho -{\rho_q}) (-1+{\rho_q}) (-{\rho_q}+\rho  (5+2 {\rho_q}))\nonumber\\
 &\hspace{0pt} +2 \beta  (-1+{\rho_q}) {\rho_q} \left(62 \rho  {\rho_q}+8 {\rho_q}^2-\rho ^2 (632+95 {\rho_q})\right)\nonumber\\
 &\hspace{0pt}+24 h_2(\chi,\chi_q) \rho  (-1+{\rho_q}) \left(-2 {\rho_q}^2-2 \rho  {\rho_q}^2+\rho ^2 (-2+(-4+{\rho_q}) {\rho_q})\right)\nonumber\\
 &\hspace{0pt}+\rho  (-1+{\rho_q}) \left(568 \rho ^2-48 \rho  (11+8 \rho ) {\rho_q}+(48+\rho  (-96+59 \rho )) {\rho_q}^2\right) \ln [\chi ]\nonumber\\
 &\hspace{0pt}\left.-8 {\beta_q} (\rho -{\rho_q}) \left(-2 \rho  {\rho_q} (4+{\rho_q})+{\rho_q}^2 (4+{\rho_q})+\rho ^2 (-68+73 {\rho_q})\right) \ln \left[\frac{\chi +{\chi_q}}{1+\chi  {\chi_q}} \right]
\right\}
\end{align}
\begin{align}
&\bar c_{\lVVFL,\Pg}^{(1)}=\nonumber\\
 & \frac{-\rho_q}{16\pi(1-\rho_q)^2(\rho-\rho_q)^3}\left\{48 h_3(\chi,\chi_q) \beta  \rho ^2 (-1+{\rho_q})^2 {\rho_q}\right.\nonumber\\
 &\hspace{0pt}-12 h_2(\chi,\chi_q) \rho ^3 (-1+{\rho_q})^2 {\rho_q}\nonumber\\
 &\hspace{0pt}+2 \beta  (-1+{\rho_q}) {\rho_q} \left(36 \rho ^2-35 \rho ^2 {\rho_q}-2 (1+\rho ) {\rho_q}^2+3 {\rho_q}^3\right)\nonumber\\
 &\hspace{0pt} -8 \rho ^2 (-1+{\rho_q})^2 (3 {\rho_q}+\rho  (-6+5 {\rho_q})) \ln [\chi ]\nonumber\\
 &\hspace{0pt}\left.+{\beta_q} (-\rho +{\rho_q}) \left(2 \rho  {\rho_q}^2+{\rho_q}^3 (-4+3 {\rho_q})+\rho ^2 (-6+5 {\rho_q}) (-8+9 {\rho_q})\right) \ln\!\left[\frac{\chi +{\chi_q}}{1+\chi  {\chi_q}}\right] \!\right\}
\end{align}
\begin{align}
\bar c_{\lVVx2g1,\Pg}^{(1)} &= \frac{-3\rho_q}{16\pi(\rho-\rho_q)^3}\left\{ 4 h_1(\chi,\chi_q) \rho  (\rho -{\rho_q}) (2 \rho -{\rho_q})+32 \beta  \rho  {\rho_q} (-\rho +{\rho_q})\right.\nonumber\\
 &\hspace{0pt}-2 h_2(\chi,\chi_q) \rho  (\rho -{\rho_q}) (\rho +{\rho_q})\nonumber\\
 &\hspace{0pt}+4 h_3(\chi,\chi_q) \beta  \rho  (\rho -{\rho_q}) (\rho +3 {\rho_q})-\rho  (\rho -{\rho_q}) (11 {\rho_q}+\rho  (-13+4 {\rho_q})) \ln [\chi ]\nonumber\\
 &\hspace{0pt}\left.-4 {\beta_q} \rho  \left(3 \rho ^2-4 \rho  {\rho_q}+{\rho_q}^2\right) \ln \left[\frac{\chi +{\chi_q}}{1+\chi  {\chi_q}}\right] \right\}
\end{align}
\begin{align}
\bar c_{\lAAF2,\Pg}^{(1)} &= \frac{\rho_q}{128\pi(1-\rho_q)(\rho-\rho_q)^3}\left\{96 h_3(\chi,\chi_q) \beta  \rho  (-1+{\rho_q}) \left(\rho ^2+{\rho_q}^2+\rho  {\rho_q} (6+{\rho_q})\right)\right.\nonumber\\
 &\hspace{0pt}+48 h_1(\chi,\chi_q) \rho  (\rho -{\rho_q}) (-1+{\rho_q}) ((-1+{\rho_q}) {\rho_q}+\rho  (5+{\rho_q}))\nonumber\\
 &\hspace{0pt}+ 2 \beta  (-1+{\rho_q}) {\rho_q} \left(-632 \rho ^2+(62-95 \rho ) \rho  {\rho_q}+(8+9 (-2+\rho ) \rho ) {\rho_q}^2\right)\nonumber\\
 &\hspace{0pt}+24 h_2(\chi,\chi_q) \rho  (-1+{\rho_q}) \left(-6 \rho  {\rho_q}^2+2 (-1+{\rho_q}) {\rho_q}^2+\rho ^2 (-2+(-2+{\rho_q}) {\rho_q})\right)\nonumber\\
 &\hspace{0pt}+\rho  (1-{\rho_q}) \left(-48 {\rho_q}^2 (1+{\rho_q})-24 \rho  {\rho_q} (-22+(-10+{\rho_q}) {\rho_q})\right.\nonumber\\
 &\hspace{60pt}\left.+\rho ^2 (-568+{\rho_q} (312+{\rho_q} (-59+9 {\rho_q})))\right) \ln [\chi ]\nonumber\\
 &\hspace{0pt}\left.-8 {\beta_q} (\rho -{\rho_q}) \left((4-5 {\rho_q}) {\rho_q}^2+2 \rho  {\rho_q} (-4+5 {\rho_q})+\rho ^2 (-68+67 {\rho_q})\right) \!\ln\! \left[\frac{\chi +{\chi_q}}{1+\chi  {\chi_q}}\right] \!\right\}
\end{align}
\begin{align}
\bar c_{\lAAFL,\Pg}^{(1)} &= \frac{\rho_q}{128\pi(1-\rho_q)(\rho-\rho_q)^3}\left\{ 192 h_3(\chi,\chi_q) \beta  \rho ^2 (-1+{\rho_q}) {\rho_q} (2+{\rho_q})\right.\nonumber\\
 &\hspace{0pt}+48 h_1(\chi,\chi_q) \rho  (\rho -{\rho_q}) (-1+{\rho_q}) {\rho_q} (3 \rho +{\rho_q})\nonumber\\
 &\hspace{0pt}+48 h_2(\chi,\chi_q) \rho  (-1+{\rho_q}) {\rho_q} \left(\rho ^2 (-1+{\rho_q})-4 \rho  {\rho_q}+{\rho_q}^2\right)\nonumber\\
 &\hspace{0pt}+2 \beta  (-1+{\rho_q}) {\rho_q} \left(-288 \rho ^2-184 \rho ^2 {\rho_q}+(16+9 (-2+\rho ) \rho ) {\rho_q}^2\right)\nonumber\\
 &\hspace{0pt}-\rho  (-1+{\rho_q}) \left(-384 \rho ^2+8 \rho  (24+35 \rho ) {\rho_q}-112 (-3+\rho ) \rho  {\rho_q}^2\right.\nonumber\\
 &\hspace{70pt}\left.+3 (-4+\rho ) (4+3 \rho ) {\rho_q}^3\right) \ln [\chi ]\nonumber\\
 &\hspace{0pt}\left.-32 {\beta_q} (\rho -{\rho_q}) \left(2 \rho  {\rho_q}^2-{\rho_q}^3+\rho ^2 (-12+11 {\rho_q})\right) \ln \left[\frac{\chi +{\chi_q}}{1+\chi  {\chi_q}}\right]
 \right\}
\end{align}
\begin{align}
\bar c_{\lAAx2g1,\Pg}^{(1)} &= \bar c_{\lVVx2g1,\Pg}^{(1)}
\end{align}

Near threshold we can approximate the above expressions and find
\begin{equation}
\bar c_{\vec \kappa,\Pg}^{(1),\tThr} = c_{\vec \kappa,\Pg}^{(0),\text{thr}} \frac{1}{\pi^2} C_A\left(\bar a_{\vec \kappa,\Pg}^{(1,1)}\ln(\beta) + \bar a_{\vec \kappa,\Pg}^{(1,0)}\right)\;,
\end{equation} 
with
\begin{align}
\bar a^{(1,1)}_{\vec \kappa,\Pg} &= -\frac 1 2
\end{align}
and
\begin{align}
\bar a^{(1,0)}_{\lVVF2,\Pg} &= \bar a^{(1,0)}_{\lVVx2g1,\Pg} = \bar a^{(1,0)}_{\lAAF2,\Pg} = \bar a^{(1,0)}_{\lAAFL,\Pg} = \bar a^{(1,0)}_{\lAAx2g1,\Pg}\nonumber\\
 &= -\frac 1 4 \ln\left(\frac{16\chi_q}{(1+\chi_q)^2}\right)+\frac 1 2\\
\bar a^{(1,0)}_{\lVVFL,\Pg} &= \bar a^{(1,0)}_{\lVVF2,\Pg} + \frac 1 6
\end{align}

\section{Next-to-leading Order Bethe-Heitler Coefficient Function $c_{\vec\kappa,\Pq}^{(1)}$}
\label{sec:Appendix:Partonic:cq1}
Near threshold we can solve the coefficient functions analytically using the methods from \cite{Hekhorn:2018ywm} and find
\begin{align}
c_{\vec \kappa,\Pq}^{(1),\tThr} &= c_{\vec \kappa,\Pg}^{(0),\text{thr}} \frac{\beta^2\rho_q}{\pi^2(\rho_q-1)} \frac{K_{\Pq\Pgg}}{6K_{\Pg\Pgg}} \left[a_{\vec \kappa,\Pq}^{(1,1)}\ln(\beta) + a_{\vec \kappa,\Pq}^{(1,0)}\right],
\end{align}
with
\begin{align}
a^{(1,1)}_{\lVVF2,\Pq} &= a^{(1,1)}_{\lVVx2g1,\Pq} = a^{(1,1)}_{\lAAF2,\Pq} = a^{(1,1)}_{\lAAFL,\Pq} = a^{(1,1)}_{\lAAx2g1,\Pq} = 1\\
a^{(1,1)}_{\lVVFL,\Pq} &= a^{(1,1)}_{\lVVF2,\Pq} - \frac 2 {3}
\end{align}
and
\begin{align}
a^{(1,0)}_{\lVVF2,\Pq} &= a^{(1,0)}_{\lAAF2,\Pq} = a^{(1,0)}_{\lAAFL,\Pq} = a^{(1,0)}_{\lAAx2g1,\Pq} = -\frac{13}{12} + \frac 3 2 \ln(2)\\
a^{(1,0)}_{\lVVFL,\Pq} &= -\frac{77}{100} + \frac 9 {10} \ln(2) \\
a^{(1,0)}_{\lVVx2g1,\Pq} &= a^{(1,0)}_{\lVVF2,\Pq} - \frac{1}{4}
\end{align}

\section{Next-to-leading Order Bethe-Heitler Scaling Coefficient Function $\bar c_{\vec\kappa,\Pq}^{(1),F}$} \label{sec:Appendix:Partonic:cqBarF1}
We use the defined $h_1$ from above and find
\begin{align}
\bar c_{\lVVF2,\Pq}^{(1),F} &= \frac{\rho_q}{36\pi(1-\rho_q)(\rho-\rho_q)^3}\left\{ 3 h_1(\chi,\chi_q) \rho  (\rho -{\rho_q}) (-1+{\rho_q}) (-2 {\rho_q}+\rho  (4+{\rho_q}))\right.\nonumber\\
 &\hspace{0pt} -2 \beta  (-1+{\rho_q}) {\rho_q} \left(-7 \rho  {\rho_q}-{\rho_q}^2+\rho ^2 (16+{\rho_q})\right)\nonumber\\
 &\hspace{0pt}+\rho  (-1+{\rho_q}) \left(-21 \rho  {\rho_q}+3 {\rho_q}^2+\rho ^2 (14+(-6+{\rho_q}) {\rho_q})\right) \ln [\chi ]\nonumber\\
 &\hspace{0pt} \left.+{\beta_q} (-\rho +{\rho_q}) \left(-2 \rho  {\rho_q} (4+{\rho_q})+{\rho_q}^2 (4+{\rho_q})+\rho ^2 (-14+19 {\rho_q})\right) \ln \left[\frac{\chi +{\chi_q}}{1+\chi  {\chi_q}}\right] \right\}
\end{align}
\begin{align}
\bar c_{\lVVFL,\Pq}^{(1),F} &= \frac{-\rho_q}{36\pi(1-\rho_q)^2(\rho-\rho_q)^3}\left\{-4 \rho ^2 (-1+{\rho_q})^2 (\rho  (-3+{\rho_q})+3 {\rho_q}) \ln [\chi ] \right.\nonumber\\
 &\hspace{0pt} -2 \beta  (-1+{\rho_q}) {\rho_q} \left(2 \rho  {\rho_q}^2+(2-3 {\rho_q}) {\rho_q}^2+\rho ^2 (-6+5 {\rho_q})\right)\nonumber\\
 &\hspace{0pt} \left.+{\beta_q} (-\rho +{\rho_q})\! \left(2 \rho  {\rho_q}^2+{\rho_q}^3 (-4+3 {\rho_q})+\rho ^2 (12+{\rho_q} (-22+9 {\rho_q}))\right) \!\ln\!\! \left[\frac{\chi +{\chi_q}}{1+\chi  {\chi_q}}\right] \!\right\}
\end{align}
\begin{align}
\bar c_{\lVVx2g1,\Pq}^{(1),F} &= \frac{\rho_q}{36\pi(1-\rho_q)(\rho-\rho_q)^3}\left\{ -36 \beta  \rho  (\rho -{\rho_q}) (-1+{\rho_q})^2 {\rho_q}\right.\nonumber\\
 &\hspace{0pt}+6 h_1(\chi,\chi_q) \rho  (-1+{\rho_q})^2 \left(2 \rho ^2-3 \rho  {\rho_q}+{\rho_q}^2\right)\nonumber\\
 &\hspace{0pt} -3 \rho  (\rho -{\rho_q}) (-1+{\rho_q})^2 (\rho  (-6+{\rho_q})+7 {\rho_q}) \ln [\chi ]\nonumber\\
 &\hspace{0pt}\left.-6 {\beta_q} \rho  (-1+{\rho_q})^2 \left(3 \rho ^2-4 \rho  {\rho_q}+{\rho_q}^2\right) \ln \left[\frac{\chi +{\chi_q}}{1+\chi  {\chi_q}}\right] \right\}
\end{align}
\begin{align}
&\bar c_{\lAAF2,\Pq}^{(1),F} =\nonumber\\
 & \frac{\rho_q}{36\pi(1-\rho_q)(\rho-\rho_q)^3}\left\{ -3 h_1(\chi,\chi_q) \rho  (\rho -{\rho_q}) (-1+{\rho_q}) (\rho  (-4+{\rho_q})-2 (-1+{\rho_q}) {\rho_q}) \right.\nonumber\\
 &\hspace{0pt} +\beta  (-1+{\rho_q}) {\rho_q} \left(\rho ^2 (-32+{\rho_q})+\rho  (14-3 {\rho_q}) {\rho_q}+2 {\rho_q}^2\right)\nonumber\\
 &\hspace{0pt} +\frac{1}{2} \rho  (-1+{\rho_q}) \left(6 {\rho_q}^2 (1+{\rho_q})+3 \rho  {\rho_q} (-14+(-6+{\rho_q}) {\rho_q})-\rho ^2 \left(-28+{\rho_q}^2\right)\right) \ln [\chi ]\nonumber\\
 &\hspace{0pt} \left. +{\beta_q} \left(\rho ^3 (14-13 {\rho_q})+3 \rho ^2 (-2+{\rho_q}) {\rho_q}+(4-5 {\rho_q}) {\rho_q}^3+3 \rho  {\rho_q}^2 (-4+5 {\rho_q})\right) \!\ln\! \left[\frac{\chi +{\chi_q}}{1+\chi  {\chi_q}}\right]\! \right\}
\end{align}
\begin{align}
\bar c_{\lAAFL,\Pq}^{(1),F} &= \frac{\rho_q}{72\pi(1-\rho_q)(\rho-\rho_q)^3}\left\{ 12 h_1(\chi,\chi_q) \rho  (\rho -{\rho_q}) (-1+{\rho_q}) {\rho_q}^2\right.\nonumber\\
 &\hspace{0pt}-2 \beta  (-1+{\rho_q}) {\rho_q} \left(-4 {\rho_q}^2+3 \rho  {\rho_q}^2+\rho ^2 (12+{\rho_q})\right)\nonumber\\
 &\hspace{0pt}+\rho  (-1+{\rho_q}) \left(6 {\rho_q}^3+3 \rho  {\rho_q} (-8+(-6+{\rho_q}) {\rho_q})+\rho ^2 (24+(-4+{\rho_q}) {\rho_q})\right) \ln [\chi ]\nonumber\\
 &\hspace{0pt} \left.-8 {\beta_q} \left(3 \rho ^2 {\rho_q}-3 \rho  {\rho_q}^3+{\rho_q}^4+\rho ^3 (-3+2 {\rho_q})\right) \ln \left[\frac{\chi +{\chi_q}}{1+\chi  {\chi_q}}\right] \right\}
\end{align}
\begin{align}
\bar c_{\lAAx2g1,\Pq}^{(1),F} &= \bar c_{\lVVx2g1,\Pq}^{(1),F}
\end{align}

Near threshold we can approximate the above expressions and find
\begin{align}
\bar c_{\vec \kappa,\Pq}^{(1),F,\tThr} &= -c_{\vec \kappa,\Pg}^{(0),\text{thr}} \frac{\beta^2\rho_q}{\pi^2(\rho_q-1)} \frac{K_{\Pq\Pgg}}{24K_{\Pg\Pgg}} \cdot \bar a_{\vec \kappa,\Pq}^{(1,0)}
\end{align}
with
\begin{align}
\bar a^{(1,0)}_{\lVVF2,\Pq} &= \bar a^{(1,0)}_{\lVVx2g1,\Pq} = \bar a^{(1,0)}_{\lAAF2,\Pq} = \bar a^{(1,0)}_{\lAAFL,\Pq} = \bar a^{(1,0)}_{\lAAx2g1,\Pq} = 1\\
\bar a^{(1,0)}_{\lVVFL,\Pq} &= \bar a^{(1,0)}_{\lVVF2,\Pq} - \frac 2 {3}
\end{align}

\section{Next-to-leading Order Coulomb Coefficient Function $d_{\vec\kappa,\Pq}^{(1)}$} \label{sec:Appendix:Partonic:dq1}
There are no new functions compared to \cite{Hekhorn:2018ywm} as we do find
\begin{align}
d_{\lVVF2,\Pq}^{(1)} &= d_{\lAAF2,\Pq}^{(1)} = -d_{\lVAg4,\Pq}^{(1)}\,,\\
d_{\lVVFL,\Pq}^{(1)} &= d_{\lAAFL,\Pq}^{(1)} = -d_{\lVAgL,\Pq}^{(1)}\,,\\
d_{\lVVx2g1,\Pq}^{(1)} &= d_{\lAAx2g1,\Pq}^{(1)} = d_{\lVAxF3,\Pq}^{(1)}\,. 
\end{align}
We again define the additional helper function from \cite{Hekhorn:2018ywm} for the Coulomb part by
\begin{align}
h_4(\chi,\chi') &= \DiLog\left(\frac {1+\chi'}{1+\chi}\right)- \DiLog\left(\chi \frac{1+\chi'}{1+\chi}\right) - \DiLog\left(\chi'\frac{1+\chi}{1+\chi'}\right)+ \DiLog\left(\frac{\chi'(1+\chi)}{\chi(1+\chi')} \right)\nonumber\\
 &\hspace{20pt}+ \frac1 2 \ln^2 (\chi ) +  \ln (\chi ) \ln\left[\frac{(1+\chi )(1+\chi')}{(\chi -\chi')(1-\chi  \chi')}\right]
\end{align}
and find
\begin{align}
d_{\lVVF2,\Pq}^{(1)} &= \frac{1}{5184\pi\rho}\left\{ 36 h_4(\chi,\chi') \left(-8 \rho ^2+8 \rho  {\rho'}+\left(-4+3 \rho ^2\right) {\rho'}^2-27 \rho  {\rho'}^3+27 {\rho'}^4\right) \right.\nonumber\\
 &\hspace{0pt}+\beta  \left(2 \rho ^2 (718+5 \rho )-4 \rho  (758+91 \rho ) {\rho'}+4 (622+1277 \rho ) {\rho'}^2-6456 {\rho'}^3\right)\nonumber\\
 &\hspace{0pt}+9 \rho  \left(3 \rho ^3+6 \rho ^2 {\rho'}+4 {\rho'} \left(4-27 {\rho'}^2\right)+6 \rho  \left(-4+9 {\rho'}^2\right)\right) \ln [\chi ]\nonumber\\
 &\hspace{0pt}\left.-24 {\beta'} \left(8 \rho  (7-13 {\rho'}) {\rho'}+\rho ^2 (-38+23 {\rho'})+2 {\rho'}^2 (-17+47 {\rho'})\right) \ln \left[\frac{\chi -{\chi'}}{1-\chi  {\chi'}}\right] \right\}
\end{align}
\begin{align}
d_{\lVVFL,\Pq}^{(1)} &= \frac{\rho'}{864\pi\rho}\left\{ 108 h_4(\chi,\chi') {\rho'}^2 (-\rho +{\rho'})\right.\nonumber\\
 &\hspace{0pt}+4 \beta  \left(23 \rho ^2+2 (25-93 {\rho'}) {\rho'}+19 \rho  (-2+7 {\rho'})\right)\nonumber\\
 &\hspace{0pt}\left.+18 \rho  \left(\rho ^2+3 \rho  {\rho'}-6 {\rho'}^2\right) \ln [\chi ]+24 {\beta'} (\rho -{\rho'}) (-2+11 {\rho'}) \ln \left[\frac{\chi -{\chi'}}{1-\chi  {\chi'}}\right] \right\}
\end{align}
\begin{align}
d_{\lVVx2g1,\Pq}^{(1)} &= \frac{1}{5184\pi\rho}\left\{ 36 h_4(\chi,\chi') \left(-4 {\rho'}^2+\rho  {\rho'} \left(8-3 {\rho'}^2\right)+\rho ^2 \left(-8+3 {\rho'}^2\right)\right)\right.\nonumber\\
 &\hspace{0pt} +2 \beta  \left(5 \rho ^3+\rho ^2 (718-458 {\rho'})+8 (109-75 {\rho'}) {\rho'}^2+20 \rho  {\rho'} (-53+41 {\rho'})\right)\nonumber\\
 &\hspace{0pt} +9 \rho  \left(16 {\rho'}+3 \rho  \left(-8+\rho ^2-2 \rho  {\rho'}+4 {\rho'}^2\right)\right) \ln [\chi ]\nonumber\\
 &\hspace{0pt} \left. -24 {\beta'} \left(2 \rho  (22-19 {\rho'}) {\rho'}+\rho ^2 (-38+23 {\rho'})+{\rho'}^2 (-28+25 {\rho'})\right) \ln\! \left[\frac{\chi -{\chi'}}{1-\chi  {\chi'}}\right]\! \right\}
\end{align}

\chapter{Languages, Programs and Tools}\label{sec:Appendix.Tools}
\chapterquote[ngerman]{Das ist nur ein Echo \ldots ein Echo}{der Programmierer}

In this \Appendix we explain about all the different methods and computer languages we used to obtain this work. We joined several approaches, trying to extract the best out of each tool.

\subsubsection{Pen and Paper}
Undoubtedly, pen and paper are the most important tool for studying 10 years of physics as \textit{complex} problems are discussed best with an analog paper in hand - or alternatively in front of a blackboard. The first step for this work was first drawing and then writing down the fundamental physics building blocks, the Feynman diagrams - most conveniently done by hand.

\subsubsection{FORM}
Computer are great in tackling \textit{complicated} problems, so \texttt{FORM}\cite{Vermaseren:2000nd} was used as a dedicated Computer Algebra System (CAS) to compute the raw matrix elements. Furthermore the applied MVV-scheme\cite{Moch:2015usa} is implemented most easily in \texttt{FORM}. Note that we were not able to compute all required matrix elements in the alternative programs \texttt{Tracer}\cite{Tracer} nor \HEPMath\cite{Wiebusch:2014qba}.

\subsubsection{Mathematica}
We use \MMa{} to post-process the matrix elements, i.e.\ simplification of expressions, format output to C++ code, insertion of analytical phase space integrals and all analytical integrations needed for analytical results in \AppendixRef{sec:Appendix.Partonic}. \HEPMath{} has a built-in Passarino-Veltman decomposition (see also \AppendixRef{sec:Appendix.s4.Decomposition}) that comes handy for virtual matrix elements.

\subsubsection{C++}
The use of a modern, object-oriented, modular programming language should be common sense by now: it largely reduces the \textit{development} time, which usually consumes the vast majority of the project time ($\sim$ month/years), and due to clever compiler programs it is compatible in \textit{execution} time ($\sim$ hours/days). For the computations of large scale codes and to avoid license issues \MMa{} should be avoided in production codes. C++ also offers a large and vivid sea of modules and specialised programs - here we did use
\begin{description}
\item[GNU Scientific Library (GSL)\cite{GSL}]\url{http://www.gnu.org/software/gsl/} - provides integration routines, special functions and histograms
\item[Boost] \url{http://www.boost.org/} - provides filesystem operations and python bindings
\item[Dvegas\cite{Kauer:2001sp,Kauer:2002sn}] \url{https://dvegas.hepforge.org/} - provides an advanced integration routine
\item[Relativistic Kinematics (RK)\cite{RK}] \url{http://rk.hepforge.org/} - provides accurate numerics for relativistic kinematics
\end{description}

The implementation of software design patterns should become more widely used in scientific research as they improve the structure of the code itself. Here we implemented
\begin{description}
\item[Don't repeat yourself (DRY)] use abstract base classes, that cover as much as possible (\textit{here: e.g.\ Common::AbstractLeptonProduction})
\item[Template pattern] as discussed are all projections $\vec\kappa$ mostly dealt in parallel, so unify their use
\item[Facade pattern] setup an opaque front, that hides all implementation details (\textit{here: e.g.\ InclusiveLeptonProduction})
\item[Wrapper pattern] encapsulate the access to different sources of (p)PDFs (\textit{here: PdfWrapper})
\item[Module pattern] keep inclusive and fully differential calculations in separate namespaces, while keep the common elements in yet an other (\textit{here: e.g.\ Common::})
\item[Strategy pattern] allow to implement and exchange different integration routines (\textit{here: GSL, Dvegas})
\item[State pattern] each application will only do a single integration at a time, so we can unify the treatment for the different kind of integration tasks
\item[Command pattern] encapsulate logical units of parameters\\ (\textit{here: e.g.\ Common::IntegrationConfig})
\end{description}

\begin{figure}[ht]
\begin{center}
\includegraphics[width=.99\textwidth]{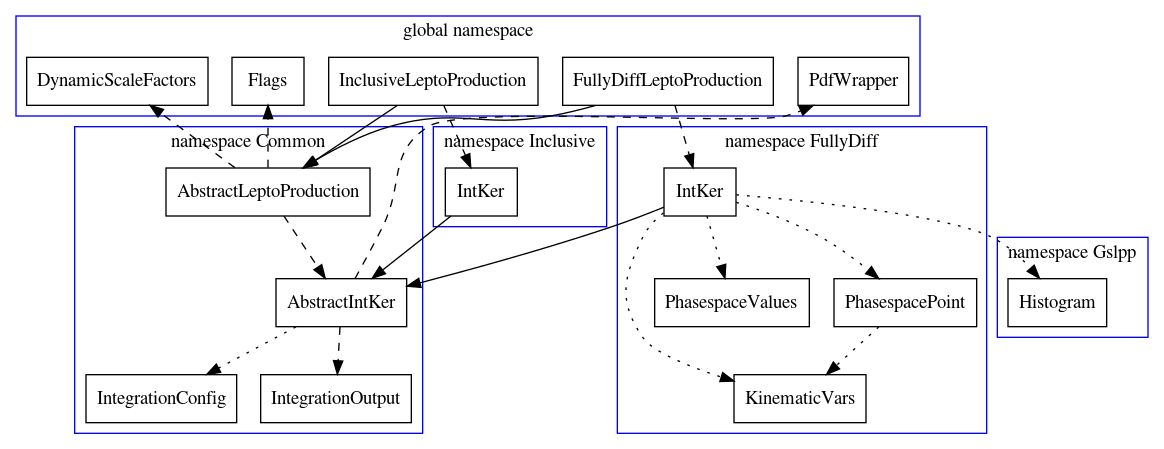}
\end{center}
\caption{class diagram of our code - solid lines represent inheritance relations, dashed lines represent member relations and dotted lines represent usage relations.} \label{fig:classDiagram}
\end{figure}
A class diagram of our code is shown in \FigureRef{fig:classDiagram}. \texttt{CodeLite}\footnote{\url{https://codelite.org/}} is a lightweight IDE providing the most important features such as auto completion and a smooth interface to \texttt{GDB}\footnote{\url{https://www.gnu.org/software/gdb/}}. A solid documentation of all code parts is strictly required and \texttt{Doxygen}\footnote{\url{http://www.doxygen.nl/}} provides a standardized visual representation thereof.

\subsubsection{Python}
We use Python to parse the \MMa{} output to C++ templates and via Boost we provide a simple Python module to access our codes. So our final run routines are written in Python and we use the \texttt{multiprocessing}\footnote{\url{https://docs.python.org/2/library/multiprocessing.html}} module to provide an easy and flexible parallelization, allowing ourselves to run the code on a High Performance Cluster, such as the bwUniCluster.

\subsubsection{LaTeX}
When writing a longer work in natural science, \LaTeXe{} is indispensable. We give here are a list of useful packages:
\begin{description}
\item[biblatex] \url{https://ctan.org/pkg/biblatex} - to manage references
\item[hepthesis] \url{https://ctan.org/pkg/hepthesis} - as a solid basis to build on
\item[siunitx] \url{https://ctan.org/pkg/siunitx} - to output any kind of numbers/units
\end{description}
Further useful programs are
\begin{description}
\item[Dot] \url{https://www.graphviz.org/doc/info/lang.html} - to draw the class diagram
\item[Hunspell] \url{http://hunspell.github.io/} - used for spell checking
\item[Latexmk] \url{http://personal.psu.edu/jcc8//software/latexmk-jcc/} - to automatically run \LaTeXe{}
\item[PyFeyn] \url{https://pyfeyn.hepforge.org/} - to draw Feynman diagrams
\end{description}

\end{appendices}
\begin{backmatter}

\bookmarksetup{startatroot}
\chapter{Acknowledgments} \label{chap:Acknowledgments}
I thank my supervisor Prof. Dr. Werner Vogelsang for the possibility to write this thesis and introducing me into the field of high energy particle physics. Ever since he gave the lecture on \foreignlanguage{ngerman}{Physik III} (Introduction to Theoretical Physics) in my third semester, I simply followed his lectures all the way up to QCD, admiring his lecture style and his open minded attitude.

I thank Prof. Dr. Thomas Gutsche for providing the second appraisel of this thesis, the thorough management of all the construction work and the interesting discussions at lunch break.

I thank Dr. Marco Stratman for the collaboration and the help for writing our publications.

The author acknowledges support by the state of Baden-Württemberg through bwHPC.

I am grateful to Marina for proof-reading and enhancing this thesis in many ways and for all the support and discussions in our office

I am grateful to all current and former members of our research group and our friendly neighbours: Frau Prof. Jäger, Valery, Sabine, Felix P., Felix K., Juliane, Fabian, Maurizio, Patrick, Dominik, Markus, Julien, Christopher, Johannes, Santiago, Marc, Peter, Tom, Pit, Patriz, Daniele, Julius, Rouven, Lukas, Matthias. And a special thanks to Felix R., who was and is a role model, to Ilkka, for his company and all the good times in Japan, to Martin, for all the good times in Saint-Jacut-de-la-Mer \foreignlanguage{italian}{e agli italiani, Margaritha, Maria Vittoria, Fausto e Gabriele, per parlare italiano con me}.

I am grateful to my physics buddy group who I literally met on the first day of studying: Paula, Benedikt, Timo and Thomas, whith whom I spent wonderful days and nights, talking about physics and much more. In particular my choir, yoga and walking mate Timo is an endless source of support, happiness and solid discussions.

\begin{otherlanguage}{ngerman}
Ohne Zweifel gilt mein größter Dank meiner Familie: meinen Eltern, für ihre Unterstützung, allen meinen Geschwistern, für ihre Unterhaltung, und meinen Großeltern, für deren Rückhalt. Insbesondere möchte ich meiner großartigen und bewundernswerten Oma danken für ihre unendliche Wärme und ihre lehrreichen Geschichten.
\end{otherlanguage}

\listoffigures
\listoftables
\printbibliography[heading=bibintoc]

\end{backmatter}
\end{document}